\documentclass[twocolumn]{aastex62}

\usepackage{color}
\usepackage{amsmath}

\newcommand{\Teff}{\mbox{$T_\mathrm{eff}$}}
\newcommand{\Mjup}{\mbox{$M_\mathrm{Jup}$}}
\newcommand{\Msun}{\mbox{$M_{\odot}$}}

\newcommand{\Rsun}{\mbox{$R_{\odot}$}}

\shorttitle{Rotation, Inclinations, and Obliquities of Substellar Host Stars}
\shortauthors{Bowler et al.}

\begin{document}

\title{Rotation Periods, Inclinations, and Obliquities of Cool Stars Hosting Directly Imaged Substellar Companions: 
Spin-Orbit Misalignments are Common}

\correspondingauthor{Brendan P. Bowler}
\email{bpbowler@astro.as.utexas.edu}

\author[0000-0003-2649-2288]{Brendan P. Bowler}
\affiliation{Department of Astronomy, The University of Texas at Austin, 2515 Speedway, Stop C1400, Austin, TX 78712, USA}

\author[0000-0001-6532-6755]{Quang H. Tran}
\affiliation{Department of Astronomy, The University of Texas at Austin, 2515 Speedway, Stop C1400, Austin, TX 78712, USA}

\author[0000-0002-3726-4881]{Zhoujian Zhang}\thanks{NASA Sagan Fellow}
\affiliation{Department of Astronomy \& Astrophysics, University of California, Santa Cruz, 1156 High St, Santa Cruz, CA 95064, USA}

\author[0000-0003-4022-6234]{Marvin Morgan}
\affiliation{Department of Astronomy, The University of Texas at Austin, 2515 Speedway, Stop C1400, Austin, TX 78712, USA}

\author{Katelyn B. Ashok}
\affiliation{Department of Astronomy, The University of Texas at Austin, 2515 Speedway, Stop C1400, Austin, TX 78712, USA}

\author[0000-0002-3199-2888]{Sarah Blunt}
\affiliation{Department of Astronomy, California Institute of Technology, Pasadena, CA, USA}

\author[0000-0002-6076-5967]{Marta L. Bryan}\thanks{NASA Sagan Fellow}
\affiliation{Department of Astronomy, 501 Campbell Hall, University of California Berkeley, Berkeley, CA 94720-3411, USA}

\author[0000-0003-0850-7749]{Analis E. Evans}
\affiliation{Department of Physics, University of Florida, 2001 Museum Road, Gainesville, FL 32611-8440, USA}

\author[0000-0003-4557-414X]{Kyle Franson}
\altaffiliation{NSF Graduate Research Fellow}
\affiliation{Department of Astronomy, The University of Texas at Austin, 2515 Speedway, Stop C1400, Austin, TX 78712, USA}

\author[0000-0001-8832-4488]{Daniel Huber}
\affiliation{Institute for Astronomy, University of Hawai‘i, 2680 Woodlawn Drive, Honolulu, HI 96822, USA}

\author{Vighnesh Nagpal}
\affiliation{Astronomy Department, University of California, Berkeley, CA 94720, USA}

\author{Ya-Lin Wu}
\affiliation{Department of Physics, National Taiwan Normal University, Taipei City, Taiwan}

\author[0000-0003-2969-6040]{Yifan Zhou}
\altaffiliation{51 Pegasi b Fellow}
\affiliation{Department of Astronomy, The University of Texas at Austin, 2515 Speedway, Stop C1400, Austin, TX 78712, USA}



\begin{abstract}

The orientation between a star's spin axis and a planet's orbital plane provides
valuable information about the system's formation and dynamical history.
For non-transiting planets at wide separations, true stellar obliquities are challenging to measure, but 
lower limits on spin-orbit orientations can be determined from the difference between
the inclination of the star's rotational axis and the companion's orbital plane ($\Delta i$).
We present results of a uniform analysis of rotation periods, 
stellar inclinations, and obliquities of cool stars (SpT $\gtrsim$ F5) hosting directly imaged 
planets and brown dwarf companions.
As part of this effort, we have acquired new $v \sin i_*$ values for 22 host stars with the high-resolution Tull spectrograph at
the Harlan J. Smith telescope.
Altogether our sample contains 62 host stars with rotation periods, most of which are newly measured using light
curves from the \emph{Transiting Exoplanet Survey Satellite}.
Among these, 53 stars have inclinations determined from projected rotational and equatorial velocities,
and 21 stars predominantly hosting brown dwarfs have constraints on $\Delta i$.
Eleven of these (52$^{+10}_{-11}$\% of the sample) are likely misaligned, while  
the remaining ten host stars are consistent with spin-orbit alignment.
As an ensemble, the minimum obliquity distribution between 10--250~AU is more 
consistent with a mixture of isotropic and aligned systems than either extreme scenario alone---pointing to direct cloud collapse, 
formation within disks bearing primordial alignments and misalignments, or architectures processed by dynamical evolution.
This contrasts with stars hosting directly imaged planets, which show a preference for low obliquities.
These results reinforce an emerging distinction between the orbits of long-period brown dwarfs and giant planets in
terms of their stellar obliquities and orbital eccentricities.

\end{abstract}



\section{Introduction} \label{sec:intro}

Measurements of stellar obliquity---the orientation between a star's spin axis and a planet's orbital angular momentum
vector---provide valuable clues about how planets form, migrate, and dynamically interact 
(\citealt{Winn:2015jt}; \citealt{Albrecht:2022aa}).
In the absence of external perturbers, the axisymmetric collapse of a molecular cloud core 
is expected to result in mutual alignment between the stellar spin and the protoplanetary disk's axis of rotation,
which would then be inherited by any planets that form in the disk.

Many processes can disrupt this alignment and produce 
a wide range of stellar and planetary obliquities during and after 
the phase of planet formation.
These mechanisms can act on the star, the planet, or the protoplanetary disk and
include the influence of passing stars (\citealt{Heller:1993aa}; \citealt{Batygin:2020dl});
secular torques from wide stellar and substellar companions 
(\citealt{Fabrycky:2007jh}; \citealt{Batygin:2012ig}; \citealt{Anderson:2016aa}; \citealt{Lai:2018cb});
and gravitational scattering with other planets (\citealt{Chatterjee:2008gd}; \citealt{Ford:2008jo}).
Primordial misalignments of protoplanetary disks (both ``intact'' and ``broken'') appear to be
common, indicating that some planets probably form with substantial spin-orbit angles 
(e.g., \citealt{Bate:2010fz}; \citealt{Huber:2013aa}; \citealt{Marino:2015hg}; \citealt{Ansdell:2020fo}; \citealt{EpsteinMartin:2022aa}).
Even the Sun is misaligned by about 6$\degr$ with respect to the solar system's invariable plane, but
 the origin of this enigmatic offset  
continues to be debated (e.g., \citealt{Adams:2010bb}; \citealt{Bailey:2016iu}; \citealt{Spalding:2019do}). 

Large transiting planets have provided a wealth of information about stellar obliquities, largely through individual and collective
studies of Rossiter-McLaughlin measurements 
(\citealt{Rossiter:1924aa}; \citealt{McLaughlin:1924vz}; \citealt{Queloz:2000ui}; \citealt{Gaudi:2007vi}).
It is now clear that non-zero obliquities are common among field stars hosting short-period planets (\citealt{Fabrycky:2009ke}; \citealt{Schlaufman:2010db}; \citealt{Louden:2021jz}; \citealt{Munoz:2018jq}), 
and especially those with temperatures above the Kraft break at 
$\approx$6250~K (\citealt{Winn:2010dr}; \citealt{Schlaufman:2010db}).
As an ensemble, however,
most compact multiplanet systems tend to have lower stellar obliquities 
(\citealt{Albrecht:2013uq}; \citealt{Morton:2014in}; \citealt{Winn:2017ip}).
A complex picture is emerging in which several mechanisms are probably responsible for tilting and in some contexts 
realigning the orbits of close-in planets over many different timescales
(e.g., \citealt{Morton:2011jc};  \citealt{Albrecht:2012hk}; \citealt{Dawson:2014bj}).

Less, however, is known about stellar obliquities for systems hosting long-period companions.
At wide separations, stellar obliquities can be used to provide clues about the formation and 
orbital evolution of companions amenable to direct imaging (\citealt{Bowler:2016jk}).
In particular, planets and brown dwarfs have been discovered at separations spanning tens to 
thousands of AU and may originate from a combination of outward gravitational scattering 
(\citealt{Boss:2006ge}; \citealt{Veras:2009br}; \citealt{Bailey:2019js});
\emph{in situ} formation in a disk via pebble accretion or disk instability (\citealt{Durisen:2007wg}; \citealt{Lambrechts:2012gr}); 
dynamical capture (\citealt{Perets:2012cv}); or 
direct collapse 
from a fragmenting molecular cloud core  (\citealt{Bate:2012hy}).
There is growing evidence that most directly imaged planets within about one hundred AU are formed in a disk
based on their eccentricity distributions (\citealt{Bowler:2020hk}), mass functions (\citealt{Wagner:2019iy}),
and demographic trends (\citealt{Nielsen:2019cb}; \citealt{Vigan:2021dc}).
Spin-orbit angles (between the stellar spin axis and companion orbital plane) 
and spin-spin angles (between the spin axes of the host star and companion)
can provide another unique perspective on the origin and orbital evolution of this population.


\begin{figure*}
    \gridline{\fig{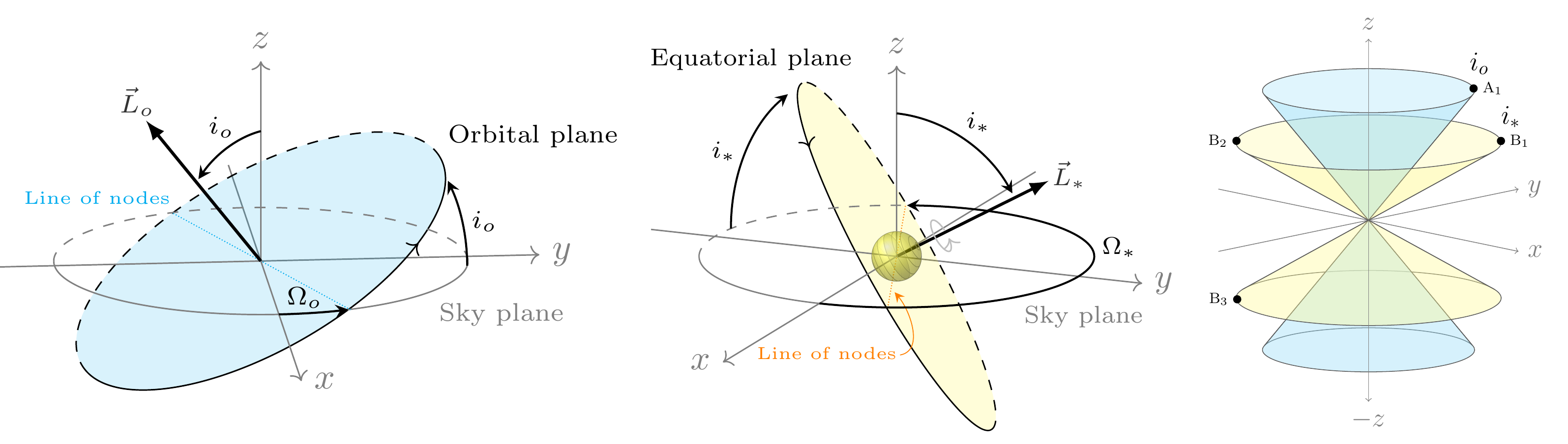}{1.0\textwidth}{}}
     \vskip -.33 in
  \caption{Orbital and rotational geometry 
  relevant for obliquity constraints of stars hosting imaged companions.  The observer is looking down from the $z$-axis.  The $x$-$y$ plane is
  the plane of the sky, and the star is centered on the origin.  The $x$-axis points north and the $y$-axis points east.
  \emph{Left}: The inclination of the companion's orbital plane, $i_o$, intersects the sky along the line of nodes, which is oriented by 
  $\Omega_o$ from true north to the ascending node.  In this configuration the orbital angular momentum vector $\overrightarrow{L_o}$ points in the hemisphere facing the observer,
  with $\overrightarrow{L_o}$ pointing toward the observer along the $z$ axis when $i_o$=0$\degr$.
  \emph{Middle:} The star's orientation in space determines the inclination of its equatorial plane, $i_*$, and the position angle of its line of nodes.  
  The sense of the stellar spin sets its rotational angular momentum
  vector $\overrightarrow{L_*}$ and the longitude of ascending node $\Omega_*$.
  \emph{Right:} If only the inclinations $i_o$ and $i_*$ are measured, but not their true (or relative) orientations in the plane of the sky ($\Omega_o$ and $\Omega_*$), 
  then each angular momentum vector can fall anywhere along nested cones opening toward or away from the observer.  
  If $\Omega_o$ is known (point \texttt{A$_1$}, for example), but not $\Omega_*$, then the obliquity can be as small as $|i_o$--$i_*|$ (at \texttt{B$_1$}),
  can reach $i_o$ + $i_*$ for prograde orbits (\texttt{B$_2$}), or can be as large as $\pi - |i_o - i_*|$ (\texttt{B$_3$})
  for retrograde orbits.
 \label{fig:geometry} } 
 \end{figure*}

Measuring stellar obliquities is challenging at wide separations.
The Rossiter-McLaughlin effect becomes less practical as a tool to study projected spin-orbit orientations
because the geometric probability of a transit is lower, transit events are less frequent, and transit durations become longer;
this explains why few systems with orbital periods beyond a few tens of days have had stellar 
obliquity measurements (\citealt{Albrecht:2022aa}).
Our strategy in this study is to use stellar rotation periods, projected rotational velocities, and stellar radii to infer the spin-orientation of
stars hosting imaged substellar companions.  The goal is to uniformly establish stellar inclinations for the full
sample of host stars, then determine stellar obliquities for the subset of these with measurements of the orbital inclination from orbit monitoring.
As a byproduct of this analysis, we also present rotation periods for 62 host stars, which can be used to refine the ages of these systems
using gyrochronology.  

This paper is organized as follows.  In Section~\ref{sec:geometry} we summarize the geometry and overall framework to constrain inclinations
and obliquities for host stars of imaged companions. 
A description of our sample, 
new high-resolution spectroscopy, and \emph{Transiting Exoplanet Survey Satellite} (\emph{TESS}; \citealt{Ricker:2015fy}) light curves 
 are detailed in Section~\ref{sec:obs}.
Section~\ref{sec:results} includes our 
$v \sin{i_*}$ measurements, periodicity analysis, stellar line-of sight inclinations, and obliquity constraints.
The interpretation of our results and a discussion of sample biases can be found in Section~\ref{sec:discussion}.
Our conclusions are summarized in Section~\ref{sec:summary}.

\section{Stellar Obliquity Geometry for Directly Imaged Companions} \label{sec:geometry}

The problem of observationally measuring the angle between the stellar spin axis and a companion's orbital plane, $\psi$, 
has a long history in the context of stellar binaries and the coplanarity of disks (e.g., \citealt{Hale:1994gv}; \citealt{Watson:2011gf}).
For clarity we briefly review the basic geometric
setup of this problem as it applies to directly imaged companions.

Figure~\ref{fig:geometry} shows the relevant orbital elements of a
visual binary companion (left diagram) and the orientation of the spin axis of its host star (middle diagram).
The observer is looking down along the $z$ axis toward the $x$-$y$ sky plane.
The orbital plane of an imaged planet is characterized by its inclination $i_o$ with respect to the sky plane,
which is equal to the angle between the $z$ axis and the orbital angular momentum vector $\overrightarrow{L_o}$,
and the longitude of ascending node $\Omega_o$, defined from celestial north (here the $x$ axis) increasing
eastward (toward the $y$ axis) to the line of nodes.  $i_o$ spans 0--$\pi$  and $\Omega_o$ spans 0--2$\pi$.
In general, orbit monitoring with relative astrometry results in an ambiguity between $\Omega_o$ and $\Omega_o$ + 180$\degr$,
which can be distinguished with radial velocities (RVs) of the companion (e.g., \citealt{Snellen:2018aa}; \citealt{Ruffio:2019aa}; \citealt{Ruffio:2021aa}) 
or its host star (e.g., \citealt{Crepp:2012eg}; \citealt{Bowler:2018gy}).

The setup is similar for the spinning star: its equatorial plane is inclined by $i_*$ from the sky plane,
equal to the angle between the observer's line of sight and the star's rotational angular momentum vector $\overrightarrow{L_*}$.
The direction in which $\overrightarrow{L_*}$ points is determined by its orientation on the sky, $\Omega_*$.
With few exceptions for direct measurements of stellar oblateness (e.g., \citealt{VanBelle:2012aa}) and 
resolved rotation (\citealt{Kraus:2020aa}), $\Omega_*$ is generally unknown.

The relationship between $i_o$, $\Omega_o$, $i_*$, and $\Omega_*$ can be derived by an observer-to-orbit coordinate system transformation
(\citealt{Fabrycky:2009ke}) or via the spherical law of cosines (\citealt{Glebocki:1997ul}):
\begin{equation}
\cos{\psi} = \cos{i_o} \cos{i_*} + \sin{i_o} \sin{i_*} \cos(\Omega_o - \Omega_*)
\end{equation}

\noindent For transiting planets, $i_o$ is known ($\approx$90$\degr$) and the projected spin-orbit orientation 
$\Delta \Omega$ = $\Omega_o - \Omega_*$ (often denoted as $\lambda$)
can be measured with the Rossiter-McLaughlin effect (\citealt{Holt:1893aa}; \citealt{Rossiter:1924aa}; \citealt{McLaughlin:1924vz}).  In this case
$i_*$ in general is unknown, and $\cos{\psi}$ $\approx$ $\sin{i_*} \cos(\lambda)$, so $\lambda$ is a lower limit
on the true obliquity.  

For directly imaged planets, $i_o$ can be measured through orbit monitoring.  In some cases the inclination of the host star can be determined 
through asteroseismology (e.g., \citealt{Zwintz:2019aa}) or the projected rotational velocity technique, which relies on knowing $v \sin{i_*}$, the rotation period $P_\mathrm{rot}$, 
and the stellar radius $R_*$ (\citealt{Shajn:1929aa}).
This latter approach is the focus of this study and is possible for spotted stars 
whose rotational modulations manifest as periodic light curve variations. 


\begin{figure*}
     \vskip -1.9 in
    \resizebox{7.1in}{!}{\includegraphics{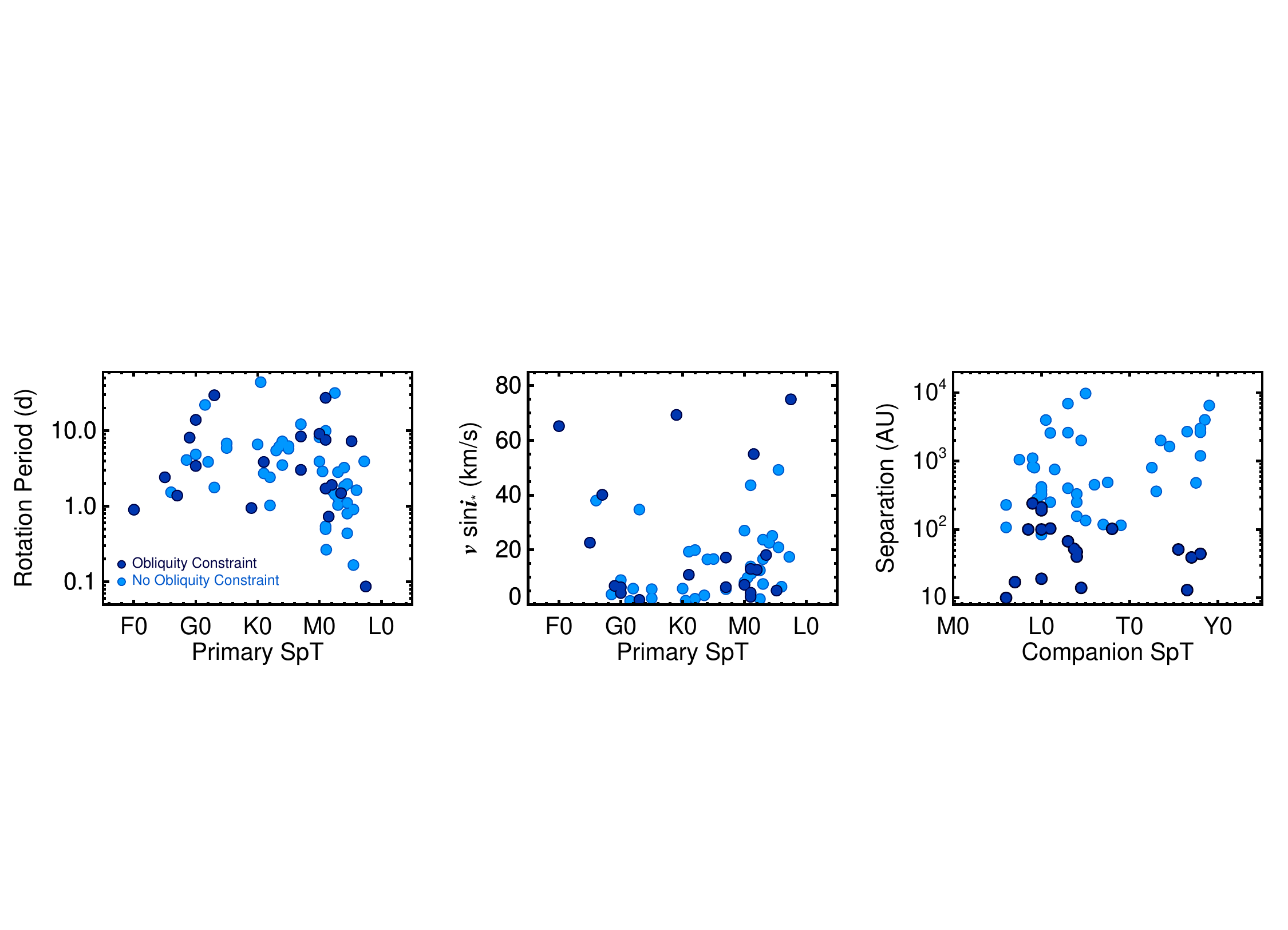}}
     \vskip -1.45 in
  \caption{Overview of the host star spectral types and rotation periods (left), projected rotational velocities (middle), and companion spectral types and separations (right).  Systems with stellar obliquity constraints are plotted in dark blue.  Companions with obliquity constraints in this study reside within $\approx$250~AU where orbital motion is more readily detectable.
 \label{fig:overview} } 
 \end{figure*}

There is an important distinction between simple coplanarity of the orbital and stellar equatorial planes,
versus genuine alignment of the orbital and rotational angular momentum vectors between the companion and the host star---the primary angle
of interest in this study.
When considering only coplanarity (which could include spin-orbit ``anti-alignment'', or retrograde orbits)
and both $i_o$ and $i_*$ are known (but not $\Delta \Omega$), the sum and absolute difference between the orbital and stellar inclinations 
provides boundaries on the true spin-orbit alignment:

\begin{equation}
 | i_o - i_* |  \le \psi_\mathrm{coplanar} \le  i_o + i_*.
\end{equation}

On the other hand, if one is concerned with angular momentum alignment in the system, and only $i_o$ and $i_*$ are known, there exists a broader range of 
possibilities for $\psi$ because the measurement of $i_*$ does not contain information about the sense of the spin.
Depending on the direction of rotation, the spin angular momentum vector may be pointing toward the observer 
or away from the observer into the plane of the sky.  In this case the constraint on $\psi$ becomes:

\begin{equation}
| i_o - i_* | \le \psi_\mathrm{true} \le \pi - | i_o - i_* |.
\end{equation}

\noindent Regardless, if $i_o$ and $i_*$ differ then $\psi$ is non-zero and the system's orbital and rotational planes are misaligned by at least $| i_o - i_* |$. 
However, if $i_o$ and $i_*$ are identical then the system is consistent with being aligned but the true obliquity may be much larger.

This is illustrated in the right-most diagram in Figure~\ref{fig:geometry}.  If only $i_o$ and $i_*$ are known, the projection of the orbital and rotational 
angular momentum vectors will trace out nested cones with a degeneracy between the cones opening toward and away from the observer.
If $i_o$ is fixed at point \texttt{A$_1$}, for example, and a stellar inclination $i_*$ is measured, the minimum value of $\psi$ will occur when $\Omega_*$=$\Omega_o$
at point \texttt{B$_1$}.  If $\Omega_*$=$\Omega_o$ + 180$\degr$ (\texttt{B$_2$}), $\psi$ can reach $i_o$ + $i_*$.  But if $\overrightarrow{L_*}$
points away from the observer (residing on the lower cone opening in the $-z$ direction), 
$\Delta \Omega$ = 180$\degr$ (\texttt{B$_3$}), and $\psi$ can reach a value as high as 180$\degr$ -- $| i_o - i_* |$.
For instance, if $i_o$ = $i_*$ = 30$\degr$, the orbital and rotational planes can be perfectly coincident ($\psi$ = 0$\degr$) or misaligned by up to 60$\degr$, and the
spin-orbit angle can be misaligned by up to 120$\degr$.  
This study focuses on measurements of the minimum misalignment, $\Delta i$ = $| i_o - i_* |$,
for stars hosting directly imaged substellar companions.

\section{Observations} \label{sec:obs}

\subsection{Sample of Substellar Hosts} \label{sec:sample}

Our sample originates from a compilation of 177 stars hosting imaged substellar companions spanning a wide range of orbital separations.
The list includes discoveries based on high-contrast imaging as well as wide common-proper motion companions 
identified from seeing-limited surveys.  The compilation is assembled predominantly from the more focused lists in \citet{Deacon:2014ey}, \citet{Bowler:2016jk}, 
and \citet{Bowler:2020hk}, as well as additional systems identified more recently.  
The properties of the original parent sample are broad.  Host star masses span the hydrogen burning limit up to supernova progenitors (\citealt{Squicciarini:2022aa});
companions range in mass from 2 to about 75~\Mjup \ and orbital separations from a few AU to thousands of AU (Figure~\ref{fig:overview}).

\begin{deluxetable*}{lcccccccc}
\renewcommand\arraystretch{0.9}
\tabletypesize{\footnotesize}
\setlength{ \tabcolsep } {.1cm}
\tablewidth{0pt}
\tablecolumns{9}
\tablecaption{Tull Spectrograph Radial Velocities and Projected Rotational Velocities\label{tab:tullspec}}
\tablehead{
    \colhead{Name}  &  \colhead{UT Obs. Date}  &  \colhead{Exp. Time}  & \colhead{RV}           & \colhead{$\sigma_{\mathrm{RV}}$}  & \colhead{$v \sin i$}    & \colhead{$\sigma_{v \sin i}$} &   \colhead{Standard}    & \colhead{SpT} \\
                    &  \colhead{(YYYY-MM-DD)}       &  \colhead{(s)}        & \colhead{(m s$^{-1}$)} & \colhead{(m s$^{-1}$)}            & \colhead{(km s$^{-1}$)} & \colhead{(km s$^{-1}$)} &   \colhead{or Model}    & \colhead{or $T_\mathrm{eff}$}
        }
\startdata
1RXS J034231.8+121622   &  2022-02-21  &  1200   & 34.7  &   2.1  &   4.9   &  3.3    &    HD 119850   &  M1.5  \\
1RXS J034231.8+121622    &   2022-02-21  &  1200  &  35.0  &   1.6  &     5.2  &   3.2   &     HD 119850  &   M1.5  \\
1RXS J160929.1--210524   &  2019-06-26   & 1200  &  --8.4  &   1.1  &   8   &  2  &      HD 166620   &  K2  \\
1RXS J160929.1--210524    &  2019-06-26  &  1200 &   --8.2  &   1.0  &   8.3  &   1.7  &      HD 166620   &   K2  \\
2MASS J22362452+4751425   &   2019-06-17  &  1200   &  $\cdots$  & $\cdots$  & 6.8   &  1.4  &   PHOENIX  &    4100 K \\
2MASS J22362452+4751425   &   2019-06-27   & 1200   & --22.9   &  0.7  &   6.8  &   2.0  &    HD 173701   &  K0  \\
2MASS J23513366+3127229   &   2019-07-31  &  1200  &  $\cdots$  & $\cdots$ &15  &   3   &    PHOENIX  &   3600 K \\
2MASS J23513366+3127229   &   2019-07-31  &  1200   & $\cdots$  & $\cdots$ & 15 &   4    &   PHOENIX  &    3600 K \\
51 Eri  &   2021-10-16   &  120  &  22.3  &   3.3  &  69   &  3   &  HD 207978  &  F2 \\
51 Eri  &  2022-02-21    & 300   &  22.0   &  2.4   & 69  &   6  &     HIP 29396  &  F0 \\
G 196-3  &    2019-04-01  &  1200  &  0.0  &   0.7  &  16.8   &  1.7  &      GJ388   &    M4.5  \\
Gl 229  &  2021-10-16  &   800   &  4.8  &   0.5  &   4.5  &   1.4  &       HD 245409   &   K7 \\
Gl 504  &   2019-03-31  &   300  & $\cdots$ &$\cdots$  &   9.2  &  1.7  &  PHOENIX  &   5900 K \\
Gl 504  &   2019-03-31  &   900  & $\cdots$ & $\cdots$ &  8.8  &   0.9  &  PHOENIX  &  5900 K \\
Gl 504   &  2019-03-31  &   900  & $\cdots$ & $\cdots$ &  9.0  &   1.0  &  PHOENIX  &   5900 K \\
Gl 504  &  2019-03-31   &  900   & $\cdots$  & $\cdots$ &  8.7  &   1.0  &  PHOENIX  &  5900 K \\
Gl 504  &  2022-02-21   &  300  &  --27.7  &   0.6   &  7.6  &   0.2  &  HD 141004  &  G0  \\
Gl 758  &   2019-06-15   &  300  & --22.2   &  0.4  &   5.1   &  0.4  &  HD 173701  & K0  \\
Gl 758   &  2019-06-17   &  600  &  $\cdots$  & $\cdots$ & 5.4   &  2.3  &  PHOENIX  &  5300 K \\
Gl 758  &   2019-06-27   &  300  &  $\cdots$  & $\cdots$ & 6.2   &  1.8   &  PHOENIX  &  5300 K \\
GSC 6214-210  &   2019-06-26  &  1200  &   --6.1  &   0.8   &  6.8  &   0.5  &   HD 166620  &  K2 \\
GSC 6214-210  &    2019-06-26  &  1200  &  --6.0  &   0.9  &   6.6  &   0.5   &  HD 166620  &  K2 \\
GU Psc   &  2019-07-31  &  1200  & $\cdots$  &  $\cdots$  &  23    & 4     &  PHOENIX  &  3400 K \\
GU Psc   &  2019-07-31  &  1200   & $\cdots$  &  $\cdots$  &  25   &  5    &   PHOENIX  &  3400 K \\
HD 1160  &  2019-06-15 &  600  & $\cdots$  &  $\cdots$  & 96  &  10  &  PHOENIX  &  9800 K \\
HD 1160 &  2019-06-17  &   600  & $\cdots$  &  $\cdots$  &     97  &   7    &  PHOENIX  & 9800 K  \\
HD 1160  &  2019-07-31  &   300  &  $\cdots$  &  $\cdots$  &  95   &  7    &   PHOENIX  & 9800 K  \\
HD 19467  &  2021-10-16  &   600  &   6.1  &   0.2  &   3.3  &     0.4      &    HD 187923   &      G0    \\
HD 206893  &    2019-06-15  &   600  & --13.0   &  2.3  &  33  &   3   &    HD 182572  &  G8 \\
HD 206893  &  2019-06-17  &   600  &  $\cdots$  & $\cdots$ &  34  &   3   &   PHOENIX  &  6600 K  \\
HD 4747   &  2021-10-16  &   600  &   9.6         &         0.3     &   3.1  &   0.3   &        HD 4628        &     K2  \\
HD 49197 &   2019-05-12   &  600  &   9.8  &   0.6  &  23.1  &   1.4   &     HD 122652   &   F8  \\
HD 984  &  2019-06-17  &   600  & $\cdots$  &  $\cdots$  &   38.7  &   2.5  &  PHOENIX  & 6300 K  \\
HR 7672   &    2019-06-15   &  300  &   5.2  &   0.6  &   5.8  &   2.1  &  HD 173701   & K0  \\
HR 7672   &    2019-06-17   &  600  &   $\cdots$  & $\cdots$  &  6.4  &   1.5  &  PHOENIX  &  5900 K \\
HR 7672   &    2019-06-27   &  300  &   5.3  &   0.4   &  4.1  &   1.8  &   HD 182488  & G8 \\
HR 8799   &  2019-06-15  &   300  & $\cdots$  & $\cdots$ & 43  &   5  &   PHOENIX  &  7200 K \\
HR 8799   &  2019-06-16   &  300  & $\cdots$  & $\cdots$ &  52   &  7   &   PHOENIX  &    7200 K \\
HR 8799   &  2019-07-31  &    60  & $\cdots$  & $\cdots$ & 38  &   5  &    PHOENIX  &    7200 K \\
$\kappa$ And  &    2019-06-15  &   300  & $\cdots$  & $\cdots$ &  182  &  24  &  PHOENIX  &  10800 K \\
$\kappa$ And  &    2019-06-17  &   300  & $\cdots$  & $\cdots$ &  183  &  35   & PHOENIX  & 10800 K \\
$\kappa$ And  &    2019-07-31  &    45   & $\cdots$  & $\cdots$ &  173  &  28   &  PHOENIX  & 10800 K \\
Ross 458  &     2019-04-01 &   1200  & --12.3  &   0.9  &  11.5  &     1.3  &        GJ388  &   M4.5  \\
ROXs 12  &   2020-07-20   & 1200  &  --5.4   &  1.3  &   8.2  &   2.4  &     HD 157881  &    K5  \\
ROXs 12  &  2020-07-20  &  1200  &  --5.8  &   1.9  &   8.5  &   2.6   &     HD 157881  &    K5  \\
\enddata
\end{deluxetable*}

From this list we identify stars with rotation periods and $v \sin i_*$ measurements---either new in this work or previously published.
When considering rotation periods, we restrict our analysis to spectral types of F5 or later to reduce confusion between 
rotation periods and stellar pulsations  (e.g., \citealt{Sepulveda:2022aa}).\footnote{Note that this is not an explicit cut on mass because
spectral type generally evolves during the pre- and post-main sequence phases with changing \Teff.  
However, for mid-F stars the difference is modest.
On the main sequence, F5 corresponds to an effective temperature 
of $\approx$6510~K and a mass of about 1.4~\Msun \ (\citealt{Drilling:2000vo}; \citealt{Pecaut:2013ej}).  
Similarly, for the young ($\approx$18~Myr) equal-mass F5 binary AK Sco,
\citet{Czekala:2015vb} found a total dynamical mass of 2.49 $\pm$ 0.10~\Msun,
implying individual masses of about 1.25~\Msun.}
This is especially true of the intermediate-mass $\gamma$ Dor pulsators 
which span late-A to mid-F spectral types and have characteristic $g$-mode oscillation frequencies 
of a few hours to a few days (\citealt{Kaye:1999aa}; \citealt{Aerts:2021aa}).
Light curves and periodograms of stars in the remaining sample are further visually inspected for signs of pulsations.

Two exceptions are made to this spectral type cut.
HIP 21152 is an F5 member of the Hyades with a recently discovered brown dwarf companion 
(\citealt{Bonavita:2022aa}; \citealt{Kuzuhara:2022aa}; \citealt{Franson:2022bb}).  Its light curve periodogram
shows significant peaks that are not integer harmonics of its strongest signal, and are therefore not likely to be
caused by rotational modulation.  We therefore remove this star from our sample.

51 Eri is a young F0 member of the $\beta$~Pic moving group with an imaged planet at 11~AU (\citealt{Macintosh:2015fw}).
\citet{Sepulveda:2022aa} found the host to be a $\gamma$~Dor pulsator, casting doubt on previous rotation periods determined
from light curves.  However, they also find a plausible core rotation rate of 0.9$^{+0.3}_{-0.1}$ d, so we include 
this star in our sample but recognize that detailed asteroseismic analysis using a longer time series
is needed to more reliably constrain the rotation 
properties of this massive star.

Finally, all light curves with low-amplitude modulations are carefully examined on a case-by-case basis
for signs that the variations might be instrumental rather than astrophysical in nature.
Optical artifacts from bright sources and scattering from Earthshine can result in low-level
changes that may mimic real signals.  Our criteria for retaining light curves are that the signals
must show clear and consistent periodic brightness changes that are evident from visual inspection, 
rotation periods must be shorter than half of the \emph{TESS} sector baseline ($\lesssim$13 d for a single sector), 
and amplitudes must be substantially higher than the
expected sensitivity floor for the target star brightness.

\begin{figure}
  \vskip -0 in
  \hskip 0. in
  \resizebox{3.3in}{!}{\includegraphics{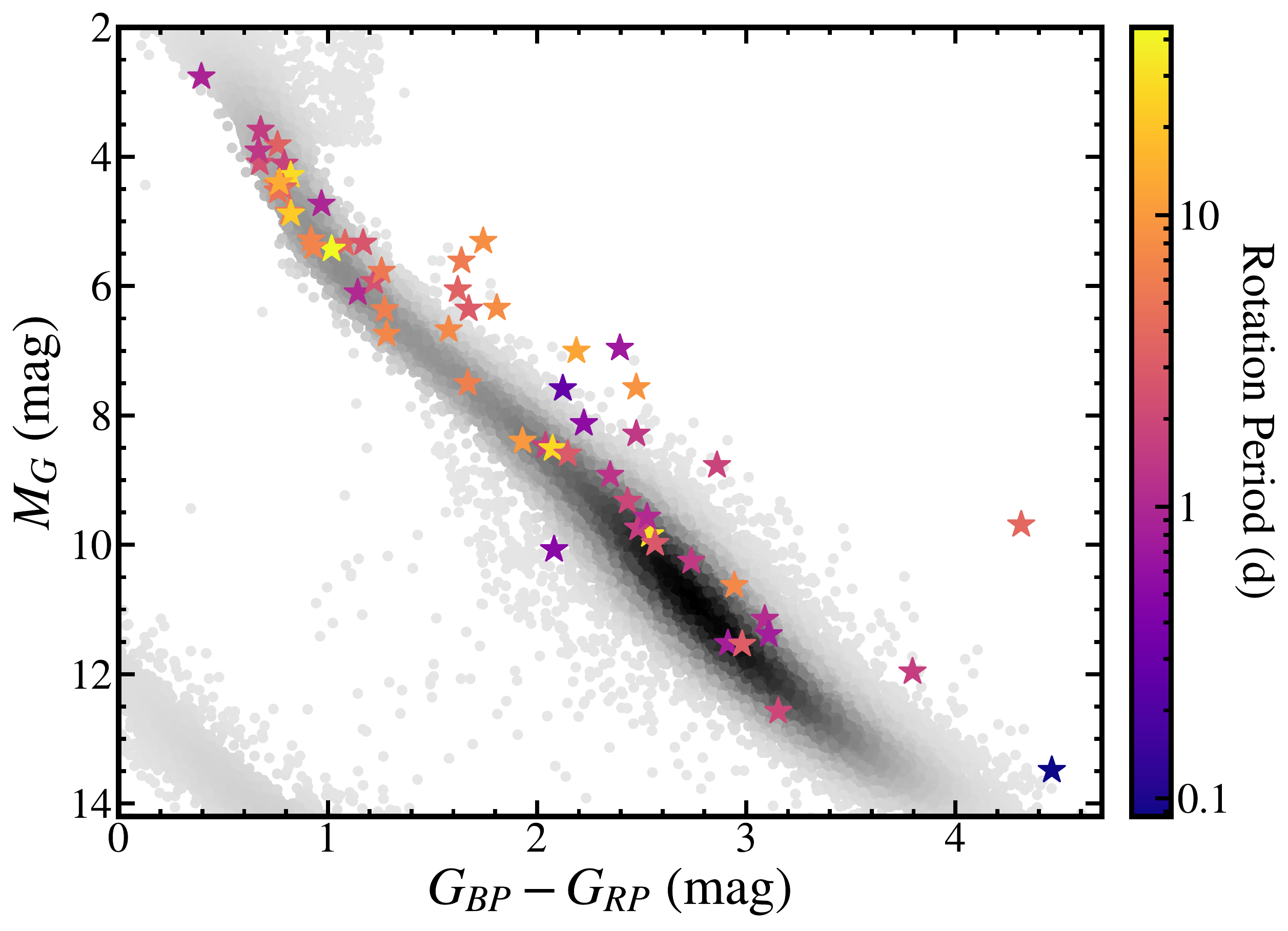}}
  \vskip -.1 in
  \caption{\emph{Gaia} $G_{BP}$--$G_{RP}$ color-magnitude diagram of stars hosting imaged substellar companions 
  in this study.  With the exception of 51 Eri, the bluest star in the figure, we focus on cool stars of F5 and later to avoid
  confusion between stellar rotation and pulsations in our light curve analysis.
  Host stars are color-coded by rotation period and range from 0.08--44~d.  Long rotation periods are either determined from
  multiple \emph{TESS} sectors or are adopted from previous measurements in the literature.   \label{fig:cmd} } 
 \end{figure}

After implementing these cuts, our final sample for this study comprises 62 host stars with rotation periods,
53 of which also have projected rotational velocities and stellar radius determinations.  
These measurements are detailed in the following subsections and are summarized in Table~\ref{tab:hosttable} and Table~\ref{fig:overview}.
A \emph{Gaia} color-magnitude diagram of host stars in our final sample is shown in Figure~\ref{fig:cmd}.


\subsection{High-Resolution Spectroscopy with the Tull Spectrograph} \label{sec:ts23}

High-resolution optical spectra for 22 targets from our broader sample were obtained 
with the Tull Coud\'{e} Spectrograph (\citealt{Tull:1995tn}) at McDonald Observatory's 2.7-m Harlan J. Smith telescope
between March 2019 and February 2022.  The 1$\farcs$2 slit and E2 grating were used with the TS23 setup for all observations, which 
resulted in a resolving power of $R = \lambda / \delta \lambda \approx 60,000$.
56 \'{e}chelle orders were simultaneously captured with the TK3 Tektronix CCD from 3870~\AA \ to 10500~\AA \ 
in (largely non-overlapping) segments ranging from 60~\AA \ to 200~\AA.  
On most nights we also observed several bright, slowly rotating RV and $v \sin i_*$ standards from \citet{Chubak:2012tv} and \citet{Soubiran:2018fz} 
spanning F through M spectral types.
A summary of the observations can be found Table~\ref{tab:tullspec}.


\begin{figure*}
  \vskip -0.2 in
  \hskip 0.1 in
  \resizebox{7in}{!}{\includegraphics{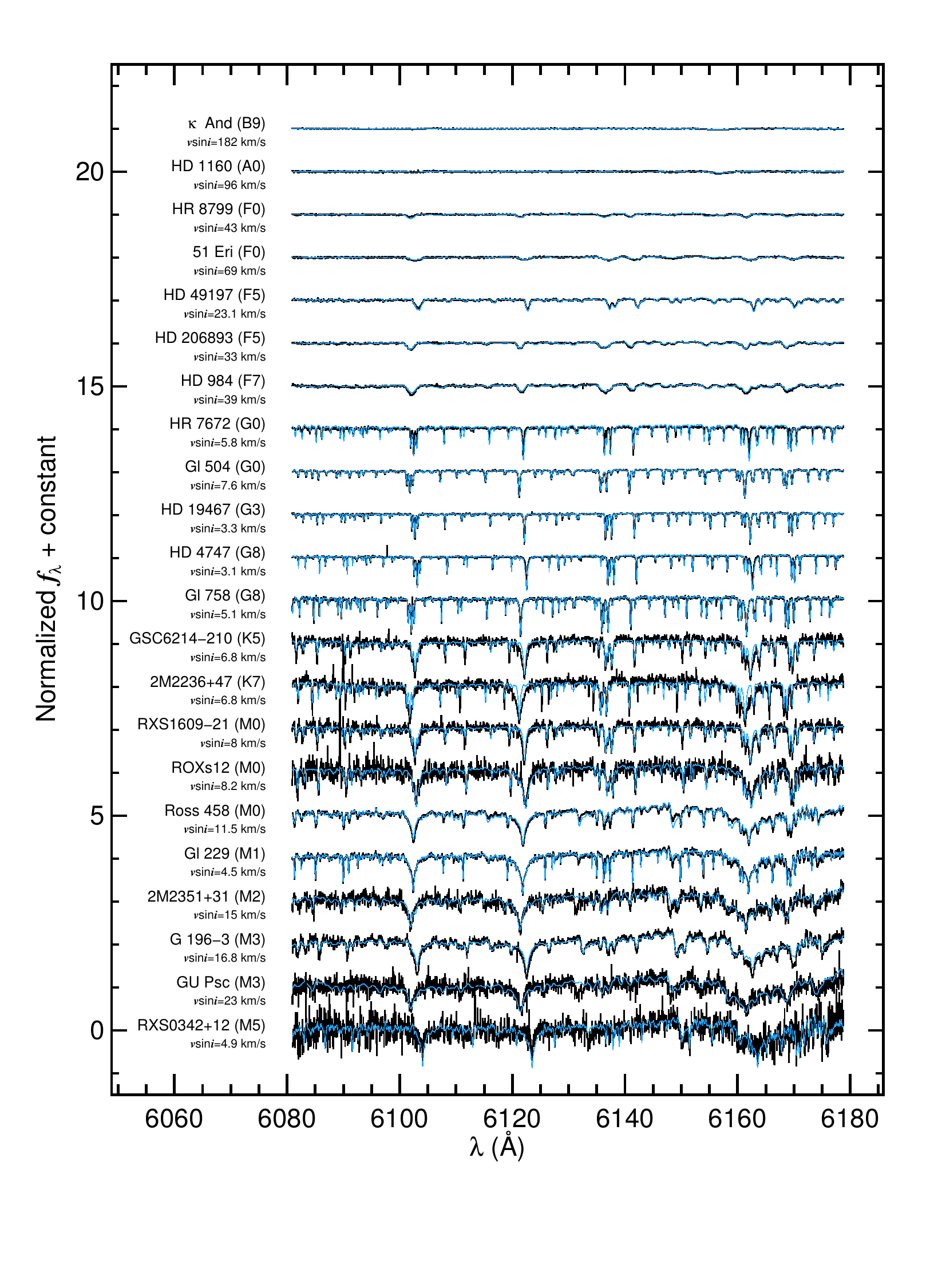}}
  \vskip -0.9 in
  \caption{Example of a single order from our observations with the Tull high-resolution optical spectrograph (black).  
  Stars are organized by decreasing spectral type from late-B to mid-M.  
  Standard stars are overplotted in blue and have been broadened and
  shifted to match the measured radial and projected rotational velocities of the substellar host stars in our sample.    \label{fig:ts23} } 
 \end{figure*}

2-dimensional traces for the science observations are defined using a spectral flat field calibration frame taken on the same night.
After bias subtraction and correction for known bad pixels, each curved spectral order is resampled to a linear two-dimensional trace using sub-pixel interpolation.
Night sky lines are subtracted from each order using dispersed sky regions near the ends of the projected slit.
Orders are optimally extracted following the method of \citet{Horne:1986bg}, and fifth-order polynomial fits to the extracted flat spectra are used
to remove the blaze function.

A wavelength solution is derived for each of the 56 orders using a ThAr emission line spectrum obtained on the same night and with the same setup 
as the science observations.  A ThAr line list comprising a total of 9145 emission lines 
with documented relative line intensities was assembled spanning 3785~\AA \ to 10507~\AA \ from \citet{Lovis:2007cp} 
(for $\lambda$~$<$ 6915~\AA) and \citet{Murphy:2007aa} (for $\lambda$~$>$ 6915~\AA).  
For each order, a synthetic emission line spectrum was generated at the resolving power of the Tull spectrograph.
To automate the process of mapping pixel values to wavelengths, we iteratively solved for the coefficients of 
a third-order polynomial model by maximizing the cross-correlation function between the observed spectral order and a ``true'' 
emission spectrum using the \texttt{AMOEBA} algorithm (\citealt{Nelder:1965tk}).  With reasonable initial estimates for the coefficients, this approach
performed well and the solution was visually verified for each order on each night.

\subsection{\emph{TESS} Light Curves} \label{sec:tess}

We used the \texttt{lightkurve} (\citealt{LightkurveCollaboration:2018aa}) package to download the 2-minute cadence Science Processing Operations Center (SPOC) Pre-search Data Conditioning Simple Aperture Photometry (PDCSAP) light curve (\citealt{Smith:2012aa}; \citealt{Stumpe:2012aa}; \citealt{Stumpe:2014aa}; \citealt{Jenkins:2016aa})  from the Mikulski Archive for Space Telescopes (MAST).\footnote{\href{https://archive.stsci.edu/missions-and-data/tess}{https://archive.stsci.edu/missions-and-data/tess/}} Two targets (GJ 3305 and SDSS J130432.93+090713.7) were analyzed using their 30-minute Full Frame Images (FFIs), which were reduced by either the \emph{TESS}-SPOC (\citealt{Caldwell:2020aa}) or the MIT Quick-Look Pipeline (\citealt{Huang:2020bb}) and manually retrieved from the MAST online portal.\footnote{\href{https://mast.stsci.edu/portal/Mashup/Clients/Mast/Portal.html}{https://mast.stsci.edu/portal/Mashup/Clients/Mast/Portal.html}} 

All photometric measurements listed as \texttt{NaN} are removed. Individual \emph{TESS} sector light curves are normalized by dividing both the flux and flux uncertainties by the median flux value of that sector.  Each  complete light curve is then compiled by stitching together these normalized sectors. Flares, transits, and other photometric outliers are removed by detrending the light curve with a high-pass Savitzky-Golay filter (\citealt{Savitzky:1964aa}) and only selecting data points in the original light curve that lie within one standard deviation of the flattened light curve.  Note that because of flares and outlier photometric points, this threshold is typically larger than what it would be for pure Gaussian noise.  Our period measurements (described below) are in good agreement whether or not we include this outlier rejection step.

For each processed light curve used in this study, we produce Generalized Lomb-Scargle (GLS) periodograms (\citealt{Zechmeister:2009ii}) over the frequency range 0.0005--100.0 d$^{-1}$ (0.01--2000 days)  to search for any periodic modulation that could be attributed to rotation.  
A Gaussian is fit to the highest periodogram peak and the resulting mean and standard deviation are adopted for the period and its uncertainty, respectively. 
For stars with multiple sectors separated by data gaps, aliasing due to the spectral window function causes a ``fringe pattern'' with a broad envelope for all periodogram peaks. In such cases, the Gaussian is fit to the envelope structure in order to reflect that spread.
Each resulting light curve is visually inspected to ensure significant periodicities reflected astrophysical variations instead of spacecraft-related systematic oscillations, for example from optical artifacts or Earthshine.  
These can manifest as gradual low-amplitude variations, sharp discontinuities from spacecraft motion, 
and scattered light features which rise and fall in sync with the 13.7-day orbit.
Period measurements are considered reliable if they show 
a single strong peak, a phase-folded light curve with consistent structure among phased curves, and a rotational modulation amplitude 
that is large compared to the limiting sensitivity of \emph{TESS} for that target brightness.
Finally, we limit adopted measurements to periods $\lesssim$13~d,
which corresponds to half of the \emph{TESS} sector baseline; instrumental systematics make it challenging to recover 
lower frequency signals in \emph{TESS} light curves (e.g., \citealt{Avallone:2022aa}).

Altogether 46 stars have light curves that appear to reflect true rotational modulation.  They exhibit a wide range of 
characteristic behaviors---amplitudes with 20\%-level variations  (PDS 70); 
consistent, symmetric, sinusoid-like modulations (2MASS~J02155892--0929121, CD--35~2722, and G~196-3); 
signs of dramatic and rapid starspot evolution (HD 130948, HD 16270, HD 49197, and HD 97334);
double-peaked structure (2MASS J01225093--2439505 and PZ Tel);
and beating patterns reflecting the superposition of two frequencies, most likely reflecting binarity or differential rotation (TWA 5 Aab).
Full light curves, periodograms, and phased light curves are shown in Figures~\ref{fig:tess_1}--\ref{fig:tess_8}.  Rotation periods and 
\emph{TESS} sectors used in this analysis are listed in Table~\ref{tab:hosttable}.

LP 261-75 reflects an interesting example of an M dwarf whose rotation period would have been interpreted incorrectly
if only one \emph{TESS} sector had been available.  
The periodogram of its full light curve from Sectors 21 and 48
shows two comparably strong peaks at 1.11~d and 2.23~d (Figure~\ref{fig:tess_6}); one signal is clearly a harmonic of the true rotation period.
This confusion is evident in reported photometric rotation period measurements of this star in the literature:
\citet{Newton:2016ea} and \citet{Irwin:2018aa} find values of 2.219 d and 2.22 d, respectively, while
\citet{CantoMartins:2020aa} list a value of 1.105 $\pm$ 0.027 d.
We therefore separately analyzed each individual sector light curve.  The Sector 21 light curve shows a strong peak at 1.11~d with
smooth modulations and consistent amplitudes.  712 days later marks the beginning of the next set of observations in Sector 48.  
In this sector the regular modulations occur with a periodicity of 2.23~d.

We interpret this longer modulation as the true
rotation period.  
This ``period-doubling'' behavior is consistent with the following scenario: 
during the Sector 21 observations, two groups of starspots with comparable surface covering fractions 
may have been located on opposite sides of the spinning star with a longitudinal phase difference of $\sim$180$\degr$.
Typically this setup with two large groups of spots would produce a light curve with a double-peaked structure having two amplitudes, but here
the similar spot sizes and 180$\degr$ phase difference seem to have conspired to mimic a faster period in Sector~21, 
which only later could be differentiated from the 
true value when one group of spots evolved or disappeared.
This phenomenon is different from typical rotation period confusion from photometric light curves, which
are usually caused by sampling and windowing effects from insufficient cadence relative to the rotation 
period of the star.  This is not the case for these fine 2-minute observations from \emph{TESS}.
Instead, LP 261-75 

\startlongtable
\begin{longrotatetable}
\begin{deluxetable*}{lccccccccccccccccc}
\renewcommand\arraystretch{0.9}
\tabletypesize{\scriptsize}
\setlength{ \tabcolsep } {.1cm} 
\tablewidth{0pt}
\tablecolumns{18}
\tablecaption{Host Star Properties\label{tab:hosttable}}
\tablehead{
       \colhead{Name}         & \colhead{Host}   & \colhead{Comp.}  &  \colhead{Disc.}  & \colhead{Sep.} & \colhead{\emph{TESS}}     & \colhead{$P_\mathrm{rot}$} & \colhead{$\sigma_{P, \mathrm{tot}}$\tablenotemark{a}} &  \colhead{$P_\mathrm{rot}$}  & \colhead{$R_*$}         & \colhead{$\sigma_{R_*, \mathrm{tot}}$\tablenotemark{b}} &   \colhead{$R_*$} &  \colhead{$v \sin i$}     &  \colhead{$\sigma_{v \sin i}$}  &  \colhead{$v_\mathrm{eq}$}     &  \colhead{$\sigma_{v_\mathrm{eq}}$}  \\ 
       \colhead{}             & \colhead{SpT}    & \colhead{SpT}    &  \colhead{Ref.}       & \colhead{(AU)}       & \colhead{Sectors}         & \colhead{(d)}               & \colhead{(d)}                                        &  \colhead{Ref.}              & \colhead{($R_{\odot}$)} & \colhead{($R_{\odot}$)}                 & \colhead{Ref.} &   \colhead{(km s$^{-1}$)}  &   \colhead{(km s$^{-1}$)}      &   \colhead{(km s$^{-1}$)}  &   \colhead{(km s$^{-1}$)}      
        }   
\startdata
          1RXS J034231.8+121622    &                  M5.2    &                    L0    &                     1    &                    19     &                                            5; 42-44     &         7.3     &           0.3     &                    72     &        0.35     &           0.03     &                    86     &           5     &           2     &         2.4     &         0.2    \\
\enddata
\end{deluxetable*}
\end{longrotatetable}

\noindent  illustrates a (probably rare) instance of bias that can  arise if only a single sector is available, which
applies to 13 of the 62 host stars with rotation periods in our sample.
Note that the deep 1.882-day transit events from LP 261-75 C (\citealt{Irwin:2018aa}) were removed 
from the light curves for this analysis.  These small regular gaps do not impact the results.

\section{Results}{\label{sec:results}}

\subsection{New Projected Rotational and Radial Velocities} \label{sec:vsini}


\begin{figure}
  \vskip -0.7 in
  \hskip -0.8 in
  \resizebox{5.4in}{!}{\includegraphics{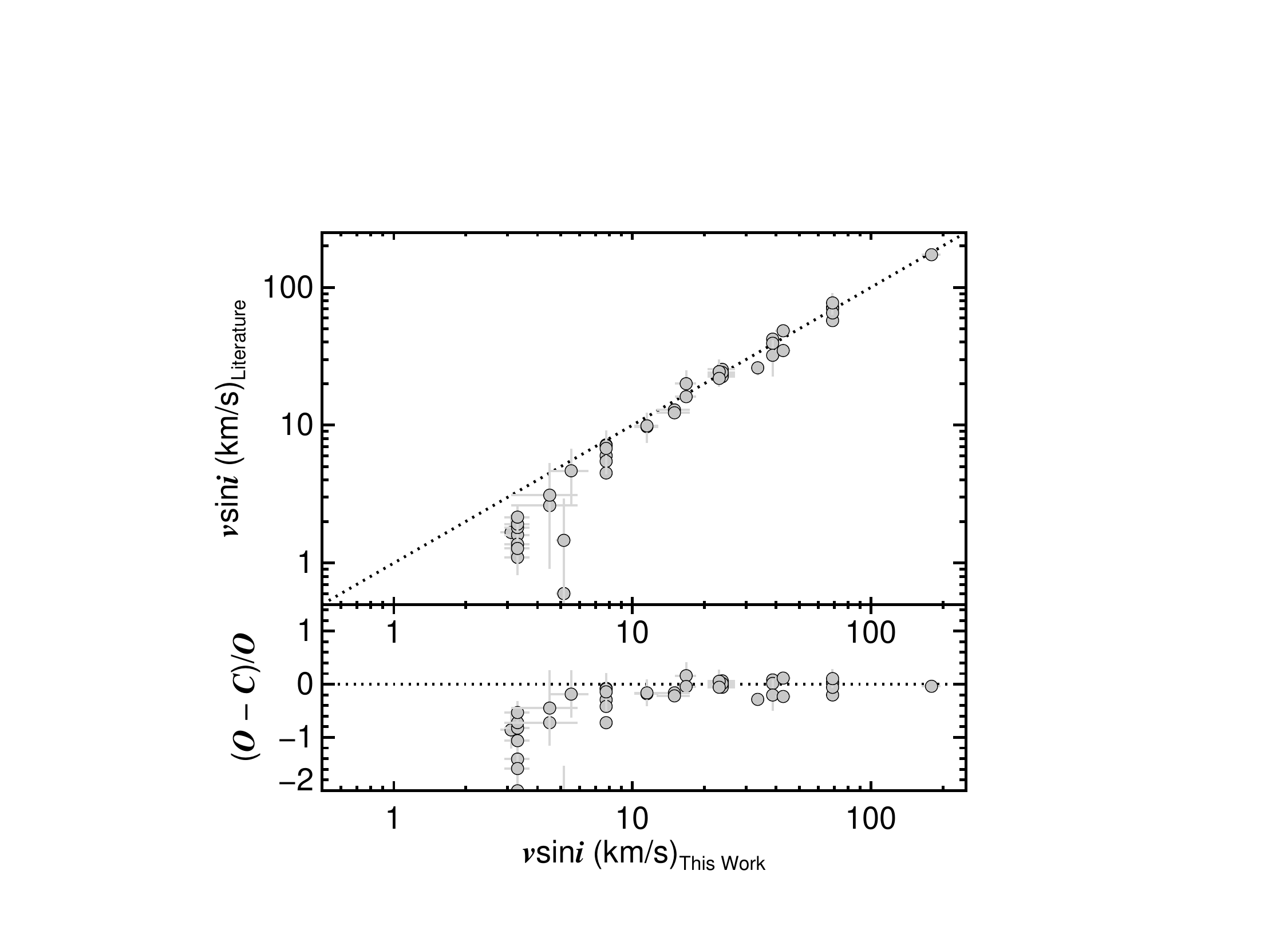}}
  \vskip -.2 in
  \caption{Comparison between $v \sin i_*$ values from this work and those from the
  literature for the same star.   The dotted line is a 1:1 relationship.  Above about 6 km s$^{-1}$,
  the agreement is good, but our $v \sin i_*$ values derived through cross correlation with
  slowly rotating standard stars are generally overestimated for low projected rotational velocities. 
  The bottom panel shows fractional residuals.  \label{fig:vsini_comp} } 
 \end{figure}

Stars that we observed with the Tull Spectrograph range in spectral type from B9 to M5.
Projected rotational velocities are determined by either applying a rotational broadening kernel to spectra of 
slowly rotating RV standards, or by broadening an appropriate model spectrum from the
PHOENIX-ACES grid (\citealt{Husser:2013ca}).
Orders with low SNR, strong telluric lines, or few absorption lines (for early-type stars) were avoided.  
This typically resulted in 20--40 good orders.
When using the empirical standards as a reference template, we first cross correlate every standard spectrum observed on the same 
night with each order of the science spectrum.  The resulting individual cross-correlation functions (CCFs) for a particular standard 
are summed and the reference spectrum that results in the highest CCF value is adopted for the broadening analysis.
Each order of the template spectrum is then sequentially broadened through a fine grid of $v \sin i_*$ values spanning 0 to 200 km/s 
with a rotational kernel following \citet{Gray:2005va}.
A new CCF is computed for each $v \sin i_*$ value, and the broadening kernel resulting in the highest CCF peak for that order is
added in quadrature with the measured $v \sin i_*$ value of the standard star.  This is typically
obtained from the 
compilation of projected rotational velocities by \citet{Glebocki:2005aa}; most of the standards we used have $v \sin i_*$ values
below 3 km~s$^{-1}$ for G, K, and M dwarfs and below 10 km~s$^{-1}$ for F stars.
The final adopted projected rotational velocity is then derived from the robust mean and standard deviation 
of the $v \sin i_*$ measurements for each individual order following the procedure in \citet{Beers:1990aa}.
Uncertainties generally scale with the $v \sin i_*$ value but most are measured at the $\sigma_{v \sin i_*}$/$v \sin i_*$ $\sim$ 5--30\% level.

This procedure also results in an instantaneous RV relative to the RV standard.  A barycentric
correction is  applied following the approach in \citet{stumpff:1980aa}, then an absolute RV is computed using
the RV of the standard.  The total uncertainty is determined from the quadrature sum of our measured 
relative RV and the published uncertainty for the standard star.  
Although these measurements are not the primary focus of this study, we report RVs 
alongside $v \sin i_*$ values in Table~\ref{tab:tullspec}.

On nights when an appropriate slowly rotating standard was not observed, or for early-type stars for which fewer
standards are available, we used solar-metallicity synthetic spectra from the PHOENIX-ACES model atmosphere grid to 
determine projected rotational velocities.  
Model effective temperatures were chosen to the  
nearest 100~K (for $T_{\mathrm{eff}}$ $<$ 7000~K) 
or 200~K (for $T_{\mathrm{eff}}$ $\ge$ 7000~K)  using the SpT--$T_{\mathrm{eff}}$ scale from \citet{Pecaut:2013ej}.  
The models were first smoothed with a Gaussian kernal to the resolving power of the Tull spectrograph, 
trimmed to the wavelength range of the 
individual orders, and resampled onto the same wavelength grid
as the science spectra.  The same procedure for
assessing rotational broadening is carried out for the models as for the standard star observations described above.
Fewer orders were generally used for the B, A, and F stars as many orders for these early-type stars lacked lines altogether.
$v \sin i_*$ values are determined separately for each order and then a robust mean and standard deviation are computed.
Model effective temperatures are listed with the $v \sin i_*$ results in Table~\ref{tab:tullspec}.
Examples of spectra and broadened templates (both empirical and from model spectra) 
for a single order are shown in Figure~\ref{fig:ts23}.

We compiled previously determined $v \sin i_*$ values for targets in Table~\ref{tab:tullspec} 
to assess how our measurements compare with those in the literature (see Appendix~\ref{app:vsini} and

\startlongtable
\begin{longrotatetable}
\begin{deluxetable*}{lcccccccccc}
\renewcommand\arraystretch{0.9}
\tabletypesize{\footnotesize}
\setlength{ \tabcolsep } {.2cm} 
\tablewidth{0pt}
\tablecolumns{11}
\tablecaption{Stellar Inclinations and Minimum Obliquities\label{tab:inclinations}}
\tablehead{
 \colhead{Name} &  \multicolumn{3}{c}{$i_*$} &  \multicolumn{2}{c}{$i_o$}  &   \multicolumn{3}{c}{$\Delta i_*$} &  \colhead{$P(\Delta i > 10\degr)$}  &  \colhead{Misaligned?\tablenotemark{b}}  \\
 \cline{2-4}
 \cline{5-6}
 \cline{7-9}
                        & \colhead{MAP\tablenotemark{a}}          &   \colhead{Median}        &  \colhead{95.4\% C.I.}          & \colhead{Median}        & \colhead{Ref.}    & \colhead{MAP}                & \colhead{Median}            &  \colhead{95.4\%  C.I.}   &   &  \\ 
 \colhead{}             & \colhead{($\degr$)}                     & \colhead{($\degr$)}       &  \colhead{($\degr$)}            & \colhead{($\degr$)}     &                   & \colhead{($\degr$)}          & \colhead{($\degr$)}         &  \colhead{($\degr$)}   &  &
        }   
\startdata
          1RXS J034231.8+121622    &        90.0    &          66.5$^{+   23.4  }_{-   10.8  }$  &               27.6--90.0                  &   79.6$^{+   6.8  }_{-  5.5 }$    &   1           &    2.5            &        26.7$^{  +12.4 }_{-  26.7 }$  &   0.0--76.0   &  0.803     &  No Evidence \\
          1RXS J160929.1-210524    &        90.0    &          71.8$^{+   18.2  }_{-    7.3  }$  &               47.5--90.0                  &   $\cdots$                        &   $\cdots$    &    $\cdots$       &    $\cdots$                          &   $\cdots$    &  $\cdots$  &  $\cdots$   \\
          2MASS J01033563-5515561  &         8.9    &           8.9$^{+    0.9  }_{-    1.3  }$  &                6.9--11.0                  &   $\cdots$                        &   $\cdots$    &    $\cdots$       &    $\cdots$                          &   $\cdots$    &  $\cdots$  &  $\cdots$   \\
          2MASS J01225093-2439505  &        90.0    &          85.9$^{+    4.1  }_{-    2.0  }$  &               78.1--90.0                  &  103.4$^{+   7.7  }_{-  9.2 }$    &   2           &    9.5            &        14.2$^{   +5.8 }_{-  13.7 }$  &   0.0--49.2   &  0.677     &  No Evidence \\
          2MASS J02155892-0929121  &        66.4    &          69.1$^{+   15.1  }_{-   12.6  }$  &               48.3--89.9                  &   $\cdots$                        &   $\cdots$    &    $\cdots$       &    $\cdots$                          &   $\cdots$    &  $\cdots$  &  $\cdots$   \\
          2MASS J02192210-3925225  &        51.7    &          54.4$^{+    6.9  }_{-   10.0  }$  &               39.2--77.8                  &   $\cdots$                        &   $\cdots$    &    $\cdots$       &    $\cdots$                          &   $\cdots$    &  $\cdots$  &  $\cdots$   \\
          2MASS J04372171+2651014  &        77.3    &          73.9$^{+   15.7  }_{-    6.3  }$  &               55.4--89.9                  &   $\cdots$                        &   $\cdots$    &    $\cdots$       &    $\cdots$                          &   $\cdots$    &  $\cdots$  &  $\cdots$   \\
          2MASS J16103196-1913062  &        90.0    &          68.7$^{+   21.3  }_{-    9.1  }$  &               36.9--90.0                  &   $\cdots$                        &   $\cdots$    &    $\cdots$       &    $\cdots$                          &   $\cdots$    &  $\cdots$  &  $\cdots$   \\
          2MASS J23513366+3127229  &        90.0    &          77.7$^{+   12.3  }_{-    5.0  }$  &               61.3--90.0                  &   127.1$^{+  12.0  }_{-  15.9 }$  &   1           &    39.5           &        38.6$^{  +17.3 }_{-  22.5 }$  &   0.0--75.0   &  0.920     &  Likely     \\
                        51 Eri     &        44.1    &          53.5$^{+   13.2  }_{-   19.3  }$  &               31.9--87.6                  &   144$^{+8}_{-8}$\tablenotemark{c} &   3          &    0.0            &        54.4$^{  +28.3 }_{-  54.2 }$  &   0.0--110.5  &  0.854     &  No Evidence \\
                        AB Pic     &        54.4    &          58.5$^{+    7.7  }_{-   12.0  }$  &               43.1--84.5                  &   90$^{+12}_{-12}$\tablenotemark{c} & 4         &    32.4           &        30.1$^{  +15.5 }_{-  16.4 }$  &   0.0--55.3   &  0.888     &  Likely      \\
           ASAS J212528-8138.5     &        44.0    &          45.1$^{+    4.1  }_{-    5.5  }$  &               36.2--56.4                  &   $\cdots$                        &   $\cdots$    &    $\cdots$       &    $\cdots$                          &   $\cdots$    &  $\cdots$  &  $\cdots$   \\
                      BD+21 55     &         3.7    &           3.8$^{+    1.2  }_{-    1.6  }$  &                1.1-- 6.6                  &   $\cdots$                        &   $\cdots$    &    $\cdots$       &    $\cdots$                          &   $\cdots$    &  $\cdots$  &  $\cdots$   \\
                    CD-35 2722     &        52.2    &          54.4$^{+    5.6  }_{-    8.5  }$  &               41.5--74.1                  &   150.8 $^{+  20.7  }_{-  14.5 }$ &   1           &    35.5           &   54.8$^{  +36.0 }_{-  54.6 }$       &   0.0--118.2  &  0.903     &  Likely     \\
                        FU Tau     &        75.0    &          75.8$^{+   12.8  }_{-    6.3  }$  &               60.7--89.9                  &   $\cdots$                        &   $\cdots$    &    $\cdots$       &    $\cdots$                          &   $\cdots$    &  $\cdots$  &  $\cdots$   \\
                     FW Tau AB     &        43.6    &          44.9$^{+    4.4  }_{-    5.9  }$  &               35.4--57.6                  &   $\cdots$                        &   $\cdots$    &    $\cdots$       &    $\cdots$                          &   $\cdots$    &  $\cdots$  &  $\cdots$   \\
                       G 196-3     &        56.1    &          60.0$^{+    8.8  }_{-   12.2  }$  &               43.9--86.1                  &   $\cdots$                        &   $\cdots$    &    $\cdots$       &    $\cdots$                          &   $\cdots$    &  $\cdots$  &  $\cdots$   \\
                      G 204-39     &        90.0    &          64.0$^{+   25.9  }_{-   11.7  }$  &               23.0--90.0                  &   $\cdots$                        &   $\cdots$    &    $\cdots$       &    $\cdots$                          &   $\cdots$    &  $\cdots$  &  $\cdots$   \\
                    GJ 3305 AB     &         6.9    &           7.4$^{+    1.0  }_{-    1.7  }$  &                4.9--10.9                  &   $\cdots$                        &   $\cdots$    &    $\cdots$       &    $\cdots$                          &   $\cdots$    &  $\cdots$  &  $\cdots$   \\
                        Gl 229     &        90.0    &          79.4$^{+   10.5  }_{-    5.0  }$  &               59.7--90.0                  &   5.50$^{+0.16}_{-0.16}$\tablenotemark{c} &   5   &    85.4           &        84.5$^{  +15.3 }_{-  16.2 }$  &   52.6--114.8 &  1.00      &  Yes        \\
                        Gl 504     &        18.4    &          18.5$^{+    0.7  }_{-    0.9  }$  &               16.9--20.1                  &   141.4$^{+  7.5  }_{-  10.7 }$   &   1           &    120.5          &        97.3$^{  +29.7 }_{-  82.7 }$  &   0.0--140.7  &  0.910     &  Likely     \\
                        GQ Lup     &        38.3    &          40.9$^{+    5.9  }_{-    8.3  }$  &               28.2--62.2                  &   61.4$^{+   7.2  }_{-  5.9 }$    &   6           &    77.7           &   38.9$^{  +44.3 }_{-  24.3 }$       &   0.0--92.3   &  0.896     &  Likely     \\
               GSC 06214-00210     &        45.2    &          48.5$^{+    6.6  }_{-   10.1  }$  &               33.9--73.9                  &   114.4$^{+  8.9  }_{-  12.2 }$   &   7           &    20.5           &    39.5$^{  +29.7 }_{-  31.7 }$      &   0.0--94.6   &  0.865     &  Likely     \\
               GSC 08047-00232     &        90.0    &          79.0$^{+   10.9  }_{-    4.8  }$  &               63.2--90.0                  &   $\cdots$                        &   $\cdots$    &    $\cdots$       &    $\cdots$                          &   $\cdots$    &  $\cdots$  &  $\cdots$   \\
                        GU Psc     &        90.0    &          79.1$^{+   10.8  }_{-    4.9  }$  &               62.4--90.0                  &   $\cdots$                        &   $\cdots$    &    $\cdots$       &    $\cdots$                          &   $\cdots$    &  $\cdots$  &  $\cdots$   \\
                     HD 116402     &        60.7    &          65.4$^{+    9.5  }_{-   12.9  }$  &               50.3--89.1                  &   $\cdots$                        &   $\cdots$    &    $\cdots$       &    $\cdots$                          &   $\cdots$    &  $\cdots$  &  $\cdots$   \\
                     HD 129683     &        54.7    &          59.0$^{+    9.5  }_{-   13.4  }$  &               41.9--86.5                  &   $\cdots$                        &   $\cdots$    &    $\cdots$       &    $\cdots$                          &   $\cdots$    &  $\cdots$  &  $\cdots$   \\
                     HD 130948     &        90.0    &          75.1$^{+   14.7  }_{-    6.4  }$  &               53.4--90.0                  &   105$^{+4}_{-4}$\tablenotemark{c} &   8          &    0.0            &        17.7$^{   +8.5 }_{-  17.5 }$  &   0.0--47.2   &  0.706     &  No Evidence \\
                      HD 16270     &        36.4    &          37.9$^{+    4.2  }_{-    6.1  }$  &               28.4--50.7                  &   $\cdots$                        &   $\cdots$    &    $\cdots$       &    $\cdots$                          &   $\cdots$    &  $\cdots$  &  $\cdots$   \\
                      HD 19467     &        49.4    &          54.9$^{+    9.9  }_{-   14.4  }$  &               36.9--85.2                  &   129.8$^{+   6.6  }_{-   6.6 }$  &   9           &    0.0            &        40.7$^{  +28.5 }_{-  40.5 }$  &   0.0--90.3   &  0.745     &  No Evidence \\
                     HD 203030     &        60.0    &          65.3$^{+   10.3  }_{-   13.7  }$  &               49.5--89.6                  &   $\cdots$                        &   $\cdots$    &    $\cdots$       &    $\cdots$                          &   $\cdots$    &  $\cdots$  &  $\cdots$   \\
                       HD 3651     &        90.0    &          77.0$^{+   12.9  }_{-    5.9  }$  &               55.5--90.0                  &   $\cdots$                        &   $\cdots$    &    $\cdots$       &    $\cdots$                          &   $\cdots$    &  $\cdots$  &  $\cdots$   \\
                      HD 37216     &        74.4    &          65.2$^{+   24.8  }_{-    9.6  }$  &               34.1--90.0                  &   $\cdots$                        &   $\cdots$    &    $\cdots$       &    $\cdots$                          &   $\cdots$    &  $\cdots$  &  $\cdots$   \\
                      HD 49197     &        75.0    &          72.8$^{+   16.7  }_{-    6.7  }$  &               53.6--90.0                  &   97.6$^{+   4.7  }_{-   8.6 }$   &   1           &    4.5            &        18.4$^{   +8.1 }_{-  18.0 }$  &   0.0--52.1   &  0.734     &  No Evidence \\
                       HD 8291     &        20.9    &          22.2$^{+    7.7  }_{-    9.8  }$  &                5.7--41.1                  &   $\cdots$                        &   $\cdots$    &    $\cdots$       &    $\cdots$                          &   $\cdots$    &  $\cdots$  &  $\cdots$   \\
                      HD 97334     &        25.3    &          25.8$^{+    3.0  }_{-    4.1  }$  &               19.2--33.5                  &   $\cdots$                        &   $\cdots$    &    $\cdots$       &    $\cdots$                          &   $\cdots$    &  $\cdots$  &  $\cdots$   \\
                        HD 984     &        72.8    &          74.3$^{+   12.1  }_{-    8.8  }$  &               58.2--89.9                  &   120.8$^{+1.7}_{-1.7}$\tablenotemark{c}  &   10   &    12.6           &        30.8$^{  +16.6 }_{-  21.4 }$  &   2.2--58.0   &  0.838     &  Likely     \\
                     HIP 70319     &        41.2    &          47.8$^{+   15.6  }_{-   21.0  }$  &               22.2--88.2                  &   $\cdots$                        &   $\cdots$    &    $\cdots$       &    $\cdots$                          &   $\cdots$    &  $\cdots$  &  $\cdots$   \\
                        HN Peg     &        57.9    &          62.5$^{+    8.5  }_{-   12.5  }$  &               47.1--87.1                  &   $\cdots$                        &   $\cdots$    &    $\cdots$       &    $\cdots$                          &   $\cdots$    &  $\cdots$  &  $\cdots$   \\
                       HR 7672     &        90.0    &          73.7$^{+   16.2  }_{-    7.0  }$  &               49.7--90.0                  &   97.4$^{+0.4}_{-0.4}$\tablenotemark{c} &   11    &    0.0            &        16.6$^{   +7.6 }_{-  16.2 }$  &   0.0--43.2   &  0.697     &  No Evidence \\
                  $\kappa$ And     &        30.1\tablenotemark{d}  &  30.1$^{+  4.0 }_{- 4.0 }$ &               22.1--38.1                  &   139.1$^{+13.8}_{-18.4}$         &   1           &    0.0            &        72.4$^{   +31.0 }_{- 72.4 }$  &   0.0--133.3  &  0.822     &  No Evidence \\
                       ksi UMa     &        19.6    &          20.7$^{+    6.4  }_{-    7.9  }$  &                7.1--36.3                  &   $\cdots$                        &   $\cdots$    &    $\cdots$       &    $\cdots$                          &   $\cdots$    &  $\cdots$  &  $\cdots$   \\
                       L 34-26     &        90.0    &          80.2$^{+    9.7  }_{-    4.3  }$  &               65.2--90.0                  &   $\cdots$                        &   $\cdots$    &    $\cdots$       &    $\cdots$                          &   $\cdots$    &  $\cdots$  &  $\cdots$   \\
                       LkCa 15     &        90.0    &          79.8$^{+   10.1  }_{-    4.6  }$  &               63.7--90.0                  &   $\cdots$                        &   $\cdots$    &    $\cdots$       &    $\cdots$                          &   $\cdots$    &  $\cdots$  &  $\cdots$   \\
                        PDS 70     &        54.7    &          60.4$^{+    9.9  }_{-   14.7  }$  &               43.8--88.2                  &   132.8$^{+3.0}_{-3.0}$\tablenotemark{c} &   12   &    1.4            &        43.2$^{  +24.1 }_{-  43.1 }$  &   0.0--85.9   &  0.804     &  No Evidence \\
                PM J02133+3648     &        80.6    &          77.9$^{+   11.7  }_{-    4.9  }$  &               63.5--90.0                  &   $\cdots$                        &   $\cdots$    &    $\cdots$       &    $\cdots$                          &   $\cdots$    &  $\cdots$  &  $\cdots$   \\
                        PZ Tel     &        90.0    &          78.9$^{+   11.0  }_{-    4.7  }$  &               63.7--90.0                  &    93.5$^{+   1.4  }_{-   2.0 }$  &   1           &    0.0            &         9.2$^{   +4.1 }_{-   9.0 }$  &   0.0--24.0   &  0.561     &  No Evidence \\
                   Ross 458 AB     &        85.0    &          77.4$^{+   12.5  }_{-    4.9  }$  &               61.9--89.9                  &   $\cdots$                        &   $\cdots$    &    $\cdots$       &    $\cdots$                          &   $\cdots$    &  $\cdots$  &  $\cdots$   \\
                       ROXs 12     &        90.0    &          82.3$^{+    7.6  }_{-    3.5  }$  &               69.8--90.0                  &   135.0$^{+  23.1  }_{-  19.8 }$  &   13          &    45.6           &        45.0$^{  +22.9 }_{-  24.0 }$  &   0.2--82.3   &  0.949     &  Likely     \\
              RX J1602.8-2401B     &        68.7    &          69.9$^{+   16.8  }_{-   10.2  }$  &               48.9--89.9                  &   $\cdots$                        &   $\cdots$    &    $\cdots$       &    $\cdots$                          &   $\cdots$    &  $\cdots$  &  $\cdots$   \\
                      SR 12 AB     &        90.0    &          67.7$^{+   22.3  }_{-   10.0  }$  &               31.3--90.0                  &   $\cdots$                        &   $\cdots$    &    $\cdots$       &    $\cdots$                          &   $\cdots$    &  $\cdots$  &  $\cdots$   \\
                     TWA 5 Aab     &        63.4    &          68.2$^{+   12.1  }_{-   13.9  }$  &               50.2--89.9                  &   150.8$^{+  12.3  }_{-  13.3 }$  &   1           &    42.5           &        60.9$^{  +29.5 }_{-  29.7 }$  &   10.6--109.7 &  0.982     &  Yes        \\
               TYC 8984-2245-1     &        69.2    &          72.2$^{+   11.8  }_{-   11.0  }$  &               55.4--89.9                  &   $\cdots$                        &   $\cdots$    &    $\cdots$       &    $\cdots$                          &   $\cdots$    &  $\cdots$  &  $\cdots$   \\
          VHS J125601.92-125723.9  &        49.1    &          58.9$^{+   17.5  }_{-   21.2  }$  &               32.9--89.9                  &   26$^{+14}_{-14}$\tablenotemark{c} &   14        &    35.5           &        63.1$^{  +38.7 }_{-  44.0 }$  &   0.0--122.3  &  0.933     &  Likely     \\
                     Wolf 1130     &        26.6    &          27.1$^{+    3.6  }_{-    4.4  }$  &               19.4--35.5                  &   $\cdots$                        &   $\cdots$    &    $\cdots$       &    $\cdots$                          &   $\cdots$    &  $\cdots$  &  $\cdots$   \\
\enddata
\tablerefs{
(1) \citet{Bowler:2020hk}; 
(2) \citet{Bryan:2020ji};
(3) \citet{Dupuy:2022bb};
(4) \citet{PalmaBifani:2022aa}; 
(5) \citet{Feng:2022zz};
(6) \citet{Stolker:2021aa};
(7) \citet{Pearce:2019iv};
(8) Wang et al. (in prep.);
(9) \citet{Maire:2020iu};
(10) \citet{Franson:2022bl};
(11) \citet{Brandt:2019ey};
(12) \citet{Wang:2020jb};
(13) \citet{Bryan:2016eo};
(14) \citet{Dupuy:2022aa}.
}
\tablenotetext{a}{Maximum a posteriori probability.}
\tablenotetext{b}{Systems are classified as misaligned (``Yes'') if the probability that $\Delta i$ values are greater than 10$\degr$ is $\ge$95\% \emph{and} the MAP value of the $\Delta i$ posteriors is greater than 10$\degr$.
If $P(\Delta i > 10\degr)$ $\ge$ 80\% and the $\Delta i$ MAP value is $>10\degr$ then the system is classified as being ``Likely'' misaligned.  If neither is satisfied, the system is consistent with spin-orbit alignment (``No Evidence'').}
\tablenotetext{c}{Our normal approximation to reported $i_o$.}
\tablenotetext{d}{The stellar inclination of $\kappa$ And is taken from the solar-metallicity model fits to oblateness measurements in \citet{Jones:2016hg}.  The slightly asymmetric reported uncertainties are averaged to 4$\degr$ for these purposes.}
\end{deluxetable*}
\end{longrotatetable}

\noindent Section~\ref{sec:vsinicomp}).
Results are shown in Figure~\ref{fig:vsini_comp}.  Above about 6 km s$^{-1}$, the agreement between
our measurement and previous values is generally good; the Pearson correlation coefficient is 0.99 and the 
rms of the relative residuals is 20\%.  
Below 6 km s$^{-1}$, our measurements using the order-by-order CCF approach generally overestimates reported values.
This is likely because we are not carrying out a detailed line profile analysis of high-SNR spectra focused on select individual lines.
We are limited by previous determinations of broadening in standard stars, which result in a lower limit on our achieved
value, and we do not take into account any additional broadening effects such as micro- and macro-turbulence that could 
contribute to the shape and strength of absorption lines.  
We therefore caution that low $v \sin i_*$ values from our analysis may be somewhat overestimated.
However, only 8 of our 45 measurements are below 6 km s$^{-1}$, and among stars that 
also have rotation periods (which enable an inclination analysis), all but one (1RXS~J034231.8+121622) 
have previous $v \sin i_*$ measurements
in the literature.  These previous measurements are taken into account in the final adopted $v \sin i_*$ value 
(Appendix~\ref{app:vsini} and Table~\ref{tab:hosttable}).

\subsection{Interpreting Periodic Modulations as Rotation Periods} \label{sec:prot}

Starspots are the main source of large-scale periodic photometric variability in intermediate- and low-mass stars.
In the Sun, differential rotation gives rise to surface rotation periods of about 25 days
at the equator and 34 days at the poles.
The locations of sunspots vary during the solar activity cycle but are generally confined to latitudes of about $\pm$30$\degr$.  However, for 
other stars with different convection zone depths, rotation rates, masses, compositions, and evolutionary states, the behavior of differential rotation, starspot
locations, and spot filling factors are expected to fundamentally differ from the Sun (\citealt{schussler:1996aa}; \citealt{Barnes:2005aa}; \citealt{Berdyugina:2005aa}).
If spots are positioned at non-equatorial latitudes, differential rotation will result in a rotationally modulated signal
that deviates from the equatorial rotation period.  This can bias measurements of the stellar equatorial velocity and inclination 
using the projected rotational velocity, whereby $i_*$ = $\sin^{-1} ( P_\mathrm{rot} v \sin i_* / (2 \pi R_*) )$.

Absolute shear $\Delta \Omega$ is a common metric to quantify differential rotation.  It represents the difference
in angular frequency at a star's equator and its pole:

\begin{equation}{\label{eqn:domega}}
\Delta \Omega = \Omega_\mathrm{Equator} - \Omega_\mathrm{Pole} =  2 \pi \Big(\frac{1}{P_\mathrm{min}} -  \frac{1}{P_\mathrm{max}} \Big).
\end{equation}

\noindent \citet{Reinhold:2013iz} and \citet{Reinhold:2015ep} analyzed thousands of (typically old) stars from the \emph{Kepler} mission and found that
absolute shear as traced by multiple periodogram peaks can span a wide range of values for a given rotation period---a proxy for age for Sun-like stars---and 
effective temperature,
but does not vary strongly as a function of these parameters, at least below
about 6200~K ($\approx$F8).  
The typical range of $\Delta \Omega$ values spans 0.01--0.1 rad d$^{-1}$.
At higher effective temperatures the absolute shear increases to values greater than 1 rad d$^{-1}$.
For comparison, the Sun's absolute shear is 0.07 rad d$^{-1}$.

Because differential rotation is likely to play an important role in interpreting the periodicity of light curves from stars in our sample,
we inflate the rotation period uncertainties computed from periodograms to more accurately reflect the unknown latitude 
being tracked by dominant starspot groups.
We define an error term associated with the star's shear, $\sigma_{P, \mathrm{shear}}$ that is designed to conservatively
reflect half the difference between the maximum (polar) and minimum (equatorial) rotation periods:

\begin{displaymath}
\sigma_{P, \mathrm{shear}} \approx \frac{P_\mathrm{max}  - P_\mathrm{min} }{2}.
\end{displaymath}

Solving for $P_\mathrm{min}$ and inserting it into Equation~\ref{eqn:domega} allows us to relate
$\sigma_{P, \mathrm{shear}}$, $P_\mathrm{max}$ (for which we adopt our measured photometric rotation period), and $\Delta \Omega$:

\begin{equation}
\sigma_{P, \mathrm{shear}} \approx \frac{1}{2} \Big( P_\mathrm{max} - \Big( \frac{\Delta \Omega}{2 \pi} +  \frac{1}{P_\mathrm{max}} \Big)^{-1} \Big)
\end{equation}

\noindent The Sun, for instance, would have $\sigma_{P, \mathrm{shear}}$ of 2.5~d---about 10\% of its actual equatorial velocity.
Assuming the same absolute shear, a younger Sun with $P_\mathrm{rot} = 10$~d would have $\sigma_{P, \mathrm{shear}}$ = 0.5~d, a 5\% relative uncertainty.
To estimate $\sigma_{P, \mathrm{shear}}$ for stars in our sample, we assume a constant 
solar-like absolute shear of $\Delta \Omega$ = 0.07 rad d$^{-1}$.

Our final adopted rotation period uncertainty is the quadrature sum of the uncertainty from the periodogram analysis
and from potential differential rotation:

\begin{equation}
\sigma_{P, \mathrm{tot}}  = \sqrt{ \sigma_{P, \mathrm{per}}^2  + \sigma_{P, \mathrm{shear}}^2}.
\end{equation}

The median rotation period precision, $\sigma_{P, \mathrm{tot}} / P_\mathrm{rot}$, is 2\% 
with a range of 0.2\%--20\% across our entire sample of 64 stars.

\subsection{Compilation of Projected Rotational Velocities} \label{sec:vsinicomp}

$v \sin i_*$ values are compiled from our own measurements and those in the literature.
There is a wide range in the quality of published projected rotational velocities based on the type of observations
(for example, medium versus high spectral resolution) and the approach to the measurement itself 
(such as detailed treatment of broadening mechanisms).  Some studies report uncertainties, but many do not.
Furthermore, many measurements of $v \sin i_*$ for the same star are formally inconsistent.

Our strategy for this study is to incorporate as much information as possible while also balancing the
reliability of various measurements made over the past several decades.  
Robust uncertainties are especially important to ensure the accuracy of our stellar inclination posterior distributions. 
If we obtained more than one $v \sin i_*$  measurement from our own Tull spectrograph observations, 
then these are combined into a single value through a weighted mean and weighted standard deviation.
If more than one value was identified in the literature, and uncertainties are reported,
then these are treated as separate independent measurements.
In cases where measurement uncertainties are not reported, we compute their mean and standard
deviation and treat this as an additional measurement.
All of these are then combined into a single (presumably more accurate) $v \sin i_*$  value which we adopt for this 
study using weighted mean and standard deviation.
This relies on reported uncertainties having been reliably estimated so that they avoid 
unjustifiably driving the weighted mean value to that measurement. 
We therefore also set an upper limit of 1\% on the relative precision of $\sigma_{v \sin i_*}$/$v \sin i_*$
for any single measurement.
Details can be found in Appendix~\ref{app:vsini}, and individual measurements 
and adopted $v \sin i_*$ values are listed in Table~\ref{tab:vsinis}.

\subsection{Stellar Radii} \label{sec:radii}

Most stellar radius estimates have been adopted from the
Revised \emph{TESS} Input Catalog (TIC; \citealt{Stassun:2019aa})
to ensure they are determined in a self-consistent fashion.
These values are based on the Stefan-Boltzmann relation
and incorporate \emph{Gaia}-based distances, extinction-corrected
$G$-band magnitudes, and $G$-band bolometric corrections.
\citet{Stassun:2019aa} found that the resulting radii were in good agreement (typically within 7\%) with
values for the same stars measured through asteroseismology by \citet{Huber:2017aa}.  
We therefore inflate uncertainties from the TIC catalog by adding 7\% errors in quadrature with the
quoted uncertainties to reflect a more realistic spread in individual radius estimates.
TIC radii are adopted for 47 of the 64 stars in our full sample.

For the remaining stars, radii are either taken from other literature sources or individually determined in this study.
This is the case when TIC radii are not listed, or if we suspect binarity may be severely impacting the inferred values.
When determining radii in this study, we make use of the Stefan-Boltzmann relation,
\begin{equation}
M_\mathrm{bol} = - 10 \log \big(T_\mathrm{eff} / T_{\odot} \big) - 5 \log \big( R / R_{\odot} \big) +  M_\mathrm{bol,\odot},
\end{equation}
and adopt a solar bolometric magnitude of 
$M_\mathrm{bol,\odot}$ = 4.755 mag and an
effective temperature of $T_\mathrm{eff, \odot}$ = 5772~K (\citealt{Mamajek:2012ga}; \citealt{Pecaut:2013ej}).

To decompose the apparent magnitudes of visual binaries into their individual constituent brightnesses, we make use of 
the contrast $\Delta m$ in magnitudes
and the integrated-light apparent magnitude $m$.  The apparent magnitude of the primary $m_A$ in each system is then 

\begin{equation}{\label{eqn:decompose}}
m_A = m + 2.5 \log (1 + 10^{-\Delta m / 2.5} ),
\end{equation}
and the apparent magnitude of the secondary is $m_B = \Delta m + m_A$.
Details for individual systems can be found in Appendix~\ref{sec:notes}, and 
radii are listed in Table~\ref{tab:hosttable}.


\begin{figure*}
  \hskip -0.3 in
  \resizebox{8in}{!}{\includegraphics{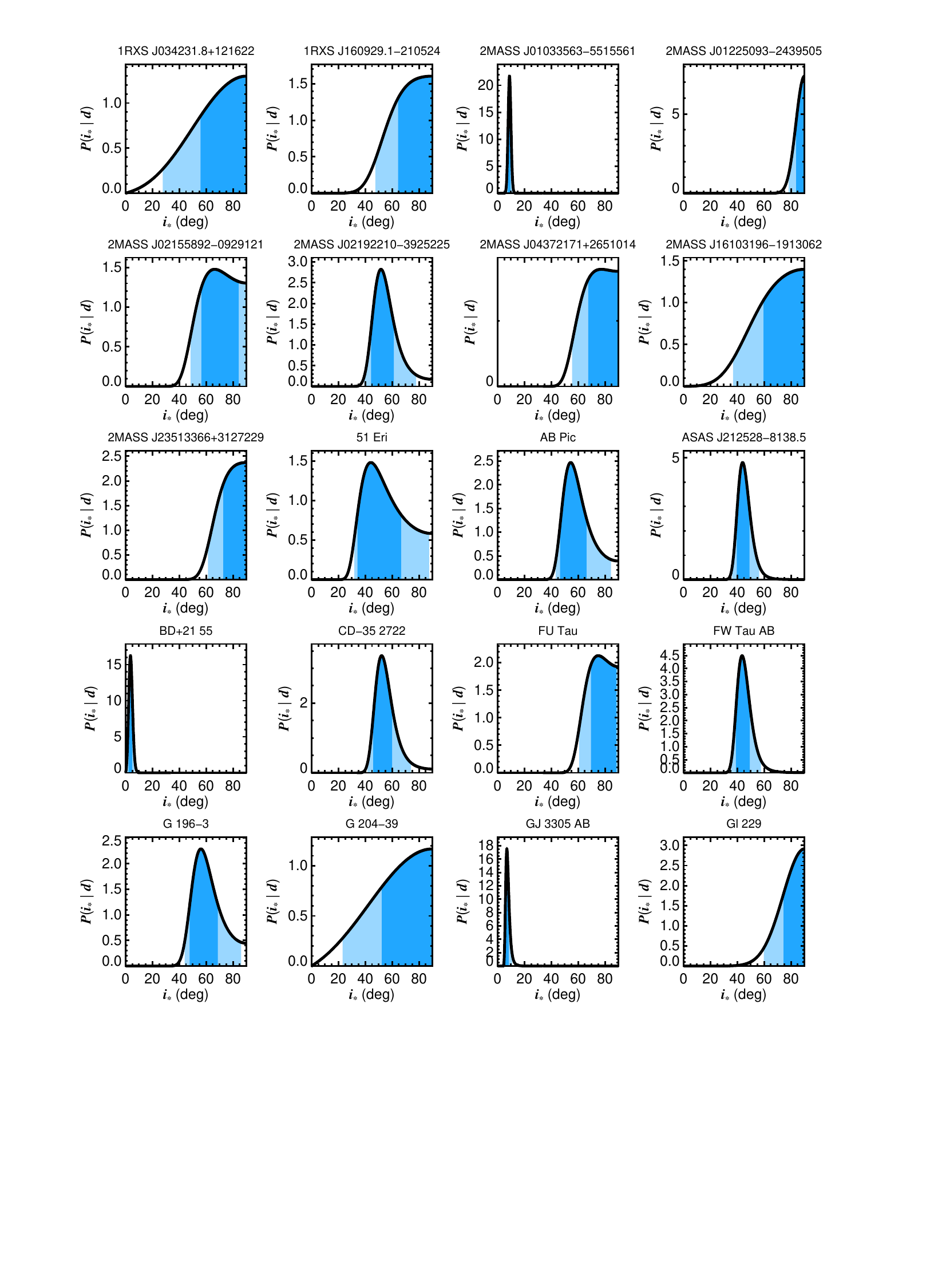}}
  \vskip -2.4 in
  \caption{Posterior distributions for line-of-sight stellar inclinations determined in this study.  Blue shaded regions show
   1-$\sigma$ and 2-$\sigma$ credible intervals.  Stars viewed equator-on have inclinations of 90$\degr$, while pole-on orientations have inclinations of 0$\degr$.
   Note that the direction of the angular momentum vector in the sky plane is usually unknown, as is its orientation toward or away from the 
   observer.  For comparison, an isotropic prior distribution is shown in Figure~\ref{fig:incl3}.
      \label{fig:incl1} } 
 \end{figure*}

\begin{figure*}
  \hskip -0.3 in
  \resizebox{8in}{!}{\includegraphics{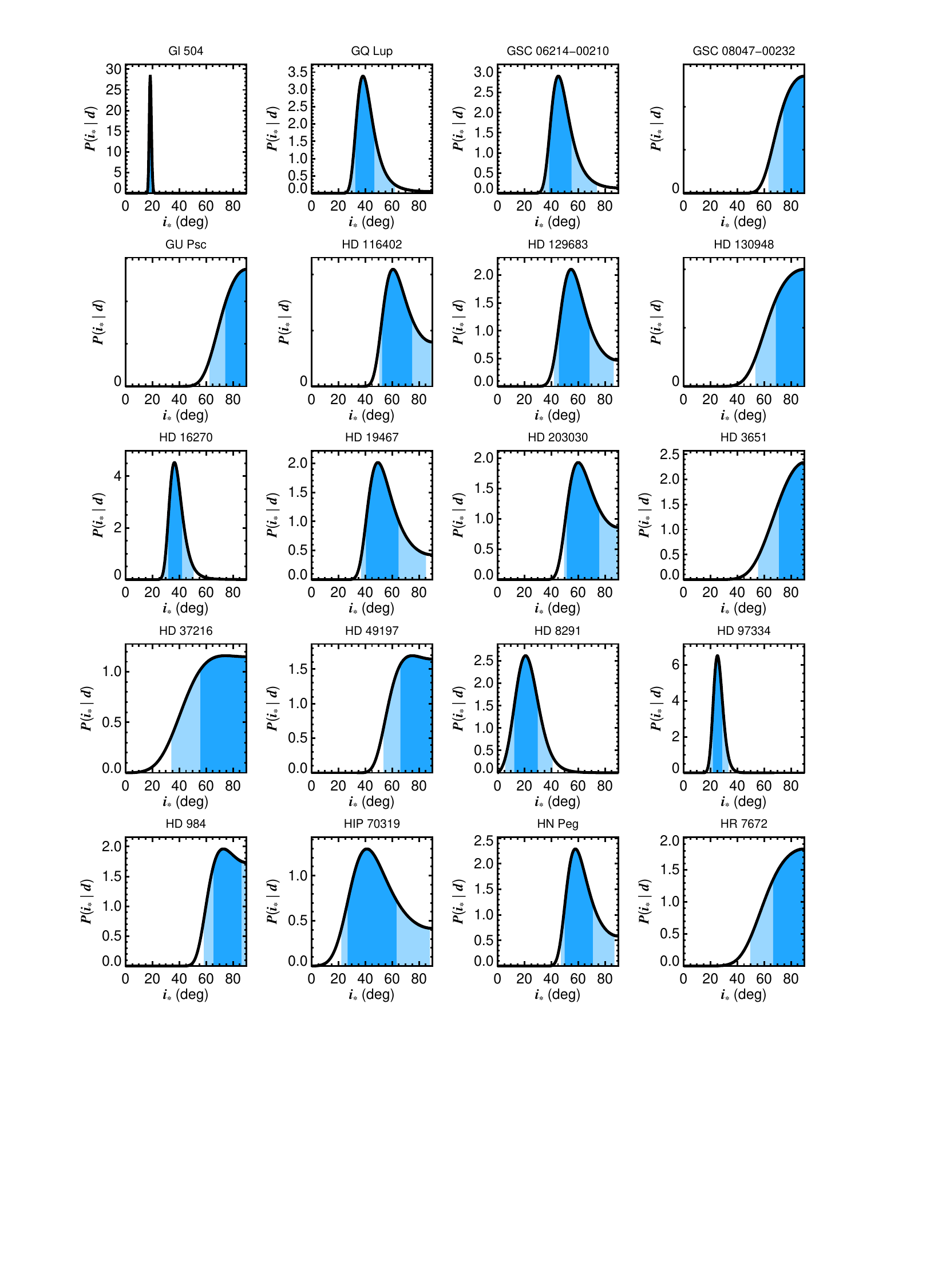}}
  \vskip -2.4 in
  \caption{Posterior distributions for line-of sight stellar inclinations.  See Figure~\ref{fig:incl1} for details.  
  For comparison, an isotropic prior distribution is shown in Figure~\ref{fig:incl3}.   \label{fig:incl2} } 
 \end{figure*}

\begin{figure*}
  \hskip -0.3 in
  \resizebox{8in}{!}{\includegraphics{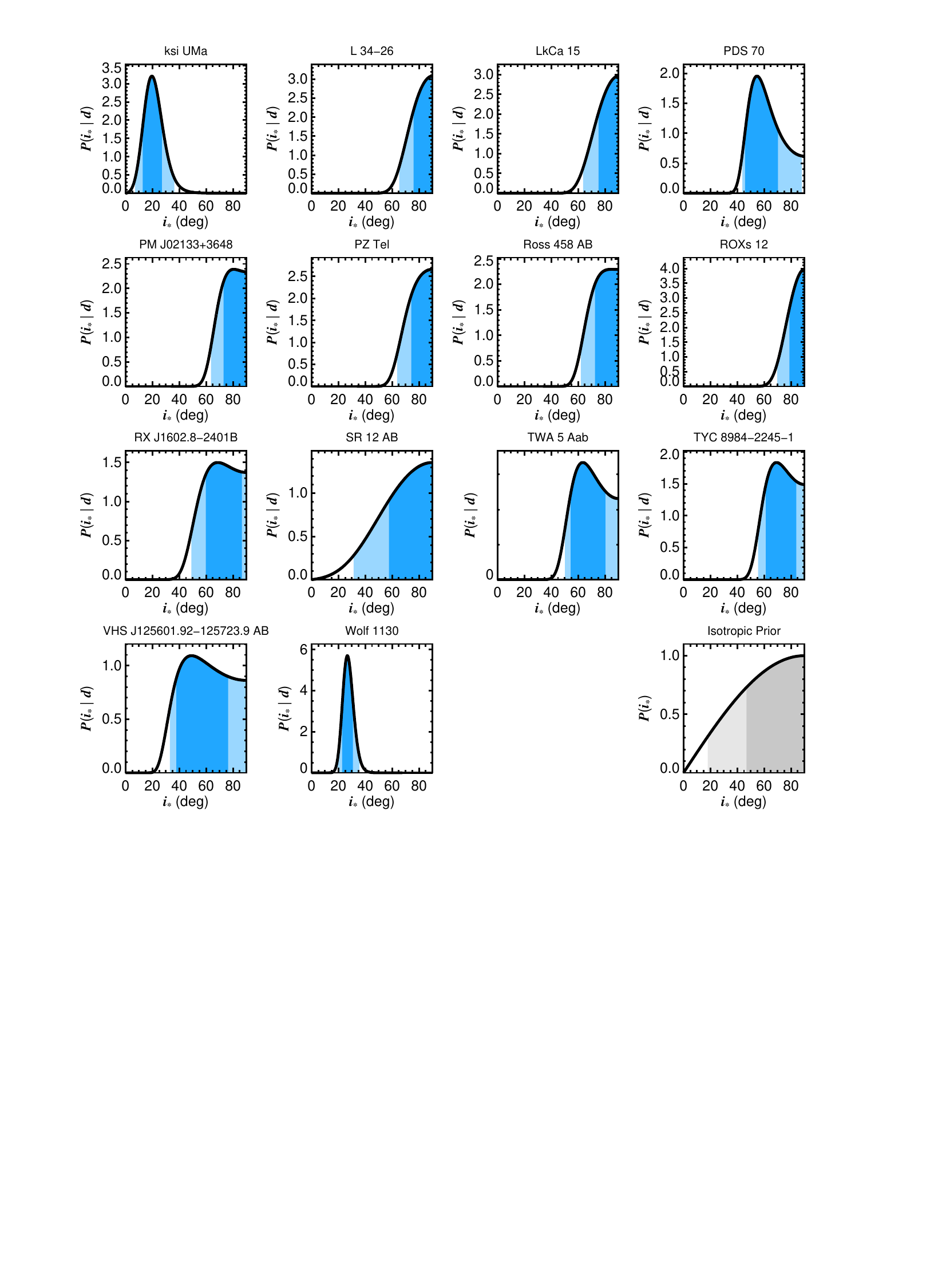}}
  \vskip -4. in
  \caption{Posterior distributions for line-of sight stellar inclinations.  See Figure~\ref{fig:incl1} for details.  For comparison, an isotropic prior distribution is shown in the bottom right panel.   \label{fig:incl3} } 
 \end{figure*}

\subsection{Stellar Inclinations} \label{sec:inclinations}

Our approach to constrain $i_*$ relies on the Bayesian probabilistic framework developed by  \citet{Masuda:2020dp}, which takes into account the correlation 
between the projected and equatorial rotational velocities $v \sin i_*$ and $v_\mathrm{eq}$.
Assuming normally distributed measurement errors for $v \sin{i_*}$, $P_\mathrm{rot}$, and  $R_*$; 
sufficiently precise measurements of the rotation period (at the $\lesssim$20\% level); and
uniform priors for all three parameters, we show in Appendix~\ref{app:istar} that 
the posterior distribution of $i_*$ can be expressed as:

\begin{equation}{\label{eqn:i}}
P(i_*\mid P_\mathrm{rot}, R_*, v\sin i_*) \propto  \sin i_* \times \frac{e^{- \frac{\big(v \sin i_* - \frac{2\pi R_*}{P_\mathrm{rot}}\sin i_* \big)^2}{2\big(\sigma_{v\sin i_*}^2 + \sigma_{v_\mathrm{eq}}^2 \sin^2 i_* \big)}} }{\sqrt{\sigma_{v\sin i_*}^2 + \sigma_{v_\mathrm{eq}}^2 \sin^2 i_*}},
\end{equation}

where

\begin{equation}{\label{eqn:veqerr}}
\sigma_{v_\mathrm{eq}} =  \frac{2\pi R_*}{P_\mathrm{rot}} \sqrt{\Big(\frac{\sigma_{R_*}}{R_*}\Big)^2 + \Big(\frac{\sigma_{P_\mathrm{rot}}}{P_\mathrm{rot}}\Big)^2 }.
\end{equation}

The resulting line-of-sight inclination posteriors for all 53 host stars are shown in Figures~\ref{fig:incl1}--\ref{fig:incl3},
and summary statistics are listed in Table~\ref{tab:inclinations}.  Constraints on the stellar inclination vary dramatically from star to star
depending on the precision of the input rotation period, projected rotational velocity, and stellar radius.
In some cases when these constraints are poor, the data add very little information and the posterior on $i_*$ resembles 
that for an isotropic prior, $\sin i_*$.  Most inclination uncertainties span $\approx$10$\degr$--40$\degr$, but 
in some cases the posterior is very well constrained to within 1--2 degrees.  
As expected from random orientations, the majority of 
inclination angles peak at high values of $i_*$ $>45\degr$, with many consistent with equator-on orientations of $i_* = 90\degr$.
BD+21 55 has the most pole-on orientation with $i_* = 3.8^{+1.2}_{-1.6} \degr$.

For cases where the host star is a binary, it is assumed that the measurements of $P_\mathrm{rot}$ and $v \sin i_*$ are for
the brighter component, and therefore that the inclination usually reflects the more massive member of the system.  In these cases,
and especially when the mass ratio is near unity, these constraints should be treated with caution because it is possible that the companion
could impact the input values of period, radius, or projected rotational velocity. 

The inferred equatorial velocity should always be higher than the projected rotational velocity for any
line-of-sight inclinations that depart from 90$\degr$.
In several instances in Table~\ref{tab:hosttable}, the $v \sin i_*$ value is larger than $v_\mathrm{eq}$, but for most of these systems
they are consistent to within 1--2~$\sigma$.
However, there are a few notable exceptions including
2MASS J01225093--2439505 ($v_\mathrm{eq}$ = 12.4 $\pm$ 1.0 km s$^{-1}$; $v \sin i_*$ = 18.1 $\pm$ 0.5 km s$^{-1}$) 
and Gl 229 ($v_\mathrm{eq}$ = 1.0 $\pm$ 0.2 km s$^{-1}$; $v \sin i_*$ = 2.8 $\pm$ 0.4 km s$^{-1}$).
This discrepancy may be due to overestimated rotational broadening measurements, 
underestimated radius estimates, or periods that are 
overestimated.  Longer-than-expected rotation periods could originate from stars with solar-like dynamos that have both strong differential rotation and 
spots located at non-equatorial latitudes such that the observed modulations are not tracking equatorial rotation regions.
2MASS J01225093--2439505 and Gl 229 are individually discussed in more detail in Appendix~\ref{sec:notes}.
For this work, we treat the posterior distributions of $i_*$ at face value when $v_\mathrm{eq} < v \sin i_*$ 
because, when this occurs, the Bayesian framework we make use of yields qualitatively similar results to $v_\mathrm{eq} = v \sin i_*$;
both imply an equator-on orientation with a posterior ``pressed up'' against $i_* = 90\degr$.

\subsection{Orbital Inclinations} \label{sec:radii}

Probability distributions for orbital inclinations are assembled from the literature for companions with orbit fits to 
various combinations of relative astrometry, absolute astrometry, and radial velocities.  This limits the available sample of obliquity constraints 
to 21 companions at separations where orbital motion is apparent on timescales of years to a few decades.
For our sample the range of separations spans 10--250~AU, with most companions residing between 10--100~AU.
\footnote{Note that in addition to 20 systems from our main sample with $i_*$ constraints determined using rotation periods, radii, and projected
rotational velocities, we also include  in our obliquity analysis $\kappa$~And---a rapidly rotating B9 star hosting a substellar companion at $\approx$55~AU (\citealt{Carson:2013fw}).
 \citet{Jones:2016hg} measured a stellar inclination of 30$^{+3}_{-5}$\degr and a polar position angle of 63$^{+5}_{-1}$\degr based on interferometric observations
 and a solar metallicity model.  However, the sense of stellar rotation (clockwise or counter-clockwise on the sky) is unknown, so the rotational angular momentum vector is degenerate across the sky plane.
\citet{Bowler:2020hk} determined an orbital inclination of 139$^{+13}_{-18}$\degr and a longitude of ascending node of 72$^{+21}_{-16}$\degr for $\kappa$~And~B.
 Averaging the asymmetric uncertainties implies a true spin-orbit angle of either $\psi$ = 109$^{+16}_{-16}$\degr or $\psi$ = 70$^{+15}_{-17}$\degr depending on 
 whether the stellar inclination is $\approx$30\degr and the polar position angle is $\approx$63\degr, or whether the stellar inclination is $\approx$180\degr -- 30\degr = 150\degr
 and the position angle is $\approx$63\degr + 180\degr = 243\degr.  Regardless, it appears that $\kappa$~And~B is significantly misaligned on a prograde
 or retrograde orbit around its host star.  However, for a fair incorporation with the rest of our sample, which does not possess information about the orientation of the star in the sky plane,
 we neglect the polar position angle and only include information about the rotational and orbital inclinations in our subsequent analysis of the $\kappa$~And system.
 }

When possible, actual orbital inclination posterior distributions are used in our analysis, but in some instances we make use of 
our own approximations (for example, assumptions of normality) 
based on the reported summary statistics such as the mean and standard deviation.
Orbital inclination posteriors for eight systems are taken from the uniform analysis in \citet{Bowler:2020hk}, which makes use
of the \texttt{orbitize!} orbit-fitting package (\citealt{Blunt:2017eta}; \citealt{Blunt:2020bb}):
1RXS J034231.8+121622 B, 2MASS J23513366+3127229 B, CD-35 2722 B, Gl 504 B, HD 49197 B, $\kappa$ And B, PZ Tel B, and TWA 5 B.
The orbits of two other companions were separately fit using the same package and are directly adopted for this study: 
2M0122--2439 b (\citealt{Bryan:2020ji}) and ROXs 12 B (\citealt{Bryan:2016eo}).
Eight systems have reported posteriors that are approximately normal, so we adopt Gaussian distributions for the following:
51 Eri b (\citealt{Dupuy:2022bb}), 
AB Pic b (\citealt{PalmaBifani:2022aa}),
Gl 229 B (\citealt{Feng:2022zz}), 
HD 984 B (\citealt{Franson:2022bl}),
HD 19467 B (\citealt{Maire:2020iu}),
HD 130948 BC (Wang et al. in prep; T. Dupuy, 2022, priv. communication),
HD 7672 B (\citealt{Brandt:2019ey}),
PDS 70 b (\citealt{Wang:2020jb}),
and VHS J125601.92-125723.9 b (\citealt{Dupuy:2022aa}).
Finally, for two systems the inclination distributions are digitally extracted:\footnote{Histograms are extracted using \textit{WebPlotDigitizer} at \url{https://automeris.io/WebPlotDigitizer/}.}
GQ Lup B (\citealt{Stolker:2021aa}) and GSC 6214-210 B (\citealt{Pearce:2019iv}).
The distribution medians and 68.3\% credible intervals are summarized in Table~\ref{tab:inclinations}.

\begin{figure*}
  \vskip -0.2 in
  \hskip -.5 in
  \resizebox{7.6in}{!}{\includegraphics{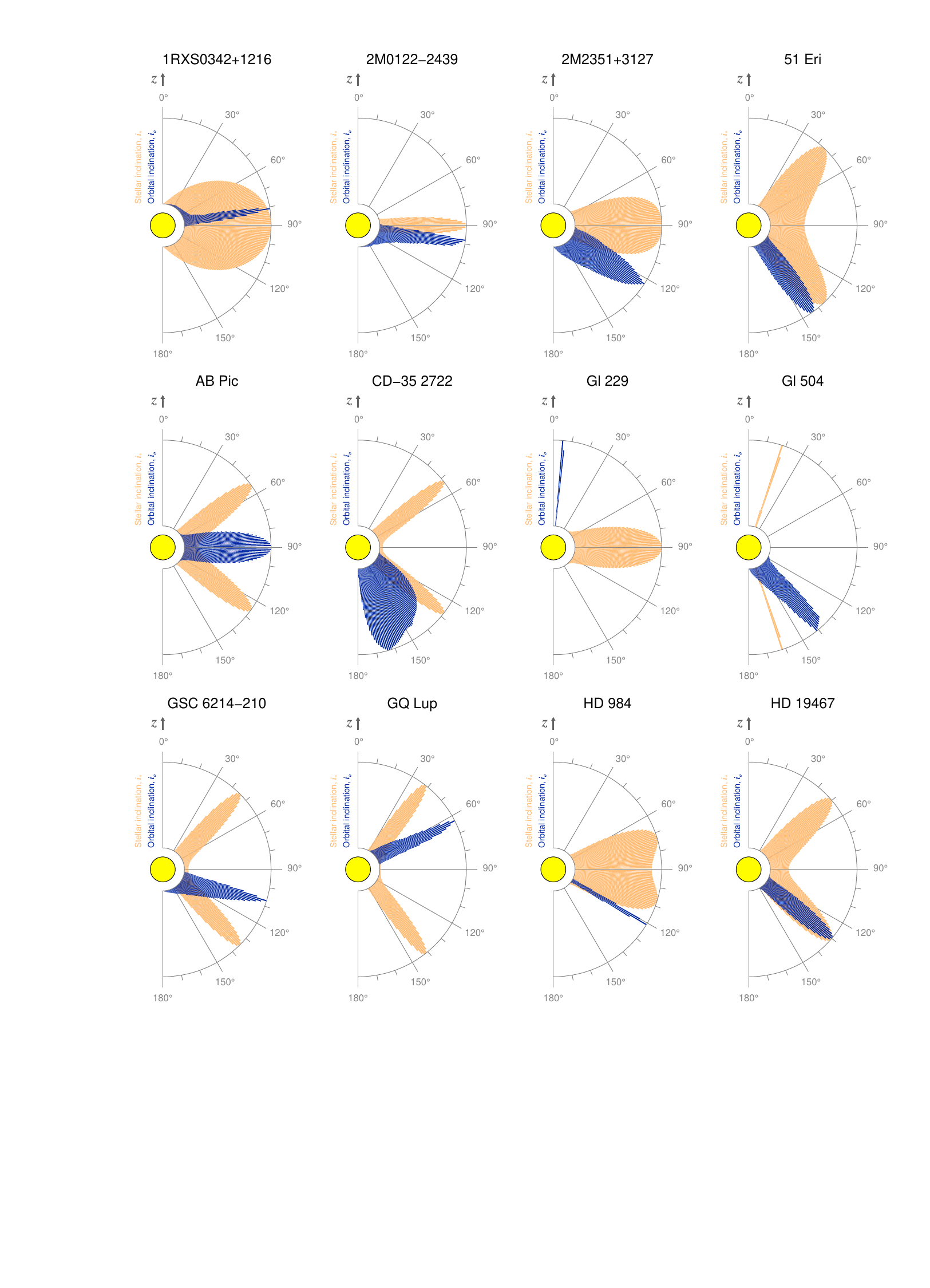}}
  \vskip -2.1 in
  \caption{Line-of-sight stellar inclinations ($i_*$, orange) compared with the orbital inclinations of substellar companions ($i_o$, blue).  
  In these ``boomerang diagrams'', stellar inclinations are
  mirrored about the $x-y$ sky plane ($i=90\degr$) because angular momentum vectors may point toward or away
  from the observer looking down along the $z$-axis (as depicted in the bottom right panel of Figure~\ref{fig:di2}).  
  If $i_o$ agrees with $i_*$ (e.g., HD~49197), this implies that the host and companion
  are consistent with spin-orbit alignment;  however, the true spin-orbit angle may nevertheless be non-zero in these cases.  
  If $i_o$ and $i_*$ disagree (e.g., Gl 504), 
  the system is misaligned by at least the difference in the inclination angles.  See Sections~\ref{sec:geometry} and \ref{sec:obliquities} for details.    
       \label{fig:di1} } 
 \end{figure*}

\begin{figure*}
  \vskip -.2 in
  \hskip -.5 in
  \resizebox{7.6in}{!}{\includegraphics{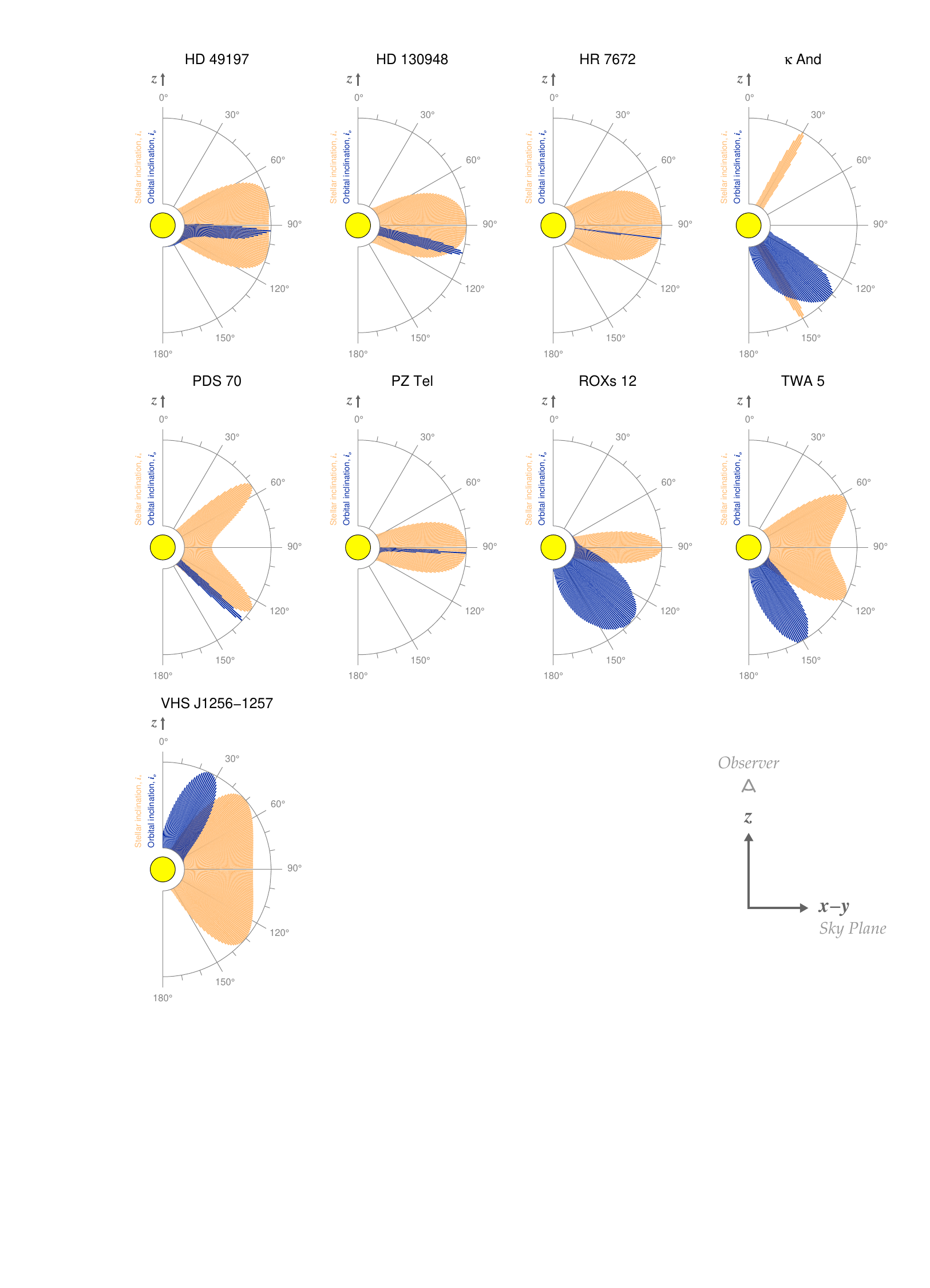}}
  \vskip -2.1 in
  \caption{Line-of-sight stellar inclinations ($i_*$, orange) compared with the orbital inclinations of substellar companions ($i_o$, blue).  
  See Figure~\ref{fig:di1} caption for details.  The bottom right panel depicts the orientation of the $z$ axis, 
  pointing toward the observer, and the $x-y$ sky plane.   \label{fig:di2} } 
 \end{figure*}

\subsection{Minimum Stellar Obliquities} \label{sec:obliquities}

Figures~\ref{fig:di1} and \ref{fig:di2} show the posterior distributions of $i_*$ and $i_o$ visualized in polar coordinates.
Here the observer is looking down along the $z$-axis and the $x$-$y$ sky plane is located at an inclination of 90$\degr$.
Because the orientations of stellar inclinations in the sky plane are unconstrained, as are the senses of their spin directions 
(clockwise or counterclockwise from the observer's perspective), the stellar rotational angular momentum vectors 
may point toward the observer or away into the plane of the sky.  When assessing spin-orbit alignment---mutual agreement 
of orbital and angular angular momentum vectors---the stellar inclination distribution is therefore mirrored about $i=90\degr$.
In these polar plots the results often look like boomerang throwing sticks, so we refer to these figures as ``boomerang diagrams.''

\begin{figure}
  \vskip -.1 in
  \hskip -1.6 in
  \resizebox{5.6in}{!}{\includegraphics{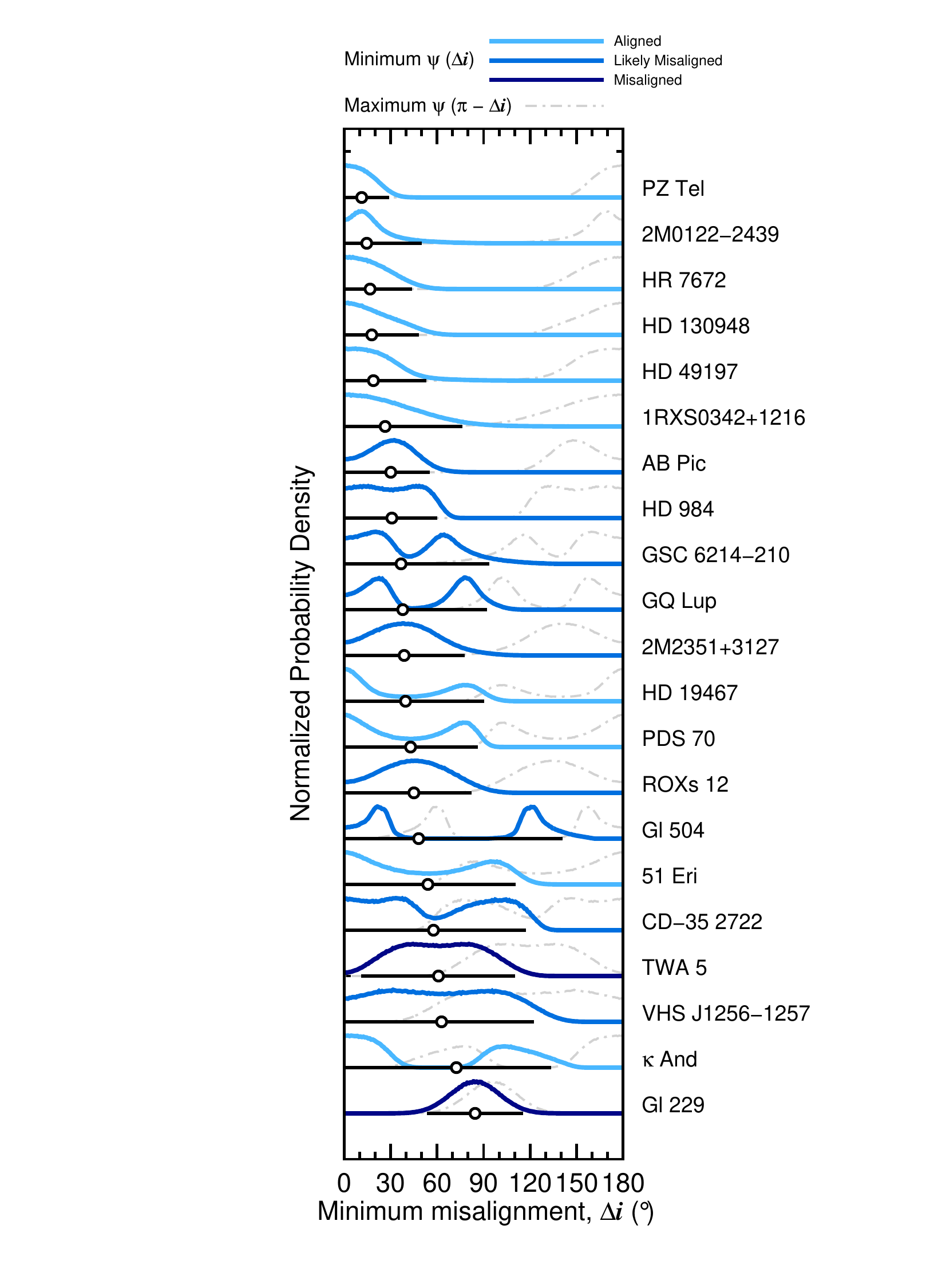}}
  \vskip -.35 in
  \caption{Distribution functions of the minimum misalignment value, $\Delta i$, equal to 
  the absolute difference between the stellar and orbital inclinations.
  Median values and 95.4\% credible intervals are plotted as open circles and dark blue horizontal lines.  
  Light blue distributions are formally
  consistent with alignment. TWA 5 and Gl 229 are significantly misaligned (dark blue), and nine other systems are likely misaligned (blue).
  See Section~\ref{sec:obliquities} and Table~\ref{tab:inclinations} for details.  The maximum misalignment ($\pi - \Delta i$) is
  shown in light gray.  For Gl 229, this implies it is on a near-polar orbit.
     \label{fig:dicomp} } 
 \end{figure}

The consistency between $i_*$ and $i_o$ is quantified by sampling from these distributions and evaluating whether their
absolute difference, $\Delta i$, deviates from 0$\degr$.
A $\Delta i$ value of 0$\degr$ is consistent with alignment, $0\degr < \Delta i < 90\degr$ is consistent with a prograde orbit,
$\Delta i$ = 90$\degr$ is consistent with a polar orbit, and $90\degr < \Delta i$~$\le$~$180\degr$ is consistent with a retrograde orbit.
However, in all cases true obliquity angles ($\psi$) can be larger.
The resulting distributions of $\Delta i$ for all 21 systems are sorted by their median values in Figure~\ref{fig:dicomp}.
These differenced distributions take on a range of shapes: some are pressed up against $i=0\degr$;
some are bimodal, a reflection of the degeneracy of stellar inclination about the sky plane; and
others significantly depart from alignment.
Many distributions are quite broad and could be consistent with alignment, polar orbits, or retrograde motion.
Table~\ref{tab:inclinations} summarizes the $\Delta i$ maximum a posteriori (MAP) values, median values, and credible intervals.

We define two criteria to classify a system as being misaligned based on these $\Delta i$ distributions.
Most of the distribution power must be located at values of $\Delta i$ away from 0$\degr$.  But this alone would not be a sufficient metric because
a flat distribution from 0--180$\degr$, for instance, could satisfy this criterion but $\Delta i$ would be unconstrained.
So we also require that the MAP value is non-zero and beyond a given threshold.  
The characteristic uncertainty in stellar inclination measurements is about 10$\degr$, so we adopt this as a reasonable 
value to distinguish alignment from misalignment.
Systems for which $P(\Delta i > 10\degr)$ $\ge$ 95\% and where $\Delta i$ MAP values are beyond 10$\degr$ are deemed ``misaligned.''
Those with $P(\Delta i > 10\degr)$ $\ge$ 80\% and $\Delta i$ MAP $>$ 10$\degr$ are labeled as ``likely misaligned'';
most of their posterior power lay beyond that threshold.
If neither are satisfied, the system is considered to be consistent with alignment.  
Finally, in addition to the minimum true obliquity distributions, we also overplot $\pi - \Delta i$, the maximum values of $\psi$, 
in Figure~\ref{fig:dicomp}.  In most cases these maximum values are near 180$\degr$, showing the broad range of potential
angular momentum orientations in each system.

Two companions in the sample are on significantly misaligned orbits.  
The orbital plane of TWA 5 B is offset with respect to the rotation axis of at least one component of the host binary TWA 5 Aab 
by at least 61~$\pm$~30$\degr$.  The  95.4\% credible interval of $\Delta i$ spans 11--110$\degr$.
As with most systems, this implies that prograde, polar, and retrograde orbits are allowed. 
Note, however, that because TWA 5 Aab is a near-equal mass resolved binary (\citealt{Macintosh:2001tr}), it is possible that the
stellar inclination could be impacted by the blended light curve and unresolved $v \sin i_*$ measurement. 
 
Gl 229 B orbits its host in a nearly 
face-on orientation ($i_o = 5.50 \pm 0.16 \degr$; \citealt{Feng:2022zz})\footnote{Note that \citet{Brandt:2021cc} found a similar 
orbital inclination of $i_o = 7.7^{+7.6}_{-4.4} \degr$ for Gl 229 B.  The uncertainties differ substantially between these two
studies despite using similar datasets.  For this analysys we adopt the inclination posterior from \citet{Feng:2022zz}, but results for 
Gl 229 would be similar if the broader posterior from \citet{Brandt:2021cc}  was chosen.}, while we find the star is viewed nearly equator-on.  
The resulting minimum misalignment value is $\Delta i$ = 85$^{+15}_{-16} \degr$.  
Although the direction of the stellar rotation is not known, this does not
impact the interpretation of the spin-orbit misalignment because in either scenario (whether it rotates clockwise or counterclockwise
from the observer's perspective), the inclination is confined close to the sky plane.
As a result, the maximum value of $\psi$ ($\pi - \Delta i$) is close to the minimum value, so $\psi \approx 90\degr$.  
Gl 229 B therefore appears to be on a polar orbit around Gl 229 A.  This is the only system with 
this type of potentially well-constrained geometry in the sample.

We caution that this interpretation for the Gl 229 system is especially sensitive to the $v \sin i_*$ value of Gl 229 A
because its equatorial velocity is so small (1.0 $\pm$ 0.2 km s$^{-1}$).
It is difficult to reliably determine projected rotational velocities below about 2 km s$^{-1}$, so if the true $v \sin i_*$ value is, for instance, 0.5 km s$^{-1}$,
that would be challenging to measure.  Our adopted $v \sin i_*$ value is elevated above $v_\mathrm{eq}$, indicating that
the projected rotational velocity may be overestimated---likely because it is near the systematics-dominated floor for broadening measurements.
See Section~\ref{sec:notes} for a detailed discussion of this system.
Additional in-depth rotational broadening analysis would help to clarify the inclination and spin-axis orientation of this system.
For the purposes of this study we rely on these available measurements at face value, so a pole-on orbital orientation
appears to be the most likely angular momentum architecture for this system, although this could change if the actual $v \sin i_*$ value of the
host star is less than $\approx$1 km s$^{-1}$.

Nine companions are classified as being likely misaligned with the spin axes of their host stars: 
2MASS J23513366+3127229~B, AB Pic b, CD--35~2722~B, Gl~504~B, 
GSC~6214-210~B, GQ~Lup~B, HD~984~B, ROXs~12~B, and VHS~J125601.92--125723.9~b.
There are a variety of reasons these systems are less confidently misaligned than for TWA 5 B and Gl 229 B.
The primary contributing factors are the precision of the $v \sin i_*$ measurement, 
which impacts the width or the stellar inclination posterior, and the constraint on the orbital inclination.
Improvements to both of these parameters will help narrow the resulting $\Delta i$ distributions to more definitively establish
the angular momentum geometry.

Altogether, this implies that a total of 11 out of 21 systems (52$^{+10}_{-11}$\% of the sample) show
signatures of offsets between the rotational and orbital angular momentum vectors.  
Note that this measurement assumes that host star inclinations are equally likely to point toward or away from the observer
(creating the boomerang structures in Figures~\ref{fig:di1}--\ref{fig:di2}).
If mutual alignment is \emph{a priori} expected (from particular disk-based formation channels, for instance), 
and this non-uniform prior probability were to be imposed on the stellar inclination distributions
so as to make them asymmetric about 90$\degr$,
this would impact the resulting $\Delta i$ distributions and the implied misalignment fraction.
As expected with Bayesian probabilistic inference, the answer and interpretation will depend
to some degree on the priors.  We simply
emphasize here that the priors for the orientation of the rotational angular momentum vectors are
identical and are not conditioned on whether the orbital inclination distribution points
toward or away from the observer.
We discuss the 
implications of this prevalence of misalignments in Section~\ref{sec:discussion}.

The remaining ten systems are consistent with alignment:
1RXS J034231.8+121622, 2MASS~J01225093--2439505, 
51~Eri, HD 130948, HD 19467, HD 49197, HR 7672, $\kappa$~And\footnote{For $\kappa$ And, we adopt
the stellar inclination of this oblate B9 host star from fits to interferometric measurements by \citet{Jones:2016hg}.}, PDS 70, and PZ Tel.
However, as is evident in Figure~\ref{fig:dicomp}, the complementary distribution to $\Delta i$ 
for the maximum value of $\psi$ implies that
there are a wide range of potential true obliquities each system can have.
It is challenging to interpret individual systems with $\Delta i$ distributions consistent with 0$\degr$ because
these represent minimum misalignments; their true obliquities can be much larger.  In many cases obliquities
spanning any value from 0--180$\degr$ are consistent with the observations.  
Instead we focus the following discussion on misaligned (and likely misaligned) systems
and we develop simple population-level models to compare with our observed family of $\Delta i$ distributions.

\section{Discussion}{\label{sec:discussion}}

\subsection{Comparison with Previous Obliquity Constraints} \label{sec:prevconstraints}

Several previous studies have explored spin-orbit alignment for individual systems with imaged substellar companions
similar to the method we employ here:
\begin{itemize}
\item \citet{Bowler:2017kt} found that the orbit of ROXs~12~B is likely to be 
misaligned with the spin axis of its host star, with $P(\Delta i > 10\degr)$ = 0.94.  This is similar to the misalignment probability of 0.95
derived in this study.

\item \citet{Bonnefoy:2018ch} carried out the same analysis for Gl 504 and note possible signs of misalignment, with $P(\Delta i > 10\degr)$ = 0.78.
We infer a higher probability of 0.91, most likely because of the more precise $v \sin i_*$ we adopt as well as the updated orbit constraint.

\item \citet{Maire:2020iu} found that the brown dwarf companion HD 19467 B is consistent with alignment.  We reach a similar conclusion,
with $P(\Delta i > 10\degr)$ = 0.75, which is below the threshold of 80\% that we adopt for classifying systems as being ``likely misaligned.''

\item \citet{Bryan:2020ji} performed a detailed analysis of spin-orbit and spin-spin alignments 
in the 2MASS J01225093--2439505 system.  Their results support a low stellar obliquity for the host star, consistent with our findings here.

\item \citet{Schwarz:2016fl}, \citet{Wu:2017kd}, and \citet{Stolker:2021aa} 
analyzed the orbital configuration of GQ~Lup~B with respect to the host
star's spin axis, the companion's spin axis, and the orientation of the circumstellar disk.  Although the orbit is not
yet well constrained, both studies conclude that non-zero stellar obliquity is possible.  We find that spin-orbit misalignment is
likely in the GQ~Lup system, with $P(\Delta i > 10\degr)$ = 0.896.
\end{itemize}
Altogether our results from this study are in good agreement with previous work focused on individual systems.

\subsection{Implications of Misalignments} \label{sec:implications}

Most of our 
sample is comprised of brown dwarf companions (Figure~\ref{fig:violin}), but there are two stars hosting giant planets
(PDS 70 and 51 Eri) as well as a handful of objects whose masses and origin are more ambiguous
(such as VHS~J1256--1257 b, 2M0122--2439 b, AB Pic b, Gl 504 B, and GSC~6214-210~B).  
Because the overall sample size is modest, and individual constraints are fairly broad, we consider our sample as
belonging to one substellar population of (predominantly) brown dwarf companions for this analysis and discussion about 
 formation channels.

Brown dwarf companions are expected to form like stellar binaries from the turbulent fragmentation of molecular cloud cores
or through disk fragmentation 
(e.g., \citealt{Low:1976wt}; \citealt{Bate:2002iq}; \citealt{Bate:2009br}; 
\citealt{Stamatellos:2009fw}; \citealt{Jumper:2013ct}).  In principle, these formation sites---in cores, 
filamentary structures, or in disks---can result in similar
orbital and rotational axes for the host and companion through a localized conservation of angular momentum during collapse.
However, a variety of mechanisms have been proposed that can disrupt this alignment
during or after the star formation process.
This includes non-axisymmetric collapse from variable accretion, interactions with nearby protostars, or inhomogeneities in the 
parent cloud core (e.g., \citealt{Tremaine:1991aa}; \citealt{Bate:2010fz}; \citealt{Fielding:2015aa}; \citealt{Offner:2016gl}); 
interactions with the protoplanetary disk (e.g., \citealt{Lai:2011bb}; \citealt{Batygin:2013fr}; \citealt{EpsteinMartin:2022aa});
or dynamical encounters with companions or passing stars at older ages (\citealt{Batygin:2012ig}; \citealt{Anderson:2017aa}).
There have been few large-scale empirical tests of these various models owing to the difficulty in determining obliquities
for stars hosting long-period brown dwarf companions.

Observationally, close stellar binaries have been shown to be well aligned in many individual instances (e.g., \citealt{Sybilski:2018aa}),
although there are several notable examples of spin-orbit misalignments (e.g., \citealt{Albrecht:2009df}).
Ensemble obliquity measurements of binary systems have yielded tentative or 
inconclusive results (\citealt{Hale:1994gv}; \citealt{Justesen:2020jz}).
Recently, however, \citet{Marcussen:2022aa} found that most close binaries in their sample of 43 systems 
are consistent with alignment.

The most significant statistical result from this analysis is that spin-orbit misalignments appear to be common among
stars hosting wide substellar companions.  
The (minimum) fraction of systems likely to be misaligned, 52$^{+10}_{-11}$\%, is higher than  
for cool stars hosting hot Jupiters ($\sim$5\%: \citealt{Winn:2010dr}; \citealt{Albrecht:2022aa}) and 
close-in small planets (\citealt{Campante:2016jq}; \citealt{Winn:2017ip}).
It also appears to be higher than the misalignment fraction for stars hosting debris disks 
($\lesssim$10\%: \citealt{LeBouquin:2009aa}; \citealt{Watson:2011gf}; \citealt{Greaves:2014aa}),
a comparison that may be more relevant because the spatial scales of tens to hundreds of AU are
closer to the population of companions considered in this study.
Since debris disks are expected to trace the sites of planetesimals and planet formation, 
this disagreement between the incidence of misaligned debris disks and misaligned brown dwarf companions 
may indicate that companions in our sample did not predominantly form in disks;
if this were the case, the alignment frequencies would be expected to be similar.  
Less is known about the overall coplanarity of systems that contain both 
spatially resolved debris disks and substellar companions.  
HD 2562~B (\citealt{Konopacky:2016dk}; \citealt{Maire:2018aa}) 
and HD 206893~B (\citealt{Milli:2017fs}; \citealt{Delorme:2017aa}; \citealt{Marino:2020aa}; \citealt{WardDuong:2021ev}) 
are notable examples of brown dwarf companions that appear to be aligned with
their exterior debris disks---suggesting a disk-based origin for these companions---whereas 
planets orbiting HR 8799, $\beta$~Pic, and HD 106906 show a range of
orientations with respect to exterior and interior disks (e.g., \citealt{Dawson:2011eu}; \citealt{Bailey:2014et}; \citealt{Wilner:2018aa}; \citealt{Nguyen:2021aa}).
A larger sample would help clarify the prevalence of aligned and misaligned 
orientations.

An alternative explanation is
that subsequent post-formation scattering with a second object in the system 
(so as to alter the initial obliquity distribution) may be more common for stars
hosting wide substellar companions, perhaps because multiple massive objects formed in the same disk.
With a few exceptions (\citealt{Haffert:2019ba}; \citealt{Lagrange:2019bi}; \citealt{Bohn:2020ge}; \citealt{Hinkley:2022aa}), however,
most follow-up searches have not identified additional giant planets and brown dwarfs
which could represent potential scatterers (e.g., \citealt{Bryan:2016eo}).

It is also possible that some widely separated substellar companions could have been captured at an early age while still
embedded in a dense cluster.  Close encounters with passing stars can liberate planets and brown dwarfs (e.g, \citealt{Parker:2012aa}; \citealt{DaffernPowell:2022aa});
these free-floating planets, and other isolated brown dwarfs formed during the star formation process,
could then become captured onto very wide orbits by other stars, especially during the cluster dispersal phase (e.g., \citealt{Perets:2012cv}; \citealt{Parker:2022aa}).  
Without any significant realignment mechanism at these large orbital distances, 
the stellar obliquity distribution in such cases would be expected to be isotropic.  
The misaligned systems in our sample
are therefore also consistent with this capture process.

There is no simple interpretation of this high misalignment fraction.  If circumstellar disks are predominantly aligned with the spin axes of their host
stars, this result points to a diversity of formation channels or dynamical processing in some systems.  Alternatively, the observed distribution of spin-orbit orientations
could reflect the primordial distribution of aligned and misaligned disks.  Systems with spin-orbit misalignment are also consistent with the early capture of free-floating giant planets and brown dwarfs in dense clusters.  A better understanding of the initial conditions for disks around cool stars
would provide additional context to interpret results from this study.

\subsection{Stellar Obliquities in Systems with Directly Imaged Planets} \label{sec:imagedplanets}

Alignment rates are less clear among systems with directly imaged planets, largely because long-period giant planets tend to reside around
early-type stars whose rotation rates cannot be accurately determined through photometric monitoring in the same way as
can be inferred for cool stars (\citealt{Sepulveda:2022aa}).  
However, in some cases stellar inclinations and true three-dimensional orientations can be constrained using 
asteroseismology or spatially resolved interferometry.  Below we summarize orbital and rotational angular momentum
alignment in five planetary systems with previous 
obliquity constraints: $\beta$ Pic, HR 8799, 51 Eri, HD 206893, and PDS 70.
Note that among these, only 51 Eri and PDS 70 are in our sample of 21 stars with obliquity constraints;
the rest have had previous spin-orbit measurements obtained using a variety of alternative techniques.

\citet{Kraus:2020aa} analyzed the spatial displacement of the A6 host star $\beta$~Pic in the Br$\gamma$ absorption line using 
VLTI/GRAVITY spectro-interferometry.  When combined with the asteroseismic stellar inclination from \citet{Zwintz:2019aa},
they found that $\beta$~Pic is well aligned (to within $\le$3~$\pm$~5$\degr$) 
with the orbit of $\beta$ Pic b and the outer debris disk.
More recently, a second giant planet in the system, $\beta$~Pic~c (\citealt{Lagrange:2019bi}), was directly imaged
and shown to be consistent with this coplanar geometry 
(\citealt{Nowak:2020cc}; \citealt{Lagrange:2020do}; \citealt{Brandt:2021aa}; \citealt{Lacour:2021aa}).

\begin{figure*}
  \vskip -.4 in
  \hskip -0.6 in
  \resizebox{8.1in}{!}{\includegraphics{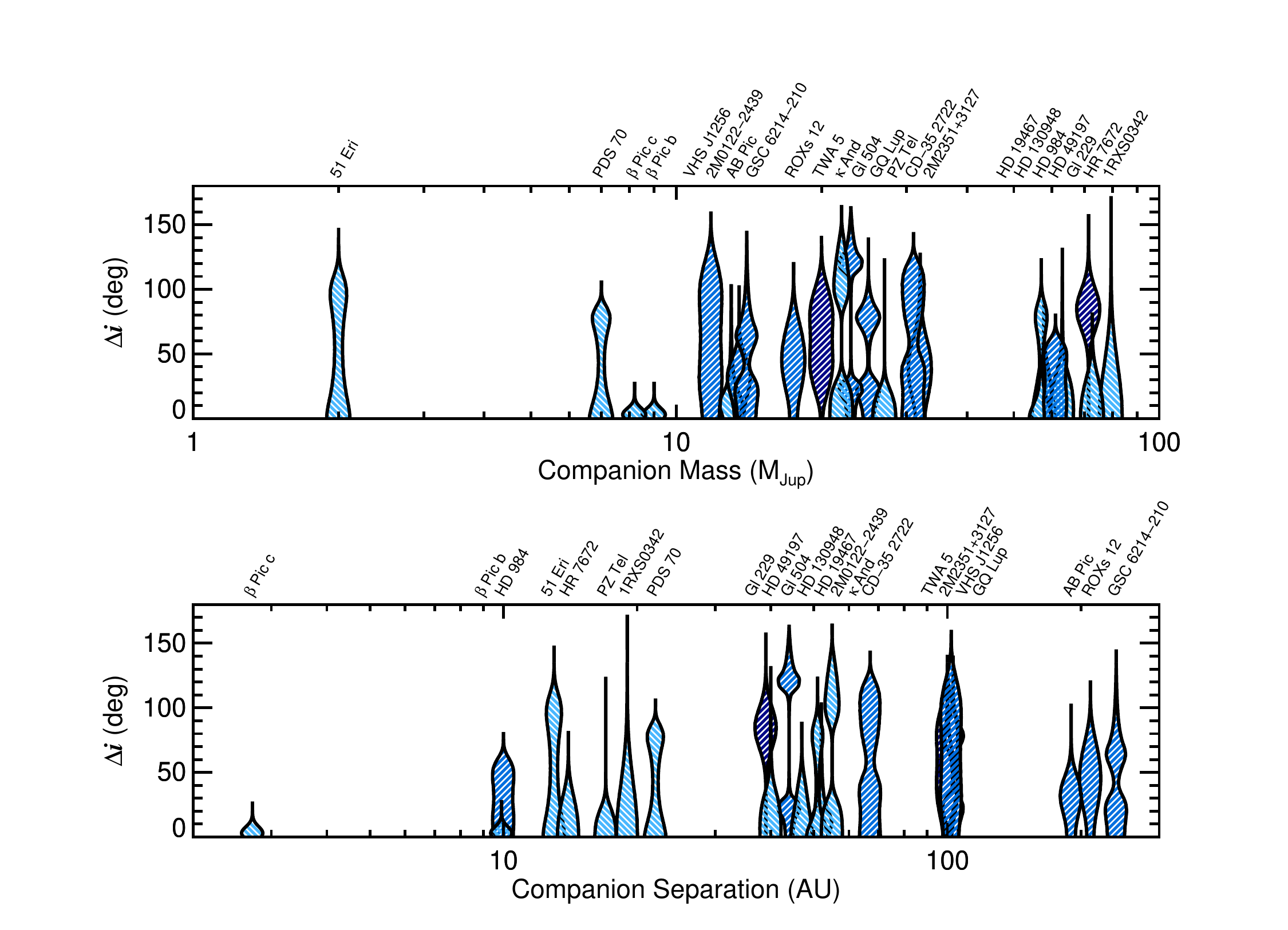}}
  \vskip -.1 in
  \caption{\emph{Top:} Minimum stellar obliquity ($\Delta i$) as a function of companion mass.  Color coding for each 
  violin plot follows the scheme in Figure~\ref{fig:dicomp}: light blue is consistent with alignment, while 
  blue and dark blue colors denote likely spin-orbit misalignment.  All planets below 
   10~\Mjup \ with stellar obliquity constraints are consistent with spin-orbit alignment,
  although the sample remains small.  The brown dwarf
  companions show a mix of aligned and misaligned systems with no apparent trend with companion mass.
  $\beta$~Pic~b and c have been added 
  based on their \emph{true} stellar obliquity from \citet{Kraus:2020aa} and dynamical masses from \citet{Nowak:2020cc}.
  Note that, in general, the uncertainties in model-inferred masses (which apply to most systems in this figure) can
  be very large and typically range from a few \Mjup~to tens of \Mjup. \emph{Bottom:} Same as the upper panel but
  for orbital separation.  With the exception of HD 984 B, which is likely (though not definitively) misaligned, 
  all companions within 20 AU are consistent with spin-orbit alignment.
     \label{fig:violin} } 
 \end{figure*}

The orbital inclinations (and longitudes of ascending node) of the four planets orbiting 
HR 8799 (\citealt{Marois:2008ei}; \citealt{Marois:2010gp}) have progressively 
improved over time with continued orbit monitoring.  
The precision of the constraint depends on assumptions about resonant orbits, stability, and coplanarity, but 
in the unconstrained case the four orbital planes lie between about 20--40$\degr$ 
(e.g., \citealt{Wang:2018aa}; \citealt{Sepulveda:2022by}); in the most restricted case, \citet{Zurlo:2022aa} found an inclination
of 26.9$^{+0.16}_{-0.17}\degr$ for the planetary system.  
\citet{Wright:2011wu} measured an inclination of $\gtrsim$40$\degr$ for the host star using oscillation frequencies from 
high-cadence radial velocities, which would imply a spin-orbit misalignment by at least 10$\degr$.
However, \citet{Sodor:2014aa} concluded that asteroseismic analysis of HR 8799 as an A5/F0 $\gamma$ Dor variable 
is challenging because of resonances and amplitude variations among the pulsation frequencies.   
The inclination and obliquity of HR 8799 are therefore not yet reliably established,
although an estimate of the core rotation period by \citet{Sepulveda:2023aa}
implies a stellar inclination in good agreement with the orbits of its four planets.
Furthermore, the exterior cold disk in this system 
appears to be coplanar with the planets' orbits (e.g., \citealt{Faramaz:2021aa}).

\citet{Maire:2019aa} derived a stellar inclination of the young F0 planet host 51 Eri (\citealt{Macintosh:2015fw}) using
the equatorial and projected rotational velocity method.  The equatorial velocity was based on a rotation period of 0.65~$\pm$~0.03 d computed
using \emph{Hipparcos} photometry.  They conclude that the stellar inclination (39--51$\degr$ or 129--141$\degr$) and orbital 
inclination of 51 Eri b (126--147$\degr$) are consistent with orbital and rotational alignment.
However, \citet{Sepulveda:2022aa} recently analyzed the \emph{TESS} light curve of 51 Eri and identify it as a $\gamma$ Dor pulsating star.
The variability previously attributed to rotational modulation is more likely caused by pulsation modes.
Although the surface rotation rate is not constrained, the authors estimate a core rotation period of 0.9$^{+0.3}_{-0.1}$~d and use this to 
deduce a stellar inclination of $\sim$62$\degr$.
We make use of this core rotation period estimate here; based on our stellar inclination of 54$^{+13}_{-19}\degr$ we conclude
there is no evidence of misalignment. 

HD 206893 hosts a brown dwarf companion (\citealt{Milli:2017fs}) 
and giant planet (\citealt{Hinkley:2022aa}) on orbits consistent with being coplanar.  
\citet{Delorme:2017aa} presented ground-based photometric monitoring of the F5 host star and found clear modulations
with a period very close to 1 day.  They interpret this signal of 0.996 $\pm$ 0.03 d as most likely originating from a rotation period.
They then use the equatorial and rotational velocities to infer a stellar inclination of 30 $\pm$ 5$\degr$,
which (given the symmetric nature of the distribution about 90$\degr$), is consistent with 
the orbital inclination of 146 $\pm$ 3$\degr$ for HD 206893 B and 150 $\pm$ 3$\degr$ for c  (\citealt{Hinkley:2022aa}).
This is also consistent with measurements of 40~$\pm$~3$\degr$ (\citealt{Marino:2020aa}) and 45~$\pm$~4$\degr$ (\citealt{Nederlander:2021aa}) 
for the orientation of the debris disk.
However, \citet{Delorme:2017aa} note that the photometric periodicity they measured could be attributed to stellar pulsation modes if HD 206893
is a $\gamma$ Dor variable star.  This is certainly possible as F5 is near the boundary of the onset of this instability strip 
(\citealt{Kaye:1999aa}; \citealt{Aerts:2021aa}).  While the $\approx$1-day rotation period is plausible, 
additional monitoring of this system would be valuable to more robustly establish the nature of the periodic variations.
For the time being, we therefore cautiously interpret the angular momentum geometry of this system as being 
potentially---but not securely---aligned.\footnote{We do not include this star in our analysis because of the unclear nature of the
photometric modulations.}

PDS 70 hosts two accreting giant planets nested inside a transition disk 
(\citealt{Keppler:2018dd}; \citealt{Haffert:2019ba}; \citealt{Zhou:2021ky}).
The orbits of PDS 70 b and c have been constrained to 133$^{+4}_{-3}\degr$ and 132~$\pm$~3$\degr$, respectively,
assuming non-crossing orbits and near-coplanarity (\citealt{Wang:2021aa}).
\citet{Thanathibodee:2020fe} measure a rotation period of 3.03~$\pm$~0.06~d for PDS 70 from the \emph{TESS} light curve
and infer a stellar rotation axis of 50~$\pm$~8$\degr$ (equal to 130~$\pm$~8$\degr$)---in good agreement with the 
orbits.  We find a similar result consistent with spin-orbit alignment: 
$i_*$ = 60$^{+10}_{-15}\degr$, $\Delta i$ = 43$^{+24}_{-43}$, and $P(\Delta i > 10\degr)$ = 0.804.
These orbital and rotational orientations are also in agreement with those of the transition disk (\citealt{Keppler:2019aa}). 
The PDS 70 system therefore appears to be in a state of organized angular momentum alignment.

In summary, a variety of constraints are now available on the true obliquities or minimum misalignments for a handful of stars hosting directly 
imaged planets.  The $\beta$~Pic system is unambiguously aligned in $\psi$.  
PDS 70 is consistent with spin-orbit alignment in $\Delta i$.  
51 Eri and HD 206893 are potentially aligned with their companions.
The stellar inclination and obliquity of HR 8799 is not reliably constrained.

In Figure~\ref{fig:violin} we compare the minimum misalignment constraint for each system as a function
of companion mass and separation.  $\beta$~Pic b and c are included based on their \emph{true} obliquity constraint
from \citet{Kraus:2020aa} and planet properties from \citet{Nowak:2020cc}.  The sample size is modest, but we
note that all four of the companions in three systems with masses below 10~\Mjup \ are consistent with spin-orbit alignment,
and, similarly, seven of the eight companions with separations less than 20~AU are consistent with alignment.  At higher masses
and wider separations, there is a mixture of aligned and misaligned systems with no clear correlation with increasing brown dwarf
mass or orbital separation.  

Altogether, this points to an emerging trend that systems with long-period giant planets have low stellar obliquities, although 
the small sample size of (primarily early-type) host stars with inclination measurements remains very small.
This would, however, have important implications for the formation of these companions.
A uniform analysis of eccentricities by \citet[see also \citealt{Nagpal:2022aa}]{Bowler:2020hk} revealed that imaged planets
between 5--100~AU have low eccentricities, whereas brown dwarf companions have a broader range of eccentricities.
If both of these trends are validated with larger samples, this would reinforce a distinction in the orbits, obliquities,
and formation of these two populations: long-period planets are most consistent with a disk-based origin with minimal
dynamical evolution, whereas brown dwarf companions as a population are more aligned with formation from turbulent cloud fragmentation.

\begin{figure*}
  \hskip -.3 in
  \resizebox{8.5in}{!}{\includegraphics{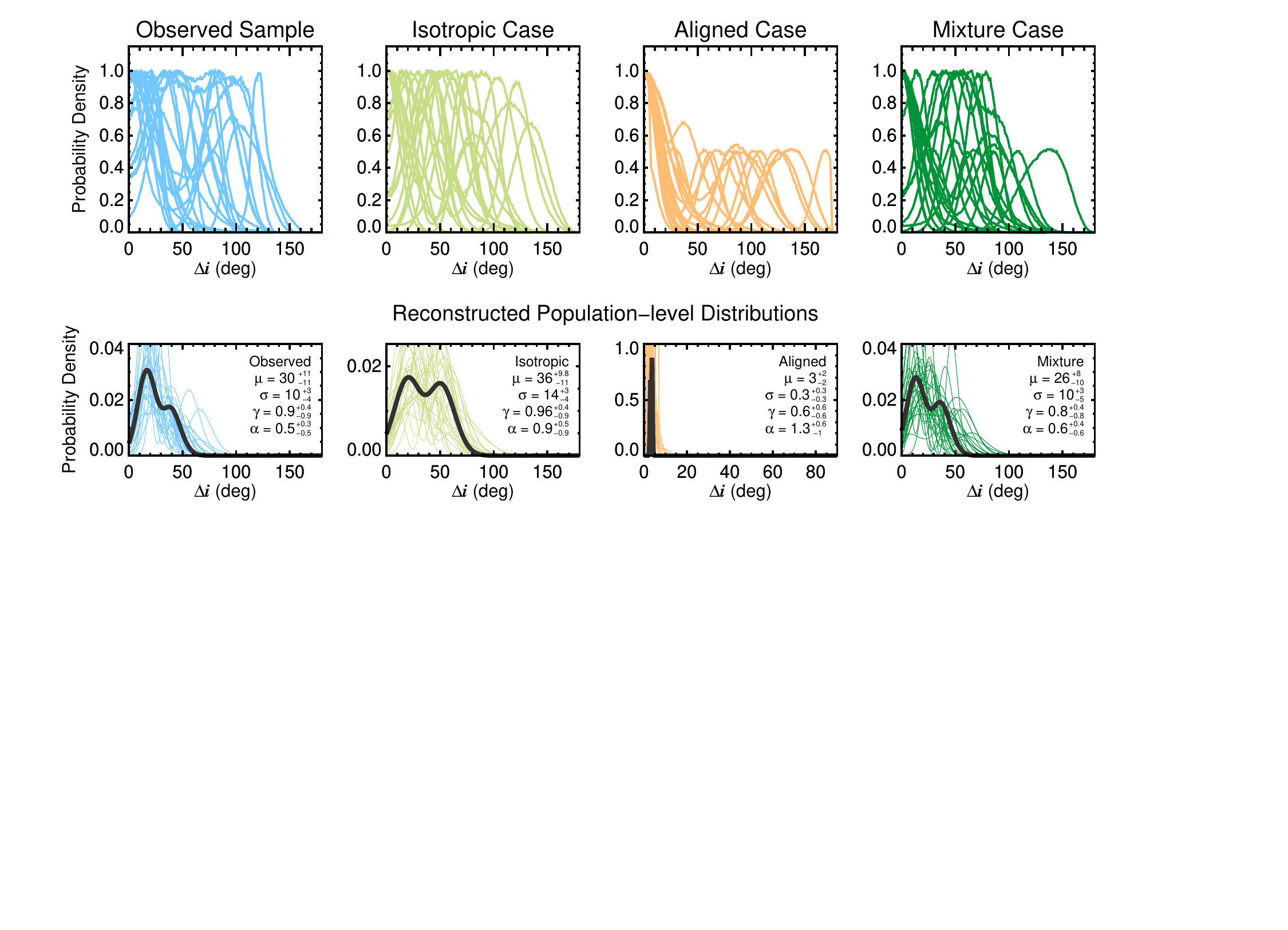}}
  \vskip -2.8 in
  \caption{\emph{Top}:  $\Delta i$ distributions for our observed sample of substellar companions (left) compared with one synthetic realization for
  an isotropic stellar inclination distribution (middle left), perfect spin-orbit alignment (middle right), and a random mixture of synthetic isotropic and aligned
  distributions in equal proportions (right).  
  The observed sample is qualitatively most similar to the isotropic or mixture cases, but differs substantially from a purely aligned scenario.
  \emph{Bottom}: Reconstructed underlying distributions for each sample using hierarchical Bayesian modeling.  Here the flexible LGB distribution is
  used for our population-level model, and constraints on hyperparameter posteriors are listed in each panel.  The population-level distribution for the observed sample closely resembles a mixture of isotropic and aligned systems.  The thick curve shows the LGB distribution using median hyperparameter values, while thin curves show 30 draws from the MCMC chains. Note that the scale has been adjusted in the aligned panel (middle right).}    \label{fig:hbm} 
 \end{figure*}

\subsection{The Underlying Distribution of $\Delta i$ with Hierarchical Bayesian Modeling} \label{sec:hbm}

Although it is difficult to make a meaningful constraint on the underlying true obliquity distribution, 
we can nevertheless compare our observed sample of $\Delta i$ distributions 
with extreme cases of isotropic stellar orientations and completely aligned spin-orbit orientations,
as well as a mixture of these two scenarios.
We generate three synthetic datasets for these purposes.  

For the isotropic case, we randomly 
draw a stellar inclination from a $\sin i_*$ distribution from 0 to $\pi$/2.  Assuming the measurement of 
$i_*$ follows a Gaussian with a standard deviation of 10$\degr$, which is a characteristic level of uncertainty
from this study, we then mirror the distribution across 90$\degr$ to reflect the 
unknown orientation of the spin angular momentum vector.  For each system we use the measured
$i_o$ and our new $i_*$ distributions to compute a realization of a sample of $\Delta i$ distributions
assuming isotropic stellar line-of-sight inclinations.

For the aligned case, the procedure is the same as the isotropic case, except instead of selecting
a randomly drawn mean value of $i_*$ we use the MAP value of $i_o$.  If this lies between 0 to $\pi$/2 then
$i_*$ is mirrored onto the $\pi/2$ to $\pi$ range, and vice versa.  The absolute difference then provides
a distribution of minimum obliquities assuming perfect spin-orbit alignment and an unknown
orientation of $\overrightarrow{L_*}$ to mimic the information available to an observer.

For the mixture case, we randomly draw 10 unique distributions 
from the isotropic sample described above and nine from the aligned sample to create an approximately 
even mixture of 19 systems.   This represents an intermediate scenario between the two
extreme examples of completely aligned and independent angular momentum architectures.

The observed and synthetic limiting cases are shown in the upper panels of Figure~\ref{fig:hbm}.  
$\Delta i$ distributions for the isotropic case are broadly located between about 0--130$\degr$.
As expected, the aligned case is clustered at $\Delta i$ = 0$\degr$ with many individual
distributions being bimodal because $i_*$ is often bimodal.
Altogether, the observed sample better resembles the isotropic or mixed cases with many distributions
peaking at non-zero $\Delta i$ values.

To quantify this agreement, we employ hierarchical Bayesian modeling to reconstruct the
underlying population-level distributions of the real and synthetic samples.  Our general strategy
follows the framework described by \citet{Hogg:2010gh}, in which an importance sampling approximation
of the likelihood function enables a phased approach to the problem: individual-level posteriors can be generated first
and subsequently sampled to constrain hyperparameters of an assumed population-level model.

Here we adopt a flexible model introduced by \citet{Iriarte:2021aa} called the Lambert
generalized bimodal (LGB) distribution, which was developed to introduce skewness to the generalized
bimodal distribution.  The LGB distribution has four parameters: the location parameter $\mu$
(defined for all real numbers),
the scale parameter $\sigma$ (defined for positive numbers), and the shape parameters
$\gamma$ (defined over $[$0, 2)) and $\alpha$ (defined over (0, $e$), where $e$ is Euler's constant).
Depending on these parameters, the LGB distribution can be unimodal, bimodal, symmetric, or
asymmetric; examples of the diversity of shapes can be seen in Figure 1 of \citet{Iriarte:2021aa}.
$\gamma$ controls the bimodality of the distribution, and $\alpha$ determines the relative heights of the 
bimodal peaks or unimodal asymmetry.

This choice of the underlying model is motivated by the bimodal nature of many $\Delta i$ distributions,
which tend to produce power at secondary peaks, as well as the versatility and modest number of hyperparameters
of the LGB distribution.  It is not, however, otherwise physically motivated, and certainly other distributions
or mixture models could be used in its place to capture the behavior of the (unknown)  distribution from which this
sample was drawn.
The exact shape and inferred parameter values are less important for this exercise than a comparison of the results from the observed
sample with the isotropic, aligned, and mixture cases.

The LGB probability distribution function for $\Delta i$ 
takes on the following form:

\begin{eqnarray}
f(\Delta i | \mu, \sigma, \gamma, \alpha) = \frac{\gamma + z^2}{\sigma (\gamma + 1)} \phi(z) \alpha^{\Phi(z) - \frac{z}{\gamma+1}\phi(z)}  \\
\nonumber \times \Big( 1 - \ln{\alpha} \big( 1 - \Phi(z) + \frac{z}{\gamma + 1} \phi(z) \big)   \Big)
\end{eqnarray}

\noindent Where $z$ = $(\Delta i - \mu) / \sigma$, $\phi(z)$ is the standard normal distribution, $e^{-0.5 z^2}/\sqrt{2 \pi}$, and
$\Phi(z)$ is the cumulative distribution function of the standard normal: $\big(1 + \mathrm{erf}(z/\sqrt{2}) \big)/2$.

Hierarchical Bayesian modeling of all three samples is carried out in a similar fashion as 
in \citet{Bowler:2020hk} for the case of the two-parameter Beta distribution, except here the posteriors
of four hyperparameters are sampled.  We implement a Metropolis-Hastings Markov Chain Monte Carlo (MCMC) 
algorithm (\citealt{Metropolis:1953vj}; \citealt{Hastings:1970wm})
with tunable Gaussian jump parameters and Gibbs sampling.  
Uniform hyperpriors are adopted for all four hyperparameters: 
$\mathcal{P}(\mu)$ $\sim$ $\mathcal{U}(-1000\degr, 1000\degr)$, so as to allow the location parameter to drift beyond the range of the data,
$\mathcal{P}(\sigma)$ $\sim$ $\mathcal{U}(0\degr, 1000\degr)$, 
$\mathcal{P}(\gamma)$ $\sim$ $\mathcal{U}(0, 2)$, and
$\mathcal{P}(\alpha)$ $\sim$ $\mathcal{U}(0, e)$.\footnote{ \citet{Nagpal:2022aa} analyzed 
the eccentricity distribution of long-period substellar companions and found that hyperpriors on the
two Beta distribution shape parameters can impact the hyperparameter posteriors in counterintuitive ways.
Specifically, a uniform hyperprior can result in a biased eccentricity distribution, and the wider the range
of the shape parameter hyperprior, the more narrow the resulting posteriors.  
However, in this case for the LGB distribution, the parameters $\mu$ and $\sigma$ directly map the location
and spread of the distribution function, and the two shape parameters $\gamma$ and $\alpha$ 
by definition have limited ranges. 
We therefore do not expect uniform hyperpriors to bias our results in the same way as they would for the Beta distribution,
although more study on the effect of hyperpriors for this distribution is needed}.
A single Markov chain ran for 10$^5$ steps with parameter acceptance rates
generally falling between 20--50\%.  Trace plots are visually analyzed to assess convergence,
and a burn-in fraction of 30\% is adopted.

Results are shown in the bottom panels of Figure~\ref{fig:hbm}.
The family of hyperparameter posterior distributions is clustered between $\Delta i$ 
values of $\approx$0--50$\degr$ for the observed sample
($\mu = 30^{+11}_{-11}$, $\sigma = 10^{+3}_{-4}$, $\gamma = 0.9^{+0.4}_{-0.9}$, and $\alpha = 0.5^{+0.3}_{-0.5}$),
$\Delta i \approx$0--80$\degr$ for the isotropic case
($\mu = 36^{+10}_{-11}$, $\sigma = 14^{+3}_{-4}$, $\gamma = 1.0^{+0.4}_{-0.9}$, and $\alpha = 0.9^{+0.5}_{-0.9}$), 
$\Delta i \lesssim$10$\degr$ for the aligned case
($\mu = 3^{+2}_{-2}$, $\sigma = 0.3^{+0.3}_{-0.3}$, $\gamma = 0.6^{+0.6}_{-0.6}$, and $\alpha = 1.3^{+0.6}_{-1}$), and
$\Delta i \approx$0--50$\degr$ for the mixture case
($\mu = 26^{+8}_{-10}$, $\sigma = 10^{+3}_{-5}$, $\gamma = 0.8^{+0.4}_{-0.8}$, and $\alpha = 0.6^{+0.4}_{-0.6}$).
Quoted values represent median and 68.3\% credible intervals of the marginalized posteriors.\footnote{Although all four cases show some asymmetric structure 
in the recovered distributions, we do not necessarily ascribe this as a real feature,
as this is likely to simply be an artifact of a weakly constrained $\gamma$ parameter.}

Following \citet{Nagpal:2022aa}, we define the metric $\mathcal{M}$ 
to quantify the level of agreement between the family of underlying distributions in the observed sample 
and the three synthetic cases:

\begin{equation}
\mathcal{M} = \frac{1}{N}\sum_{n=1}^{N} \int \left| f(\Delta i)_{\mathrm{obs}, n} - f(\Delta i)_{\mathrm{model}, n}  \right| \, d\Delta i
\end{equation}

\noindent $\mathcal{M}$ is the average area of the absolute difference between random posterior draws ($n$)
from the underlying model of the observed sample, $f(\Delta i)_{\mathrm{obs}, n}$, compared with the isotropic, aligned,
or mixture models, $f(\Delta i)_{\mathrm{model}, n}$.  Lower values of $\mathcal{M}$ correspond to better agreement between the two underlying families of distributions.  
We also quantify the spread in this metric with the standard
deviation of these absolute residual areas:

{\small{
\begin{equation}
\sigma_\mathcal{M} = \sqrt{ \frac{ \sum_{n=1}^N \Big(   \int \left| f(\Delta i)_{\mathrm{obs}, n} - f(\Delta i)_{\mathrm{model}, n}  \right| \, d\Delta i  - \mathcal{M} \Big)  }{N-1} }
\end{equation}
}}

We find $\mathcal{M}$ and $\sigma_{\mathcal{M}}$ values of 0.83 $\pm$ 0.30 when comparing the observed and isotropic case, 
2.00 $\pm$ 0.46 for the observed and aligned case, and  0.67 $\pm$ 0.24 for the observed and mixture case.
The inferred distribution for the observed sample is in much better agreement with the mixture scenario
than either the pure aligned or isotropic cases.  
 Indeed, all four parameters in the LGB distribution model are consistent at the 1-$\sigma$ level with the mixture sample.  
The aligned scenario has the poorest agreement among the three simulated samples that we tested.
Although we do not attempt to constrain the relative proportion of isotropic versus aligned orbits that best fit
the observed sample, the good agreement with the equal mixture suggests that there are more aligned cases in our observations than would be
expected for a purely isotropic distribution.\footnote{It is interesting to note that the stars with
the two lowest-mass companions in the sample---51 Eri and PDS 70---are consistent with alignment.
If these were excluded, it is likely that the underlying distribution for the observed sample would broaden out to better reflect the
isotropic distribution.  That is, the mixture model might be preferred in part because our sample itself is a mixture of brown dwarf companions
and giant planets.} 

Overall, there does not appear to be a strong correlation between the orbital and rotational planes
for systems with wide substellar companions.
These results reinforce the view that misalignments are common among brown dwarf companions.
A larger sample and more precise individual constraints will help to clarify this picture in the future.

\subsection{Sample Biases} \label{sec:biases}

There are several observational biases that are important to highlight as they are likely to impact the sample selection  
from this study with varying degrees of significance.  These are individually detailed below and summarized in Section~\ref{sec:biassumary}.

\subsubsection{Light Curve Selection Biases} 

Because our strategy to constrain $i_*$ relies on rotation period measurements, any biases that impact our ability to detect
and interpret periodic photometric modulations will influence the types of host stars we can analyze.
For instance, light curves with a limited time baseline are inherently more sensitive to brighter stars with 
larger-amplitude variations and shorter rotation periods.  Some of the \emph{TESS} light curves we rely on comprise only a single 27-day sector
(see Table~\ref{tab:hosttable}), which means we would only be able to reliably establish periods on timescales of about half that baseline ($\sim$13~d).
The Sun has a small surface spot covering fraction, long equatorial rotation period (24.5~d), 
and brightness variations typically under 0.1\% (\citealt{Reinhold:2020aa}).
Similarly old G dwarfs would not be recovered in \emph{TESS}, so inclination and obliquity measurements for these types of host
stars would not be possible using methods that rely on photometric rotation periods.
As a result, there is very likely a selection bias in this study in favor of systems with brighter, younger, and later-type host stars in our sample,
as has been recently established by \citet{Masuda:2022aa} for Sun-like stars in the \emph{Kepler} field.
This age bias is already present for systems with directly imaged planets, but it is likely reinforced for stars with brown dwarf companions
as well by disfavoring older stars.

\subsubsection{Pole-on Orientations}{\label{sec:poleon}}

We also expect there to be a geometric bias associated with selecting stars based on the presence of light curve modulations.
For stars with solar-like dynamos, spot distributions located in equatorial regions would be more visible and would produce
higher amplitude variations if seen in an equator-on orientation.  Pole-on viewing angles are inherently geometrically unlikely, but for
Sun-like stars this dearth of pole-on viewing angles is likely to be reinforced.
However, high-latitude and polar spots are commonly observed on the surfaces 
of rapidly rotating active stars (\citealt{Strassmeier:2009aa}; \citealt{Yadav:2015aa}; \citealt{Roettenbacher:2016aa})
as a result of youth, spin-orbit coupling in close binaries, or low stellar masses with large convective envelopes.
If the dominant spot structure is located in a large polar cap-like spot, and the evolution of that spot is slow, 
the impact on brightness variations in a light curve
would be small for a star viewed pole-on compared to a similar spot rotating in and out of view in equatorial regions.  
There may therefore be a two-fold selection bias against pole-on orientations near $i_*$=0$\degr$ as a result of the lack of polar 
spot structures for Sun-like stars and long-lived polar spots for active stars in the sample.

This possible bias against pole-on systems can be directly tested with our sample.  Three out of 53 stars with inclination 
constraints in Table~\ref{tab:inclinations} have posterior MAP values of $i_*$ under 10$\degr$.  
For an isotropic inclination distribution, $P(i_*) = \sin i_*$,  
the probability of obtaining an orientation within 10$\degr$ of pole-on is $\int_{i_*=0\degr}^{i_*=10\degr} \sin i_* d i_*$ = 0.015.
The expectation value in a sample of 53 stars is 0.8, so we would expect about one star to be viewed this way.  This is similar
to the number of near-pole on stars in our sample.  We therefore do not find strong evidence for a 
significant bias disfavoring pole-on orientations.

\subsubsection{Differential Rotation}{\label{sec:diffrot}}

Differential rotation is expected to be common and has been observed in many stars.
Starspots located at mid- or high-latitudes can result in a systematically longer inferred rotation period than 
the true equatorial rotation period.  The severity of this bias is increased for spots located at higher latitudes and 
stars with stronger levels of shear.
An overestimated rotation period will underestimate the equatorial velocity, which in turn will overestimate the stellar inclination
when using projected rotational velocity to infer $i_*$.

For example, the active regions in the Sun are confined to latitudes of about $\pm$30$\degr$ where
the rotation period is $\approx$25.7~d (based on the latitude-dependent relations for differential rotation from \citealt{Snodgrass:1990aa}).  
This is only 5\% higher than the equatorial rotational velocity of 24.47~d, so the effect is not particularly large for old Sun-like stars
with solar-like dynamos.  However, using asteroseismology of stars in the \emph{Kepler} field, \citet{Benomar:2018aa} found that 
some solar-type stars rotate twice as fast at their equators compared to their mid-latitudes.  
A  broad distribution of absolute shear also appears to be present among  
low-mass stars, even though the overall level of differential rotation decreases with effective temperature 
(\citealt{Barnes:2005aa}; \citealt{Reinhold:2013iz}; \citealt{Reinhold:2015ep}).
Although most stars are expected to follow a solar-like differential rotation law with shorter rotation periods at the equator 
and longer periods at the poles, antisolar rotation (with poles rotating faster than the equator) and cylindrical rotation
may also be possible (\citealt{Brun:2017aa}).
Altogether, there is considerable uncertainty in the detailed picture of differential rotation and the overall impact of  
spot distributions on light curves.  While we have made conservative assumptions about the uncertainties associated with our
photometric rotation periods (see Section~\ref{sec:prot}), 
we did not make explicit adjustments to the period values to correct them for potential systematic overestimates.

\begin{figure*}
  \vskip -.3 in
  \hskip .3 in
  \resizebox{7in}{!}{\includegraphics{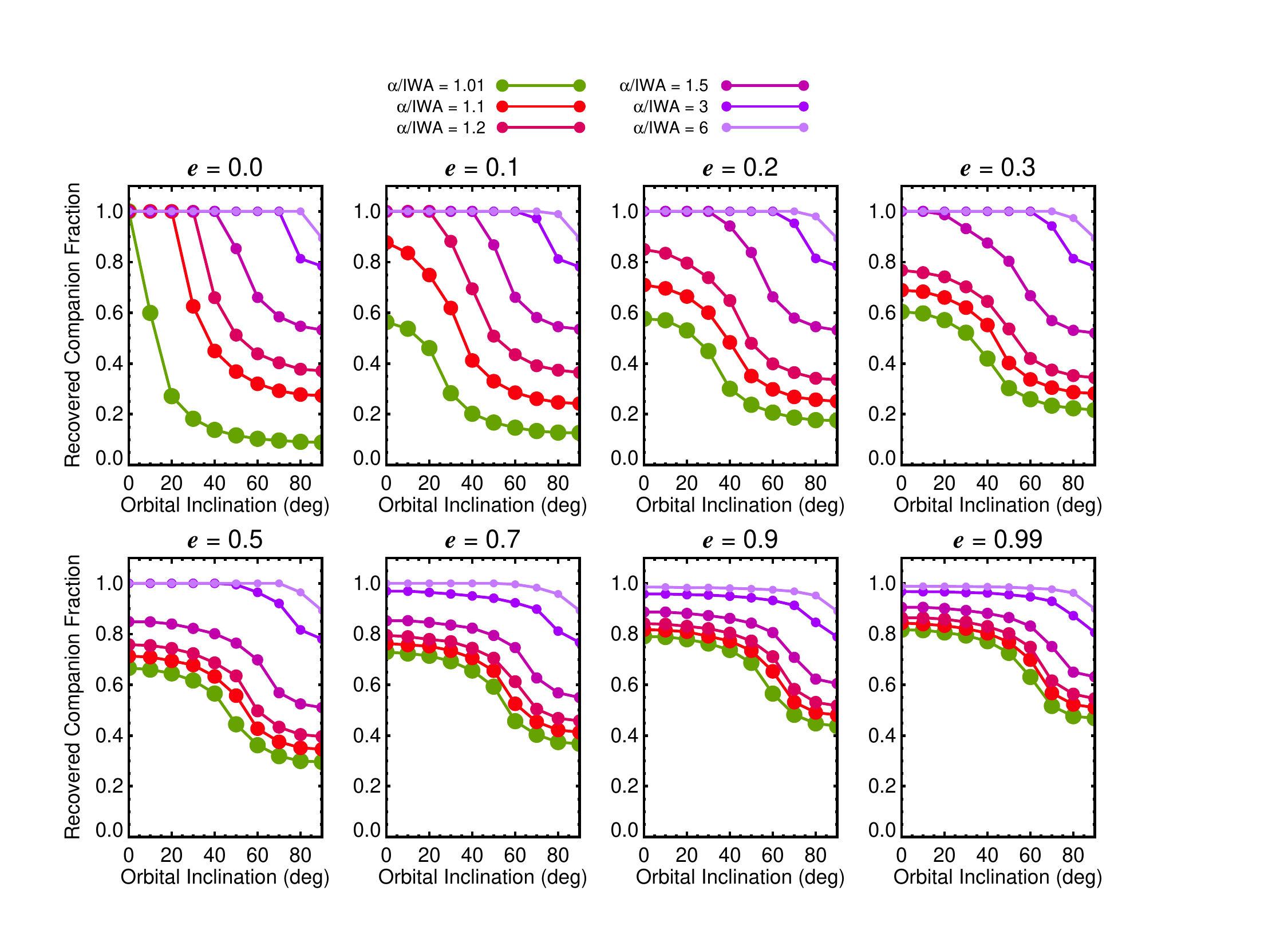}}
  \vskip -.2 in
  \caption{Assessing biases in orbital inclinations for direct imaging surveys.  The sensitivity to faint substellar companions 
  on various orbital inclinations depends on orbital eccentricity and the ratio of the angular semi-major axis, $\alpha$, 
  and the inner working angle (IWA), which establishes whether a companion is detected or obscured (by a circular coronagraph,
  for example).  Imaging is more sensitive to companions with low inclinations, wider orbits (higher values of $\alpha$/IWA), 
  and higher eccentricities.  These trends are corrected for isotropic orbital viewing geometry in Figure~\ref{fig:incsims_sini}.     \label{fig:incsims} } 
 \end{figure*}

\begin{figure*}
  \vskip -.3 in
  \hskip .3 in
  \resizebox{7in}{!}{\includegraphics{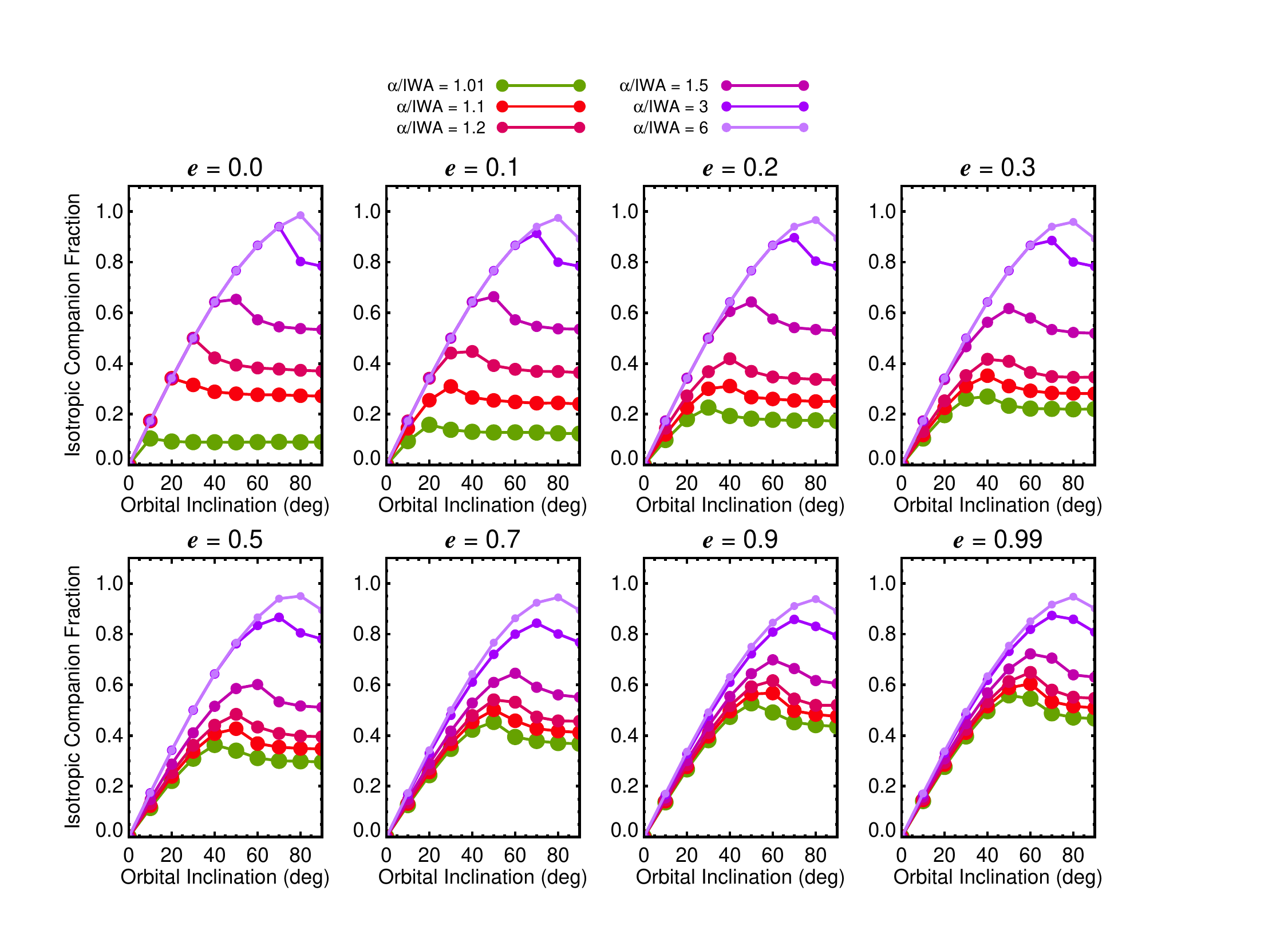}}
  \vskip -.2 in
  \caption{Same as Figure~\ref{fig:incsims} but now taking into account the relative probability of 
  observing companions with random orientations.   When accounting for this $\sin i_*$ isotropic inclination distribution,
  the sensitivity to a companion is roughly constant for a given semi-major axis from the $\sin i_*$ envelope to $i \approx 90\degr$.
  This results in a ``truncated isotropic inclination distribution'' with fewer companions having higher (more edge-on) inclinations
  and represents a more realistic outcome for the expected orbital inclination distribution of discoveries from blind
  direct imaging surveys.     \label{fig:incsims_sini} } 
 \end{figure*}

One approach to assess whether there is a bias toward high inclinations is to 
check for an over-representation of equator-on systems in our sample in the same fashion as we
did in Section~\ref{sec:poleon}.  For this analysis, we adopt MAP values of $i_*$ above 80$\degr$ as a threshold for equator-on viewing angles.
There are 19 stars in the sample with $i_* > 80\degr$.  The probability of a random orientation falling in that range is 
$\int_{i_*=80\degr}^{i_*=90\degr} \sin i_* d i_*$ = 0.174, implying an expectation value of $\approx$9 for a sample of 53 stars.
It does therefore appear that there are more equator-on stars than expected from chance alone.
To quantify the significance of this discrepancy, 
we can treat this outcome as a Bernouilli trial and compute the probability of an event occurring that is at least as extreme
as what was measured ($k$ = 19 out of $n$ = 53 systems) using binomial statistics.  Here the probability that a success occurs is $p$ = 0.174.
This statement is equivalent to computing 1 minus
the probability of observing fewer than 19 stars in the 80$\degr$ $<$ $i_*$ $<$ 90$\degr$ range:
$P(k \ge 19 | p=0.174, n=53)$ 
= $1 - P(k < 19 | p=0.174, n=53)$ 
= $1 - \sum_{k=0}^{18} {n \choose k} p^k (1-p)^{n-k}$ 
= 0.0010.
The probability of there being so many equator-on systems by chance is therefore very small.
This can be interpreted either as confirmation that the overestimated rotation period bias is impacting our sample,
or alternatively as a result of the Bayesian formalism we adopt, in which a broad or 
unconstrained $v \sin i_*$ value relative to $v_\mathrm{eq}$ will revert toward the $\sin i_*$ isotropic prior in which $i_*$ is near 90$\degr$.
Of course, both explanations could be at play here.

To examine whether poorly constrained $v \sin i_*$ values are driving the high incidence of equator-on stars, 
we compare the distribution of $\sigma_{v \sin i_*}$/$v \sin i_*$ values from the full sample in Table~\ref{tab:hosttable}
to the subset of 19 stars with $i_* > 80 \degr$.  If the bias in line-of-sight stellar inclinations is caused by poor constraints
on projected rotational velocities, we would expect the subsample to be shifted to higher values of $\sigma_{v \sin i_*}$/$v \sin i_*$.
However, this is not what we find: the median precision of $v \sin i_*$ for the full sample is 3\% (with a standard deviation of 13\%), and
for the subsample is 5\% (with a standard deviation of 13\%).  This suggests that overestimated rotation periods, likely caused by
differential rotation, are artificially increasing some of the stellar inclinations in our sample.

\subsubsection{Inclination Discovery Bias}  

Companions found through high-contrast imaging are only visible outside of an inner working angle (IWA), which represents
the closest angular separation at which a detection can be made for a given contrast.  The IWA is often defined by the edge of a
coronagraph or the radial extent of residual features after primary subtraction.
For a star at a given distance and a companion on a circular orbit with a given semi-major axis located just outside of the IWA,
a face-on orbital orientation will be more readily discoverable in a blind survey than a companion on an inclined orbit.
The same companion on an edge-on orbit will spend the least time outside the IWA.
This ``inclination discovery bias'' will impact systems with angular semi-major axes ($\alpha$) 
close to the IWA, or $\alpha$/IWA $\sim$ 1 (e.g., \citealt{Janson:2010gg}).

\citet{Bowler:2020hk} examined the impact of (the reciprocal of) this ratio 
to assess whether an analogous ``eccentricity discovery bias'' was present in their sample of companions between 5--100~AU with 
orbit constraints.  They found that most systems at the time of discovery had $\alpha$/IWA values between 1.2--10,
and therefore that this discovery bias did not play a strong role in shaping the eccentricity distribution of their sample.
13 out of 21 systems with obliquity constraints in this study overlap with the sample in \citet{Bowler:2020hk} that was used
to analyze the impact of discovery bias.  The additional systems in this study are generally at separations slightly beyond the 100~AU
cutoff used in that study and generally have large $\alpha$/IWA values much greater than 1.0. 

Here we examine whether an inclination discovery bias might have impacted the orbital properties of substellar companions 
that have been discovered in high-contrast imaging surveys 
(e.g., \citealt{Bowler:2016jk}; \citealt{Nielsen:2019cb}; \citealt{Vigan:2021dc}), which in turn could influence the 
minimum obliquity distribution $\Delta i$ in this study.
We simulate the astrometric orbits of companions and randomly sample their projected separations 
on the plane of the sky as a function of orbital inclination, semi-major axis (in units of $\alpha$/IWA), and eccentricity.
For each orbit we randomly draw a longitude of ascending node from 0 to 2$\pi$, 
argument of periastron from 0 to 2$\pi$, and time of periastron passage.
The following $\alpha$/IWA values are examined: 1.01, 1.1, 1.2, 1.5, 3, and 6.
10$^5$ synthetic observations of companions on unique orbits are generated for each set of sampled parameters;
the proportion of observations for which the companion's projected separation is beyond 1 $\alpha$/IWA (constituting a
``detection'') represents the recovered companion fraction.
Results are displayed in Figure~\ref{fig:incsims}.

\begin{figure}
  \vskip -.4 in
  \hskip -2.1 in
  \resizebox{8.1in}{!}{\includegraphics{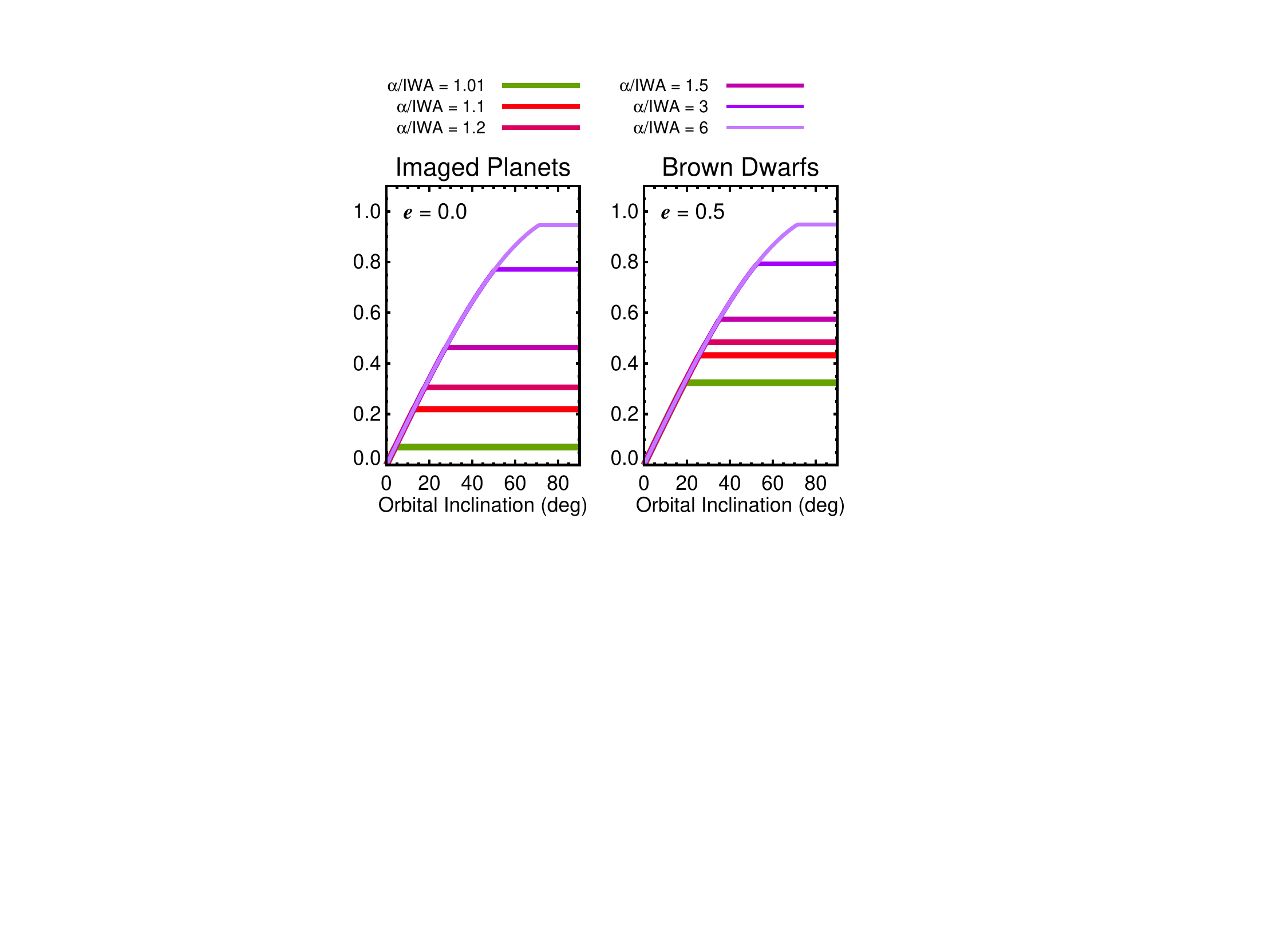}}
  \vskip -2.5 in
  \caption{Example of parameterized Bayesian priors for orbital inclinations that take into
  account the inclination discovery bias caused by the presence of a hard 
  IWA in high-contrast imaging datasets (Equations~\ref{eqn:prior1} and \ref{eqn:prior2}).  
  Note that for separations very close to the IWA, the expected orbital inclination prior deviates
  substantially from a traditional isotropic ($\sin i_o$) distribution.  For circular orbits this can resemble a uniform distribution.
  These modified isotropic distributions are unnormalized approximations to the results in Figure~\ref{fig:incsims_sini} and depend on
  the angular semimajor axis $\alpha$ and orbital eccentricity.  Because long-period 
  giant planets and brown dwarf companions have distinct eccentricity distributions 
  (\citealt{Bowler:2020hk}; \citealt{Nagpal:2022aa}), a different choice for the characteristic 
  eccentricity might be warranted depending on the properties of a newly imaged companion---for instance,
  $e=0.0$ for planets and $e=0.5$ for brown dwarfs.
      \label{fig:orbital_prior} } 
 \end{figure}

Considering the case of circular orbits, there is a steep decline sensitivity for more inclined systems
and an increase in detection sensitivity for companions on wider orbits.  The recovery rate is nearly 100\%
for $\alpha$/IWA values beyond $\approx$3, with only the most edge-on orbits experiencing some loss in sensitivity.
As soon as non-zero eccentricity is introduced, there is a general increase in sensitivity for a given $\alpha$/IWA
value, except for near-face on orbits where a slight drop is evident.
This behavior is readily explainable: companions on eccentric orbits will reach 
farther apastron distances and will spend most of their time at those locations.
But for face-on circular orbits, where detection sensitivity is maximized, even a slight eccentricity will bring part of the
orbit inside the IWA.  This trend of overall  sensitivity with increasing eccentricity persists even to unrealistically
high eccentricities above 0.9.  
There is also a clear inclination bias present: companions on more edge-on orbits are recovered at $\approx$2--3 times 
lower rates compared to companions on more face-on orbits, with this drop occurring beyond inclinations of about 40$\degr$ 
for low eccentricities and beyond 60$\degr$ for high eccentricities.

However, these results do not take into account the relative geometric probability of observing face-on versus edge-on systems.
To account for this effect, each curve should be adjusted by a factor of $\sin i_*$, as shown in Figure~\ref{fig:incsims_sini}.
The impact is to flatten the recovered companion fraction curves so that for any given semi-major axis,
there is a roughly uniform probability of observing any orbital inclination between the $\sin i_*$ envelope of isotropic orbits
to 90$\degr$.  This behavior persists to about $e$=0.5, then begins to more closely resemble an isotropic
distribution at higher eccentricities, especially for large values of $\alpha$/IWA where sensitivity to companions is nearly
complete.  In this limit the isotropic distribution is recovered.
These trends should better represent more realistic biases on the orbital inclinations of discoveries 
made through high-contrast imaging.
The actual distribution of observed orbital inclinations will depend on the underlying range of semi-major axes,
host star distances, eccentricities, and IWAs.  

\begin{figure}
  \vskip -.1 in
  \hskip -.1 in
  \resizebox{3.6in}{!}{\includegraphics{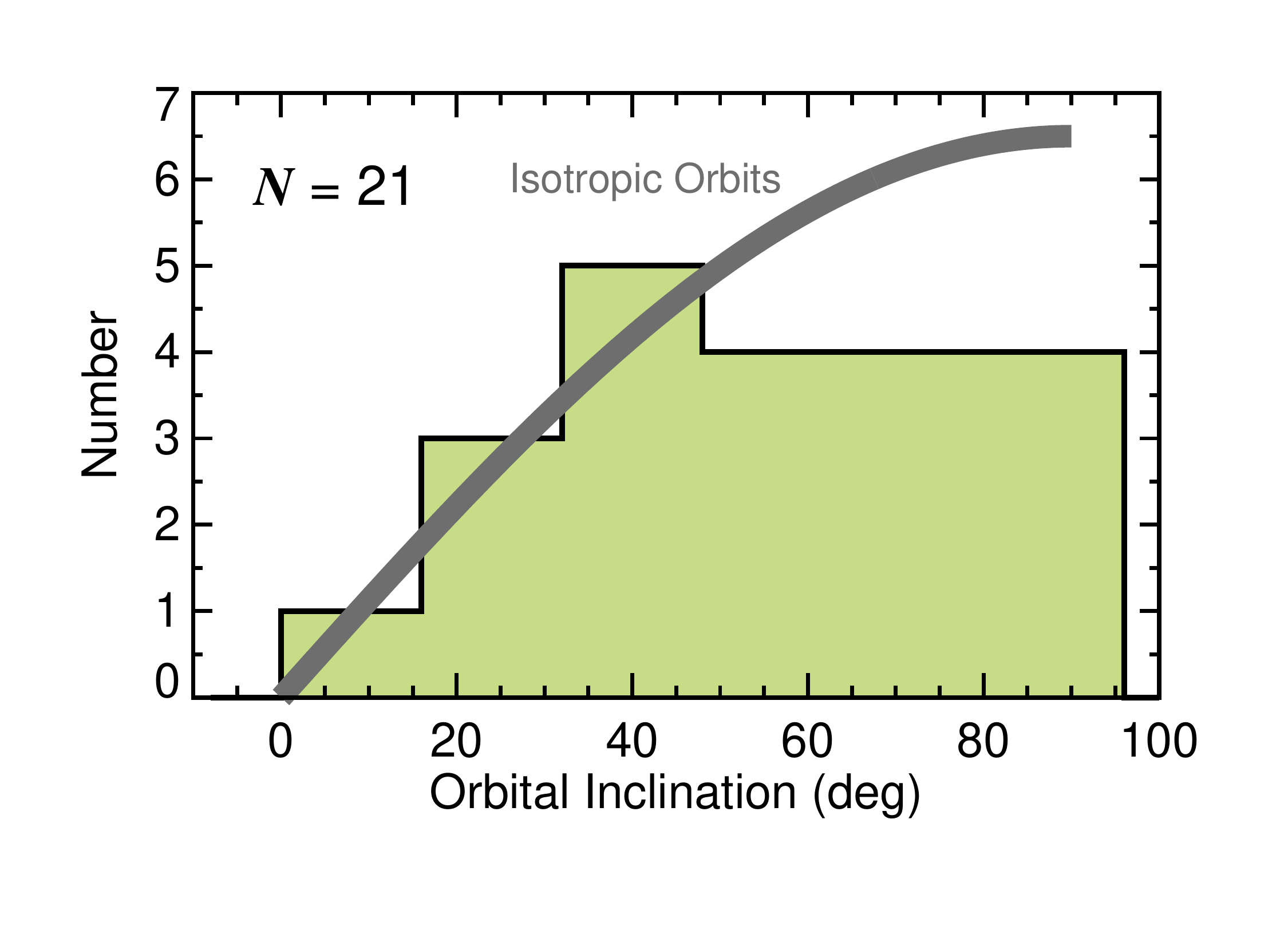}}
  \vskip -.25 in
  \caption{Distribution of orbital inclination MAP values for the 21 substellar companions with obliquity constraints 
 examined in this study.  Compared to an isotropic inclination distribution (gray), the observed sample
 has a deficit of objects with edge-on orientations between $\approx$50--90$\degr$.  
 This closely resembles predictions in Figure~\ref{fig:incsims_sini} which take into account 
 inclination discovery bias in direct imaging surveys caused by a hard IWA.    \label{fig:orbital_incs} } 
 \end{figure}

While not the focus of this study, we note that these ``truncated isotropic inclination distributions'' should be more appropriate 
to use as Bayesian priors for the orbital plane inclinations ($\mathcal{P}(i_o)$) of newly discovered companions 
located near the observational IWA than the more traditional $\mathcal{P}(i_o) = \sin i_o$ 
inclination prior that is widely adopted for orbit fitting.
Below is an approximation of these more realistic priors 
which account for the general behavior of inclination discovery bias as a function of semi-major axis
and orbital eccentricity:

\begin{equation}{\label{eqn:prior1}}
  \mathcal{P}(i_o) \propto
    \begin{cases}
      \sin i_o & \text{if $i_o$ $<$ $\gamma$}\\
      \sin \gamma & \text{otherwise}
    \end{cases}       
.
\end{equation}

\noindent $\gamma$ is the inclination in radians beyond which 
the distribution is constant for increasing values of $i_o$.
For this approximation, we implement a break from the sine curve that depends 
on both the angular semi-major axis $\alpha$, here parameterized as 
$\beta = \alpha$/IWA, and eccentricity $e$:

\begin{equation}{\label{eqn:prior2}}
\gamma(\beta, e) = 0.5  \cosh^{-1}(\beta) + 1.5 \log(1 + e/\beta^2).
\end{equation}

\noindent  Here the inverse hyperbolic cosine term is a better fit to the $\gamma$ break
values as a function of $\beta$ parameter than a polynomial.  The second term is motivated 
by a logarithmic-like scaling of $\gamma$ with eccentricity and $\beta$.  Coefficients were 
selected so as to reasonably agree with values in Figure~\ref{fig:incsims_sini} as a 
compromise between consistency with the numerical simulations and analytical simplicity.
In the limit of wide orbits ($\alpha \gtrsim 11.6$ IWA), $\gamma$ $>$ $\pi/2$ and
the standard isotropic distribution is recovered.

The choice of $e$ and $\beta$ may depend on the context of the system.
For example, $e$=0.0 may be an appropriate choice for a new directly imaged planet,
with $\beta$ corresponding to the projected separation over the IWA,
whereas a moderate eccentricity of $e\approx$0.5 may be a better choice for a brown dwarf
companion (\citealt{Bowler:2020hk}; \citealt{Nagpal:2022aa}).
Examples of these unnormalized priors are shown in Figure~\ref{fig:orbital_prior}.

How do these expectations compare with actual orbital inclinations?  
A histogram of the median orbital inclinations from Table~\ref{tab:inclinations} is shown in 
Figure~\ref{fig:orbital_incs} compared to an isotropic distribution.  
Because the sample size is modest, we have reflected the inclination values (which span 0--180$\degr$) 
about 90$\degr$ as the predicted yields are symmetric with respect to the sky plane.
The observed inclinations seems to match expectations reasonably well; there is good agreement with the isotropic case 
at low inclinations, and the overall distribution is approximately constant from $\approx$40--90$\degr$.
Altogether, this can be explained by a mixture of modest eccentricities of brown dwarfs and
the range of $\alpha$/IWA values from \citet{Bowler:2020hk}.

\subsubsection{Orbital Inclination Priors}  

Another relevant form of inclination bias is related to the challenge of fitting orbits with small phase coverage.
Most substellar companions with orbit constraints have had less than 15\% of their orbits mapped with
relative astrometry (\citealt{Bowler:2020hk}).
This makes the inferred orbit solutions and the detailed shape of the orbital element posteriors
sensitive to both random and systematic uncertainties in relative astrometry, as well as the choice
of parameter priors used in the analysis.  
Recently, \citet{FerrerChavez:2021aa} analyzed systematic biases associated with fitting orbits 
to companions with very small coverage ($\approx$0.5\%).  They found that face-on inclinations can result in
significant biases in the posterior median and mode, and that for progressively higher eccentricities, more inclined orbits
can be impacted by this bias.
A similar conclusion was found by \citet{ONeil:2019aa}; without adopting their 
observable-based priors, inclinations are broader and less accurate.
For our sample in this study, these effects can be exacerbated by the use of different approaches to
orbit fitting in the literature and parameter priors that were adopted.
It is therefore challenging to determine the exact scale of inclination bias caused by priors, 
although it is expected to be more severe for orbits with smaller fractional coverage.

\subsubsection{Slow Rotation}  

As discussed in Section~\ref{sec:obliquities}, it is challenging to reliably determine inclinations for 
slow rotators with equatorial velocities less than a few km s$^{-1}$ because this is near the
limit where traditional spectroscopic approaches to measuring rotational broadening, macroturbulence, and microturbulence can
be difficult to distinguish (e.g., \citealt{smith:1976aa}; \citealt{Gray:2005va}; \citealt{Petigura:2017bz}; \citealt{Masuda:2021aa}).  
The result of this is a tendency to interpret slow rotators as being equator-on, even 
if their actual inclinations substantially depart from 90$\degr$.
This may contribute to the overabundance
of stellar inclinations between 80--90$\degr$ discussed in Section~\ref{sec:diffrot}.
However, only seven of 53 stars with $i_*$ constraints have $v_\mathrm{eq} < 3$ km s$^{-1}$, 
so we do not expect these systematic errors to severely impact statistical results from this study.

\subsubsection{Relative Biases}{\label{sec:relativebiases}}  

In addition to overall biases that may be impacting our sample selection and inferred inclination distributions, 
\emph{relative} biases may also be present within our sample that could in principle influence the interpretation 
of the misalignment results in this study.  In particular, the apparent distinction between obliquities of stars 
hosting long-period giant planets and brown dwarf companions could instead be caused by obliquity evolution, 
for example through dynamical excitation over time.  Imaged planets 
are typically young ($\lesssim$200~Myr), whereas the systems with brown dwarfs span a much broader range of ages
because brown dwarfs remain bright enough to readily detect even at field ages of several Gyr.
If obliquities are initially low but evolve toward a broader distribution over time, perhaps through dynamical interactions,
that could manifest as a distinction between planets and brown dwarf companions as a result of their
different characteristic ages.  However, our sample only comprises a few field-age brown dwarfs---Gl 229,
HR 7672, and HD 19467---so if obliquities are evolving then this mechanism would likely need to operate
on relatively short timescales ($\lesssim$1~Gyr) corresponding to the typical age range of systems with
brown dwarfs in this study.

Other possible sources
of relative biases within our sample could include a dependence of obliquity on stellar host mass, metallicity,
or formation environment.  All of these could be explored in more detail with larger samples in the future.

\subsubsection{Summary of Biases}{\label{sec:biassumary}}  

Below we summarize the most important biases likely to impact stellar inclinations, 
orbital inclinations, and minimum obliquity values ($\Delta i$) in our sample.

\begin{itemize}

\item We expect a strong bias in favor of active stars with younger ages (faster rotation periods) 
and cooler temperatures (greater starspot covering fractions)
as a result of the finite photometric precision and limited baseline coverage of light curves.

\item Both differential rotation and slow rotation can result in a systematic overestimation of stellar inclinations.
This may explain the over-representation of equator-on orientations that is present with high confidence in our 
larger sample of 53 host stars with inclination constraints.

\item A discovery bias affecting the orbital inclination distribution 
 is expected for companions found with high-contrast imaging.
This distribution is not isotropic but instead more closely resembles a truncated $\sin i_o$ 
distribution.  This deficit of edge-on orientations is reflected in the orbital inclination distribution of companions in our sample.

\item Bayesian priors may be influencing the detailed shape of orbital inclination posteriors for 
systems with more face-on orientations and less orbital phase coverage compared to systems with well-mapped
orbits.  

\item We do not find evidence that our sample is biased against pole-on orientations as might be
expected for stars with solar-like dynamos and starspots that are confined to equatorial regions.

\item Relative age biases could impact stellar obliquities for systems with giant planets
and brown dwarf companions, but only if obliquity evolution occurs on relatively short timescales ($\lesssim$1~Gyr).

\end{itemize}

Altogether, it is difficult to confidently establish how these biases might be combining to affect individual and cumulative 
results for $\Delta i$ distributions.  This certainly warrants future consideration, especially in conjunction with a better understanding of
other unknown astrophysical
obliquity trends that we are ignoring.  These could include stellar obliquities as a function of
system age, companion mass, separation, eccentricity, or multiplicity.  We did not consider these in this study because the sample
size and individual constraints are generally modest, but in the future these relationships can be explored 
in more detail with larger samples and more precise constraints on $\Delta i$.

\subsection{Spin-Spin Opportunities} \label{sec:spin}

Most companions to the 53 host stars with stellar inclinations do not have constraints on the
inclinations of their orbital planes because the physical separations in these systems are prohibitively large.  
However, this sample may be beneficial to explore the statistical distributions of 
spin-spin orientations between the host stars ($i_*$) and companions ($i_p$).  A handful of obliquities have now been measured 
for individual brown dwarf companions and giant planets by combining rotation periods,
radii, and projected rotational velocities (e.g., \citealt{Bryan:2020ji}; \citealt{Bowler:2020ip};
\citealt{Zhou:2020cl}; \citealt{Bryan:2021aa}; \citealt{PalmaBifani:2022aa}),
but larger samples are needed to assess the prevalence of
stellar-companion spin-spin  alignment in order to 
interpret this in the context of formation secenarios (e.g., \citealt{Jennings:2021aa}).  
Companions to host stars with existing inclination constraints from this study may be promising
targets for follow-up photometric monitoring and high-resolution spectroscopy in the future.

\section{Summary}{\label{sec:summary}}

We have carried out a homogeneous analysis of the rotation periods, inclinations, and obliquities of stars hosting 
directly imaged brown dwarf and giant planet companions.  
Below we summarize the main conclusions and interpretation from this work.

\begin{itemize}

\item \emph{TESS} light curves for substellar host stars were uniformly examined in search of rotational modulations.  Our search 
was isolated to spectral types $\gtrsim$F5 to
avoid confusion with stellar pulsations.  Supplementing these rotation periods with values from the literature
yields 62 stars with period measurements, 53 of which also have $v \sin i_*$ values from our own new
observations and previous determinations.  We derived inclination posteriors using the Bayesian framework from
\citet{Masuda:2020dp} along with general analytical expressions for inclination posteriors given $P_\mathrm{rot}$,
$R_*$, and $v \sin i_*$ values in Appendix~\ref{app:istar}.

\item 21 systems with stellar inclinations also have constraints on the inclination of the companion's orbital plane.
Together these provide a measurement of $\Delta i$, the minimum value of the true spin-orbit obliquity.
Companions in this subsample are predominantly brown dwarfs located between 10--100 AU with a few 
companions extending out to 250 AU.  

\item Spin-orbit misalignments for systems with substellar companions 
appear to be common, as revealed by $\Delta i$ distributions.  Eleven of the 21 systems
in our sample (52$^{+10}_{-11}$\%) have $\Delta i$ values beyond 10$\degr$ with $\ge$80\% confidence, indicating that
misalignments are prevalent among brown dwarf companions. 
Depending on the orientation of the star in the sky plane
and the longitude of ascending node of the orbital plane, most companions can be on prograde, polar, or retrograde orbits.
This high misalignment fraction contrasts with partial or full constraints on stellar obliquities for systems with directly imaged planets,
which show an emerging pattern of angular momentum alignment between the star, planet orbital plane,
and, when present, the disk.  We also find signs of a preference for spin-orbit alignment 
for companions within 20~AU.  Brown dwarfs and companions at wide separations show a greater diversity of 
rotational and orbital angular momentum architectures.

\item When the same systems are compared with stars with random orientations, stars with rotation axes aligned with orbital 
angular momentum vectors, and a mixture of both scenarios, our hierarchical Bayesian analysis of the underlying distributions of $\Delta i$ for the observed
sample is most consistent with the even mixture and disfavors the aligned case.  However, the modest sample size, generally broad $\Delta i$ distributions,
potential for sample biases, and minimal information about the true obliquity angle $\psi$ limits more nuanced conclusions about how
obliquities might depend on age, companion mass, and separation.

\item Altogether, these results are consistent with predictions from hydrodynamic 
simulations that spin-orbit misalignments should be a common outcome for binary companions 
that form from a fragmenting molecular cloud core (e.g., \citealt{Bate:2010fz}; \citealt{Offner:2016gl}).
Although these results could also be explained by dynamical processing of systems with primordial alignments, 
 this would require another massive object to interact with in these systems.  
 Large high-contrast imaging surveys have failed to uncover a population of multiple brown dwarf companions 
 orbiting stars in non-hierarchical configurations.
Some misaligned systems may also result from the capture of isolated or ejected giant planets and brown dwarfs
 in young dense clusters.

\item We have evaluated potential biases that could be impacting the stellar and orbital inclinations in our sample.
There is evidence for a higher number of systems on edge-on orbits than would be expected from chance, which
may be a result of stars with differential rotation or slow rotation in the sample.
More detailed modeling of the impact of these effects is warranted.

\item As part of this analysis we highlight a bias in the orbital inclination distribution of companions found with high-contrast 
imaging in the presence of an IWA---which may be a coronagraph or residuals from primary star subtraction.
This ``inclination discovery bias'' is most severe for objects close to the IWA and varies with semi-major axis and eccentricity.
Its effect is to disfavor edge-on orientations and deviates substantially from an
isotropic distribution in which $\mathcal{P}(i_o) \propto \sin i_o$.
The distribution of orbital inclinations for substellar companions is consistent with having been shaped by the 
presence of this effect. 
We formulate a parametric approximation for this bias, which may be useful as a more realistic Bayesian prior for
orbit fits to new discoveries.

\end{itemize}

Stellar obliquities provide a rich resource to probe the formation and dynamical histories 
of planets and brown dwarf companions.
Results from this study reinforce an emerging distinction between imaged giant planets
and brown dwarf companions in terms of their orbits (\citealt{Bowler:2020hk}),
demographics (\citealt{Nielsen:2019cb}; \citealt{Vigan:2021dc}), and mass functions (\citealt{Wagner:2019iy}):
brown dwarf companions are most consistent with formation in a stellar-like pathway from fragmenting turbulent clouds than 
within aligned disks.
In the future, we expect that new discoveries from dynamically informed 
surveys (e.g., \citealt{Franson:2022bb}; \citealt{Hinkley:2022aa}), 
continued astrometric orbit monitoring of known systems (e.g., \citealt{Zurlo:2022aa}), 
precision astrometry from VLTI/GRAVITY (\citealt{Nowak:2020cc}; \citealt{Wang:2021aa}),
and radial velocities of companions from instruments
such as the Keck Planet Imager and Characterizer (\citealt{Mawet:2016aa}) will be especially fruitful to build a larger sample of systems well-characterized spin-orbit constraints.
In addition, spectro-interferometry capable of spatially resolving offsets from stellar rotation 
will enable true obliquity measurements for imaged planets orbiting early-type stars (\citealt{Kraus:2022aa}).
Ultimately, refined orbital inclinations and projected rotational velocities will facilitate more precise 
population-level obliquity constraints to carry out direct
tests for differences in the obliquity distributions 
between long-period brown dwarf companions and directly imaged planets.

\facility{Smith (Tull Spectrograph), TESS}

\acknowledgments

We thank the referee for their timely and constructive feedback, as well as 
Simon Albrecht, Ansgar Reiners, Aldo Sepulveda, and Josh Winn for helpful discussions and comments on this study.
B.P.B. acknowledges support from the National Science Foundation grant AST-1909209, NASA Exoplanet Research Program grant 20-XRP20$\_$2-0119, and the Alfred P. Sloan Foundation.
Q.H.T. and B.P.B. acknowledge the support from a NASA FINESST grant (80NSSC20K1554).
K.F. acknowledges support from the National Science Foundation Graduate Research Fellowship Program under Grant No. DGE 2137420.
D.H. acknowledges support from the Alfred P. Sloan Foundation and the National Aeronautics and Space Administration (80NSSC21K0652, 80NSSC21K0784).
Y.Z. acknowledges support from the Heising-Simons Foundation 51 Pegasi b Fellowship.
Support for this work was provided by NASA through the NASA Hubble Fellowship grant HST-HF2-51522.001-A awarded by the Space Telescope Science Institute, which is operated by the Association of Universities for Research in Astronomy, Inc., for NASA, under contract NAS5-26555.
This work has benefited from \emph{The UltracoolSheet}, maintained by Will Best, Trent Dupuy, Michael Liu, Rob Siverd, and Zhoujian Zhang, and developed from compilations by \citet{Dupuy:2012bp}, \citet{Dupuy:2013ks}, \citet{Liu:2016co}, \citet{Best:2018kw}, and \citet{Best:2021gm}.
This research has made use of the VizieR catalogue access tool, CDS,
 Strasbourg, France (DOI: 10.26093/cds/vizier). The original description 
 of the VizieR service was published in 2000, A\&AS 143, 23.

\clearpage
\newpage

\appendix
\section{Analytical Formalism to Infer Stellar Inclinations}{\label{app:istar}}

\subsection{Measurements of $v \sin i_*$}{\label{app:vsini_meas}}

\citet{Masuda:2020dp} provide a thorough guide to infer the posterior probability distribution function of stellar inclination ($i_*$) 
given measurements or estimates of a rotation period ($P_\mathrm{rot}$), stellar radius ($R_*$), 
and projected rotational velocity ($v \sin i_*$) following a Bayesian formalism.  
The equatorial velocity $v_\mathrm{eq}$ is simply 2$\pi$$R_*$/$P_\mathrm{rot}$.
If $i_*$, $R_*$, $P_\mathrm{rot}$, and $v \sin i_*$ are independent then $i_*$ is  
$\sin^{-1} (\frac{P_\mathrm{rot} \, v \sin i_*}{2 \pi R_*})$.  However, one of the important insights recognized by \citet{Masuda:2020dp} is
that $v_\mathrm{eq}$ and $v \sin i_*$ are not independent, so accurately determining the probability distribution of $i_*$ must take into account the correlation between these parameters.

Under the reasonable assumption of independence between the datasets used to measure $v_\mathrm{eq}$ and $v \sin i_*$ (so that their
likelihood functions are separable),
and assuming $v_\mathrm{eq}$ and $i_*$ are similarly independent (so that their priors are separable), the probability distribution for
$\cos i_*$ given a set of data $d$ is:

\begin{equation}
P(\cos i_*\mid d) \propto \mathcal{P}(\cos i_*) \int \mathcal{L}_{v_\mathrm{eq}}(v_\mathrm{eq}) \, \mathcal{L }_{v\sin i_*}(v_\mathrm{eq} \sqrt{1 - \cos^2i_*}) \, \mathcal{P}_{v_\mathrm{eq}}(v_\mathrm{eq}) \, d v_\mathrm{eq}.
\end{equation}

\noindent Here $\mathcal{L}_{v_\mathrm{eq}}$ is the likelihood function for the equatorial velocity, $\mathcal{L }_{v\sin i_*}$ is the likelihood function
for the projected rotational velocity, 
$\mathcal{P}(\cos i_*)$ is the Bayesian prior on the cosine of the inclinations---equal to 1 for isotropic orbits---and $\mathcal{P}_{v_\mathrm{eq}}(v_\mathrm{eq})$ is the prior
on the equatorial velocity.  If both priors are constant then 

\begin{equation}{\label{eqn:cosi_upriors}}
P(\cos i_* \mid d) \propto \int \mathcal{L}_{v_\mathrm{eq}}(v_\mathrm{eq}) \, \mathcal{L }_{v\sin i_*}(v_\mathrm{eq} \sqrt{1 - \cos^2i_*})  \, d v_\mathrm{eq}.
\end{equation}
 
Since $v\sin i_*$ is typically measured from high-resolution spectroscopy, it is usually reported with Gaussian uncertainties so its likelihood function
will follow a normal distribution in $v_\mathrm{eq} \sin i_*$:

\begin{equation}
\mathcal{L }_{v\sin i_*}(v_\mathrm{eq} \sqrt{1 - \cos^2i_*}) = \frac{1}{\sqrt{2 \pi} \sigma_{v\sin i_*} } e^{-\frac{1}{2}\big(\frac{v_\mathrm{eq} \sqrt{1 - \cos^2i_*} - v\sin i_*}{\sigma_{v\sin i_*}}\big)^2}.
\end{equation}

If the equatorial velocity is
taken from measurements (or estimates) of $R_*$ and $P_\mathrm{rot}$ then the likelihood function for $v_\mathrm{eq}$ may not be 
straightforward to express analytically.  While the stellar radius and rotation period may be normally distributed such that $R_* \sim \mathcal{N}(R_*, \sigma_{R_*}^2)$ 
and $P_\mathrm{rot} \sim \mathcal{N}(P_\mathrm{rot}, \sigma_{P_\mathrm{rot}}^2)$,
the ratio of two independent normally distributed random variables of the form 2$\pi$$R_*$/$P_\mathrm{rot}$ itself does \emph{not} generally follow a normal
distribution.  
An analytical expression has been derived but it is not succinct, even in the case when both random variables are uncorrelated 
(\citealt{Hinkley:1969aa}; \citealt{Oliveira:2015be}).

However, there are certain cases when this normal ratio distribution can be approximated with a Gaussian distribution.
For independent random variables $X \sim \mathcal{N}(\mu_x, \sigma_x^2)$ and $Y \sim \mathcal{N}(\mu_y, \sigma_y^2)$,
\citet{DiazFrances:2013hs} propose a second-order expansion for the variance of $Z=X/Y$ 
of the form $\sigma_z^2 = \beta^2\big(\delta_x^2 + \delta_y^2 \big)$,
where $\beta=\mu_x/\mu_y$, $\delta_x=\sigma_x/\mu_x$, and $\delta_y=\sigma_y/\mu_y$.
Smaller values of $\delta_x$ and $\delta_y$ produce better agreement between the normal approximation and true distribution function.
In general, the normal approximation is good if $\delta_y$ is small ($\lesssim$0.2).
For the application 
to $v_\mathrm{eq}$ this corresponds to a rotation period measurement at the $\lesssim$20\% level,
which is satisfied for all targets we consider in this study.

Under these assumptions the likelihood function for $v_\mathrm{eq}$ becomes
\begin{equation}{\label{eqn:veq_likelihood}}
\mathcal{L }_{v_\mathrm{eq}}(v_\mathrm{eq}) \approx \frac{1}{ \sqrt{2 \pi} \sigma_{v_\mathrm{eq}}} e^{-\frac{1}{2}\big(\frac{v_\mathrm{eq} - \frac{2\pi R_*}{P_\mathrm{rot}}}{\sigma_{v_\mathrm{eq}}}\big)^2},
\end{equation}

where

\begin{equation}{\label{eqn:veqerr}}
\sigma_{v_\mathrm{eq}} =  \frac{2\pi R_*}{P_\mathrm{rot}} \sqrt{\Big(\frac{\sigma_{R_*}}{R_*}\Big)^2 + \Big(\frac{\sigma_{P_\mathrm{rot}}}{P_\mathrm{rot}}\Big)^2 }
\end{equation}

Then the expression for the posterior in $\cos i$ is

\begin{displaymath}
P(\cos i_* \mid d) \propto \int \frac{1}{\sqrt{2 \pi} \sigma_{v\sin i_*} } e^{-\frac{1}{2}\big(\frac{v_\mathrm{eq} \sqrt{1 - \cos^2i_*} - v\sin i_*}{\sigma_{v\sin i_*}}\big)^2} \, \frac{1}{ \sqrt{2 \pi} \sigma_{v_\mathrm{eq}}} e^{-\frac{1}{2}\big(\frac{v_\mathrm{eq} - \frac{2\pi R_*}{P_\mathrm{rot}}}{\sigma_{v_\mathrm{eq}}}\big)^2} \, d v_\mathrm{eq} 
\end{displaymath}

\begin{equation}{\label{eqn:constants}}
P(\cos i_* \mid d) \propto \frac{1}{ 2 \pi  \sigma_{v_\mathrm{eq}}  b  \sqrt{1-\cos^2 i_*}} \int e^{-\frac{1}{2}\big(\frac{v_\mathrm{eq}  - a}{b}\big)^2} \,  e^{-\frac{1}{2}\big(\frac{v_\mathrm{eq} - c}{\sigma_{v_\mathrm{eq}}}\big)^2} \, d v_\mathrm{eq}
\end{equation}

\noindent where $a = v\sin i_* / \sqrt{1 - \cos^2i_*}$, $b = \sigma_{v\sin i_*} / \sqrt{1 - \cos^2i_*}$, and $c = 2\pi R_*/P_\mathrm{rot}$.
The integrand in Equation~\ref{eqn:constants} takes the form of the product of two normal distributions, which itself is a normal distribution: 
\begin{equation}{\label{eqn:productnormal}}
\mathcal{N}(\mu_1,\sigma_1^2)\times\mathcal{N}(\mu_2,\sigma_2^2) =  \frac{e^{- \frac{(\mu_1 - \mu_2)^2}{2(\sigma_1^2 + \sigma_2^2)} }}{\sqrt{2 \pi (\sigma_1^2 + \sigma_2^2)}} \times \mathcal{N}(\mu_{12},\sigma_{12}^2); 
\end{equation}

\noindent here $\mu_{12}$ and $\sigma_{12}^2$ are the mean and variance of the resulting distribution:
\begin{eqnarray}
\mu_{12} = \frac{\mu_1/\sigma_1^2 + \mu_2/\sigma_2^2}{1/\sigma_1^2 + 1/\sigma_2^2}, \\
\sigma_{12}^2 = \frac{\sigma_1^2 \sigma_2^2}{\sigma_1^2 + \sigma_2^2}.
\end{eqnarray}

Making use of Equations~\ref{eqn:constants} and~\ref{eqn:productnormal}, the posterior distribution for $\cos i_*$ can be rewritten as
\begin{displaymath}
P(\cos i_* \mid d)\propto 
\frac{1}{\sqrt{1 - \cos^2 i_*}} \,  \frac{e^{- \frac{(a - c)^2}{2(b^2 + \sigma_{v_\mathrm{eq}}^2)} }}{\sqrt{2 \pi (b^2 + \sigma_{v_\mathrm{eq}}^2 )}} 
\int \mathcal{N}(\mu_{12},\sigma_{12}^2) \,  d v_\mathrm{eq}
\end{displaymath}
\begin{equation}
P(\cos i_* \mid d)\propto \frac{1}{\sqrt{1 - \cos^2 i_*}} \, \frac{e^{- \frac{(a - c)^2}{2(b^2 + \sigma_{v_\mathrm{eq}}^2)} }}{\sqrt{b^2 + \sigma_{v_\mathrm{eq}}^2 }} 
\end{equation}

Substituting in the values of $a$, $b$, and $c$ gives the following expression for the posterior probability of the cosine of the stellar inclination:

\begin{equation}{\label{eqn:cosi}}
P(\cos i_* \mid d)\propto  \frac{e^{- \frac{\big(v \sin i_* - \frac{2\pi R_*}{P_\mathrm{rot}}\sqrt{1 - \cos^2 i_*} \big)^2}{2\big(\sigma_{v\sin i_*}^2 + \sigma_{v_\mathrm{eq}}^2 (1 - \cos^2 i_*) \big)}} }{\sqrt{\sigma_{v\sin i_*}^2 + \sigma_{v_\mathrm{eq}}^2 (1 - \cos^2 i_*)}}.
\end{equation}

\noindent The uncertainty in the equatorial velocity $v_\mathrm{eq}$ is given in Equation~\ref{eqn:veqerr}.  

Finally, it can be useful to represent the stellar inclination constraint  in terms of inclination $i_*$ itself.  This can be achieved using a variable transformation 
of the form $f(i_*) \, di_* = g(y) \, dy$.  Here $y$ is a function of $i_*$, $y=y(i_*) = \cos i_*$, and $g(y)$ is the posterior distribution
for $\cos i_*$, $P(\cos i_* \mid d)$.
Therefore $f(i_*) = P(\cos i_* \mid d) \, | d \cos i_* / di_* |$ = $P(\cos i_* \mid d) \sin i_*$.  The posterior of the stellar
inclination in radians becomes
\begin{equation}{\label{eqn:i}}
P(i_* \mid d) \propto  \sin i_* \times \frac{e^{- \frac{\big(v \sin i_* - \frac{2\pi R_*}{P_\mathrm{rot}}\sin i_* \big)^2}{2\big(\sigma_{v\sin i_*}^2 + \sigma_{v_\mathrm{eq}}^2 \sin^2 i_* \big)}} }{\sqrt{\sigma_{v\sin i_*}^2 + \sigma_{v_\mathrm{eq}}^2 \sin^2 i_*}}.
\end{equation}

Under the assumptions outlined above, \noindent Equations \ref{eqn:cosi} and \ref{eqn:i} provide the full (unnormalized) expressions for the distribution function of a stellar inclination
and its cosine given measurements of $v \sin i_*$, $P_\mathrm{rot}$, and $R_*$.

\begin{figure*}
  \vskip -0.7 in
  \hskip 0.3 in
  \resizebox{6.5in}{!}{\includegraphics{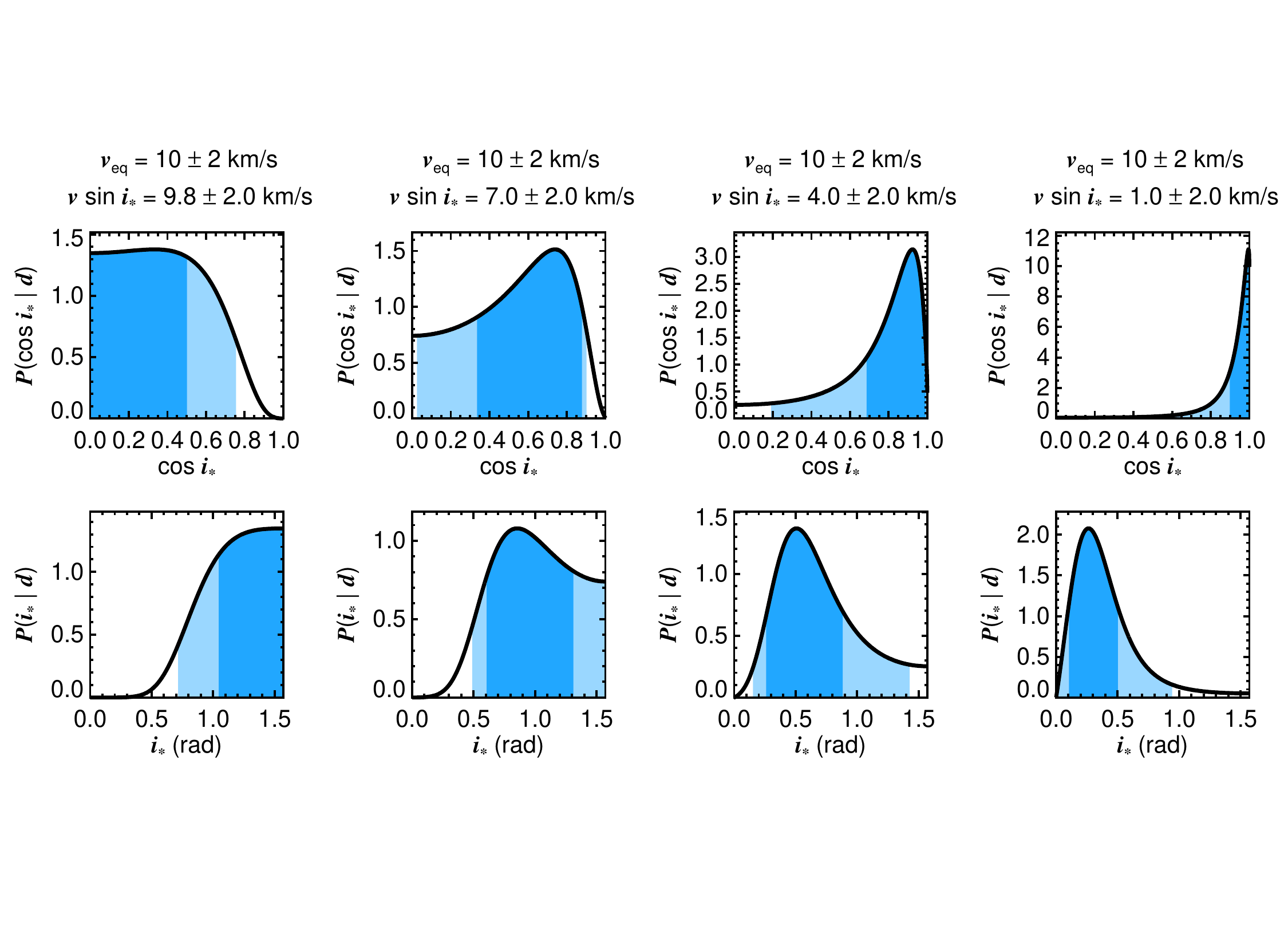}}
  \vskip -0.8 in 
  \caption{Posterior distributions for $\cos i_*$ and $i_*$ for the same examples from \citet{Masuda:2020dp}  using our Equations~\ref{eqn:cosi} and \ref{eqn:i}. Shaded regions correspond to 68.3\% and 95.4\% credible intervals.  \label{fig:masuda_inclinations} } 
\end{figure*}

In Figure~\ref{fig:masuda_inclinations} we test these relations with the same examples used in \citet{Masuda:2020dp}.
Here we assume $R_*$ = 1.0 $\pm$ 0.1 \Rsun \ and $P_\mathrm{rot}$ = 5.059 $\pm$ 0.88~d, which is tailored to give $v_\mathrm{eq}$ = 10 $\pm$ 2 km s$^{-1}$.
Compared to their Figure~1, our results are in good agreement 
for two cases of projected rotational velocities: 
$v \sin i_*$ = 9.8 $\pm$ 2.0 km s$^{-1}$ and $v \sin i_*$ = 4.0 $\pm$ 2.0 km s$^{-1}$. 

\subsection{Upper Limits on $v \sin i_*$}{\label{app:vsini_ul}}

In some cases only an upper limit on $v \sin i_*$ is available.  This can occur if absorption lines are not fully resolved by an instrument 
or if a star is rotating slowly.  Here we derive analytical expressions for this scenario in a similar fashion as was carried out for normally distributed measurements 
of $v \sin i_*$ in Appendix~\ref{app:vsini_meas}.  

Following Equation 11 of \citet{Masuda:2020dp}, the posterior distribution of $\cos i_*$ can be written as an integral over $v \sin i$:

\begin{equation}{\label{eqn:pcosi_vsini}}
P(\cos i_* \mid d) \propto \frac{\mathcal{P}(\cos i_*)}{\sqrt{1 - \cos i_*^2}} \int \mathcal{L}_{v_{\mathrm{eq}}}\Big( \frac{v \sin i_*}{\sqrt{1 - \cos i_*^2}} \Big) \, \mathcal{L}_{v \sin i_*}( v \sin i_*) \, \mathcal{P}_{v_\mathrm{eq}} \Big( \frac{v \sin i_*}{\sqrt{1 - \cos i_*^2}}  \Big) \, d \big(v\sin i_*\big)
\end{equation}

\noindent For an upper limit $l$ on the projected rotational velocity, its likelihood function is simply a uniform distribution from  0 km s$^{-1}$ to $l$ km s$^{-1}$:  
 $\mathcal{L }_{v\sin i_*}( v \sin i_*)$ $\sim$ $\mathcal{U}(v \sin i_*=0,v \sin i_*=l)$ km s$^{-1}$.  This is equivalent to placing limits on the definite integral over $v \sin i_*$ from 0 to $l$.
 The likelihood function for the equatorial velocity, $ \mathcal{L}_{v_\mathrm{eq}}(v \sin i_*/ \sqrt{1 - \cos i_*^2})$, is approximately normally distributed following
 Equation~\ref{eqn:veq_likelihood} if the rotation period is measured to a precision of $\lesssim$20\%.
Adopting an isotropic prior for  $\mathcal{P}(\cos i_*)$ and a uniform prior for $v_\mathrm{eq}$, 
Equation~\ref{eqn:pcosi_vsini} becomes

\begin{equation}{\label{eqn:vsini_gauss}}
P(\cos i_* \mid d) \propto \int_{v \sin i_* = 0}^{v \sin i_* = l} \frac{1}{ \sqrt{2 \pi} s } e^{-\frac{1}{2}\big(\frac{v \sin i_* - m}{s}\big)^2}  \, d \big(v\sin i_*\big)
\end{equation}

\noindent where $m = \frac{2\pi R_*}{P_\mathrm{rot}}\sqrt{1 - \cos i_*^2}$, $s = \sigma_{v_\mathrm{eq}}\sqrt{1 - \cos i_*^2}$, and $\sigma_{v_\mathrm{eq}} =  \frac{2\pi R_*}{P_\mathrm{rot}} \sqrt{\big(\frac{\sigma_{R_*}}{R_*}\big)^2 + \big(\frac{\sigma_{P_\mathrm{rot}}}{P_\mathrm{rot}}\big)^2 }$.
Equation~\ref{eqn:vsini_gauss} is an integral over a Gaussian, the solution of which takes the following form:

\begin{equation}{\label{eqn:cosi_ul}}
P(\cos i_* \mid d) \propto \mathrm{erf}\big(\frac{l - m}{\sqrt{2} \, s}  \big) + \mathrm{erf}\big(\frac{m}{\sqrt{2} \, s}  \big)   
\end{equation}

\noindent and simplifies to

\begin{equation}{\label{eqn:cosi_ul}}
P(\cos i_* \mid d) \propto \mathrm{erf}\Big(\frac{l -  \frac{2\pi R_*}{P_\mathrm{rot}}\sqrt{1 - \cos i_*^2}}{\sqrt{2} \, \sigma_{v_\mathrm{eq}}\sqrt{1 - \cos i_*^2}}  \Big) + \mathrm{erf}\Big(\frac{\sqrt{2} \, \pi R_*}{\sigma_{v_\mathrm{eq}} P_\mathrm{rot} }  \Big)  
\end{equation}

\noindent 
Therefore, for an upper limit on $v \sin i_*$ of $l$ km s$^{-1}$, the posterior distribution for $\cos i_*$ is simply proportional to the sum of two error functions.

Similarly, for the posterior of the inclination $i_*$, a variable transformation from $\cos i_*$ to $i_*$ gives

\begin{equation}{\label{eqn:i_ul}}
P(i_* \mid d) \propto \sin i_* \times \Big( \mathrm{erf}\Big(\frac{l - \frac{2\pi R_*}{P_\mathrm{rot}} \sin i_*}{\sqrt{2} \, \sigma_{v_\mathrm{eq}} \sin i_*}  \Big) + \mathrm{erf}\Big(\frac{\sqrt{2} \, \pi R_*}{\sigma_{v_\mathrm{eq}} P_\mathrm{rot} }  \Big)   \Big).
\end{equation}

Several examples using Equations~\ref{eqn:cosi_ul} and \ref{eqn:i_ul} 
are shown in Figure~\ref{fig:masuda_inclinations_ul} following \citet{Masuda:2020dp}.  
As in Appendix~\ref{app:vsini_meas}, we use $R_*$ = 1.0 $\pm$ 0.1 \Rsun \ and $P_\mathrm{rot}$ = 5.059 $\pm$ 0.88~d for this exercise.
For a star with $v_\mathrm{eq}$ = 10 $\pm$ 2 km s$^{-1}$, 
we vary the upper limit of a hypothetical $v \sin i_*$ measurement 
from 2.5 km s$^{-1}$ to 30 km s$^{-1}$.  As the upper limit moves beyond the equatorial rotational velocity,
the posterior distribution of $\cos i_*$ tends to unity and for $i_*$ it tends to an isotropic distribution.

\begin{figure*}
  \vskip -0.7 in
  \hskip 0.3 in
  \resizebox{6.5in}{!}{\includegraphics{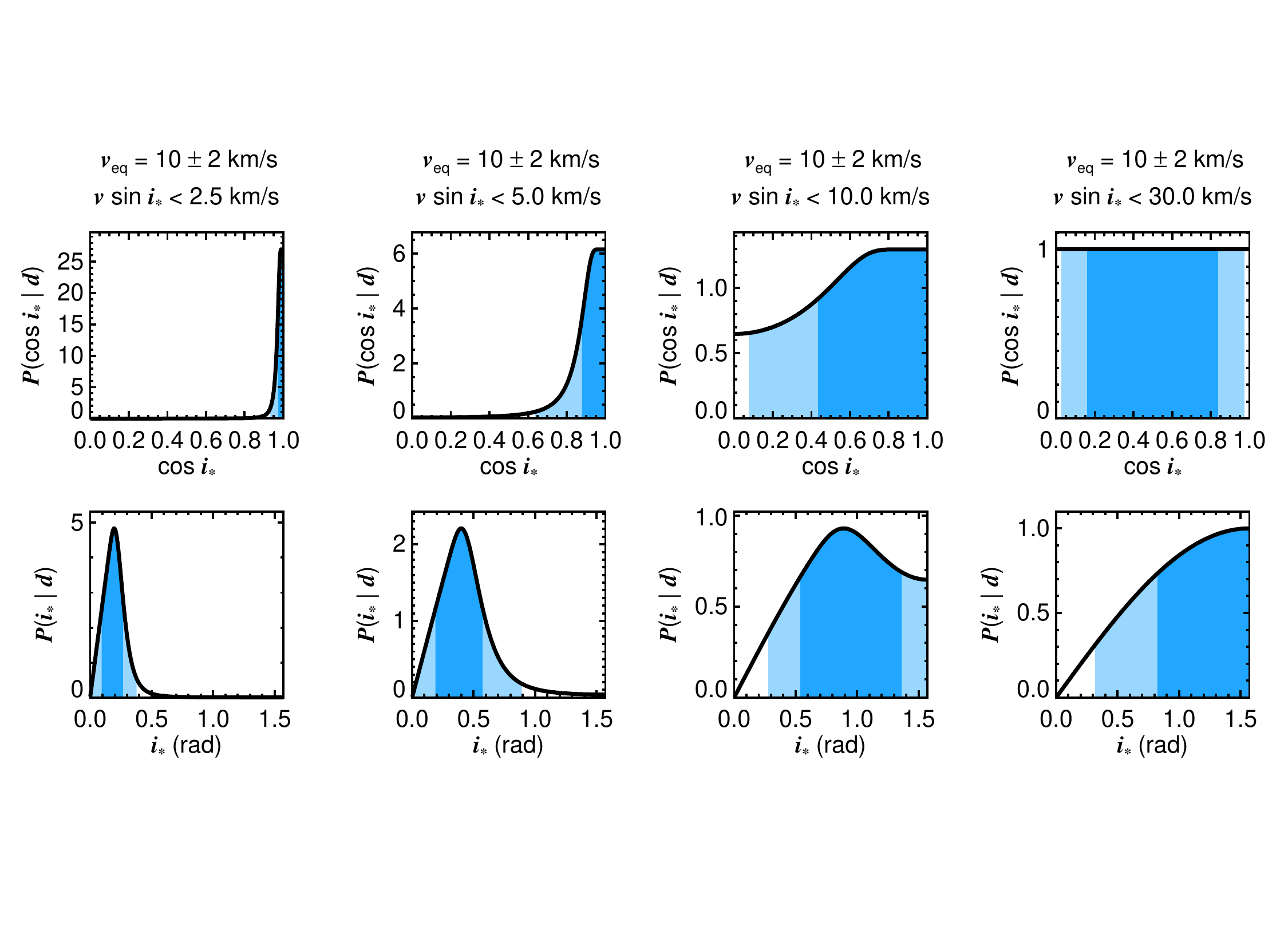}}
  \vskip -0.8 in 
  \caption{Same as Figure~\ref{fig:masuda_inclinations} but for examples of upper limits on $v \sin i_*$ from \citet{Masuda:2020dp} using our Equations~\ref{eqn:cosi_ul} and \ref{eqn:i_ul}. Shaded regions correspond to 68.3\% and 95.4\% credible intervals.  \label{fig:masuda_inclinations_ul} } 
\end{figure*}

\section{Notes on Individual Stars} \label{sec:notes}

\textbf{1RXS J034231.8+121622}
This young mid-M dwarf is a likely member of the Hyades cluster (\citealt{Kuzuhara:2022aa}).
Its brown dwarf companion was found by \citet{Bowler:2015ja} and the latest orbit for the pair is determined in \citet{Bowler:2020hk}.
Our $v \sin i_*$ measurement of 5.1 $\pm$ 2.3 km s$^{-1}$ agrees with the upper limit of 7 km s$^{-1}$ from \citet{LopezValdivia:2019aa}.
Likewise, our rotation period of 7.3 $\pm$ 0.3 d agrees with the value of 7.44~d from \citet{Newton:2016ea}.

\textbf{2MASS J01033563--5515561:}
This low-mass hierarchical triple system comprises a close visual binary (AB) with an unresolved spectral type of M5--M6 and a wide 12--14~\Mjup \ accreting L-dwarf companion, 2MASS J01033563--5515561 b (\citealt{Delorme:2013bo}; \citealt{Eriksson:2020aa}; \citealt{Betti:2022aa}).
The light curve from two \emph{TESS} sectors shows a strong periodic signal at 0.1664 $\pm$ 0.0003 d (3.994 $\pm$ 0.007 hr),
which is consistent with fast rotation expected as a low-mass member of the $\approx$45~Myr Tuc-Hor young moving group (\citealt{Kraus:2014ur}).
It is unclear which member of the binary is producing the observed modulations, but their near-equal flux ratios implies that they share similar physical properties.

To estimate their radii, we adopt an effective temperature of 2980~$\pm$~75~K and $J$-band bolometric correction of 1.99 mag for young stars from 
\citet{Herczeg:2015bp} for 2MASS J01033563--5515561 A, which corresponds to young M5 stars.  Assuming the B component is slightly cooler with a
spectral type of M5.5, this corresponds to $T_\mathrm{eff}$ = 2920 $\pm$ 75 K and BC$_J$=2.01 mag.  
The absolute $J$-band magnitude of 2MASS J01033563--5515561 A is $M_J$ = 7.36~$\pm$~0.05~mag (\citealt{Delorme:2013bo}), 
which corresponds to a bolometric magnitude
of $M_\mathrm{bol}$ = 9.35~$\pm$~0.05 mag, a luminosity of $\log{L_\mathrm{bol}/L_{\odot}}$ = --1.84~$\pm$~0.02 dex, 
and a radius of $R_*$ = 0.45~$\pm$~0.03 $R_{\odot}$.
The properties are similar for B: $M_J$ = 7.56~$\pm$~0.05~mag (\citealt{Delorme:2013bo}), 
$M_\mathrm{bol}$ = 9.57~$\pm$~0.05 mag, $\log{L_\mathrm{bol}/L_{\odot}}$ = --1.93~$\pm$~0.02 dex, and a radius of $R_*$ = 0.43~$\pm$~0.02 $R_{\odot}$.  

For the inclination analysis we assume the properties of A, although given the consistent radii the results would be similar for B.
We find a stellar inclination of $i_*$=8.9$^{+0.9}_{-1.3}\degr$.  In the future this measurement may be compared 
with the orbital angular momentum of the AB binary and, in principle, both the spin and orbit of the companion.

\textbf{2MASS J01225093--2439505:}
This M3.5 star is a member of the AB~Dor young moving group and hosts an imaged L4 substellar 
companion with a mass near the deuterium-burning limit (\citealt{Bowler:2013ek}; \citealt{Hinkley:2015gk}).

2MASS J01225093--2439505 has an inferred equatorial rotational velocity of 12.4 $\pm$ 1.0 km s$^{-1}$, which is
inconsistent with the projected rotational velocity of 18.1 $\pm$ 0.5 km s$^{-1}$.  
The equatorial rotational velocity is 
based on a radius of 0.37~$\pm$ 0.03~\Rsun \ from \citet{Stassun:2019aa} and a rotation period of 1.493 $\pm$ 0.017 d from our
analysis of the \emph{TESS} light curve spanning two sectors.  Note that this rotation period is nearly identical to the value 
of 1.49 $\pm$ 0.02 d determined by 
\citet{Bryan:2020ji} from a single \emph{TESS} sector.  
The inconsistency between $v_\mathrm{eq}$ and $v \sin i_*$ implies that either the radius is too small, 
the rotation period is overestimated, the $v \sin i_*$ value is overestimated, or
some combination of these.  

The \emph{TESS} light curve covers about 26 period cycles and shows regular single-mode modulations in 
Sector 3 followed by double-peaked modulations in Sector 30 with different peak amplitudes.  Because of this double-peaked structure,
which seems to trace two dominant groups of starspots at different longitudes, it is unlikely that the 
signal itself is an alias of a true underlying signal (as was discussed for LP 261-75 in Section~\ref{sec:tess}).
However, it is possible that mid-latitude starspots coupled with strong differential rotation could cause light curve modulations
that are longer than the equatorial period, which might account for the discrepancy between $v_\mathrm{eq}$ and $v \sin i_*$.

It is also possible that the radius is underestimated.  
If the radius is the sole origin of the discrepancy between $v_\mathrm{eq}$ and $v \sin i_*$, 
a size of $\approx$0.53~\Rsun \ would be needed to reconcile the two values.
We recompute the radius using 
the apparent 2MASS $J$-band magnitude of 10.084 $\pm$ 0.028~mag, 
a $J$-band bolometric correction of 1.885~mag (midway between M3 and M4 from \citealt{Herczeg:2015bp}),
an effective temperature of 3300~K (\citealt{Herczeg:2015bp}), and a $Gaia$ DR3-based 
distance of 33.74 $\pm$ 0.03~pc (\citealt{GaiaCollaboration:2022aa}).
This yields a bolometric magnitude of $M_\mathrm{bol}$ = 9.33~$\pm$~0.05 mag, 
a luminosity of $\log{L_\mathrm{bol}/L_{\odot}}$ = --1.83~$\pm$~0.02 dex, 
and a radius of $R_*$ = 0.37~$\pm$~0.02 $R_{\odot}$---in good agreement with the 
TIC value from \citet{Stassun:2019aa}.

Finally, it is also possible that the $v \sin i_*$ value of 2MASS J01225093-2439505 may be overestimated.
Our adopted value of 18.1  $\pm$  0.5 km s$^{-1}$ is a weighted average of 14.2  $\pm$  3.2  km s$^{-1}$ from  \citet{Malo:2014dk}   
and  18.2  $\pm$  0.5  km s$^{-1}$ from \citet{Bryan:2020ji}.
Additional measurements of rotational broadening for this star would help clarify the nature of the discrepancy 
between $v_\mathrm{eq}$ and  $v \sin i_*$.

\textbf{FU Tau:}
FU Tau A is a young accreting late-M dwarf in the Taurus star-forming complex (\citealt{Jones:1979aa}) 
which may be driving a large molecular outflow (\citealt{Monin:2013ep}; although see \citealt{Wu:2020aa} for an alternate interpretation of the line emission).
Its wide companion, FU Tau B, was discovered by \citet{Luhman:2009cx} and is expected to have a mass near the deuterium-burning boundary.

The inferred mass and radius of FU Tau A is highly sensitive to the measured spectral type, extinction, 
effective temperature, and system age; values of 0.05--0.2~\Msun \ and 0.55--1.8 \Rsun \ have been previously estimated 
(\citealt{Luhman:2009cx}; \citealt{Stelzer:2010aa}; \citealt{Stelzer:2013aa}; \citealt{Bulger:2014ei}).
The parallax of FU Tau A in $Gaia$ DR3 (\citealt{GaiaCollaboration:2016cu}; \citealt{GaiaCollaboration:2022aa}) 
is 7.835 $\pm$ 0.075~mas, which corresponds to a distance estimate of 126.4 $\pm$ 1.0 pc (\citealt{BailerJones:2021aa}).
This is about 10\% closer than the 140~pc value that was often used in the past for Taurus members that lacked direct distance measurements, including FU Tau.
We therefore re-derive the fundamental parameters of FU Tau A following a similar procedure as in \citet{Stelzer:2013aa} using 
an (extinction-corrected) $J$-band bolometric correction and the revised distance. 

Here we use a $J$-band bolometric correction of BC$_J$ = 2.045~mag from \citet{Herczeg:2015bp}, which is intermediate between young M6 and M7 stars
and assumes a spectral type of M6.5 (\citealt{Stelzer:2013aa}; \citealt{Herczeg:2014is}) for FU Tau A.
The optical extinction of $A_V$ = 0.5~$\pm$~0.5~mag from \citet{Stelzer:2013aa} corresponds to a $J$-band extinction of $A_J$ = 0.17~$\pm$~0.11~mag
based on infrared interstellar reddening law in \citet{Tokunaga:2000tr}.
This gives an absolute $J$-band magnitude of $M_J$ = 5.10 $\pm$ 0.12~mag for FU~Tau A,
a bolometric magnitude of 
$M_\mathrm{bol}$ = 7.15~$\pm$~0.12 mag, a bolometric luminosity of $\log{L_\mathrm{bol}/L_{\odot}}$ = --0.96~$\pm$~0.05 dex, 
and a radius of $R_*$ = 1.4~$\pm$~0.1 $R_{\odot}$.  
These values rely on an effective temperature of $T_{\mathrm{eff}}$ = 2815 $\pm$ 75~K for a young M6.5 star (\citealt{Herczeg:2015bp}).

Based on the periodicity at 3.93~d we find from the \emph{TESS} light curve, we infer an inclination of $i_*$=75$^{+14}_{-5}\degr$. 
The mm disk around FU Tau was detected by \citet{Wu:2020aa} with ALMA but was not spatially resolved in that data.  
In the future, the inclination of FU Tau A's disk and FU Tau B's spin orientation may be compared with the inclination measured here.

\textbf{FW Tau:}
FW Tau is a binary pair of young low-mass stars in the Taurus star-forming complex.
The pair has equal flux ratio, implying similar physical properties, and shows significant
orbital motion since 1989 (\citealt{Schaefer:2014it}).
\citet{Bowler:2014dk} found a spectral type of M6~$\pm$~1 and modest extinction of $A_V$=0.4$^{+1.3}_{-0.4}$~mag.
A faint companion was found with HST imaging by \citet{White:2001ic} and confirmed to be comoving by \citet{Kraus:2014tl},
however follow-up studies suggest FW Tau C may be a brown dwarf or a low-mass star with an edge-on disk 
(\citealt{Bowler:2014dk}; \citealt{Kraus:2015fx}; \citealt{Caceres:2015hg}; \citealt{Wu:2017aa}; \citealt{Mora:2020aa}).

We compute the radii of FW Tau A and B by assuming each component is identical in luminosity and effective temperature.  No parallax is available in $Gaia$ DR3 for
FW Tau, presumably because of its binary nature, so we adopt a characteristic distance of 140~$\pm$~10~pc for Taurus (\citealt{Kenyon:1994aa}). 
Using a bolometric correction of 2.03~mag and $T_\mathrm{eff}$ = 2860 $\pm$ 75~K from \citet{Herczeg:2015bp},
an optical extinction of $A_V$ = 0.4~mag (implying a $J$-band extinction of $A_J$ = 0.11~mag following \citealt{Tokunaga:2000tr}), and an apparent magnitude decomposed
into equal-brightness individual magnitudes of $J$ = 11.1~$\pm$ 0.05~mag, we find
$M_J$ = 5.25 $\pm$ 0.16~mag, 
$M_\mathrm{bol}$ = 7.28~$\pm$~0.16 mag, $\log{L_\mathrm{bol}/L_{\odot}}$ = --1.01~$\pm$~0.06 dex, and a radius of $R_*$ = 1.28~$\pm$~0.11 $R_{\odot}$.  

Our stellar rotation period of 0.91 d agrees well with the value of 0.906 d found by \citet{Chen:2020aa} using $g$-band photometry
from the Zwicky Transient Factory.  They also report an $r$-band period of 0.475~d, which is probably an alias of the true rotation period.
Combined with the $v \sin i_*$ of 49.2 $\pm$ 0.7 km s$^{-1}$ from \citet{Kounkel:2019cp}, the resulting
posterior stellar inclination distribution peaks at $i_*$=44$\degr$ and the 68\% credible interval spans 39--49$\degr$.
Although the orbit of AB is not yet well constrained (\citealt{Schaefer:2014it}), its inclination and the properties of FW Tau C 
such as its disk inclination and spin inclination can
eventually be compared with this stellar inclination to shed light on the formation of this hierarchical multiple system.

\textbf{Gl 229:}
This nearby (5.7614 $\pm$ 0.0006 pc; \citealt{GaiaCollaboration:2022aa})  M1 star hosts the first T dwarf companion to have been directly imaged 
(\citealt{Nakajima:1994ea}; \citealt{Oppenheimer:1995wl}; \citealt{Oppenheimer:2014et}).
With over 25 years of relative and absolute astrometry, the orbit and dynamical mass of Gl 229 B have now been exquisitely measured (\citealt{Brandt:2020ce}; \citealt{Brandt:2021cc}; \citealt{Feng:2022zz}).

The equatorial velocity of 1.0 $\pm$ 0.2 km s$^{-1}$ that we infer for Gl 229 is slightly (but significantly) smaller than its
$v \sin i_*$ value of 2.8 $\pm$ 0.4 km s$^{-1}$, which is based on a compilation of projected rotational velocities from the literature (see Table~\ref{tab:vsinis}).
Most of the literature values do not include estimates of the measurement uncertainties, so these are grouped into a single 
value following the procedure outlined in Section~\ref{sec:vsinicomp}.
Note that our adopted value is dominated by the measurement of 2.6 $\pm$ 0.4 km s$^{-1}$ from \citet{Hojjatpanah:2019aa}, which is
more heavily weighted because of its small reported uncertainties.

Most of these published $v \sin i_*$ measurements indicate that Gl 229 is a slow rotator, 
which is consistent with its old age (2--6~Gyr; \citealt{Brandt:2020ce}), lack of activity signatures, and long
rotation period of 27 d (\citealt{SuarezMascareno:2016aa}).
Several $v \sin i_*$ measurements fall below 2 km s$^{-1}$ (\citet{Glebocki:2005aa};  \citealt{Reiners:2007aa}; \citealt{Reiners:2022aa}), 
which at this level becomes very challenging to determine
because it can be difficult to separate contributions to line broadening from rotation, microturbulence, and macroturbulence.  
The resulting obliquity constraint for Gl 229 may be suffer from a systematic bias because of an overestimated $v \sin i_*$ value.

The disagreement could also stem from the radius estimate or the photometric rotation period measurement.
The radius 
of 0.55 $\pm$ 0.06~\Rsun \ we adopt is similar to several other estimates such as 0.58 $\pm$ 0.05 \Rsun \ from \citet{Houdebine:2016aa} 
and 0.52 $\pm$ 0.02 \Rsun \ from \citet{Schweitzer:2019aa}.
An independent rotation period measurement would also be valuable for this star to compare with the 27~d value found by \citet{SuarezMascareno:2016aa}.

\textbf{SR 12 AB:} 
SR 12 AB (2MASS J16271951–2441403, ROXs 21) is a close resolved binary (\citealt{Simon:1987ga}) 
and a well-established member of the $\rho$ Ophiuchus cloud complex (e.g., \citealt{Bouvier:1992ww}).
\citet{Kuzuhara:2011ic} discovered a widely separated companion, SR 12 c, with a mass near the deuterium-burning limit.
SR 12 c was later found to have an accretion disk (\citealt{SantamariaMiranda:2017aa}; \citealt{Martinez:2022aa}) with a dust mass of about 
one lunar mass based on a sub-mm detection with ALMA (\citealt{Wu:2022aa}).

Here we derive the radius of SR 12 A using the Stefan-Boltzmann relation, but results are expected to be similar if the
photometric rotation originates from SR 12 B.  
\citet{Schaefer:2018aa} measured a flux ratio in the narrow-band Keck/NIRC2 $J_\mathrm{cont}$ filter of 0.951 $\pm$ 0.006, or 0.055 $\pm$ 0.006 mag.  
The apparent integrated-light $J$-band magnitude of SR 12 is 9.424~$\pm$ 0.023 mag (\citealt{Cutri:2003tp}),
which decomposes into apparent magnitudes of $J_A$ = 10.149 $\pm$ 0.023~mag and $J_B$ = 10.204 $\pm$ 0.023~mag.
A parallax of 8.9034 $\pm$ 0.4288~mas ($d$ = 112 $\pm$ 5~pc; \citealt{BailerJones:2018aa}) is listed in \emph{Gaia} DR2, 
but no parallax is present in DR3, likely because orbital motion of the binary is evident in the updated astrometry (the excess astrometric noise parameter
is 7.121~mas but no RUWE value is listed).
We therefore instead use the \emph{Gaia}-based distance of 138.4~$\pm$ 2.6~pc inferred by \citet{OrtizLeon:2018aa}
for other members of the L1688 cloud.
Adopting a spectral type of M0 from \citet{Wilking:2005aq} implies an effective temperature of $T_\mathrm{eff}$ = 3900~$\pm$ 125~K (\citealt{Herczeg:2015bp}).
Here we assume the $J$-band extinction of $A_J$ = 0.5 $\pm$ 0.2 mag from \citet{Kuzuhara:2011ic}.
This implies an extinction-corrected absolute magnitude of $M_J$ = 3.95 $\pm$ 0.21~mag, a bolometric magnitude of 
$M_\mathrm{bol}$ = 5.61~$\pm$~0.21 mag, a bolometric luminosity of $\log{L_\mathrm{bol}/L_{\odot}}$ = --0.34~$\pm$~0.08 dex, 
and a radius of $R_*$ = 1.48~$\pm$~0.17 $R_{\odot}$.  

For this study we adopt the rotation period of 3.918 d found by \citet{Rebull:2018aa} from \emph{K2},
which is similar to the period of 3.927~d from \citet{Kiraga:2012wj} and 3.516~d from \citet{Grankin:2008aa}.
Note that \citet{Jayasinghe:2018aa} report a period of 7.849 d using light curves from the All-Sky Automated Survey for Supernovae (ASAS-SN);
this is about twice the value from \citet{Rebull:2018aa} so one is probably an alias of the true period.  
The cadence of ASAS-SN is every few days, whereas $K2$ is much finer ($\approx$30~min) and therefore should be more reliable.

Combining our radius estimate with the rotation period and adopted $v \sin i_*$ value of 27~$\pm$ 11 km s$^{-1}$ (Table~\ref{tab:vsinis})
yields a broad inclination constraint that only slightly deviates from the $\sin i_*$ prior.  Much of this is caused by the large uncertainty in projected
rotational velocity.  Orbital motion of AB has been detected but is not yet well determined (\citealt{Schaefer:2018aa});
in the future the binary orbital plane and the spin inclination of the wide companion (\citealt{Bryan:2020ex}) 
can be compared with this host inclination to test for alignment.

\textbf{TWA 5 Aab:} 
TWA 5Aab is a M2 member of the $\approx$10~Myr TW Hydrae association.  
The brown dwarf companion TWA~5 B was found by \citet{Webb:1999kf} and \citet{Lowrance:1999ck}, and the host was
later resolved into a close visual binary by \citet{Macintosh:2001tr}.

\citet{Kohler:2013im} measured a dynamical mass of 0.9~$\pm$~0.1~\Msun \ for the total mass of TWA 5Aab
and a mass ratio of 1.3, implying individual masses of about 0.4~\Msun \  and 0.5~\Msun \ of the Ab and Aa components, 
respectively---albeit with large uncertainties.
The radius of TWA 5 is listed as 0.40 $\pm$ 0.07~\Rsun \ in \citet{Stassun:2018jr}, but this value differs substantially from theoretical expectations,
perhaps because TWA 5 is a close binary.
For example, solar-metallicity MIST models (\citealt{Dotter:2016fa}; \citealt{Choi:2016kf}) predict a 
radius of 0.88~\Rsun \ for a 0.5~\Msun \ star at 10 Myr.
We instead adopt the weighted mean of radius estimates from \citet{Stassun:2019aa} for two other coeval 
stars in the TWA association with the same spectral type of M2, TWA 7 (0.917 $\pm$ 0.119~\Rsun)
and TWA 10 (0.860 $\pm$ 0.114~\Rsun).  Following the discussion in Section~\ref{sec:radii}, we also add a 7\%
uncertainty in quadrature with the original error estimates.
This yields 0.89~$\pm$~0.10~\Rsun.


\section{\emph{TESS} light curves}{\label{app:tess}}

\begin{figure*}
  \hskip -0.8 in
  \gridline{\fig{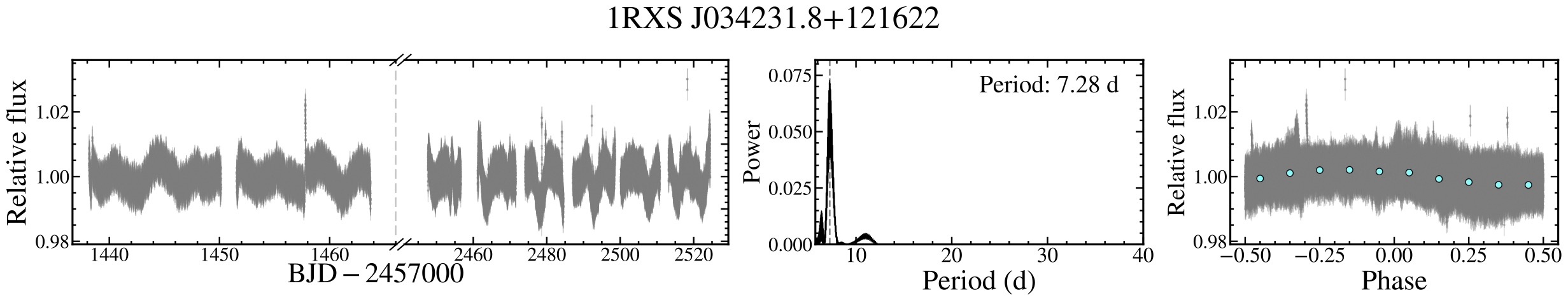}{0.9\textwidth}{}}
  \vskip -.3 in
  \gridline{\fig{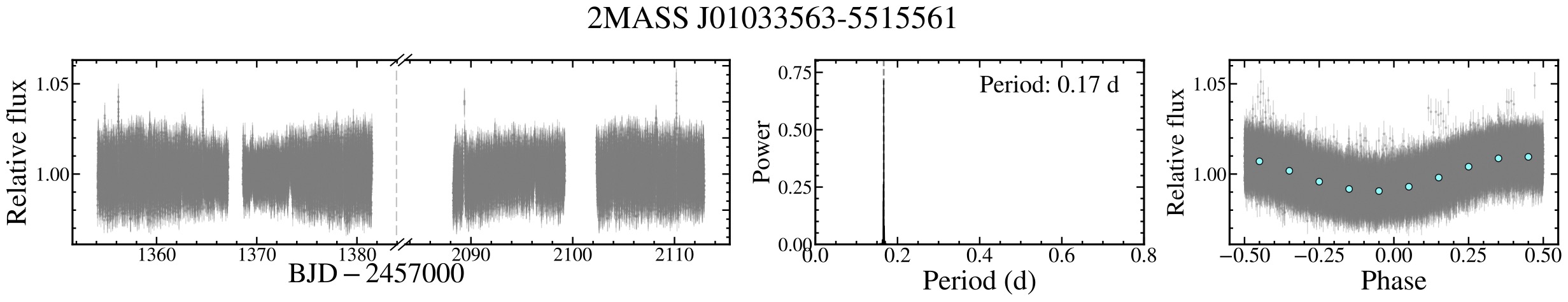}{0.9\textwidth}{}}     
  \vskip -.3 in
  \gridline{\fig{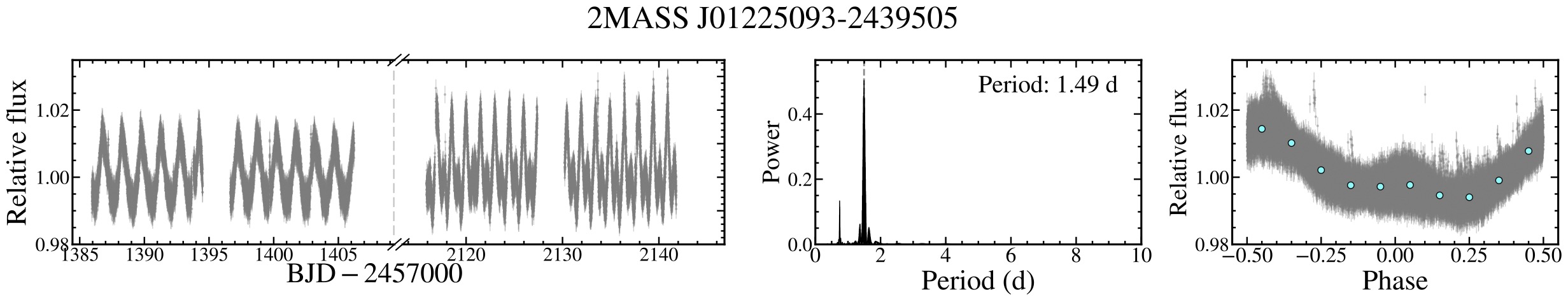}{0.9\textwidth}{}}     
  \vskip -.3 in
  \gridline{\fig{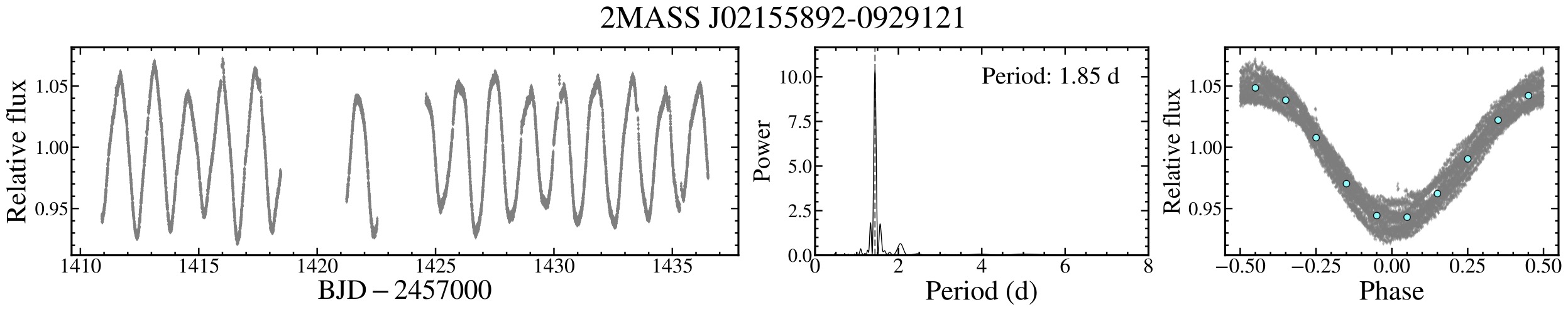}{0.9\textwidth}{}}                
  \vskip -.3 in
  \gridline{\fig{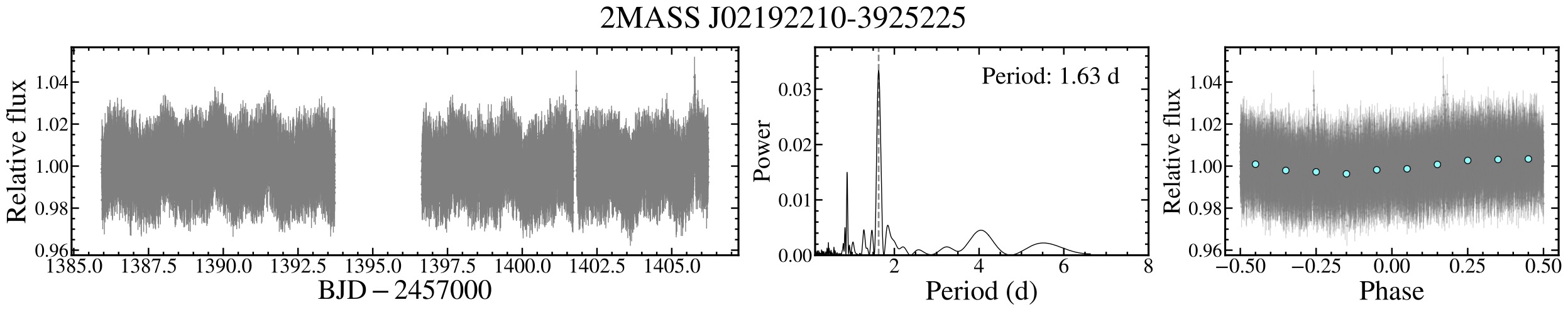}{0.9\textwidth}{}}                
  \vskip -.3 in
  \gridline{\fig{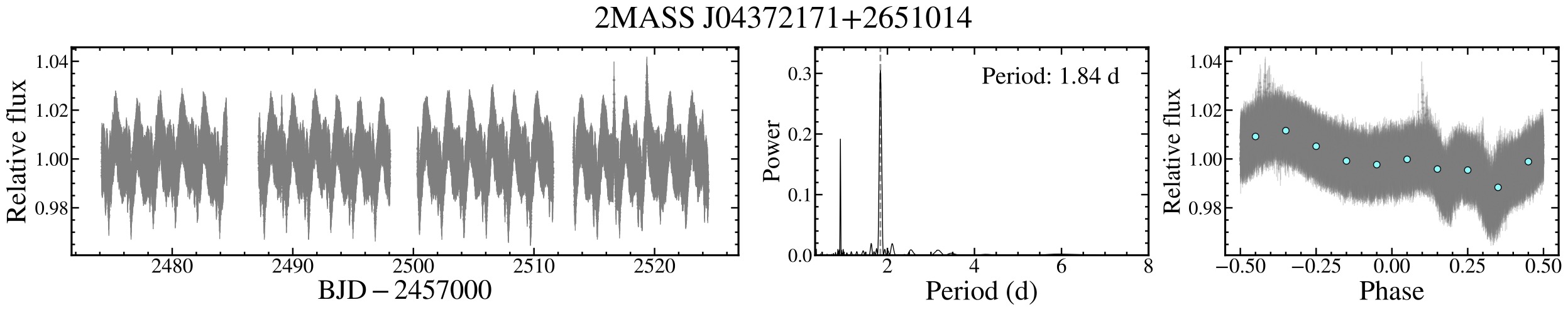}{0.9\textwidth}{}} 
  \vskip -.3 in
  \caption{Normalized \emph{TESS} light curves and Generalized Lomb-Scargle periodograms for 1RXS~J0342+1216, 2MASS~J01033563--5515561, 2MASS~J01225093--2439505, 2MASS~J02155892--0929121, 2MASS~J02192210--3925225, and 2MASS~J04372171+2651014.   
  For each light curve the strongest periodogram peak is marked with a dotted vertical line.  
  Phased light curves are shown in the right panel with median binned phases denoted by cyan circles. 
  \label{fig:tess_1} } 
\end{figure*}

\begin{figure*}
  \hskip -0.8 in
  \gridline{\fig{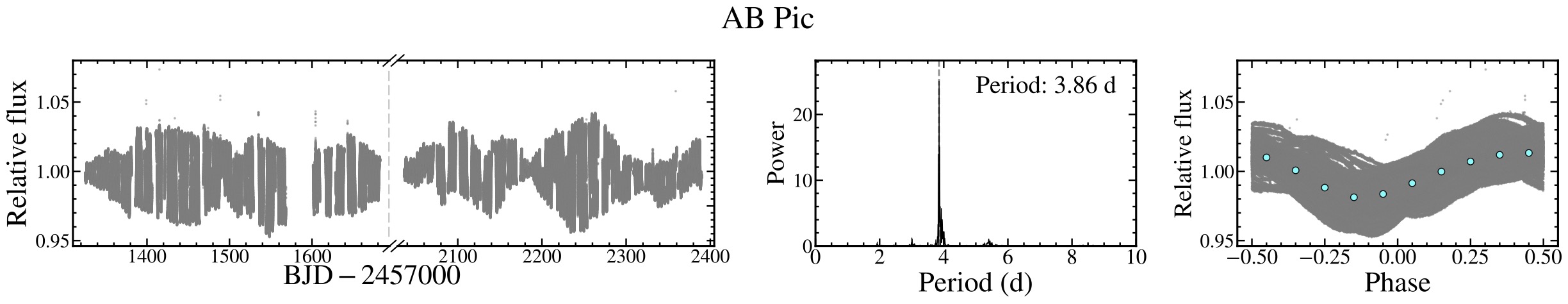}{0.9\textwidth}{}}
  \vskip -.3 in
  \gridline{\fig{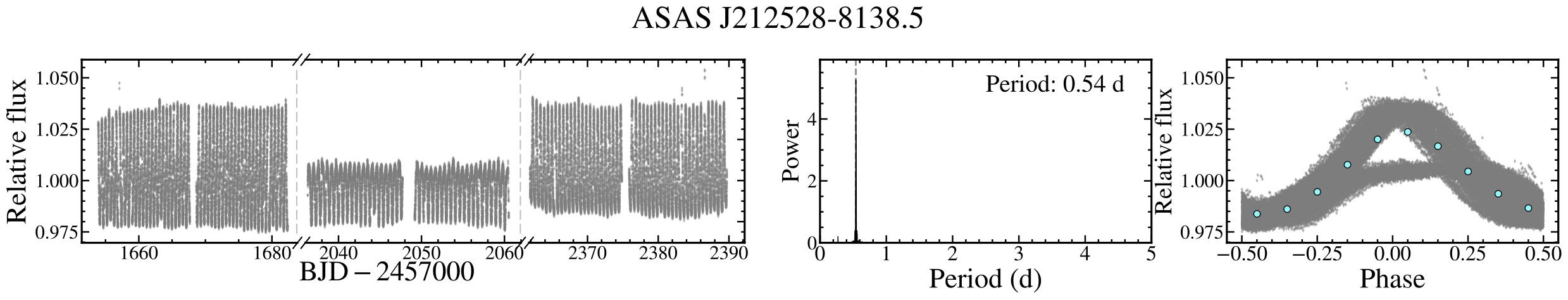}{0.9\textwidth}{}} 
  \vskip -.3 in
  \gridline{\fig{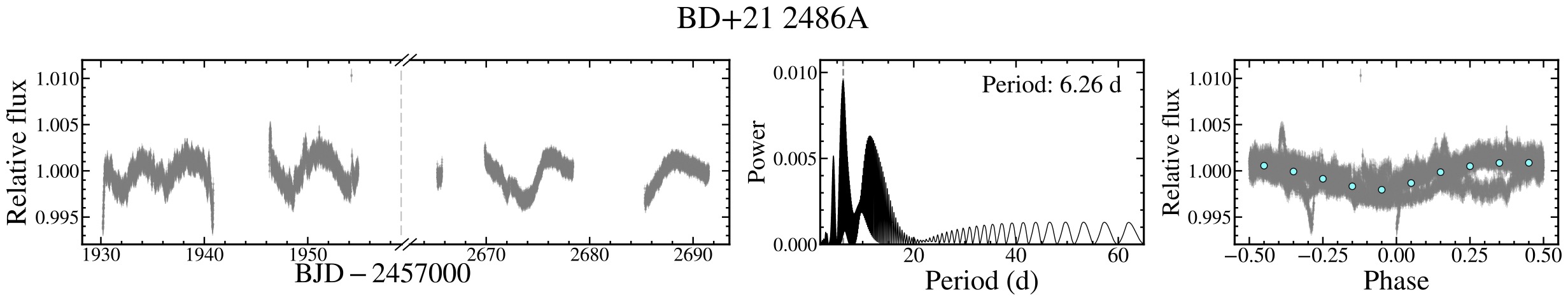}{0.9\textwidth}{}}     
  \vskip -.3 in
  \gridline{\fig{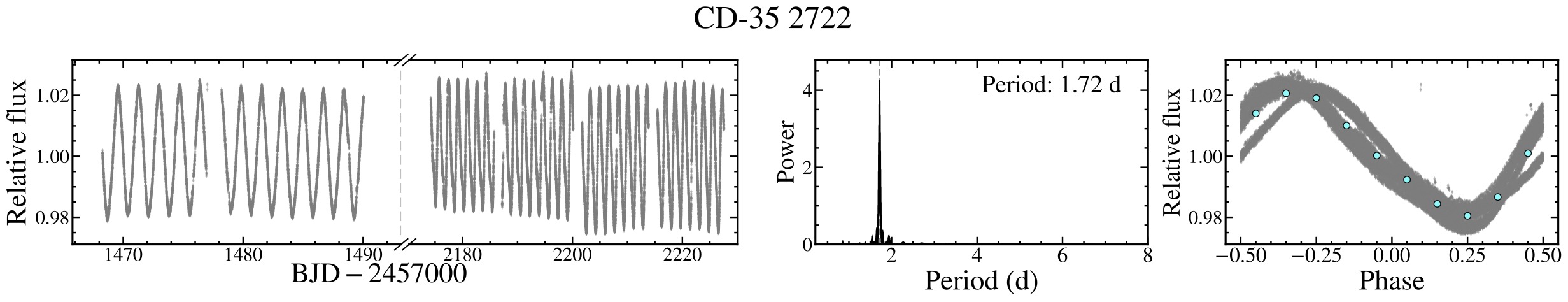}{0.9\textwidth}{}}                
  \vskip -.3 in
  \gridline{\fig{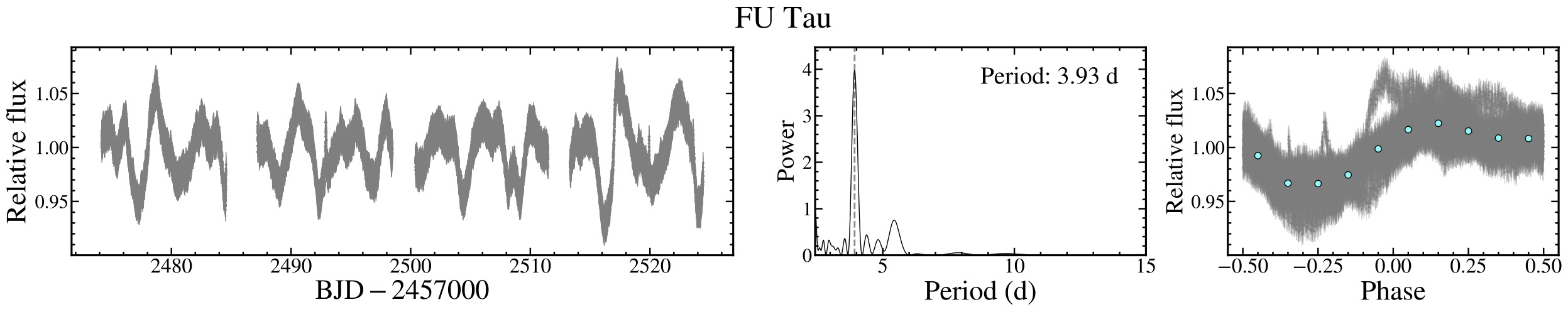}{0.9\textwidth}{}}
  \vskip -.3 in
  \gridline{\fig{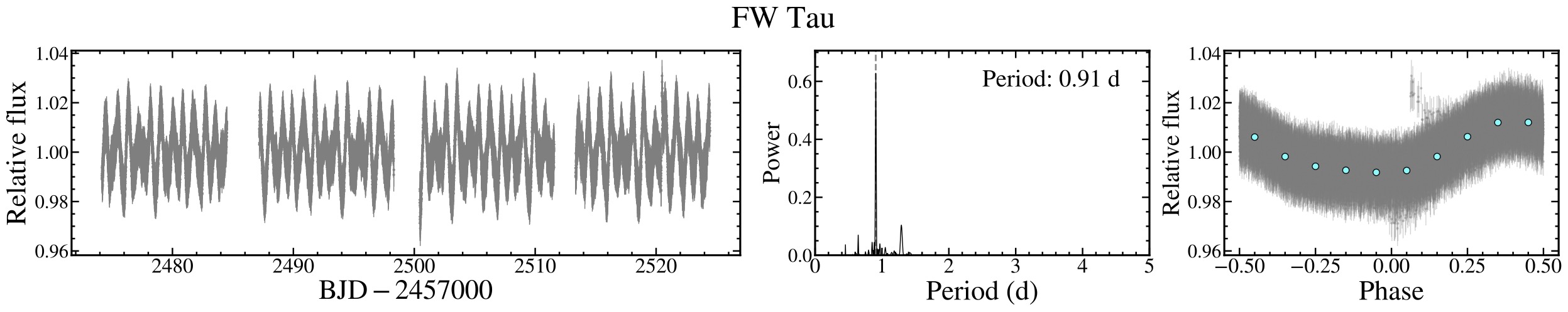}{0.9\textwidth}{}} 
  \vskip -.3 in
  \caption{\emph{TESS} light curves and periodograms for AB Pic, ASAS J212528--8138.5, BD+21 2486 AB, CD--35 2722, FU Tau, and FW Tau. 
  \label{fig:tess_2} } 
\end{figure*}

\begin{figure*}
  \hskip -0.8 in
  \gridline{\fig{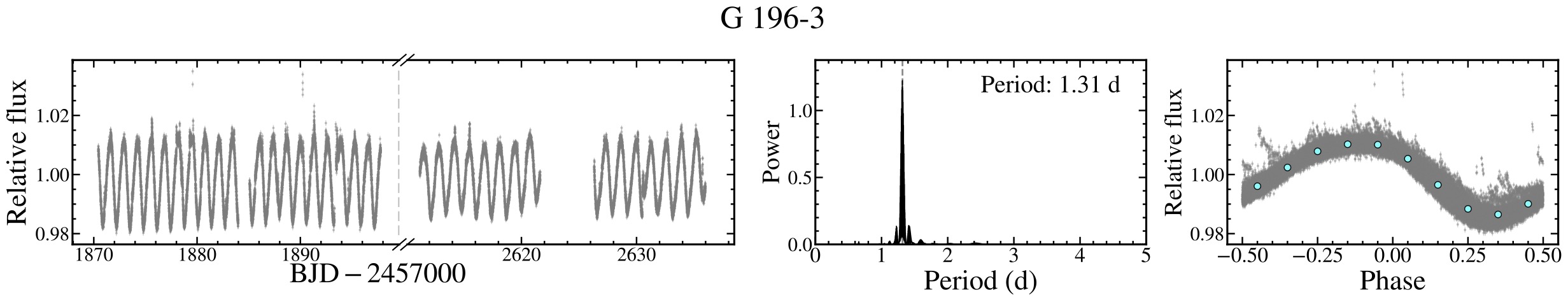}{0.9\textwidth}{}}     
  \vskip -.3 in
  \gridline{\fig{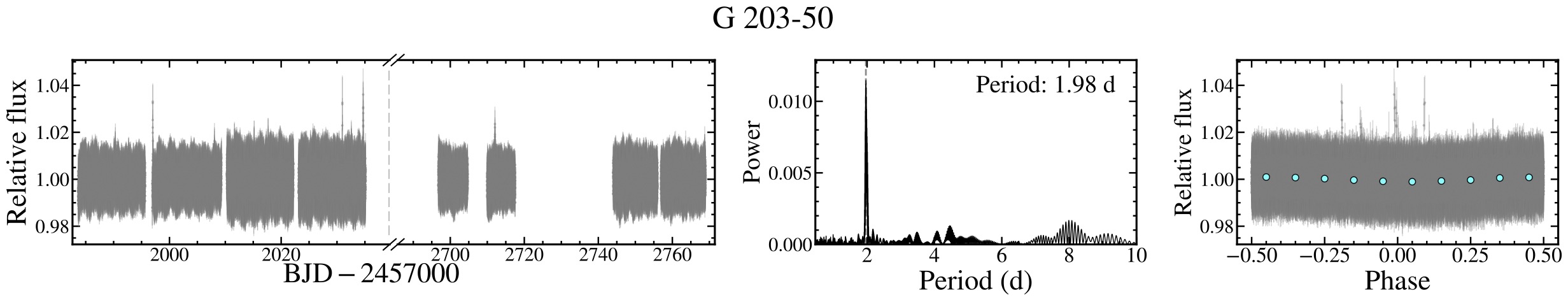}{0.9\textwidth}{}}                
  \vskip -.3 in
  \gridline{\fig{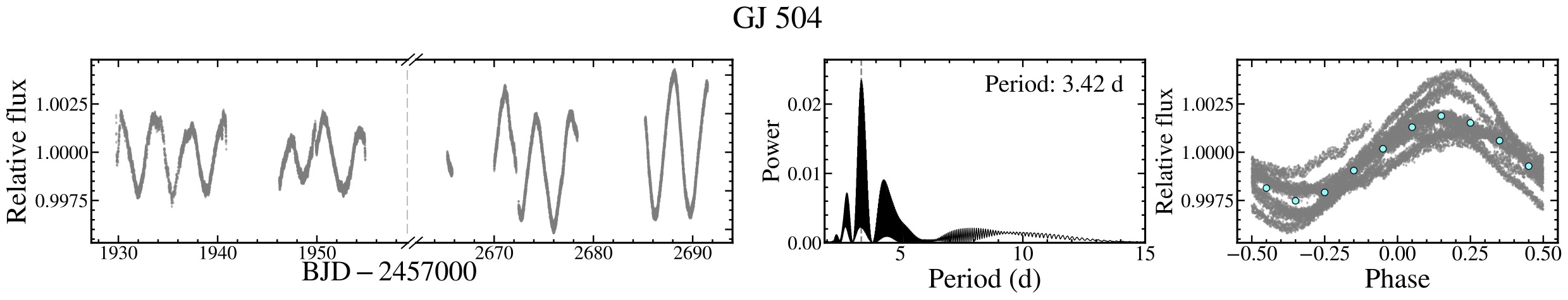}{0.9\textwidth}{}}                
  \vskip -.3 in
  \gridline{\fig{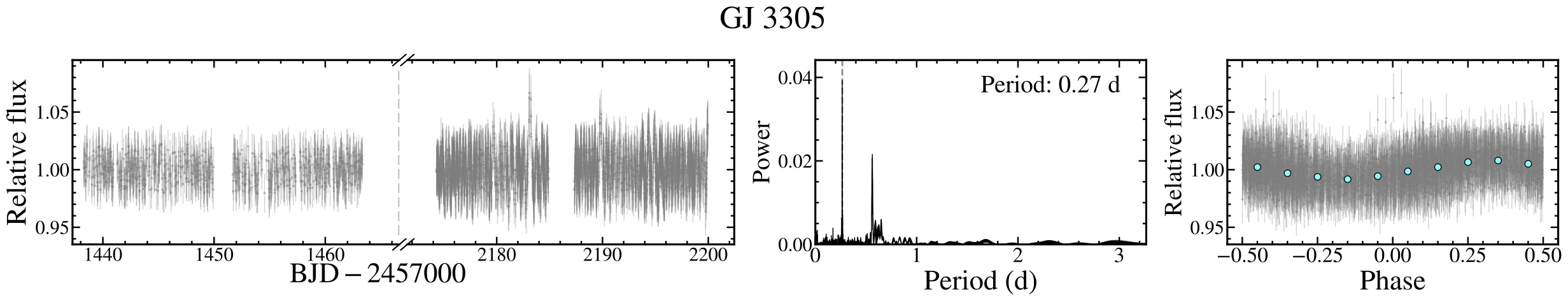}{0.9\textwidth}{}}                
  \vskip -.3 in
  \gridline{\fig{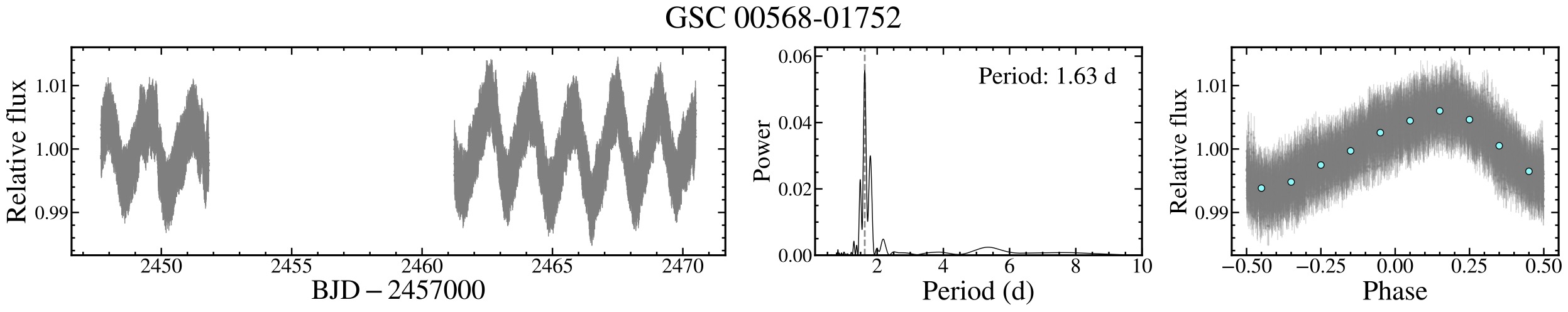}{0.9\textwidth}{}} 
  \vskip -.3 in
  \gridline{\fig{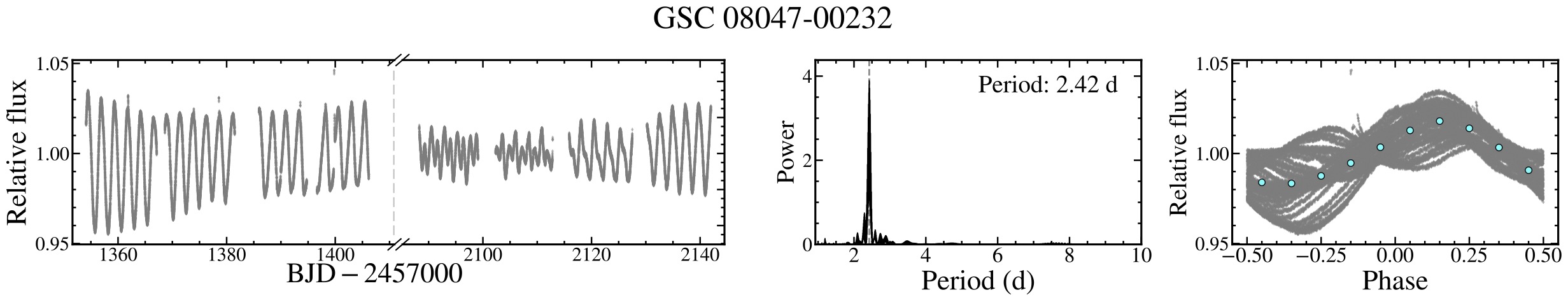}{0.9\textwidth}{}}
  \vskip -.3 in
  \caption{\emph{TESS} light curves and periodograms for G 196-3, G 203-50, GJ 504, GJ 3305 AB, GSC 00568-01752, and GSC 08047-00232. 
  \label{fig:tess_3} } 
\end{figure*}

\begin{figure*}
  \hskip -0.8 in
  \gridline{\fig{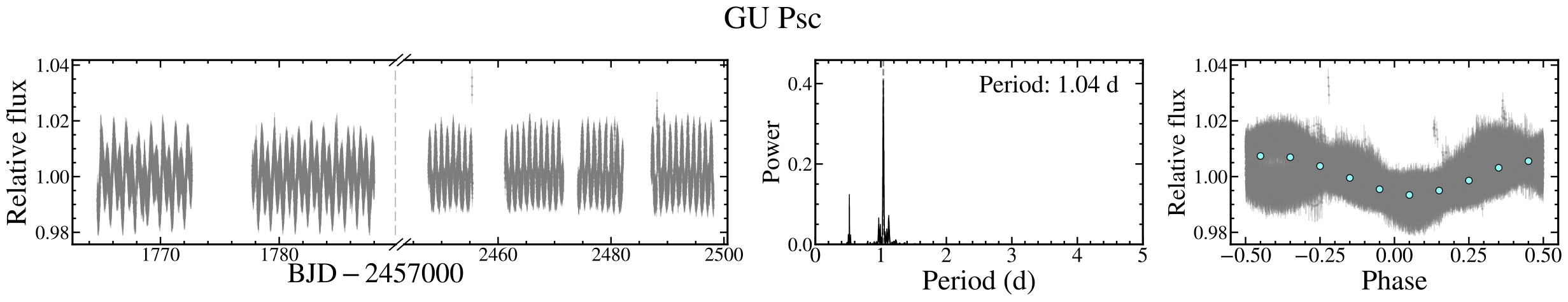}{0.9\textwidth}{}} 
  \vskip -.3 in
  \gridline{\fig{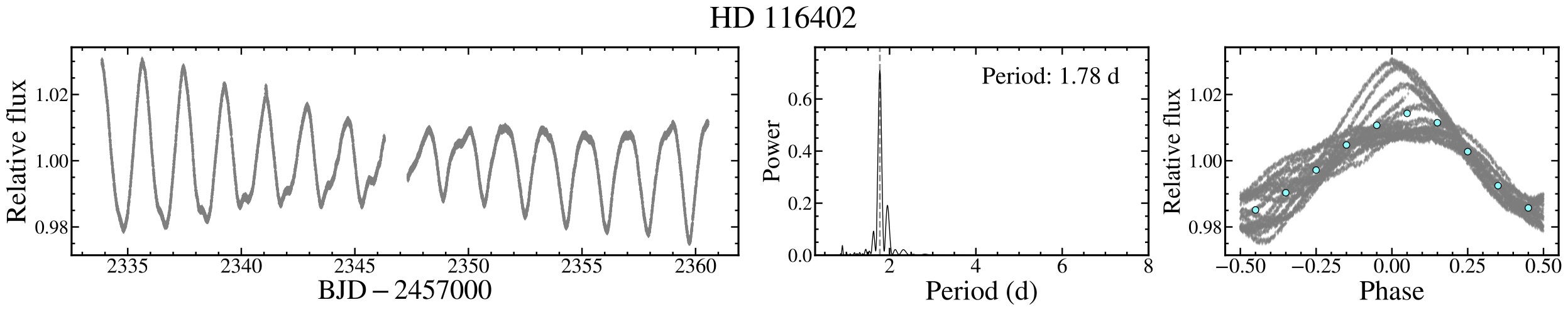}{0.9\textwidth}{}}                
  \vskip -.3 in
  \gridline{\fig{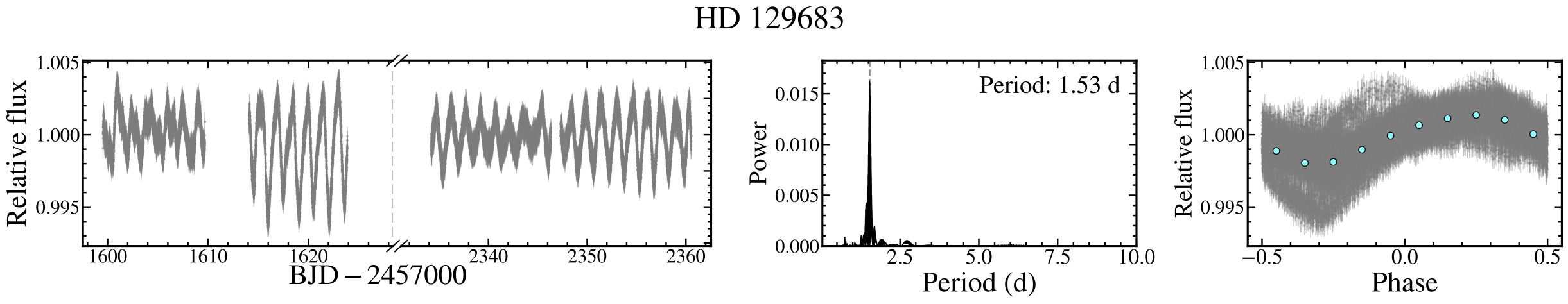}{0.9\textwidth}{}}                
  \vskip -.3 in
  \gridline{\fig{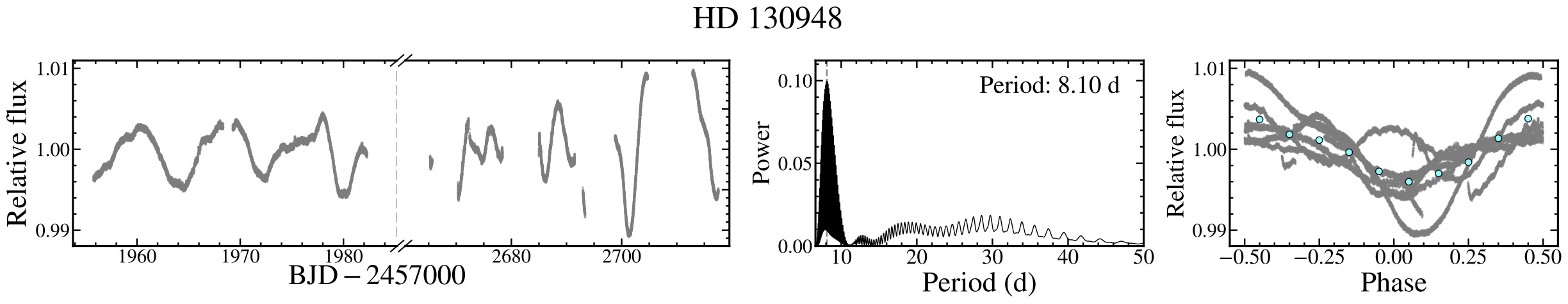}{0.9\textwidth}{}} 
  \vskip -.3 in
  \gridline{\fig{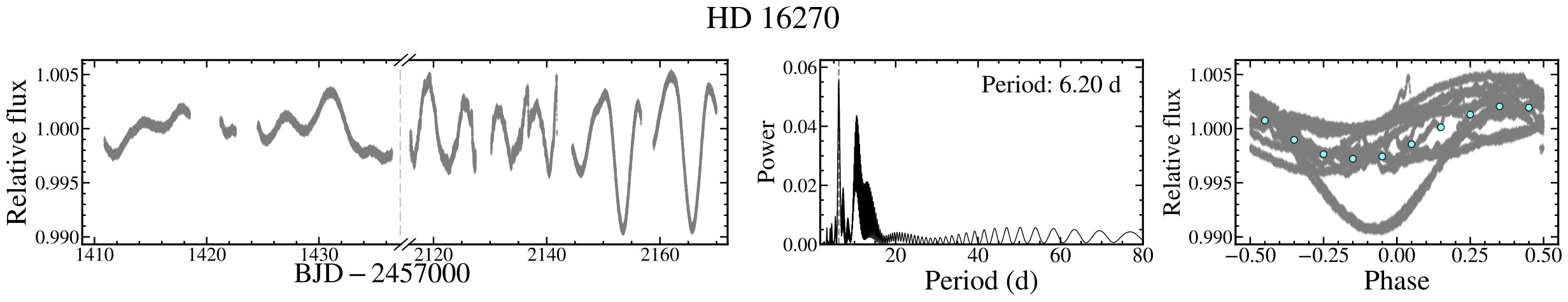}{0.9\textwidth}{}}                
  \vskip -.3 in
  \gridline{\fig{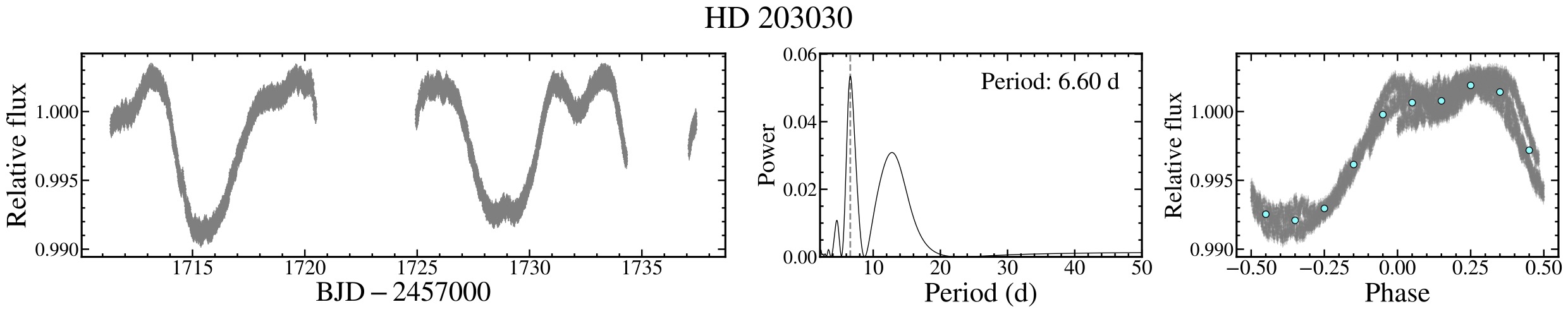}{0.9\textwidth}{}}                
  \vskip -.3 in
  \caption{\emph{TESS} light curves and periodograms for GU Psc, HD 116402, HD 129683, HD 130948, HD 16270, and HD 203030. 
  \label{fig:tess_4} } 
\end{figure*}

\begin{figure*}
  \hskip -0.8 in
  \gridline{\fig{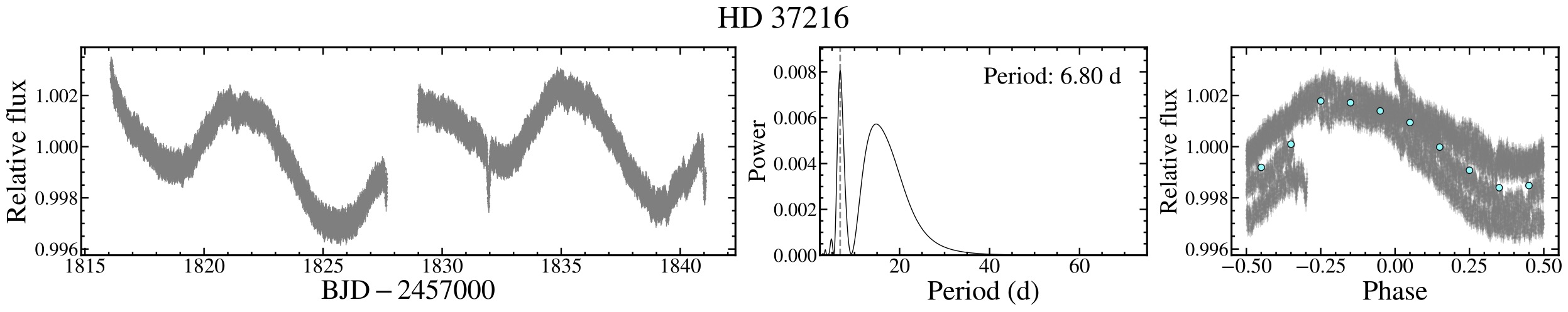}{0.9\textwidth}{}} 
  \vskip -.3 in
  \gridline{\fig{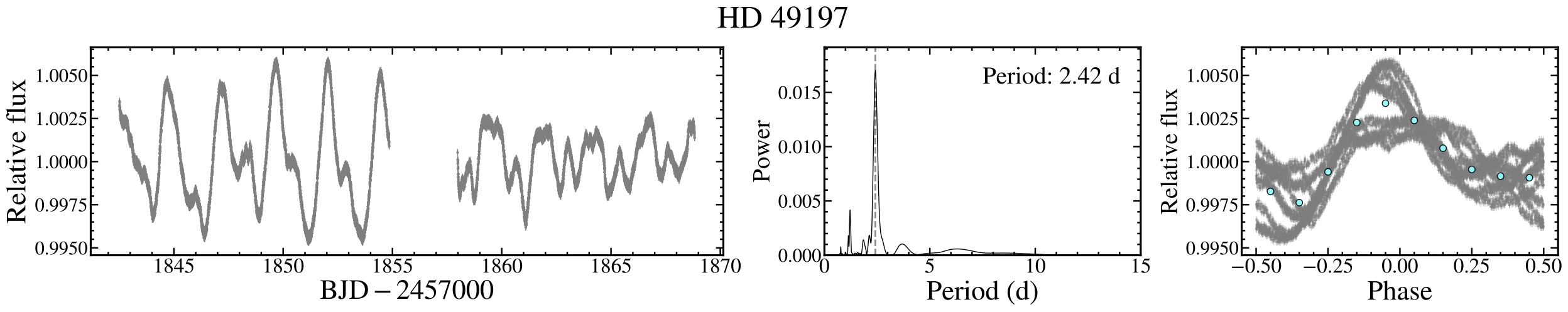}{0.9\textwidth}{}}
  \vskip -.3 in
  \gridline{\fig{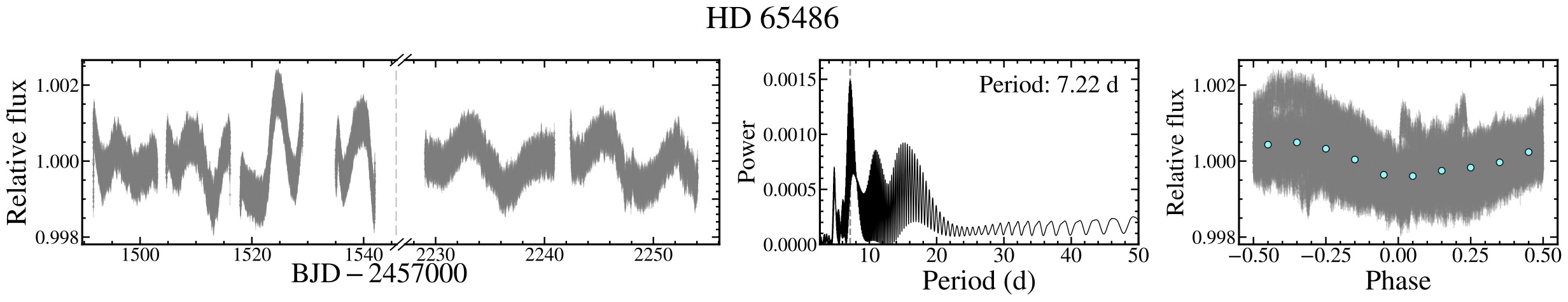}{0.9\textwidth}{}} 
  \vskip -.3 in
  \gridline{\fig{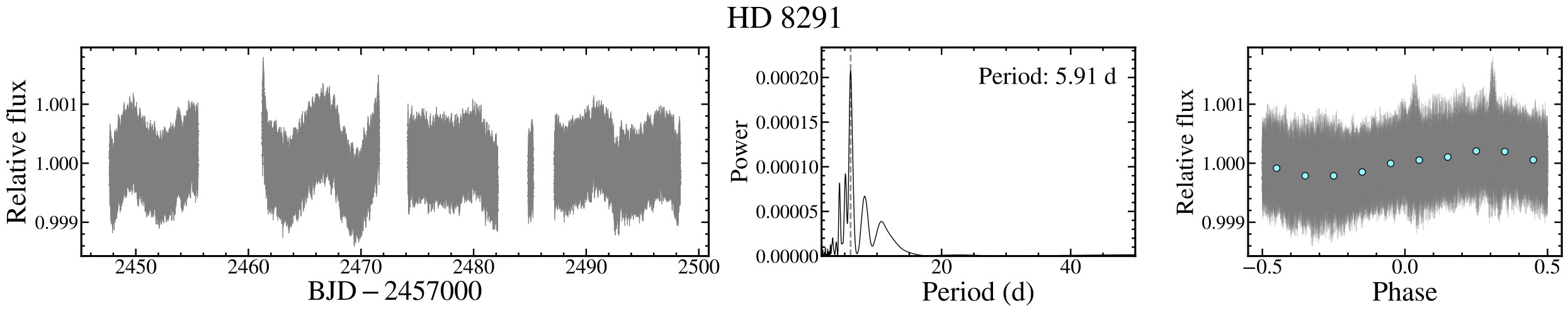}{0.9\textwidth}{}}     
  \vskip -.3 in
  \gridline{\fig{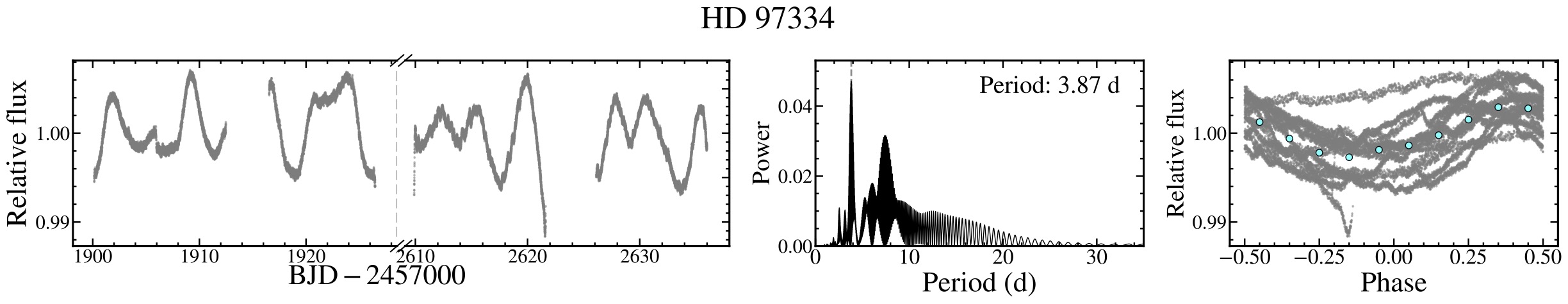}{0.9\textwidth}{}}                
  \vskip -.3 in
  \gridline{\fig{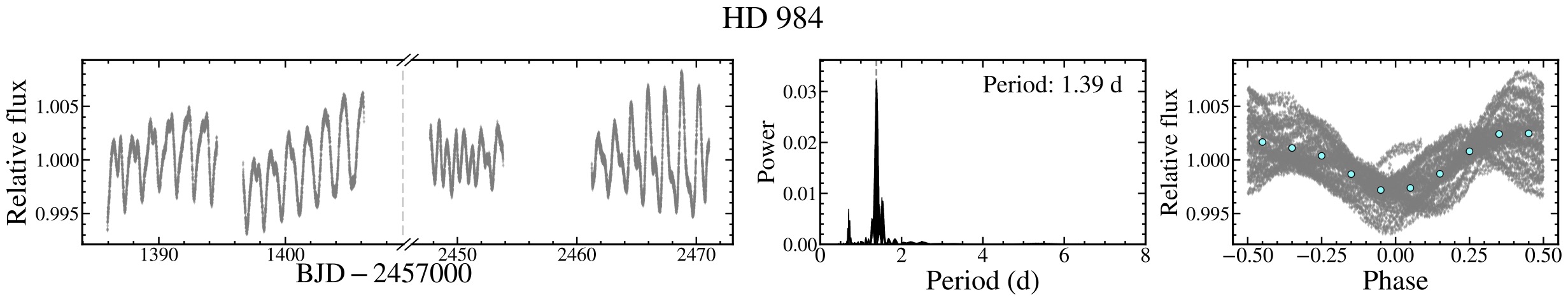}{0.9\textwidth}{}}                
  \vskip -.3 in
  \caption{\emph{TESS} light curves periodograms for HD 37216, HD 49197, HD 65486, HD 8291, HD 97334, and HD 984. 
  \label{fig:tess_5} } 
\end{figure*}

\begin{figure*}
  \hskip -0.8 in
  \gridline{\fig{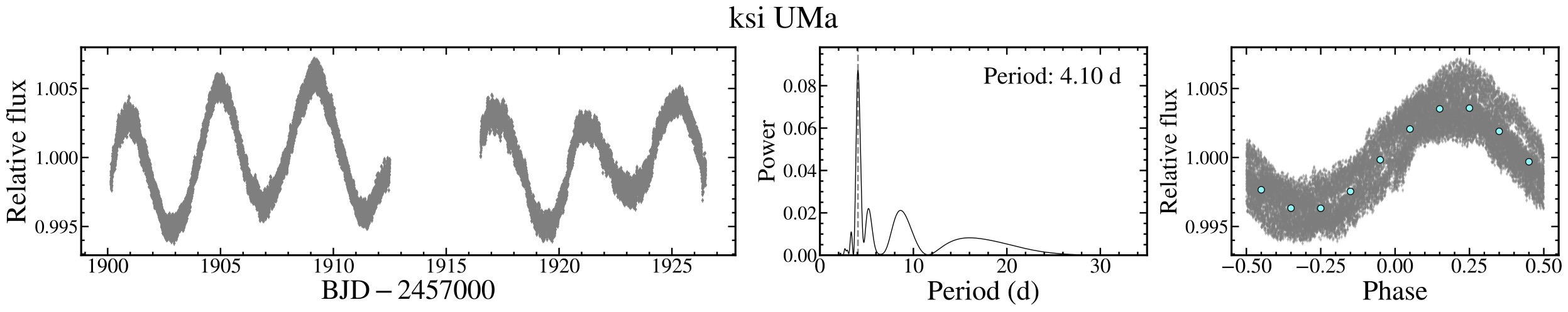}{0.9\textwidth}{}} 
  \vskip -.3 in
  \gridline{\fig{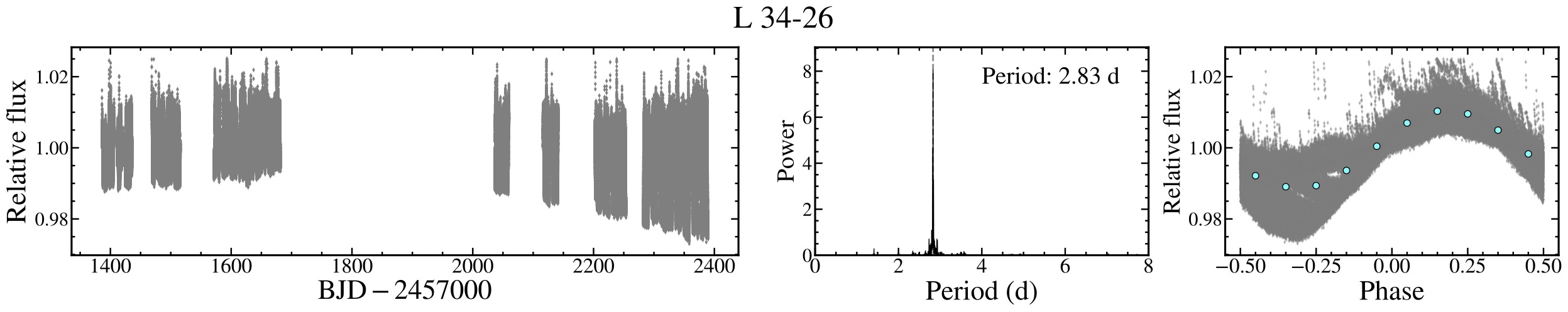}{0.9\textwidth}{}}
  \vskip -.3 in
  \gridline{\fig{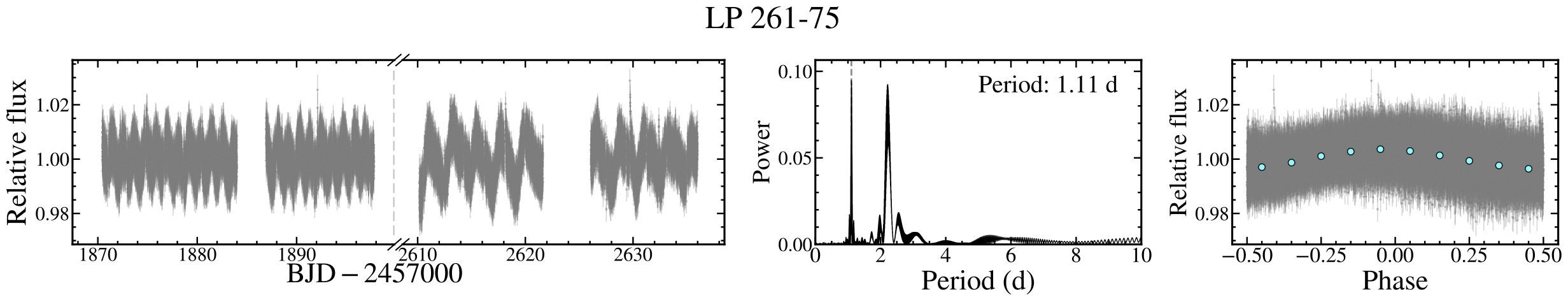}{0.9\textwidth}{}} 
  \vskip -.3 in
  \gridline{\fig{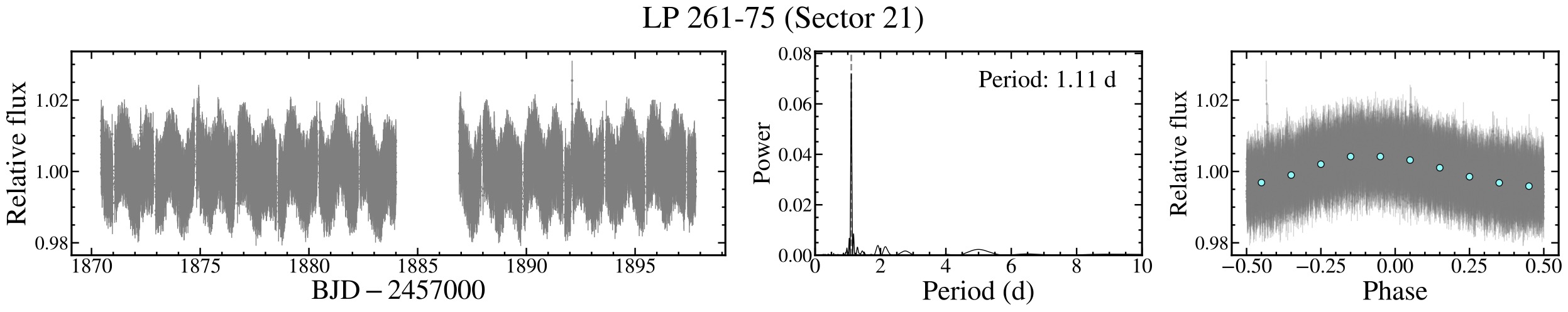}{0.9\textwidth}{}}     
  \vskip -.3 in
  \gridline{\fig{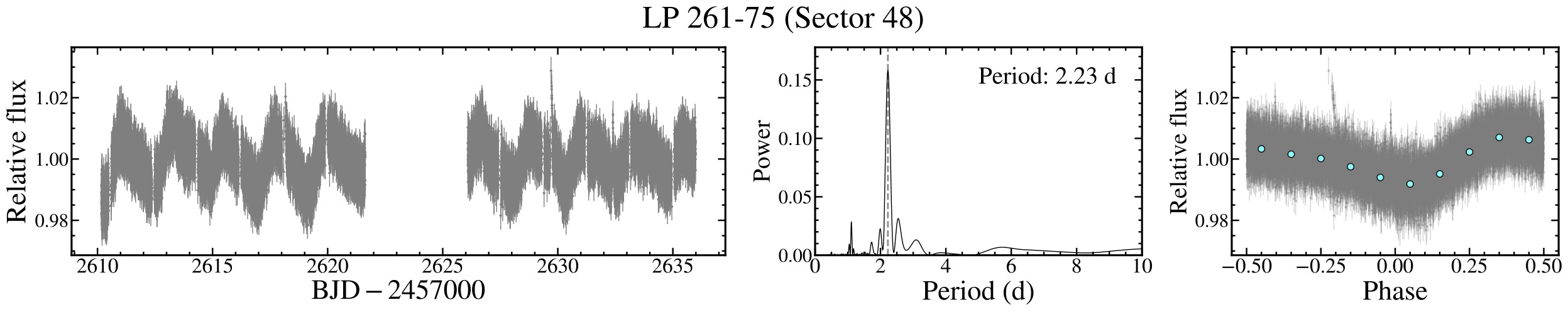}{0.9\textwidth}{}}                
  \vskip -.3 in
  \gridline{\fig{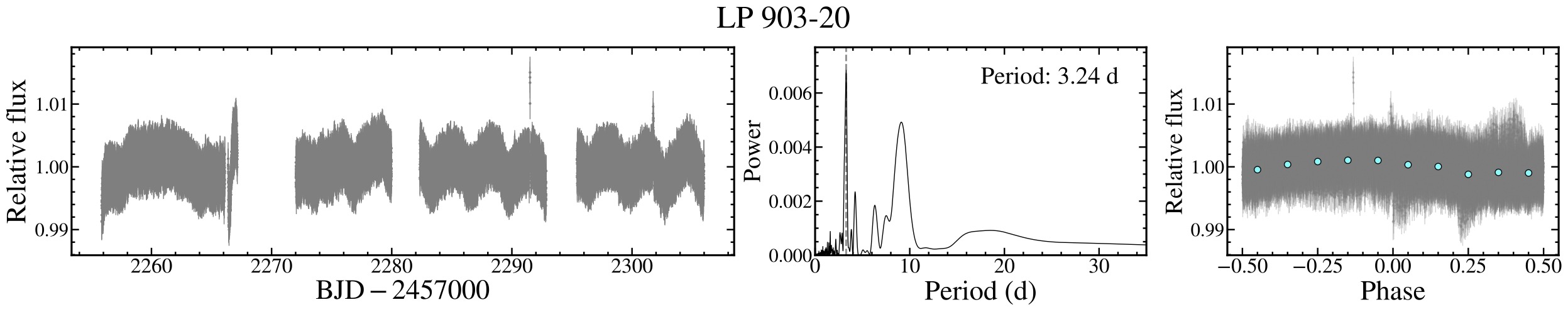}{0.9\textwidth}{}} 
  \vskip -.3 in
  \caption{\emph{TESS} light curves and periodograms for ksi UMa, L 34-26, LP 261-75 (Sectors 21 and 48), LP 261-75 (Sector 21), LP 261-75 (Sector 48), and LP 903-20.  
  \label{fig:tess_6} } 
\end{figure*}

\begin{figure*}
  \hskip -0.8 in
  \gridline{\fig{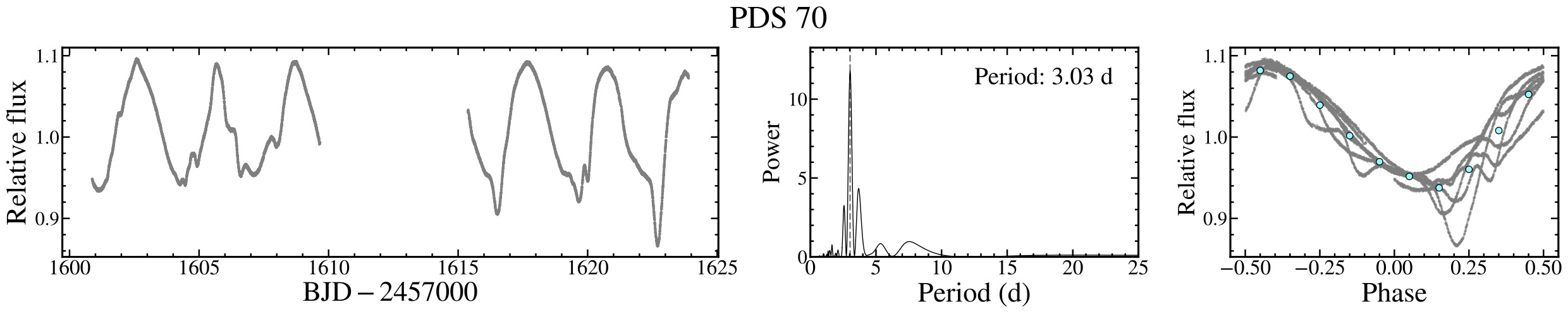}{0.9\textwidth}{}}
  \vskip -.3 in
  \gridline{\fig{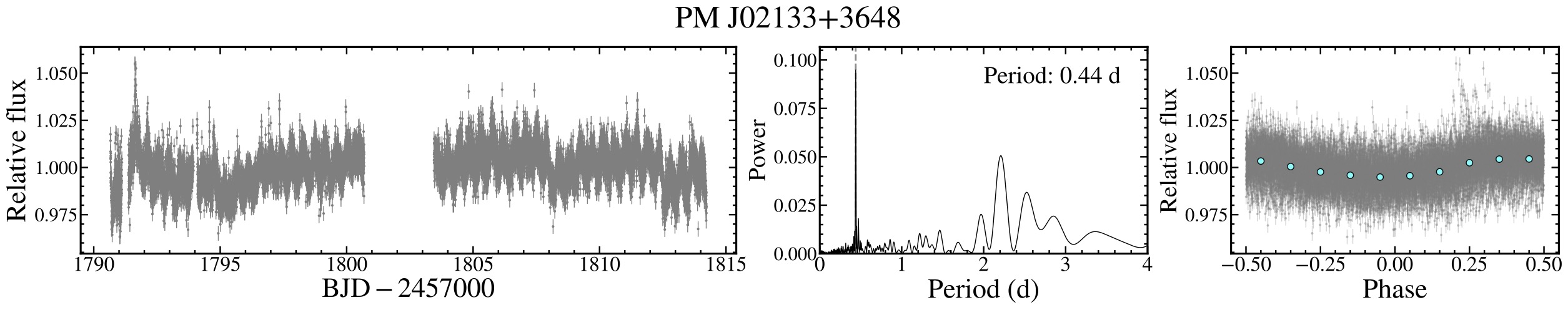}{0.9\textwidth}{}} 
  \vskip -.3 in
  \gridline{\fig{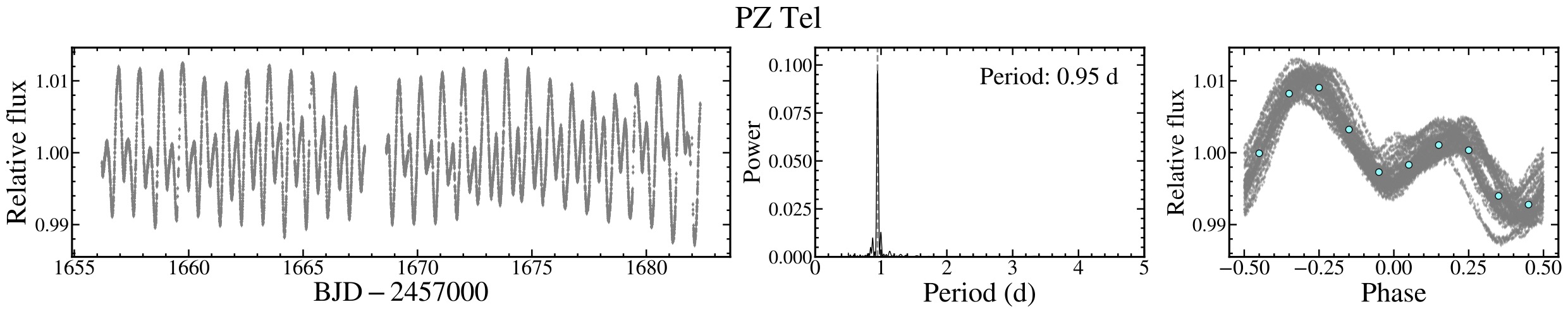}{0.9\textwidth}{}}     
  \vskip -.3 in
  \gridline{\fig{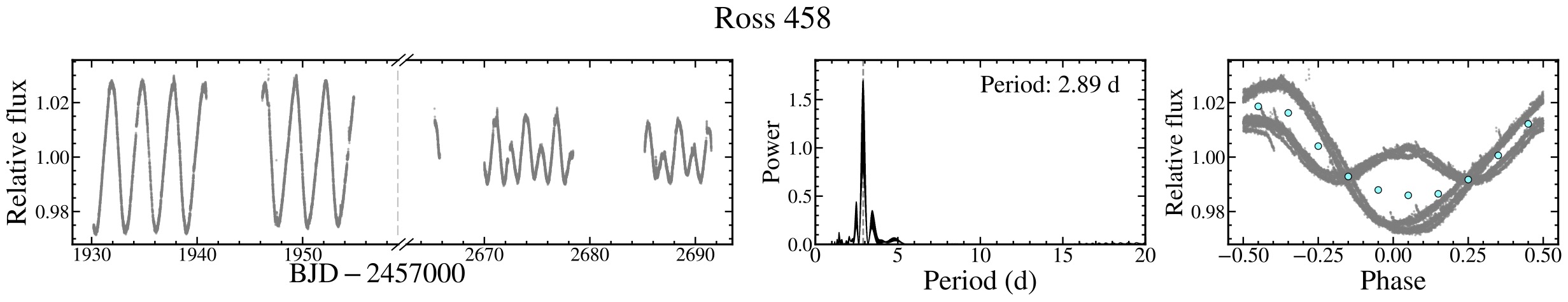}{0.9\textwidth}{}}                
  \vskip -.3 in
  \gridline{\fig{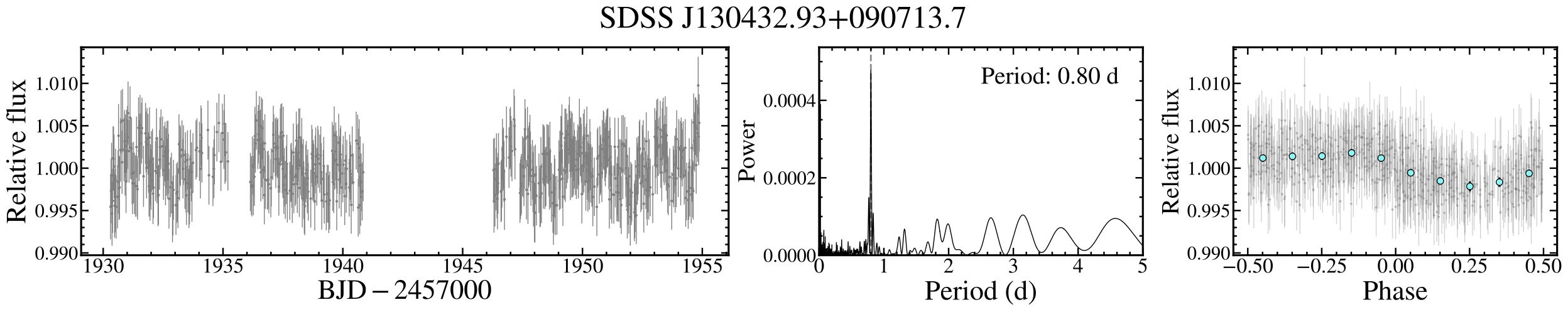}{0.9\textwidth}{}}                
  \vskip -.3 in
  \gridline{\fig{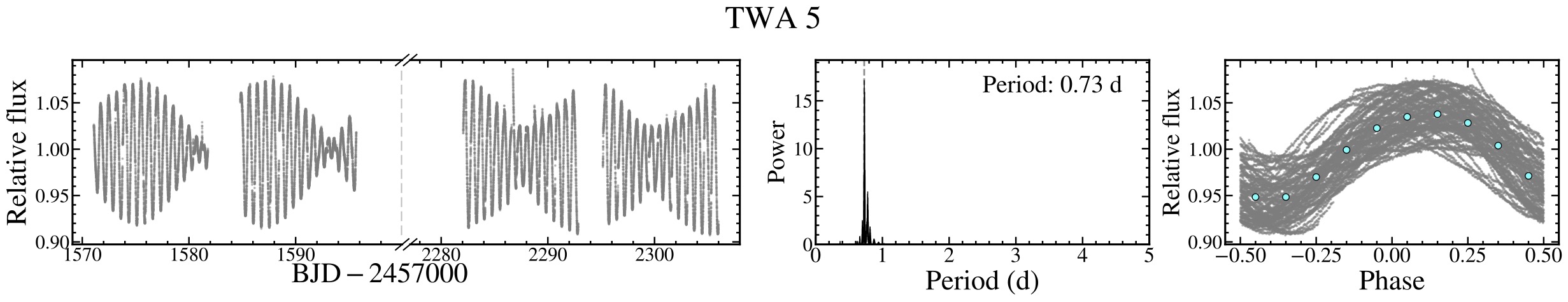}{0.9\textwidth}{}} 
  \vskip -.3 in
  \caption{\emph{TESS} light curves and periodograms for PDS 70, PM J02133+3648, PZ Tel, Ross 458 AB, SDSS J130432.93+090713.7, and TWA 5 Aab. 
  \label{fig:tess_7} } 
\end{figure*}

\begin{figure*}
  \hskip -0.8 in
  \gridline{\fig{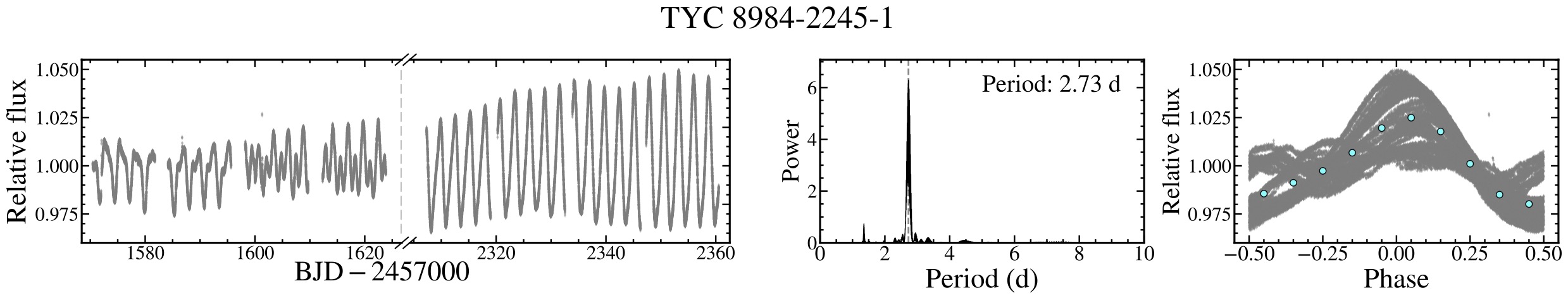}{0.9\textwidth}{}}
  \vskip -.3 in
  \gridline{\fig{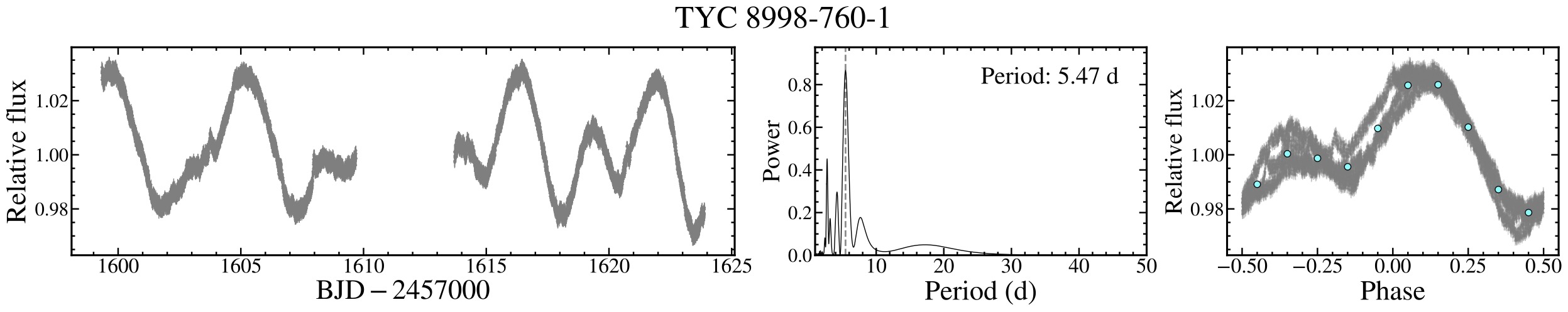}{0.9\textwidth}{}} 
  \vskip -.3 in
  \gridline{\fig{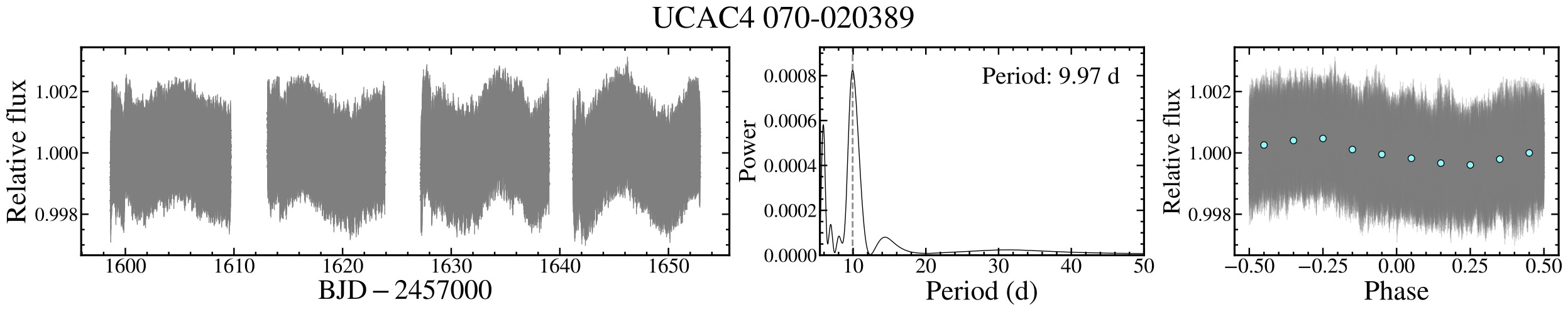}{0.9\textwidth}{}}     
  \vskip -.3 in
  \gridline{\fig{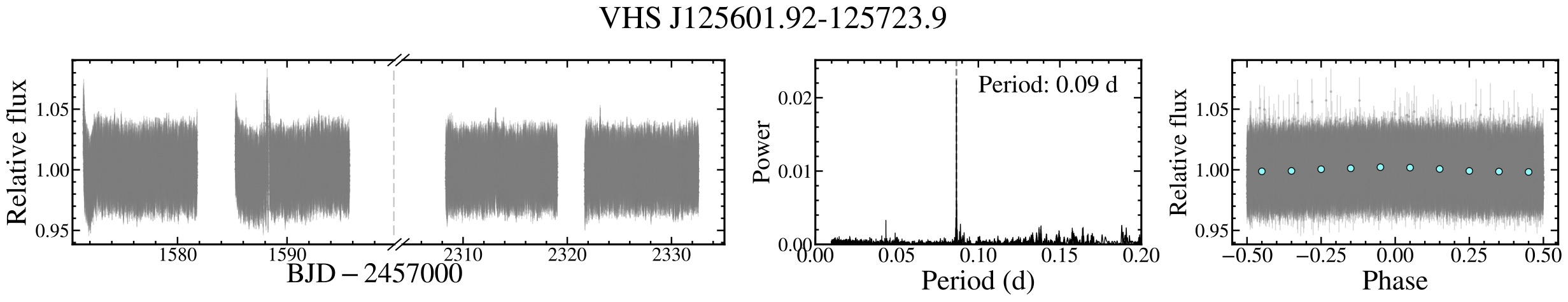}{0.9\textwidth}{}}     
  \vskip -.3 in
  \gridline{\fig{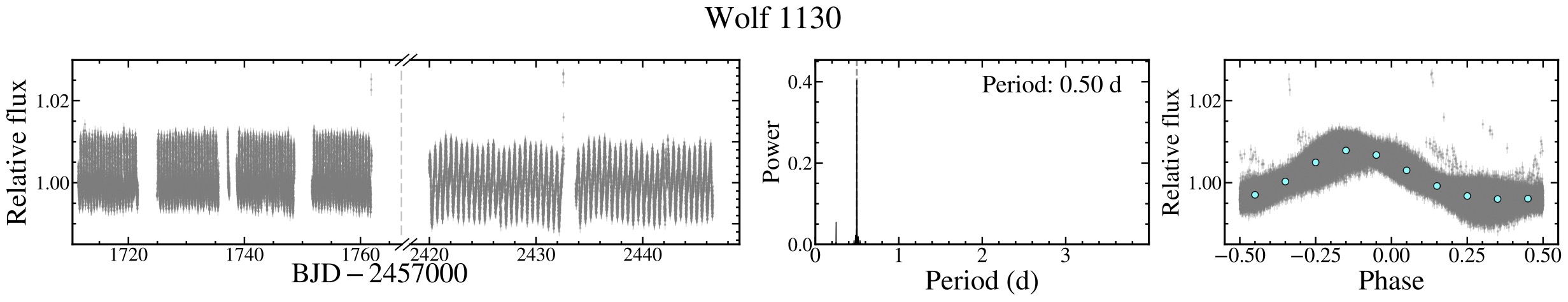}{0.9\textwidth}{}}     
  \vskip -.3 in
  \caption{\emph{TESS} light curves and periodograms for TYC 8984-2245-1, TYC 8998-760-1, UCAC4 070-020389, VHS~J125601.92--125723.9 AB, and Wolf~1130. 
  \label{fig:tess_8} } 
\end{figure*}

Figures~\ref{fig:tess_1}--\ref{fig:tess_8} show \emph{TESS} light curves for host stars in our sample with periodic variations consistent with rotational modulation.
Full light curves are shown in the left panels with breaks between non-contiguous sectors.  The middle panels depict the 
Generalized Lomb-Scargle periodogram with the most prominent peak highlighted.  The corresponding phased light curves and binned phased curves are displayed in the 
right panels.  Starspot evolution is apparent for many targets.  

Individual \emph{TESS} sectors and rotation period measurements can be found in Table~\ref{tab:hosttable}.  A full description of the light curve
extraction and reduction is summarized in Section~\ref{sec:tess}.  The observations used in this analysis can be accessed from MAST 
via \dataset[https://doi.org/10.17909/t9-r086-e880]{https://doi.org/10.17909/t9-r086-e880} and \dataset[https://doi.org/10.17909/t9-wpz1-8s54]{https://doi.org/10.17909/t9-wpz1-8s54}.

\newpage

\section{Projected Rotational Velocities}{\label{app:vsini}}

$v \sin i_*$ values for substellar host stars with rotation periods are compiled from literature and our own new observations 
in Table~\ref{tab:vsinis}.  When available, $N$ independent measurements
for the same star are combined into a single adopted value by computing a weighted mean and standard deviation: 

\begin{equation}
\mu = \frac{\sum_{i=1}^N \mu_i/\sigma_i^2 }{\sum_{i=1}^N 1/\sigma_i^2}
\end{equation}

and

\begin{equation}
\sigma = \sqrt{\frac{1}{\sum_{i=1}^N 1/\sigma_i^2 }}.
\end{equation}

In some cases, projected rotational velocities are reported in the literature at precisions below 1\%.  These 
would dominate over other measurements with more typical uncertainties (5--20\%) when computing a weighted mean.
Because of the difficulty in separating the effects of microturbulence, macroturbulence, and rotational broadening, we 
conservatively set a floor of $\sigma_{v \sin i_*}$/ $v \sin i_*$ = 0.01 when computing the weighted mean.

In many instances $v \sin i_*$ values are reported without associated uncertainties.
In these cases, when more than one are identified, we merge these into a single measurement by computing a robust mean
and standard deviation following \citet{Beers:1990aa}.

\startlongtable
\begin{deluxetable*}{lcl}
\renewcommand\arraystretch{0.9}
\tabletypesize{\footnotesize}
\setlength{ \tabcolsep } {.1cm} 
\tablewidth{0pt}
\tablecolumns{3}
\tablecaption{Adopted Projected Rotational Velocities.\label{tab:vsinis}}
\tablehead{
     & \colhead{$v \sin i_*$} &    \\
       \colhead{Name}  & \colhead{(km s$^{-1}$)}  &  \colhead{Reference\tablenotemark{a}} 
        }   
\startdata
1RXS J034231.8+121622 & 5.1  $\pm$   2.3   & This work   \\
\hline
\textbf{1RXS J034231.8+121622}  &   \textbf{5.1  $\pm$  2.3}   &     \textbf{Adopted}\tablenotemark{c} \\
\hline
1RXS J160929.1--210524  &  8.2  $\pm$ 1.3  &  This work  \\
\hline
\textbf{1RXS J160929.1--210524}  &   \textbf{8.2  $\pm$ 1.3 }   &     \textbf{Adopted}\tablenotemark{c} \\
\hline
2MASS J01033563--5515561   &    35.2  $\pm$   7.0   &  \citet{Malo:2014dk}        \\
2MASS J01033563--5515561   &    19.2  $\pm$   2.4   &  \citet{Kraus:2014ur}      \\
\hline
\textbf{2MASS J01033563--5515561}  &   \textbf{20.9  $\pm$  2.3}   &     \textbf{Adopted}\tablenotemark{c} \\
\hline
2MASS J01225093--2439505   &   14.2  $\pm$  3.2  &  \citet{Malo:2014dk}   \\
\enddata
\tablenotetext{a}{$v \sin i_*$ values from this study listed as ``This work'' represent the weighted mean of individual measurements from Table~\ref{tab:tullspec}.}
\tablenotetext{b}{Robust mean and standard deviation of literature in values without quoted uncertainties.}
\tablenotetext{c}{Weighted mean of $v \sin i_*$ measurements after upwardly revising the relative precision of 
individual uncertainties to be 1\% of their quoted values.}
\end{deluxetable*}


\begin{thebibliography}{}
\expandafter\ifx\csname natexlab\endcsname\relax\def\natexlab#1{#1}\fi

\bibitem[{Abt \& Morrell(1995)}]{Abt:1995aa}
Abt, H.~A., \& Morrell, N.~I. 1995, Astrophysical Journal Supplement, 99, 135

\bibitem[{Adams(2010)}]{Adams:2010bb}
Adams, F.~C. 2010, Annu. Rev. Astro. Astrophys., 48, 47

\bibitem[{Aerts(2021)}]{Aerts:2021aa}
Aerts, C. 2021, Reviews of Modern Physics, 93, 015001

\bibitem[{Albrecht {et~al.}(2009)Albrecht, Reffert, Snellen, \&
  Winn}]{Albrecht:2009df}
Albrecht, S., Reffert, S., Snellen, I. A.~G., \& Winn, J.~N. 2009, Nature, 461,
  373

\bibitem[{Albrecht {et~al.}(2013)Albrecht, Winn, Marcy, Howard, Isaacson, \&
  Johnson}]{Albrecht:2013uq}
Albrecht, S., Winn, J.~N., Marcy, G.~W., {et~al.} 2013, ApJ, 771, 11

\bibitem[{Albrecht {et~al.}(2012)Albrecht, Winn, Johnson, Howard, Marcy,
  Butler, Arriagada, Crane, Shectman, Thompson, Hirano, Bakos, \&
  Hartman}]{Albrecht:2012hk}
Albrecht, S., Winn, J.~N., Johnson, J.~A., {et~al.} 2012, The Astrophysical
  Journal, 757, 18

\bibitem[{Albrecht {et~al.}(2022)Albrecht, Dawson, \& Winn}]{Albrecht:2022aa}
Albrecht, S.~H., Dawson, R.~I., \& Winn, J.~N. 2022, Publications of the
  Astronomical Society of the Pacific, 134, 082001

\bibitem[{Allers {et~al.}(2010)Allers, Liu, Dupuy, \& Cushing}]{Allers:2010cg}
Allers, K.~N., Liu, M.~C., Dupuy, T.~J., \& Cushing, M.~C. 2010, The
  Astrophysical Journal, 715, 561

\bibitem[{Anderson {et~al.}(2017)Anderson, Lai, \& Storch}]{Anderson:2017aa}
Anderson, K.~R., Lai, D., \& Storch, N.~I. 2017, Monthly Notices of the Royal
  Astronomical Society, 467, 3066

\bibitem[{Anderson {et~al.}(2016)Anderson, Storch, \& Lai}]{Anderson:2016aa}
Anderson, K.~R., Storch, N.~I., \& Lai, D. 2016, MNRAS, 456, 3671

\bibitem[{Ansdell {et~al.}(2020)Ansdell, Gaidos, Hedges, Tazzari, Kraus, Wyatt,
  Kennedy, Williams, Mann, Angelo, Duchêne, Mamajek, Carpenter, Esplin, \&
  Rizzuto}]{Ansdell:2020fo}
Ansdell, M., Gaidos, E., Hedges, C., {et~al.} 2020, MNRAS, 492, 572

\bibitem[{Artigau {et~al.}(2015)Artigau, Gagné, Faherty, Malo, Naud, Doyon,
  LaFreniere, \& Beletsky}]{Artigau:2015if}
Artigau, E., Gagné, J., Faherty, J., {et~al.} 2015, ApJ, 806, 254

\bibitem[{Avallone {et~al.}(2022)Avallone, Tayar, Saders, Berger, Claytor,
  Beaton, Teske, Godoy-Rivera, \& Pan}]{Avallone:2022aa}
Avallone, E.~A., Tayar, J.~N., Saders, J. L.~v., {et~al.} 2022, The
  Astrophysical Journal, 930, 7

\bibitem[{Bailer-Jones {et~al.}(2021)Bailer-Jones, Rybizki, Fouesneau,
  Demleitner, \& Andrae}]{BailerJones:2021aa}
Bailer-Jones, C. A.~L., Rybizki, J., Fouesneau, M., Demleitner, M., \& Andrae,
  R. 2021, The Astronomical Journal, 161, 147

\bibitem[{Bailer-Jones {et~al.}(2018)Bailer-Jones, Rybizki, Fouesneau,
  Mantelet, \& Andrae}]{BailerJones:2018aa}
Bailer-Jones, C. A.~L., Rybizki, J., Fouesneau, M., Mantelet, G., \& Andrae, R.
  2018, The Astronomical Journal, 156, 58

\bibitem[{Bailey {et~al.}(2016)Bailey, Batygin, \& Brown}]{Bailey:2016iu}
Bailey, E., Batygin, K., \& Brown, M.~E. 2016, The Astronomical Journal, 152,
  126

\bibitem[{Bailey \& Fabrycky(2019)}]{Bailey:2019js}
Bailey, N., \& Fabrycky, D. 2019, The Astronomical Journal, 158, 94

\bibitem[{Bailey {et~al.}(2014)Bailey, Meshkat, Reiter, Morzinski, Males, Su,
  Hinz, Kenworthy, Stark, Mamajek, Briguglio, Close, Follette, Puglisi,
  Rodigas, Weinberger, \& Xompero}]{Bailey:2014et}
Bailey, V., Meshkat, T., Reiter, M., {et~al.} 2014, The Astrophysical Journal,
  780, L4

\bibitem[{Baliunas {et~al.}(1996)Baliunas, Sokoloff, \& Soon}]{Baliunas:1996aa}
Baliunas, S., Sokoloff, D., \& Soon, W. 1996, ApJ, 457, L99

\bibitem[{Barnes {et~al.}(2005)Barnes, Cameron, Donati, James, Marsden, \&
  Petit}]{Barnes:2005aa}
Barnes, J.~R., Cameron, A.~C., Donati, J.-F., {et~al.} 2005, MNRAS, 357, L1

\bibitem[{Barnes {et~al.}(2000)Barnes, Cameron, James, \&
  Donati}]{Barnes:2000aa}
Barnes, J.~R., Cameron, A.~C., James, D.~J., \& Donati, J.~F. 2000, MNRAS, 314,
  162

\bibitem[{Bate(2009)}]{Bate:2009br}
Bate, M.~R. 2009, Monthly Notices of the Royal Astronomical Society, 392, 590

\bibitem[{Bate(2012)}]{Bate:2012hy}
---. 2012, Monthly Notices of the Royal Astronomical Society, 419, 3115

\bibitem[{Bate {et~al.}(2002)Bate, Bonnell, \& Bromm}]{Bate:2002iq}
Bate, M.~R., Bonnell, I.~A., \& Bromm, V. 2002, Monthly Notices of the Royal
  Astronomical Society, 332, L65

\bibitem[{Bate {et~al.}(2010)Bate, Lodato, \& Pringle}]{Bate:2010fz}
Bate, M.~R., Lodato, G., \& Pringle, J.~E. 2010, Monthly Notices of the Royal
  Astronomical Society, 401, 1505

\bibitem[{Batygin(2012)}]{Batygin:2012ig}
Batygin, K. 2012, Nature, 491, 418

\bibitem[{Batygin \& Adams(2013)}]{Batygin:2013fr}
Batygin, K., \& Adams, F.~C. 2013, ApJ, 778, 169

\bibitem[{Batygin {et~al.}(2020)Batygin, Adams, Batygin, \&
  Petigura}]{Batygin:2020dl}
Batygin, K., Adams, F.~C., Batygin, Y.~K., \& Petigura, E.~A. 2020, The
  Astronomical Journal, 159, 101

\bibitem[{Baum {et~al.}(2022)Baum, Wright, Luhn, \& Isaacson}]{Baum:2022aa}
Baum, A.~C., Wright, J.~T., Luhn, J.~K., \& Isaacson, H. 2022, The Astronomical
  Journal, 163, 183

\bibitem[{Beers {et~al.}(1990)Beers, Flynn, \& Gebhardt}]{Beers:1990aa}
Beers, T.~C., Flynn, K., \& Gebhardt, K. 1990, Astronomical Journal, 100, 32

\bibitem[{Belle(2012)}]{VanBelle:2012aa}
Belle, G. T.~v. 2012, The Astronomy and Astrophysics Review, 20, 51

\bibitem[{Benomar {et~al.}(2018)Benomar, Bazot, Nielsen, Gizon, Sekii, Takata,
  Hotta, Hanasoge, Sreenivasan, \& Christensen-Dalsgaard}]{Benomar:2018aa}
Benomar, O., Bazot, M., Nielsen, M.~B., {et~al.} 2018, Science, 361, 1231

\bibitem[{Berdyugina(2005)}]{Berdyugina:2005aa}
Berdyugina, S.~V. 2005, Living Reviews in Solar Physics, 2, 8

\bibitem[{Bernacca \& Perinotto(1970)}]{Bernacca:1970aa}
Bernacca, P.~L., \& Perinotto, M. 1970, CoAsi, 239, 1

\bibitem[{Best {et~al.}(2021)Best, Liu, Magnier, \& Dupuy}]{Best:2021gm}
Best, W. M.~J., Liu, M.~C., Magnier, E.~A., \& Dupuy, T.~J. 2021, The
  Astronomical Journal, 161, 42

\bibitem[{Best {et~al.}(2018)Best, Magnier, Liu, Aller, Zhang, Burgett,
  Chambers, Draper, Flewelling, Kaiser, Kudritzki, Metcalfe, Tonry, Wainscoat,
  \& Waters}]{Best:2018kw}
Best, W. M.~J., Magnier, E.~A., Liu, M.~C., {et~al.} 2018, The Astrophysical
  Journal Supplement Series, 234, 1

\bibitem[{Betti {et~al.}(2022)Betti, Follette, Ward-Duong, Aoyama, Marleau,
  Bary, Robinson, Janson, Balmer, Chauvin, \& Palma-Bifani}]{Betti:2022aa}
Betti, S.~K., Follette, K.~B., Ward-Duong, K., {et~al.} 2022, The Astrophysical
  Journal Letters, 935, L18

\bibitem[{Biller {et~al.}(2010)Biller, Liu, Wahhaj, Nielsen, Close, Dupuy,
  Hayward, Burrows, Chun, Ftaclas, Clarke, Hartung, Males, Reid, Shkolnik,
  Skemer, Tecza, Thatte, Alencar, Artymowicz, Boss, Pino, Gregorio-Hetem, Ida,
  Kuchner, Lin, \& Toomey}]{Biller:2010ku}
Biller, B.~A., Liu, M.~C., Wahhaj, Z., {et~al.} 2010, The Astrophysical
  Journal, 720, L82

\bibitem[{Blunt {et~al.}(2017)Blunt, Nielsen, Rosa, Konopacky, Ryan, Wang,
  Pueyo, Rameau, Marois, Marchis, Macintosh, Graham, Duchene, \&
  Schneider}]{Blunt:2017eta}
Blunt, S., Nielsen, E.~L., Rosa, R. J.~D., {et~al.} 2017, The Astronomical
  Journal, 153, 229

\bibitem[{Blunt {et~al.}(2020)Blunt, Wang, Angelo, Ngo, Cody, Rosa, Graham,
  Hirsch, Nagpal, Nielsen, Pearce, Rice, \& Tejada}]{Blunt:2020bb}
Blunt, S., Wang, J.~J., Angelo, I., {et~al.} 2020, The Astronomical Journal,
  159, 89

\bibitem[{Bohn {et~al.}(2020{\natexlab{a}})Bohn, Kenworthy, Ginski, Manara,
  Pecaut, de Boer, Keller, Mamajek, Meshkat, Reggiani, Todorov, \&
  Snik}]{Bohn:2020aa}
Bohn, A.~J., Kenworthy, M.~A., Ginski, C., {et~al.} 2020{\natexlab{a}}, Monthly
  Notices of the Royal Astronomical Society, 492, 431

\bibitem[{Bohn {et~al.}(2020{\natexlab{b}})Bohn, Kenworthy, Ginski, Rieder,
  Mamajek, Meshkat, Pecaut, Reggiani, Boer, Keller, Snik, \&
  Southworth}]{Bohn:2020ge}
---. 2020{\natexlab{b}}, The Astrophysical Journal Letters, 898, L16

\bibitem[{Bohn {et~al.}(2021)Bohn, Ginski, Kenworthy, Mamajek, Pecaut,
  Mugrauer, Vogt, Adam, Meshkat, Reggiani, \& Snik}]{Bohn:2021aa}
Bohn, A.~J., Ginski, C., Kenworthy, M.~A., {et~al.} 2021, Astronomy \&
  Astrophysics, 648, A73

\bibitem[{Bonavita {et~al.}(2022)Bonavita, Fontanive, Gratton, Mužić,
  Desidera, Mesa, Biller, Scholz, Sozzetti, \& Squicciarini}]{Bonavita:2022aa}
Bonavita, M., Fontanive, C., Gratton, R., {et~al.} 2022, Monthly Notices of the
  Royal Astronomical Society, 513, 5588

\bibitem[{Bonnefoy {et~al.}(2018)Bonnefoy, Perraut, Lagrange, Delorme, Vigan,
  Line, Rodet, Ginski, Mourard, Marleau, Samland, Tremblin, Ligi, Cantalloube,
  Mollière, Charnay, Kuzuhara, Janson, Morley, Homeier, D'Orazi, Klahr,
  Mordasini, Lavie, Baudino, Beust, Peretti, Bartucci, Mesa, Bézard,
  Boccaletti, Galicher, Hagelberg, Desidera, Biller, Maire, Allard, Borgniet,
  Lannier, Meunier, Desort, Alecian, Chauvin, Langlois, Henning, Mugnier,
  Mouillet, Gratton, Brandt, Elwain, Beuzit, Tamura, Hori, Brandner, Buenzli,
  Cheetham, Cudel, Feldt, Kasper, Keppler, Kopytova, Meyer, Perrot, Rouan,
  Salter, Schmidt, Sissa, Zurlo, Wildi, Blanchard, Caprio, Delboulbé, Maurel,
  Moulin, Pavlov, Rabou, Ramos, Roelfsema, Rousset, Stadler, Rigal, \&
  Weber}]{Bonnefoy:2018ch}
Bonnefoy, M., Perraut, K., Lagrange, A.-M., {et~al.} 2018, Astronomy and
  Astrophysics, 618, A63

\bibitem[{Boss(2006)}]{Boss:2006ge}
Boss, A.~P. 2006, The Astrophysical Journal, 637, L137

\bibitem[{Bouvier \& Appenzeller(1992)}]{Bouvier:1992ww}
Bouvier, J., \& Appenzeller, I. 1992, ApJS, 92, 481

\bibitem[{Bouy {et~al.}(2003)Bouy, Brandner, Martín, Delfosse, Allard, \&
  Basri}]{Bouy:2003eg}
Bouy, H., Brandner, W., Martín, E.~L., {et~al.} 2003, The Astronomical
  Journal, 126, 1526

\bibitem[{Bowler(2016)}]{Bowler:2016jk}
Bowler, B.~P. 2016, Publications of the Astronomical Society of the Pacific,
  128, 102001

\bibitem[{Bowler {et~al.}(2020{\natexlab{a}})Bowler, Blunt, \&
  Nielsen}]{Bowler:2020hk}
Bowler, B.~P., Blunt, S.~C., \& Nielsen, E.~L. 2020{\natexlab{a}}, Astronomical
  Journal, 159, 63

\bibitem[{Bowler {et~al.}(2014)Bowler, Liu, Kraus, \& Mann}]{Bowler:2014dk}
Bowler, B.~P., Liu, M.~C., Kraus, A.~L., \& Mann, A.~W. 2014, The Astrophysical
  Journal, 784, 65

\bibitem[{Bowler {et~al.}(2013)Bowler, Liu, Shkolnik, \& Dupuy}]{Bowler:2013ek}
Bowler, B.~P., Liu, M.~C., Shkolnik, E.~L., \& Dupuy, T.~J. 2013, The
  Astrophysical Journal, 774, 55

\bibitem[{Bowler {et~al.}(2012)Bowler, Liu, Shkolnik, Dupuy, Cieza, Kraus, \&
  Tamura}]{Bowler:2012cs}
Bowler, B.~P., Liu, M.~C., Shkolnik, E.~L., {et~al.} 2012, The Astrophysical
  Journal, 753, 142

\bibitem[{Bowler {et~al.}(2015{\natexlab{a}})Bowler, Liu, Shkolnik, \&
  Tamura}]{Bowler:2015ja}
Bowler, B.~P., Liu, M.~C., Shkolnik, E.~L., \& Tamura, M. 2015{\natexlab{a}},
  The Astrophysical Journal Supplement Series, 216, 7

\bibitem[{Bowler {et~al.}(2020{\natexlab{b}})Bowler, Zhou, Morley, Kataria,
  Bryan, Benneke, \& Batygin}]{Bowler:2020ip}
Bowler, B.~P., Zhou, Y., Morley, C.~V., {et~al.} 2020{\natexlab{b}}, The
  Astrophysical Journal Letters, 893, L30

\bibitem[{Bowler {et~al.}(2015{\natexlab{b}})Bowler, Shkolnik, Liu, Schlieder,
  Mann, Dupuy, Hinkley, Crepp, Johnson, Howard, Flagg, Weinberger, Aller,
  Allers, Best, Kotson, Montet, Herczeg, Baranec, Riddle, Law, Nielsen, Wahhaj,
  Biller, \& Hayward}]{Bowler:2015ch}
Bowler, B.~P., Shkolnik, E.~L., Liu, M.~C., {et~al.} 2015{\natexlab{b}}, The
  Astrophysical Journal, 806, 62

\bibitem[{Bowler {et~al.}(2017)Bowler, Kraus, Bryan, Knutson, Brogi, Rizzuto,
  Mace, Vanderburg, Liu, Hillenbrand, \& Cieza}]{Bowler:2017kt}
Bowler, B.~P., Kraus, A.~L., Bryan, M.~L., {et~al.} 2017, The Astronomical
  Journal, 154, 165

\bibitem[{Bowler {et~al.}(2018)Bowler, Dupuy, Endl, Cochran, MacQueen, Fulton,
  Petigura, Howard, Hirsch, Kratter, Crepp, Biller, Johnson, \&
  Wittenmyer}]{Bowler:2018gy}
Bowler, B.~P., Dupuy, T.~J., Endl, M., {et~al.} 2018, The Astronomical Journal,
  155, 159

\bibitem[{Brandt {et~al.}(2021{\natexlab{a}})Brandt, Brandt, Dupuy, Li, \&
  Michalik}]{Brandt:2021aa}
Brandt, G.~M., Brandt, T.~D., Dupuy, T.~J., Li, Y., \& Michalik, D.
  2021{\natexlab{a}}, The Astronomical Journal, 161, 179

\bibitem[{Brandt {et~al.}(2021{\natexlab{b}})Brandt, Dupuy, Li, Chen, Brandt,
  Wong, Currie, Bowler, Liu, Best, \& Phillips}]{Brandt:2021cc}
Brandt, G.~M., Dupuy, T.~J., Li, Y., {et~al.} 2021{\natexlab{b}}, The
  Astronomical Journal, 162, 301

\bibitem[{Brandt {et~al.}(2019)Brandt, Dupuy, \& Bowler}]{Brandt:2019ey}
Brandt, T.~D., Dupuy, T.~J., \& Bowler, B.~P. 2019, The Astronomical Journal,
  158, 140

\bibitem[{Brandt {et~al.}(2020)Brandt, Dupuy, Bowler, Gagliuffi, Faherty,
  Brandt, \& Michalik}]{Brandt:2020ce}
Brandt, T.~D., Dupuy, T.~J., Bowler, B.~P., {et~al.} 2020, The Astronomical
  Journal, 160, 196

\bibitem[{Brewer {et~al.}(2016)Brewer, Fischer, Valenti, \&
  Piskunov}]{Brewer:2016gf}
Brewer, J.~M., Fischer, D.~A., Valenti, J.~A., \& Piskunov, N. 2016, The
  Astrophysical Journal Supplement Series, 225, 32

\bibitem[{Browning {et~al.}(2010)Browning, Basri, Marcy, West, \&
  Zhang}]{Browning:2010hw}
Browning, M.~K., Basri, G., Marcy, G.~W., West, A.~A., \& Zhang, J. 2010, The
  Astronomical Journal, 139, 504

\bibitem[{Brun {et~al.}(2017)Brun, Strugarek, Varela, Matt, Augustson, Emeriau,
  DoCao, Brown, \& Toomre}]{Brun:2017aa}
Brun, A.~S., Strugarek, A., Varela, J., {et~al.} 2017, The Astrophysical
  Journal, 836, 192

\bibitem[{Bryan {et~al.}(2018)Bryan, Benneke, Knutson, Batygin, \&
  Bowler}]{Bryan:2018hd}
Bryan, M.~L., Benneke, B., Knutson, H.~A., Batygin, K., \& Bowler, B.~P. 2018,
  Nature Astronomy, 128, 63

\bibitem[{Bryan {et~al.}(2016)Bryan, Bowler, Knutson, Kraus, Hinkley, Mawet,
  Nielsen, \& Blunt}]{Bryan:2016eo}
Bryan, M.~L., Bowler, B.~P., Knutson, H.~A., {et~al.} 2016, ApJ, 827, 100

\bibitem[{Bryan {et~al.}(2021)Bryan, Chiang, Morley, Mace, \&
  Bowler}]{Bryan:2021aa}
Bryan, M.~L., Chiang, E., Morley, C.~V., Mace, G.~N., \& Bowler, B.~P. 2021,
  The Astronomical Journal, 162, 217

\bibitem[{Bryan {et~al.}(2020{\natexlab{a}})Bryan, Ginzburg, Chiang, Morley,
  Bowler, Xuan, \& Knutson}]{Bryan:2020ex}
Bryan, M.~L., Ginzburg, S., Chiang, E., {et~al.} 2020{\natexlab{a}}, The
  Astrophysical Journal, 905, 37

\bibitem[{Bryan {et~al.}(2020{\natexlab{b}})Bryan, Chiang, Bowler, Morley,
  Millholland, Blunt, Ashok, Nielsen, Ngo, Mawet, \& Knutson}]{Bryan:2020ji}
Bryan, M.~L., Chiang, E., Bowler, B.~P., {et~al.} 2020{\natexlab{b}}, The
  Astronomical Journal, 159, 181

\bibitem[{Bulger {et~al.}(2014)Bulger, Patience, Ward-Duong, Pinte, Bouy,
  Menard, \& Monin}]{Bulger:2014ei}
Bulger, J., Patience, J., Ward-Duong, K., {et~al.} 2014, A\&A, 570, A29

\bibitem[{Butler {et~al.}(2006)Butler, Wright, Marcy, Fischer, Vogt, Tinney,
  Jones, Carter, Johnson, McCarthy, \& Penny}]{Butler:2006dd}
Butler, R.~P., Wright, J.~T., Marcy, G.~W., {et~al.} 2006, The Astrophysical
  Journal, 646, 505

\bibitem[{Caceres {et~al.}(2015)Caceres, Hardy, Schreiber, Canovas, Cieza,
  Williams, Hales, Pinte, Menard, \& Wahhaj}]{Caceres:2015hg}
Caceres, C., Hardy, A., Schreiber, M.~R., {et~al.} 2015, The Astrophysical
  Journal Letters, 806, L22

\bibitem[{Caldwell {et~al.}(2020)Caldwell, Tenenbaum, Twicken, Jenkins, Ting,
  Smith, Hedges, Fausnaugh, Rose, \& Burke}]{Caldwell:2020aa}
Caldwell, D.~A., Tenenbaum, P., Twicken, J.~D., {et~al.} 2020, RNAAS, 4, 201

\bibitem[{Campante {et~al.}(2016)Campante, Lund, Kuszlewicz, Davies, Chaplin,
  Albrecht, Winn, Bedding, Benomar, Bossini, Handberg, Santos, Eylen, Basu,
  Christensen-Dalsgaard, Elsworth, Hekker, Hirano, Huber, Karoff, Kjeldsen,
  Lundkvist, North, Aguirre, Stello, \& White}]{Campante:2016jq}
Campante, T.~L., Lund, M.~N., Kuszlewicz, J.~S., {et~al.} 2016, ApJ, 819, 85

\bibitem[{Canto Martins {et~al.}(2020)Canto Martins, Gomes, Messias, Lira,
  Leão, Almeida, Teixeira, Chagas, Bravo, Belete, \&
  Medeiros}]{CantoMartins:2020aa}
Canto Martins, B.~L., Gomes, R.~L., Messias, Y.~S., {et~al.} 2020, The
  Astrophysical Journal Supplement Series, 250, 20

\bibitem[{Carson {et~al.}(2013)Carson, Thalmann, Janson, Kozakis, Bonnefoy,
  Biller, Schlieder, Currie, Mcelwain, Goto, Henning, Brandner, Feldt, Kandori,
  Kuzuhara, Stevens, Wong, Gainey, Fukagawa, Kuwada, Brandt, Kwon, Abe, Egner,
  Grady, Guyon, Hashimoto, Hayano, Hayashi, Hayashi, Hodapp, Ishii, Iye, Knapp,
  Kudo, Kusakabe, Matsuo, Miyama, Morino, Moro-Martin, Nishimura, Pyo, Serabyn,
  Suto, Suzuki, Takami, Takato, Terada, Tomono, Turner, Watanabe, Wisniewski,
  Yamada, Takami, Usuda, \& Tamura}]{Carson:2013fw}
Carson, J., Thalmann, C., Janson, M., {et~al.} 2013, ApJL, 763, L32

\bibitem[{Chatterjee {et~al.}(2008)Chatterjee, Ford, Matsumura, \&
  Rasio}]{Chatterjee:2008gd}
Chatterjee, S., Ford, E.~B., Matsumura, S., \& Rasio, F.~A. 2008, The
  Astrophysical Journal, 686, 580

\bibitem[{Chauvin {et~al.}(2003)Chauvin, Thomson, Dumas, Beuzit, Lowrance,
  Fusco, Lagrange, Zuckerman, \& Mouillet}]{Chauvin:2003gi}
Chauvin, G., Thomson, M., Dumas, C., {et~al.} 2003, A\&A, 404, 157

\bibitem[{Chauvin {et~al.}(2005)Chauvin, Lagrange, Zuckerman, Dumas, Mouillet,
  Song, Beuzit, Lowrance, \& Bessell}]{Chauvin:2005dh}
Chauvin, G., Lagrange, A.-M., Zuckerman, B., {et~al.} 2005, A\&A, 438, L29

\bibitem[{Chen {et~al.}(2011)Chen, Mamajek, Bitner, Pecaut, Su, \&
  Weinberger}]{Chen:2011gz}
Chen, C.~H., Mamajek, E.~E., Bitner, M.~A., {et~al.} 2011, The Astrophysical
  Journal, 738, 122

\bibitem[{Chen {et~al.}(2020)Chen, Su, \& Xu}]{Chen:2020aa}
Chen, C.~H., Su, K. Y.~L., \& Xu, S. 2020, Nature Astronomy, 4, 328

\bibitem[{Choi {et~al.}(2016)Choi, Dotter, Conroy, Cantiello, Paxton, \&
  Johnson}]{Choi:2016kf}
Choi, J., Dotter, A., Conroy, C., {et~al.} 2016, The Astrophysical Journal,
  823, 102

\bibitem[{Chubak {et~al.}(2012)Chubak, Marcy, Fischer, Howard, Isaacson,
  Johnson, \& Wright}]{Chubak:2012tv}
Chubak, C., Marcy, G., Fischer, D.~A., {et~al.} 2012, arXiv:1207.6212

\bibitem[{Crepp {et~al.}(2014)Crepp, Johnson, Howard, Marcy, Brewer, Fischer,
  Wright, \& Isaacson}]{Crepp:2014ce}
Crepp, J.~R., Johnson, J.~A., Howard, A.~W., {et~al.} 2014, ApJ, 781, 29

\bibitem[{Crepp {et~al.}(2012)Crepp, Johnson, Fischer, Howard, Marcy, Wright,
  Isaacson, Boyajian, Braun, Hillenbrand, Hinkley, Carpenter, \&
  Brewer}]{Crepp:2012eg}
Crepp, J.~R., Johnson, J.~A., Fischer, D.~A., {et~al.} 2012, The Astrophysical
  Journal, 751, 97

\bibitem[{Cruz {et~al.}(2003)Cruz, Reid, Liebert, Kirkpatrick, \&
  Lowrance}]{Cruz:2003fi}
Cruz, K.~L., Reid, I.~N., Liebert, J., Kirkpatrick, J.~D., \& Lowrance, P.~J.
  2003, The Astronomical Journal, 126, 2421

\bibitem[{Cutispoto {et~al.}(2002)Cutispoto, Pastori, Pasquini, Medeiros,
  Tagliaferri, \& Andersen}]{Cutispoto:2002aa}
Cutispoto, G., Pastori, L., Pasquini, L., {et~al.} 2002, Astronomy \&
  Astrophysics, 384, 491

\bibitem[{Cutri {et~al.}(2003)Cutri, Skrutskie, Dyk, Beichman, Carpenter,
  Chester, Cambresy, Evans, Fowler, Gizis, Howard, Huchra, Jarrett, Kopan,
  Kirkpatrick, Light, Marsh, McCallon, Schneider, Stiening, Sykes, Weinberg,
  Wheaton, Wheelock, \& Zacarias}]{Cutri:2003tp}
Cutri, R.~M., Skrutskie, M.~F., Dyk, S.~V., {et~al.} 2003, The 2MASS All-Sky
  Catalog of Point Sources, University of Massachusetts and Infrared Processing
  and Analysis Center; IPAC/California Institute of Technology

\bibitem[{Czekala {et~al.}(2015)Czekala, Andrews, Jensen, Stassun, Torres, \&
  Wilner}]{Czekala:2015vb}
Czekala, I., Andrews, S.~M., Jensen, E. L.~N., {et~al.} 2015, The Astrophysical
  Journal, 806, 154

\bibitem[{Daffern-Powell {et~al.}(2022)Daffern-Powell, Parker, \&
  Quanz}]{DaffernPowell:2022aa}
Daffern-Powell, E.~C., Parker, R.~J., \& Quanz, S.~P. 2022, Monthly Notices of
  the Royal Astronomical Society, 514, 920

\bibitem[{Dahm {et~al.}(2012)Dahm, Slesnick, \& White}]{Dahm:2012iu}
Dahm, S.~E., Slesnick, C.~L., \& White, R.~J. 2012, The Astrophysical Journal,
  745, 56

\bibitem[{Dalal {et~al.}(2021)Dalal, Kiefer, Hébrard, Sahlmann, Sousa,
  Forveille, Delfosse, Arnold, Astudillo-Defru, Bonfils, Boisse, Bouchy,
  Bourrier, Brugger, Cortés-Zuleta, Deleuil, Demangeon, Díaz, Hara, Heidari,
  Hobson, Lopez, Lovis, Martioli, Mignon, Mousis, Moutou, Rey, Santerne,
  Santos, Ségransan, Strøm, \& Udry}]{Dalal:2021aa}
Dalal, S., Kiefer, F., Hébrard, G., {et~al.} 2021, Astronomy \& Astrophysics,
  651, A11

\bibitem[{Davies {et~al.}(2014)Davies, Gregory, \& Greaves}]{Davies:2014aa}
Davies, C.~L., Gregory, S.~G., \& Greaves, J.~S. 2014, Monthly Notices of the
  Royal Astronomical Society, 444, 1157

\bibitem[{Dawson(2014)}]{Dawson:2014bj}
Dawson, R.~I. 2014, The Astrophysical Journal Letters, 790, L31

\bibitem[{Dawson {et~al.}(2011)Dawson, Murray-Clay, \&
  Fabrycky}]{Dawson:2011eu}
Dawson, R.~I., Murray-Clay, R.~A., \& Fabrycky, D.~C. 2011, The Astrophysical
  Journal, 743, L17

\bibitem[{Deacon {et~al.}(2016)Deacon, Schlieder, \& Murphy}]{Deacon:2016dg}
Deacon, N.~R., Schlieder, J.~E., \& Murphy, S.~J. 2016, Monthly Notices of the
  Royal Astronomical Society, 457, 3191

\bibitem[{Deacon {et~al.}(2012)Deacon, Liu, Magnier, Bowler, Redstone, Goldman,
  Burgett, Chambers, Flewelling, Kaiser, Morgan, Price, Sweeney, Tonry,
  Wainscoat, \& Waters}]{Deacon:2012eg}
Deacon, N.~R., Liu, M.~C., Magnier, E.~A., {et~al.} 2012, The Astrophysical
  Journal, 755, 94

\bibitem[{Deacon {et~al.}(2014)Deacon, Liu, Magnier, Aller, Best, Dupuy,
  Bowler, Mann, Redstone, Burgett, Chambers, Draper, Flewelling, Hodapp,
  Kaiser, Kudritzki, Morgan, Metcalfe, Price, Tonry, \&
  Wainscoat}]{Deacon:2014ey}
---. 2014, ApJ, 792, 119

\bibitem[{Deacon {et~al.}(2017)Deacon, Magnier, Liu, Schlieder, Aller, Best,
  Bowler, Burgett, Chambers, Draper, Flewelling, Hodapp, Kaiser, Metcalfe,
  Sweeney, Wainscoat, \& Waters}]{Deacon:2017aa}
Deacon, N.~R., Magnier, E.~A., Liu, M.~C., {et~al.} 2017, Monthly Notices of
  the Royal Astronomical Society, 467, 1126

\bibitem[{Delorme {et~al.}(2013)Delorme, Gagné, Girard, Lagrange, Chauvin,
  Naud, Lafrenière, Doyon, Riedel, Bonnefoy, \& Malo}]{Delorme:2013bo}
Delorme, P., Gagné, J., Girard, J.~H., {et~al.} 2013, A\&A, 553, L5

\bibitem[{Delorme {et~al.}(2017)Delorme, Schmidt, Bonnefoy, Desidera, Ginski,
  Charnay, Lazzoni, Christiaens, Messina, D’Orazi, Milli, Schlieder, Gratton,
  Rodet, Lagrange, Absil, Vigan, Galicher, Hagelberg, Bonavita, Lavie, Zurlo,
  Olofsson, Boccaletti, Cantalloube, Mouillet, Chauvin, Hambsch, Langlois,
  Udry, Henning, Beuzit, Mordasini, Lucas, Marocco, Biller, Carson, Cheetham,
  Covino, Caprio, Delboulbe, Feldt, Girard, Hubin, Maire, Pavlov, Petit, Rouan,
  Roelfsema, \& Wildi}]{Delorme:2017aa}
Delorme, P., Schmidt, T., Bonnefoy, M., {et~al.} 2017, Astronomy \&
  Astrophysics, 608, A79

\bibitem[{Donahue {et~al.}(1996)Donahue, Saar, \& Baliunas}]{Donahue:1996ch}
Donahue, R.~A., Saar, S.~H., \& Baliunas, S.~L. 1996, Astrophysical Journal,
  466, 384

\bibitem[{Donati {et~al.}(2008)Donati, Morin, Petit, Delfosse, Forveille,
  Aurière, Cabanac, Dintrans, Fares, Gastine, Jardine, Lignières, Paletou,
  Velez, \& Théado}]{Donati:2008gy}
Donati, J.-F., Morin, J., Petit, P., {et~al.} 2008, Monthly Notices of the
  Royal Astronomical Society, 390, 545

\bibitem[{Donati {et~al.}(2012)Donati, Gregory, Alencar, Hussain, Bouvier,
  Dougados, Jardine, Menard, Romanova, \& Collaboration}]{Donati:2012aa}
Donati, J.~F., Gregory, S.~G., Alencar, S. H.~P., {et~al.} 2012, Monthly
  Notices of the Royal Astronomical Society, 425, 2948

\bibitem[{D'Orazi {et~al.}(2017)D'Orazi, Desidera, Gratton, Lanza, Messina,
  Andrievsky, Korotin, Benatti, Bonnefoy, Covino, \& Janson}]{DOrazi:2017jp}
D'Orazi, V., Desidera, S., Gratton, R.~G., {et~al.} 2017, A\&A, 598, A19

\bibitem[{Dotter(2016)}]{Dotter:2016fa}
Dotter, A. 2016, The Astrophysical Journal Supplement Series, 222, 8

\bibitem[{Drilling \& Landolt(2000)}]{Drilling:2000vo}
Drilling, J.~S., \& Landolt, A.~U. 2000, in Allen's Astrophysical Quantities,
  4th ed., ed. A. N. Cox (New York, NY: AIP Press; Springer 2000), 381

\bibitem[{Dupuy {et~al.}(2022{\natexlab{a}})Dupuy, Brandt, \&
  Brandt}]{Dupuy:2022bb}
Dupuy, T.~J., Brandt, G.~M., \& Brandt, T.~D. 2022{\natexlab{a}}, Monthly
  Notices of the Royal Astronomical Society, 509, 4411

\bibitem[{Dupuy \& Kraus(2013)}]{Dupuy:2013ks}
Dupuy, T.~J., \& Kraus, A.~L. 2013, Science, 341, 1492

\bibitem[{Dupuy \& Liu(2012)}]{Dupuy:2012bp}
Dupuy, T.~J., \& Liu, M.~C. 2012, The Astrophysical Journal Supplement, 201, 19

\bibitem[{Dupuy {et~al.}(2022{\natexlab{b}})Dupuy, Liu, Evans, Best, Pearce,
  Sanghi, Phillips, \& Gagliuffi}]{Dupuy:2022aa}
Dupuy, T.~J., Liu, M.~C., Evans, E.~L., {et~al.} 2022{\natexlab{b}}, arXiv,
  2208.08448

\bibitem[{Durisen {et~al.}(2007)Durisen, Boss, Mayer, Nelson, Quinn, \&
  Rice}]{Durisen:2007wg}
Durisen, R.~H., Boss, A.~P., Mayer, L., {et~al.} 2007, in Protostars and
  Planets V, e. B. Reipurth, D. Jewitt, \& K. Keil (Tucson, AZ: Univ. Arizona
  Press), 607

\bibitem[{Díaz-Francés \& Rubio(2013)}]{DiazFrances:2013hs}
Díaz-Francés, E., \& Rubio, F.~J. 2013, Statistical Papers, 54, 309

\bibitem[{Eiff \& Reiners(2012)}]{AmmlerVonEiff:2012de}
Eiff, M. A.-v., \& Reiners, A. 2012, Astronomy and Astrophysics, 542, A116

\bibitem[{Eiroa {et~al.}(2013)Eiroa, Marshall, Mora, Montesinos, Absil,
  Augereau, Bayo, Bryden, Danchi, Burgo, Ertel, Fridlund, Heras, Krivov,
  Launhardt, Liseau, Löhne, Maldonado, Pilbratt, Roberge, Rodmann,
  Sanz-Forcada, Solano, Stapelfeldt, Thebault, Wolf, Ardila, Arevalo,
  Beichmann, Faramaz, González-García, Gutiérrez, Lebreton,
  Martínez-Arnáiz, Meeus, Montes, Olofsson, Su, White, Barrado, Fukagawa,
  Grün, Kamp, Lorente, Morbidelli, Müller, Mutschke, Nakagawa, Ribas, \&
  Walker}]{Eiroa:2013bk}
Eiroa, C., Marshall, J.~P., Mora, A., {et~al.} 2013, A\&A, 555, A11

\bibitem[{Epstein-Martin {et~al.}(2022)Epstein-Martin, Becker, \&
  Batygin}]{EpsteinMartin:2022aa}
Epstein-Martin, M., Becker, J., \& Batygin, K. 2022, The Astrophysical Journal,
  931, 42

\bibitem[{Eriksson {et~al.}(2020)Eriksson, Torres, Janson, Aoyama, Marleau,
  Bonnefoy, \& Petrus}]{Eriksson:2020aa}
Eriksson, S.~C., Torres, R.~A., Janson, M., {et~al.} 2020, Astronomy \&
  Astrophysics, 638, L6

\bibitem[{Fabrycky \& Tremaine(2007)}]{Fabrycky:2007jh}
Fabrycky, D., \& Tremaine, S. 2007, The Astrophysical Journal, 669, 1298

\bibitem[{Fabrycky \& Winn(2009)}]{Fabrycky:2009ke}
Fabrycky, D.~C., \& Winn, J.~N. 2009, The Astrophysical Journal, 696, 1230

\bibitem[{Faherty {et~al.}(2010)Faherty, Burgasser, West, Bochanski, Cruz,
  Shara, \& Walter}]{Faherty:2010gt}
Faherty, J.~K., Burgasser, A.~J., West, A.~A., {et~al.} 2010, The Astronomical
  Journal, 139, 176

\bibitem[{Faramaz {et~al.}(2021)Faramaz, Marino, Booth, Matrà, Mamajek,
  Bryden, Stapelfeldt, Casassus, Cuadra, Hales, \& Zurlo}]{Faramaz:2021aa}
Faramaz, V., Marino, S., Booth, M., {et~al.} 2021, The Astronomical Journal,
  161, 271

\bibitem[{Feigelson {et~al.}(2006)Feigelson, Lawson, Stark, Townsley, \&
  Garmire}]{Feigelson:2006tz}
Feigelson, E.~D., Lawson, W.~A., Stark, M., Townsley, L., \& Garmire, G.~P.
  2006, AJ, 131, 1730

\bibitem[{Feng {et~al.}(2022)Feng, Butler, Vogt, Clement, Tinney, Cui, Aizawa,
  Jones, Bailey, Burt, Carter, Crane, Dotti, Holden, Ma, Ogihara, Oppenheimer,
  O’Toole, Shectman, Wittenmyer, Wang, Wright, \& Xuan}]{Feng:2022zz}
Feng, F., Butler, R.~P., Vogt, S.~S., {et~al.} 2022, The Astrophysical Journal
  Supplement Series, 262, 21

\bibitem[{Ferrer-Chávez {et~al.}(2021)Ferrer-Chávez, Wang, \&
  Blunt}]{FerrerChavez:2021aa}
Ferrer-Chávez, R., Wang, J.~J., \& Blunt, S. 2021, The Astronomical Journal,
  161, 241

\bibitem[{Fielding {et~al.}(2015)Fielding, McKee, Socrates, Cunningham, \&
  Klein}]{Fielding:2015aa}
Fielding, D.~B., McKee, C.~F., Socrates, A., Cunningham, A.~J., \& Klein, R.~I.
  2015, Monthly Notices of the Royal Astronomical Society, 450, 3306

\bibitem[{Fischer \& Valenti(2005)}]{Fischer:2005gi}
Fischer, D.~A., \& Valenti, J. 2005, The Astrophysical Journal, 622, 1102

\bibitem[{Ford \& Rasio(2008)}]{Ford:2008jo}
Ford, E.~B., \& Rasio, F.~A. 2008, The Astrophysical Journal, 686, 621

\bibitem[{Fouqué {et~al.}(2018)Fouqué, Moutou, Malo, Martioli, Lim,
  Rajpurohit, Artigau, Delfosse, Donati, Forveille, Morin, Allard, Delage,
  Doyon, Hébrard, \& Neves}]{Fouque:2018aa}
Fouqué, P., Moutou, C., Malo, L., {et~al.} 2018, Monthly Notices of the Royal
  Astronomical Society, 475, 1960

\bibitem[{Franson {et~al.}(2022{\natexlab{a}})Franson, Bowler, Brandt, Dupuy,
  Tran, Brandt, Li, \& Kraus}]{Franson:2022bl}
Franson, K., Bowler, B.~P., Brandt, T.~D., {et~al.} 2022{\natexlab{a}}, The
  Astronomical Journal, 163, 50

\bibitem[{Franson {et~al.}(2022{\natexlab{b}})Franson, Bowler, Bonavita,
  Brandt, Chen, Samland, Zhang, Lueber, Heng, Kitzmann, Wolf, Jones, Tran,
  Gagliuffi, Biller, Chilcote, Crepp, Dupuy, Faherty, Fontanive, Groff,
  Gratton, Guyon, Jensen-Clem, Jovanovic, Kasdin, Lozi, Magnier, Muzic, Sanghi,
  \& Theissen}]{Franson:2022bb}
Franson, K., Bowler, B.~P., Bonavita, M., {et~al.} 2022{\natexlab{b}}, arXiv,
  2211.09840

\bibitem[{Frasca {et~al.}(2018)Frasca, Guillout, Klutsch, Ferrero, Marilli,
  Biazzo, Gandolfi, \& Montes}]{Frasca:2018aa}
Frasca, A., Guillout, P., Klutsch, A., {et~al.} 2018, Astronomy \&
  Astrophysics, 612, A96

\bibitem[{Gaia Collaboration {et~al.}(2022)Gaia Collaboration, Vallenari,
  Brown, \& Prusti}]{GaiaCollaboration:2022aa}
Gaia Collaboration, Vallenari, A., Brown, A. G.~A., \& Prusti, T. 2022,
  arxiv:2208.00211

\bibitem[{Gaia Collaboration {et~al.}(2016)Gaia Collaboration, Prusti,
  Bruijne, Brown, Vallenari, Babusiaux, Bailer-Jones, Bastian, Biermann, Evans,
  Eyer, Jansen, Jordi, Klioner, Lammers, Lindegren, Luri, Mignard, Milligan,
  Panem, Poinsignon, Pourbaix, Randich, Sarri, Sartoretti, Siddiqui, Soubiran,
  Valette, Leeuwen, Walton, Aerts, Arenou, Cropper, Drimmel, Høg, Katz,
  Lattanzi, O'Mullane, Grebel, Holland, Huc, Passot, Bramante, Cacciari,
  Castañeda, Chaoul, Cheek, Angeli, Fabricius, Guerra, Hernandez,
  Jean-Antoine-Piccolo, Masana, Messineo, Mowlavi, Nienartowicz,
  Ordóñez-Blanco, Panuzzo, Portell, Richards, Riello, Seabroke, Tanga,
  Thévenin, Torra, Els, Gracia-Abril, Comoretto, Garcia-Reinaldos, Lock,
  Mercier, Altmann, Andrae, Astraatmadja, Bellas-Velidis, Benson, Berthier,
  Blomme, Busso, Carry, Cellino, Clementini, Cowell, Creevey, Cuypers,
  Davidson, Ridder, Torres, Delchambre, Dell’Oro, Ducourant, Frémat,
  García-Torres, Gosset, Halbwachs, Hambly, Harrison, Hauser, Hestroffer,
  Hodgkin, Huckle, Hutton, Jasniewicz, Jordan, Kontizas, Korn, Lanzafame,
  Manteiga, Moitinho, Muinonen, Osinde, Pancino, Pauwels, Petit, Recio-Blanco,
  Robin, Sarro, Siopis, Smith, Smith, Sozzetti, Thuillot, Reeven, Viala, Abbas,
  Aramburu, Accart, Aguado, Allan, Allasia, Altavilla, Álvarez, Alves,
  Anderson, Andrei, Varela, Antiche, Antoja, Antón, Arcay, Atzei, Ayache,
  Bach, Baker, Balaguer-Núñez, Barache, Barata, Barbier, Barblan, Baroni,
  Navascués, Barros, Barstow, Becciani, Bellazzini, Bellei, García,
  Belokurov, Bendjoya, Berihuete, Bianchi, Bienaymé, Billebaud, Blagorodnova,
  Blanco-Cuaresma, Boch, Bombrun, Borrachero, Bouquillon, Bourda, Bouy,
  Bragaglia, Breddels, Brouillet, Brüsemeister, Bucciarelli, Budnik, Burgess,
  Burgon, Burlacu, Busonero, Buzzi, Caffau, Cambras, Campbell, Cancelliere,
  Cantat-Gaudin, Carlucci, Carrasco, Castellani, Charlot, Charnas, Charvet,
  Chassat, Chiavassa, Clotet, Cocozza, Collins, Collins, Costigan, Crifo,
  Cross, Crosta, Crowley, Dafonte, Damerdji, Dapergolas, David, David, Cat,
  Felice, Laverny, Luise, March, Martino, Souza, Debosscher, Pozo, Delbo,
  Delgado, Delgado, Marco, Matteo, Diakite, Distefano, Dolding, Anjos,
  Drazinos, Duran, Dzigan, Ecale, Edvardsson, Enke, Erdmann, Escolar, Espina,
  Evans, Bontemps, Fabre, Fabrizio, Faigler, Falcão, Casas, Faye, Federici,
  Fedorets, Fernández-Hernández, Fernique, Fienga, Figueras, Filippi,
  Findeisen, Fonti, Fouesneau, Fraile, Fraser, Fuchs, Furnell, Gai, Galleti,
  Galluccio, Garabato, García-Sedano, Garé, Garofalo, Garralda, Gavras,
  Gerssen, Geyer, Gilmore, Girona, Giuffrida, Gomes, González-Marcos,
  González-Núñez, González-Vidal, Granvik, Guerrier, Guillout, Guiraud,
  Gúrpide, Gutiérrez-Sánchez, Guy, Haigron, Hatzidimitriou, Haywood, Heiter,
  Helmi, Hobbs, Hofmann, Holl, Holland, Hunt, Hypki, Icardi, Irwin, Fombelle,
  Jofre, Jonker, Jorissen, Julbe, Karampelas, Kochoska, Kohley, Kolenberg,
  Kontizas, Koposov, Kordopatis, Koubsky, Kowalczyk, Krone-Martins,
  Kudryashova, Kull, Bachchan, Lacoste-Seris, Lanza, Lavigne, Poncin-Lafitte,
  Lebreton, Lebzelter, Leccia, Leclerc, Lecoeur-Taibi, Lemaitre, Lenhardt,
  Leroux, Liao, Licata, Lindstrøm, Lister, Livanou, Lobel, Löffler, López,
  Lopez-Lozano, Lorenz, Loureiro, MacDonald, Fernandes, Managau, Mann,
  Mantelet, Marchal, Marchant, Marconi, Marie, Marinoni, Marrese, Marschalkó,
  Marshall, Martín-Fleitas, Martino, Mary, Matijevic, Mazeh, McMillan,
  Messina, Mestre, Michalik, Millar, Miranda, Molina, Molinaro, Molinaro,
  Molnár, Moniez, Montegriffo, Monteiro, Mor, Mora, Morbidelli, Morel,
  Morgenthaler, Morley, Morris, Mulone, Muraveva, Musella, Narbonne, Nelemans,
  Nicastro, Noval, Ordénovic, Ordieres-Meré, Osborne, Pagani, Pagano,
  Pailler, Palacin, Palaversa, Parsons, Paulsen, Pecoraro, Pedrosa,
  Pentikäinen, Pereira, Pichon, Piersimoni, Pineau, Plachy, Plum, Poujoulet,
  Prsa, Pulone, Ragaini, Rago, Rambaux, Ramos-Lerate, Ranalli, Rauw, Read,
  Regibo, Renk, Reylé, Ribeiro, Rimoldini, Ripepi, Riva, Rixon, Roelens,
  Romero-Gomez, Rowell, Royer, Rudolph, Ruiz-Dern, Sadowski, Sellés, Sahlmann,
  Salgado, Salguero, Sarasso, Savietto, Schnorhk, Schultheis, Sciacca, Segol,
  Segovia, Ségransan, Serpell, Shih, Smareglia, Smart, Smith, Solano, Solitro,
  Sordo, Nieto, Souchay, Spagna, Spoto, Stampa, Steele, Steidelmüller,
  Stephenson, Stoev, Suess, Süveges, Surdej, Szabados, Szegedi-Elek, Tapiador,
  Taris, Tauran, Taylor, Teixeira, Terrett, Tingley, Trager, Turon, Ulla,
  Utrilla, Valentini, Elteren, Hemelryck, Leeuwen, Varadi, Vecchiato,
  Veljanoski, Via, Vicente, Vogt, Voss, Votruba, Voutsinas, Walmsley, Weiler,
  Weingrill, Werner, Wevers, Whitehead, Wyrzykowski, Yoldas, Žerjal, Zucker,
  Zurbach, Zwitter, Alecu, Allen, Prieto, Amorim, Anglada-Escudé, Arsenijevic,
  Azaz, Balm, Beck, Bernstein, Bigot, Bijaoui, Blasco, Bonfigli, Bono,
  Boudreault, Bressan, Brown, Brunet, Bunclark, Buonanno, Butkevich, Carret,
  Carrion, Chemin, Chéreau, Corcione, Darmigny, Boer, Teodoro, Zeeuw, Luche,
  Domingues, Dubath, Fodor, Frézouls, Fries, Fustes, Fyfe, Gallardo, Gallegos,
  Gardiol, Gebran, Gomboc, Gomez, Grux, Gueguen, Heyrovsky, Hoar, Iannicola,
  Parache, Janotto, Joliet, Jonckheere, Keil, Kim, Klagyivik, Klar, Knude,
  Kochukhov, Kolka, Kos, Kutka, Lainey, LeBouquin, Liu, Loreggia, Makarov,
  Marseille, Martayan, Martinez-Rubi, Massart, Meynadier, Mignot, Munari,
  Nguyen, Nordlander, Ocvirk, O'Flaherty, Sanz, Ortiz, Osorio, Oszkiewicz,
  Ouzounis, Palmer, Park, Pasquato, Peltzer, Peralta, Péturaud, Pieniluoma,
  Pigozzi, Poels, Prat, Prod’homme, Raison, Rebordao, Risquez,
  Rocca-Volmerange, Rosen, Ruiz-Fuertes, Russo, Sembay, Vizcaino, Short,
  Siebert, Silva, Sinachopoulos, Slezak, Soffel, Sosnowska, Straižys, Linden,
  Terrell, Theil, Tiede, Troisi, Tsalmantza, Tur, Vaccari, Vachier, Valles,
  Hamme, Veltz, Virtanen, Wallut, Wichmann, Wilkinson, Ziaeepour, \&
  Zschocke}]{GaiaCollaboration:2016cu}
Gaia Collaboration, Prusti, T., Bruijne, J. H. J.~d., {et~al.} 2016, A\&A,
  595, A1

\bibitem[{Gaidos {et~al.}(2021)Gaidos, Hirano, Kraus, Kuzuhara, Zhang, Lee,
  Salama, Berger, Grunblatt, Ansdell, Liu, Harakawa, Hodapp, Jacobson, Konishi,
  Kotani, Kudo, Kurokawa, Nishikawa, Omiya, Serizawa, Tamura, Ueda, \&
  Vievard}]{Gaidos:2021aa}
Gaidos, E., Hirano, T., Kraus, A.~L., {et~al.} 2021, Monthly Notices of the
  Royal Astronomical Society, 512, 583

\bibitem[{Gaudi \& Winn(2007)}]{Gaudi:2007vi}
Gaudi, B.~S., \& Winn, J.~N. 2007, The Astrophysical Journal, 655, 550

\bibitem[{Gauza {et~al.}(2015)Gauza, Bejar, Pérez-Garrido, Osorio, Lodieu,
  Rebolo, Pallé, \& Nowak}]{Gauza:2015fw}
Gauza, B., Bejar, V. J.~S., Pérez-Garrido, A., {et~al.} 2015, The
  Astrophysical Journal, 804, 96

\bibitem[{Geballe {et~al.}(2002)Geballe, Knapp, Leggett, Fan, Golimowski,
  Anderson, Brinkmann, Csabai, Gunn, Hawley, Hennessy, Henry, Hill, Hindsley,
  Ivezić, Lupton, McDaniel, Munn, Narayanan, Peng, Pier, Rockosi, Schneider,
  Smith, Strauss, Tsvetanov, Uomoto, York, \& Zheng}]{Geballe:2002kw}
Geballe, T.~R., Knapp, G.~R., Leggett, S.~K., {et~al.} 2002, The Astrophysical
  Journal, 564, 466

\bibitem[{Gizis {et~al.}(2001)Gizis, Kirkpatrick, \& Wilson}]{Gizis:2001jp}
Gizis, J.~E., Kirkpatrick, J.~D., \& Wilson, J.~C. 2001, The Astronomical
  Journal, 121, 2185

\bibitem[{Glebocki \& Gnacinski(2005)}]{Glebocki:2005aa}
Glebocki, R., \& Gnacinski, P. 2005, yCat, 3244, 0

\bibitem[{Glebocki \& Stawikowski(1997)}]{Glebocki:1997ul}
Glebocki, R., \& Stawikowski, A. 1997, A\&A, 328, 579

\bibitem[{Goldman {et~al.}(2010)Goldman, Marsat, Henning, Clemens, \&
  Greiner}]{Goldman:2010ct}
Goldman, B., Marsat, S., Henning, T., Clemens, C., \& Greiner, J. 2010, Monthly
  Notices of the Royal Astronomical Society, 405, 1140

\bibitem[{Gomes {et~al.}(2013)Gomes, Pinfield, Marocco, Day-Jones, Burningham,
  Zhang, Jones, Spaandonk, \& Weights}]{Gomes:2013eg}
Gomes, J.~I., Pinfield, D.~J., Marocco, F., {et~al.} 2013, Monthly Notices of
  the Royal Astronomical Society, 431, 2745

\bibitem[{Gonzalez {et~al.}(2010)Gonzalez, Carlson, \& Tobin}]{Gonzalez:2010ey}
Gonzalez, G., Carlson, M.~K., \& Tobin, R.~W. 2010, PASJ, 403, 1368

\bibitem[{Grandjean {et~al.}(2021)Grandjean, Lagrange, Meunier, Rubini,
  Desidera, Galland, Borgniet, Zicher, Messina, Chauvin, Sterzik, \&
  Pantoja}]{Grandjean:2021dk}
Grandjean, A., Lagrange, A.-M., Meunier, N., {et~al.} 2021, Astronomy \&
  Astrophysics, 650, A39

\bibitem[{Grankin {et~al.}(2008)Grankin, Bouvier, Herbst, \&
  Melnikov}]{Grankin:2008aa}
Grankin, K.~N., Bouvier, J., Herbst, W., \& Melnikov, S.~Y. 2008, Astronomy \&
  Astrophysics, 479, 827

\bibitem[{Gray(2005)}]{Gray:2005va}
Gray, D.~F. 2005, {The Observation and Analysis of Stellar Photospheres (3rd
  Ed.; Cambridge: Cambridge Univ. Press)}

\bibitem[{Greaves {et~al.}(2014)Greaves, Kennedy, Thureau, Eiroa, Marshall,
  Maldonado, Matthews, Olofsson, Barlow, Moro-Martín, Sibthorpe, Absil,
  Ardila, Booth, Broekhoven-Fiene, Brown, Cameron, Burgo, Francesco,
  Eislöffel, Duchêne, Ertel, Holland, Horner, Kalas, Kavelaars, Lestrade,
  Vican, Wilner, Wolf, \& Wyatt}]{Greaves:2014aa}
Greaves, J.~S., Kennedy, G.~M., Thureau, N., {et~al.} 2014, MNRAS, 438, L31

\bibitem[{Guenther {et~al.}(2005)Guenther, Neuhäuser, Wuchterl, Mugrauer,
  Bedalov, \& Hauschildt}]{Guenther:2005dr}
Guenther, E.~W., Neuhäuser, R., Wuchterl, G., {et~al.} 2005, Astronomische
  Nachrichten, 326, 958

\bibitem[{Haffert {et~al.}(2019)Haffert, Bohn, Boer, Snellen, Brinchmann,
  Girard, Keller, \& Bacon}]{Haffert:2019ba}
Haffert, S.~Y., Bohn, A.~J., Boer, J., {et~al.} 2019, Nature Astronomy, 3, 749

\bibitem[{Hale(1994)}]{Hale:1994gv}
Hale, A. 1994, The Astronomical Journal, 107, 306

\bibitem[{Hartman {et~al.}(2011)Hartman, Bakos, Noyes, Sipocz, Kovacs, Mazeh,
  Shporer, \& Pál}]{Hartman:2011cq}
Hartman, J.~D., Bakos, G.~A., Noyes, R.~W., {et~al.} 2011, The Astronomical
  Journal, 141, 166

\bibitem[{Hartmann {et~al.}(1987)Hartmann, Soderblom, \&
  Stauffer}]{Hartmann:1987aa}
Hartmann, L.~W., Soderblom, D.~R., \& Stauffer, J.~R. 1987, AJ, 93, 907

\bibitem[{Hastings(1970)}]{Hastings:1970wm}
Hastings, W. 1970, Biometrika, 57, 97

\bibitem[{Heller(1993)}]{Heller:1993aa}
Heller, C.~H. 1993, The Astrophysical Journal, 408, 337

\bibitem[{Herbig \& Bell(1988)}]{Herbig:1988aa}
Herbig, G.~H., \& Bell, K.~R. 1988, in Third Catalog of Emission-Line Stars of
  the Orion Population, ed. G. H. Herbig \& K. R. Bell, Vol. 3 (Santa Cruz, CA:
  Lick Observatory), 90

\bibitem[{Herczeg \& Hillenbrand(2014)}]{Herczeg:2014is}
Herczeg, G.~J., \& Hillenbrand, L.~A. 2014, ApJ, 786, 97

\bibitem[{Herczeg \& Hillenbrand(2015)}]{Herczeg:2015bp}
---. 2015, The Astrophysical Journal, 808, 23

\bibitem[{Herrero {et~al.}(2012)Herrero, Ribas, Jordi, Guinan, \&
  Engle}]{Herrero:2012di}
Herrero, E., Ribas, I., Jordi, C., Guinan, E.~F., \& Engle, S.~G. 2012,
  Astronomy and Astrophysics, 537, A147

\bibitem[{Hinkley(1969)}]{Hinkley:1969aa}
Hinkley, D. 1969, PASJ, 56, 635

\bibitem[{Hinkley {et~al.}(2015)Hinkley, Bowler, Vigan, Aller, Liu, Mawet,
  Matthews, Wahhaj, Kraus, Baraffe, \& Chabrier}]{Hinkley:2015gk}
Hinkley, S., Bowler, B.~P., Vigan, A., {et~al.} 2015, The Astrophysical Journal
  Letters, 805, L10

\bibitem[{Hinkley {et~al.}(2022)Hinkley, Lacour, Marleau, Lagrange, Wang,
  Kammerer, Cumming, Nowak, Rodet, Stolker, Balmer, Ray, Bonnefoy, Mollière,
  Lazzoni, Kennedy, Mordasini, Abuter, Aigrain, Amorim, Asensio-Torres,
  Babusiaux, Benisty, Berger, Beust, Blunt, Boccaletti, Bohn, Bonnet,
  Bourdarot, Brandner, Cantalloube, Caselli, Charnay, Chauvin, Chomez, Choquet,
  Christiaens, Clénet, Foresto, Cridland, Delorme, Dembet, Zeeuw, Drescher,
  Duvert, Eckart, Eisenhauer, Feuchtgruber, Galland, Garcia, Lopez, Gardner,
  Gendron, Genzel, Gillessen, Girard, Grandjean, Haubois, Heißel, Henning,
  Hippler, Horrobin, Houllé, Hubert, Jocou, Keppler, Kervella, Kreidberg,
  Lapeyrère, Bouquin, Léna, Lutz, Maire, Mang, Mérand, Meunier, Monnier,
  Mordasini, Mouillet, Nasedkin, Ott, Otten, Paladini, Paumard, Perraut,
  Perrin, Philipot, Pfuhl, Pourré, Pueyo, Rameau, Rickman, Rubini,
  Rustamkulov, Samland, Shangguan, Shimizu, Sing, Straubmeier, Sturm, Tacconi,
  Dishoeck, Vigan, Vincent, Ward-Duong, Widmann, Wieprecht, Wiezorrek, Woillez,
  Yazici, Young, Zicher, \& Collaboration}]{Hinkley:2022aa}
Hinkley, S., Lacour, S., Marleau, G.~D., {et~al.} 2022, arXiv:2208.04867

\bibitem[{Hogg {et~al.}(2010)Hogg, Myers, \& Bovy}]{Hogg:2010gh}
Hogg, D.~W., Myers, A.~D., \& Bovy, J. 2010, The Astrophysical Journal, 725,
  2166

\bibitem[{Hojjatpanah {et~al.}(2019)Hojjatpanah, Figueira, Santos, Adibekyan,
  Sousa, Delgado-Mena, Alibert, Cristiani, Hernández, Lanza, Marcantonio,
  Martins, Micela, Molaro, Neves, Oshagh, Pepe, Poretti, Rojas-Ayala, Rebolo,
  Mascareño, \& Osorio}]{Hojjatpanah:2019aa}
Hojjatpanah, S., Figueira, P., Santos, N.~C., {et~al.} 2019, Astronomy \&
  Astrophysics, 629, A80

\bibitem[{Holt(1893)}]{Holt:1893aa}
Holt. 1893, AstAp, 12, 646

\bibitem[{Horne(1986)}]{Horne:1986bg}
Horne, K. 1986, Astronomical Society of the Pacific, 98, 609

\bibitem[{Houdebine(2010)}]{Houdebine:2010fv}
Houdebine, E.~R. 2010, Monthly Notices of the Royal Astronomical Society, 407,
  1657

\bibitem[{Houdebine {et~al.}(2016)Houdebine, Mullan, Paletou, \&
  Gebran}]{Houdebine:2016aa}
Houdebine, E.~R., Mullan, D.~J., Paletou, F., \& Gebran, M. 2016, The
  Astrophysical Journal, 822, 97

\bibitem[{Huang {et~al.}(2020)Huang, Vanderburg, Pal, Sha, Yu, Fong, Fausnaugh,
  Shporer, Guerrero, Vanderspek, \& Ricker}]{Huang:2020bb}
Huang, C.~X., Vanderburg, A., Pal, A., {et~al.} 2020, RNAAS, 4, 204

\bibitem[{Huber {et~al.}(2013)Huber, Carter, Barbieri, Miglio, Deck, Fabrycky,
  Montet, Buchhave, Chaplin, Hekker, Montalban, Sanchis-Ojeda, \&
  Basu}]{Huber:2013aa}
Huber, D., Carter, J.~A., Barbieri, M., {et~al.} 2013, Science, 342, 331

\bibitem[{Huber {et~al.}(2017)Huber, Zinn, Bojsen-Hansen, Pinsonneault,
  Sahlholdt, Serenelli, Aguirre, Stassun, Stello, Tayar, Bastien, Bedding,
  Buchhave, Chaplin, Davies, García, Latham, Mathur, Mosser, \&
  Sharma}]{Huber:2017aa}
Huber, D., Zinn, J., Bojsen-Hansen, M., {et~al.} 2017, The Astrophysical
  Journal, 844, 102

\bibitem[{Husser {et~al.}(2013)Husser, Berg, Dreizler, Homeier, Reiners,
  Barman, \& Hauschildt}]{Husser:2013ca}
Husser, T.~O., Berg, S. W.-v., Dreizler, S., {et~al.} 2013, A\&A, 553, A6

\bibitem[{Ireland {et~al.}(2011)Ireland, Kraus, Martinache, Law, \&
  Hillenbrand}]{Ireland:2011id}
Ireland, M.~J., Kraus, A., Martinache, F., Law, N., \& Hillenbrand, L.~A. 2011,
  The Astrophysical Journal, 726, 113

\bibitem[{Iriarte {et~al.}(2021)Iriarte, de Castro, \& Gomez}]{Iriarte:2021aa}
Iriarte, Y.~A., de Castro, M., \& Gomez, H.~W. 2021, Symmetry, 13, 269

\bibitem[{Irwin {et~al.}(2018)Irwin, Charbonneau, Esquerdo, Latham, Winters,
  Dittmann, Newton, Berta-Thompson, Berlind, \& Calkins}]{Irwin:2018aa}
Irwin, J.~M., Charbonneau, D., Esquerdo, G.~A., {et~al.} 2018, The Astronomical
  Journal, 156, 140

\bibitem[{Janson(2010)}]{Janson:2010gg}
Janson, M. 2010, Monthly Notices of the Royal Astronomical Society, 408, 514

\bibitem[{Janson {et~al.}(2012)Janson, Jayawardhana, Girard, LaFreniere,
  Bonavita, Gizis, \& Brandeker}]{Janson:2012te}
Janson, M., Jayawardhana, R., Girard, J.~H., {et~al.} 2012, ApJL, 758, L2

\bibitem[{Jayasinghe {et~al.}(2018)Jayasinghe, Kochanek, Stanek, Shappee,
  Holoien, Thompson, Prieto, Dong, Pawlak, Shields, Pojmanski, Otero, Britt, \&
  Will}]{Jayasinghe:2018aa}
Jayasinghe, T., Kochanek, C.~S., Stanek, K.~Z., {et~al.} 2018, Monthly Notices
  of the Royal Astronomical Society, 477, 3145

\bibitem[{Jeffers {et~al.}(2018)Jeffers, Schöfer, Lamert, Reiners, Montes,
  Caballero, Cortés-Contreras, Marvin, Passegger, Zechmeister, Quirrenbach,
  Alonso-Floriano, Amado, Bauer, Casal, Alonso, Herrero, Morales, Mundt, Ribas,
  \& Sarmiento}]{Jeffers:2018vt}
Jeffers, S.~V., Schöfer, P., Lamert, A., {et~al.} 2018, Astronomy \&
  Astrophysics, 614, A76

\bibitem[{Jenkins {et~al.}(2016)Jenkins, Twicken, McCauliff, Campbell,
  Sanderfer, Lung, Mansouri-Samani, Girouard, Tenenbaum, Klaus, Smith,
  Caldwell, Chacon, Henze, Heiges, Latham, Morgan, Swade, Rinehart, \&
  Vanderspek}]{Jenkins:2016aa}
Jenkins, J.~M., Twicken, J.~D., McCauliff, S., {et~al.} 2016, Proc. SPIE, 9913,
  99133E

\bibitem[{Jenkins {et~al.}(2009)Jenkins, Ramsey, Jones, Pavlenko, Gallardo,
  Barnes, \& Pinfield}]{Jenkins:2009ea}
Jenkins, J.~S., Ramsey, L.~W., Jones, H. R.~A., {et~al.} 2009, The
  Astrophysical Journal, 704, 975

\bibitem[{Jenkins {et~al.}(2012)Jenkins, Pavlenko, Ivanyuk, Gallardo, Jones,
  Day-Jones, Jones, Ruiz, Pinfield, \& Yakovina}]{Jenkins:2012ku}
Jenkins, J.~S., Pavlenko, Y.~V., Ivanyuk, O., {et~al.} 2012, Monthly Notices of
  the Royal Astronomical Society, 420, 3587

\bibitem[{Jennings \& Chiang(2021)}]{Jennings:2021aa}
Jennings, R.~M., \& Chiang, E. 2021, Monthly Notices of the Royal Astronomical
  Society, 507, 5187

\bibitem[{Jones \& Herbig(1979)}]{Jones:1979aa}
Jones, B.~F., \& Herbig, G.~H. 1979, AJ, 84, 1874

\bibitem[{Jones {et~al.}(2016)Jones, White, Quinn, Ireland, Boyajian, Schaefer,
  \& Baines}]{Jones:2016hg}
Jones, J., White, R.~J., Quinn, S., {et~al.} 2016, The Astrophysical Journal
  Letters, 822, L3

\bibitem[{Jumper \& Fisher(2013)}]{Jumper:2013ct}
Jumper, P.~H., \& Fisher, R.~T. 2013, ApJ, 769, 9

\bibitem[{Justesen \& Albrecht(2020)}]{Justesen:2020jz}
Justesen, A.~B., \& Albrecht, S. 2020, A\&A, 642, A212

\bibitem[{Jönsson {et~al.}(2020)Jönsson, Holtzman, Prieto, Cunha,
  García-Hernández, Hasselquist, Masseron, Osorio, Shetrone, Smith,
  Stringfellow, Bizyaev, Edvardsson, Majewski, Mészáros, Souto, Zamora,
  Beaton, Bovy, Donor, Pinsonneault, Poovelil, \& Sobeck}]{Jonsson:2020aa}
Jönsson, H., Holtzman, J.~A., Prieto, C.~A., {et~al.} 2020, The Astronomical
  Journal, 160, 120

\bibitem[{Kaye {et~al.}(1999)Kaye, Handler, Krisciunas, Poretti, \&
  Zerbi}]{Kaye:1999aa}
Kaye, A.~B., Handler, G., Krisciunas, K., Poretti, E., \& Zerbi, F.~M. 1999,
  Publications of the Astronomical Society of the Pacific, 111, 840

\bibitem[{Kendall {et~al.}(2007)Kendall, Jones, Pinfield, Pokorny, Folkes,
  Weights, Jenkins, \& Mauron}]{Kendall:2007fd}
Kendall, T.~R., Jones, H. R.~A., Pinfield, D.~J., {et~al.} 2007, Monthly
  Notices of the Royal Astronomical Society, 374, 445

\bibitem[{Kenyon {et~al.}(1994)Kenyon, Dobrzycka, \& Hartmann}]{Kenyon:1994aa}
Kenyon, S., Dobrzycka, D., \& Hartmann, L. 1994, AJ, 108, 1872

\bibitem[{Keppler {et~al.}(2018)Keppler, Benisty, Müller, Henning, Boekel,
  Cantalloube, Ginski, Holstein, Maire, Pohl, Samland, Avenhaus, Baudino,
  Boccaletti, Boer, Bonnefoy, Chauvin, Desidera, Langlois, Lazzoni, Marleau,
  Mordasini, Pawellek, Stolker, Vigan, Zurlo, Birnstiel, Brandner, Feldt,
  Flock, Girard, Gratton, Hagelberg, Isella, Janson, Juhasz, Kemmer, Kral,
  Lagrange, Launhardt, Matter, Menard, Milli, Mollière, Olofsson, Perez,
  Pinilla, Pinte, quanz, Schmidt, Udry, Wahhaj, Williams, Buenzli, Cudel,
  Dominik, Galicher, Kasper, Lannier, Mesa, Mouillet, Peretti, Perrot, Salter,
  Sissa, Wildi, Abe, Antichi, Augereau, Baruffolo, Baudoz, Bazzon, Beuzit,
  Blanchard, Brems, Buey, Caprio, Carbillet, Carle, Cascone, Cheetham, Claudi,
  Costille, Delboulbé, Dohlen, Fantinel, Feautrier, Fusco, Giro, Gluck, Gry,
  Hubin, Hugot, Jaquet, Mignant, Llored, Madec, Magnard, Martinez, Maurel,
  Meyer, Möller-Nilsson, Moulin, Mugnier, Origné, Pavlov, Perret, Petit,
  Pragt, Puget, Rabou, Ramos, Rigal, Rochat, Roelfsema, Rousset, Roux,
  Salasnich, Sauvage, Sevin, Soenke, Stadler, Suarez, Turatto, \&
  Weber}]{Keppler:2018dd}
Keppler, M., Benisty, M., Müller, A., {et~al.} 2018, Astronomy and
  Astrophysics, 617, A44

\bibitem[{Keppler {et~al.}(2019)Keppler, Teague, Bae, Benisty, Henning, Boekel,
  Chapillon, Pinilla, Williams, Bertrang, Facchini, Flock, Ginski, Juhasz,
  Klahr, Liu, Müller, Pérez, Pohl, Rosotti, Samland, \&
  Semenov}]{Keppler:2019aa}
Keppler, M., Teague, R., Bae, J., {et~al.} 2019, Astronomy \& Astrophysics,
  625, A118

\bibitem[{Kiefer {et~al.}(2019)Kiefer, Hébrard, Sahlmann, Sousa, Forveille,
  Santos, Mayor, Deleuil, Wilson, Dalal, Díaz, Henry, Hagelberg, Hobson,
  Demangeon, Bourrier, Delfosse, Arnold, Astudillo-Defru, Beuzit, Boisse,
  Bonfils, Borgniet, Bouchy, Courcol, Ehrenreich, Hara, Lagrange, Lovis,
  Montagnier, Moutou, Pepe, Perrier, Rey, Santerne, Ségransan, Udry, \&
  Vidal-Madjar}]{Kiefer:2019aa}
Kiefer, F., Hébrard, G., Sahlmann, J., {et~al.} 2019, Astronomy \&
  Astrophysics, 631, A125

\bibitem[{Kiraga(2012)}]{Kiraga:2012wj}
Kiraga, M. 2012, Acta Astronomica, 62, 67

\bibitem[{Kirkpatrick {et~al.}(2001)Kirkpatrick, Dahn, Monet, Reid, Gizis,
  Liebert, \& Burgasser}]{Kirkpatrick:2001bi}
Kirkpatrick, J.~D., Dahn, C.~C., Monet, D.~G., {et~al.} 2001, The Astronomical
  Journal, 121, 3235

\bibitem[{Kirkpatrick {et~al.}(2000)Kirkpatrick, Reid, Liebert, Gizis,
  Burgasser, Monet, Dahn, Nelson, \& Williams}]{Kirkpatrick:2000gi}
Kirkpatrick, J.~D., Reid, I.~N., Liebert, J., {et~al.} 2000, The Astronomical
  Journal, 120, 447

\bibitem[{Kirkpatrick {et~al.}(2011)Kirkpatrick, Cushing, Gelino, Griffith,
  Skrutskie, Marsh, Wright, Mainzer, Eisenhardt, McLean, Thompson, Bauer,
  Benford, Bridge, Lake, Petty, Stanford, Tsai, Bailey, Beichman, Bloom,
  Bochanski, Burgasser, Capak, Cruz, Hinz, Kartaltepe, Knox, Manohar, Masters,
  Morales-Calderón, Prato, Rodigas, Salvato, Schurr, Scoville, Simcoe,
  Stapelfeldt, Stern, Stock, \& Vacca}]{Kirkpatrick:2011ey}
Kirkpatrick, J.~D., Cushing, M.~C., Gelino, C.~R., {et~al.} 2011, The
  Astrophysical Journal Supplement, 197, 19

\bibitem[{Knapp {et~al.}(2004)Knapp, Leggett, Fan, Marley, Geballe, Golimowski,
  Finkbeiner, Gunn, Hennawi, Ivezić, Lupton, Schlegel, Strauss, Tsvetanov,
  Chiu, Hoversten, Glazebrook, Zheng, Hendrickson, Williams, Uomoto, Vrba,
  Henden, Luginbuhl, Guetter, Munn, Canzian, Schneider, \&
  Brinkmann}]{Knapp:2004ji}
Knapp, G.~R., Leggett, S.~K., Fan, X., {et~al.} 2004, The Astronomical Journal,
  127, 3553

\bibitem[{Konopacky {et~al.}(2016)Konopacky, Rameau, Duchene, Filippazzo,
  Godfrey, Marois, Nielsen, Pueyo, Rafikov, Rice, Wang, Ammons, Bailey, Barman,
  Bulger, Bruzzone, Chilcote, Cotten, Dawson, Rosa, Doyon, Esposito,
  Fitzgerald, Follette, Goodsell, Graham, Greenbaum, Hibon, Hung, Ingraham,
  Kalas, LaFreniere, Larkin, Macintosh, Maire, Marchis, Marley, Matthews,
  Metchev, Millar-Blanchaer, Oppenheimer, Palmer, Patience, Perrin, Poyneer,
  Rajan, Rantakyrö, Savransky, Schneider, Sivaramakrishnan, Song, Soummer,
  Thomas, Wallace, Ward-Duong, Wiktorowicz, \& Wolff}]{Konopacky:2016dk}
Konopacky, Q.~M., Rameau, J., Duchene, G., {et~al.} 2016, The Astrophysical
  Journal Letters, 829, L4

\bibitem[{Kounkel {et~al.}(2019)Kounkel, Covey, Moe, Kratter, Suárez, Stassun,
  Román-Zúñiga, Hernández, Kim, Ramírez, Roman-Lopes, Stringfellow,
  Jaehnig, Borissova, Tofflemire, Krolikowski, Rizzuto, Kraus, Badenes,
  Longa-Peña, Chew, Barba, Nidever, Brown, Lee, Pan, Bizyaev, Oravetz, \&
  Oravetz}]{Kounkel:2019cp}
Kounkel, M., Covey, K., Moe, M., {et~al.} 2019, The Astronomical Journal, 157,
  196

\bibitem[{Kraus {et~al.}(2015)Kraus, Andrews, Bowler, Herczeg, Ireland, Liu,
  Metchev, \& Cruz}]{Kraus:2015fx}
Kraus, A.~L., Andrews, S.~M., Bowler, B.~P., {et~al.} 2015, The Astrophysical
  Journal, 798, L23

\bibitem[{Kraus \& Hillenbrand(2009)}]{Kraus:2009hc}
Kraus, A.~L., \& Hillenbrand, L.~A. 2009, The Astrophysical Journal, 703, 1511

\bibitem[{Kraus \& Ireland(2012)}]{Kraus:2012gk}
Kraus, A.~L., \& Ireland, M.~J. 2012, The Astrophysical Journal, 745, 5

\bibitem[{Kraus {et~al.}(2014{\natexlab{a}})Kraus, Ireland, Cieza, Hinkley,
  Dupuy, Bowler, \& Liu}]{Kraus:2014tl}
Kraus, A.~L., Ireland, M.~J., Cieza, L.~A., {et~al.} 2014{\natexlab{a}}, The
  Astrophysical Journal, 781, 20

\bibitem[{Kraus {et~al.}(2014{\natexlab{b}})Kraus, Shkolnik, Allers, \&
  Liu}]{Kraus:2014ur}
Kraus, A.~L., Shkolnik, E.~L., Allers, K.~N., \& Liu, M.~C. 2014{\natexlab{b}},
  AJ, 147, 146

\bibitem[{Kraus {et~al.}(2020)Kraus, Bouquin, Kreplin, Davies, Hone, Monnier,
  Gardner, Kennedy, \& Hinkley}]{Kraus:2020aa}
Kraus, S., Bouquin, J.-B.~L., Kreplin, A., {et~al.} 2020, The Astrophysical
  Journal Letters, 897, L8

\bibitem[{Kraus {et~al.}(2022)Kraus, Mortimer, Chhabra, Lu, Codron, Gardner,
  Anugu, Monnier, Bouquin, Ireland, Martinache, Defrère, \&
  Martinod}]{Kraus:2022aa}
Kraus, S., Mortimer, D., Chhabra, S., {et~al.} 2022, SPIE, 12183, 121831S

\bibitem[{Kuzuhara {et~al.}(2011)Kuzuhara, Tamura, Ishii, Kudo, Nishiyama, \&
  Kandori}]{Kuzuhara:2011ic}
Kuzuhara, M., Tamura, M., Ishii, M., {et~al.} 2011, The Astronomical Journal,
  141, 119

\bibitem[{Kuzuhara {et~al.}(2013)Kuzuhara, Tamura, Kudo, Janson, Kandori,
  Brandt, Thalmann, Spiegel, Biller, Carson, Hori, Suzuki, Burrows, Henning,
  Turner, McElwain, Moro-Martín, Suenaga, Takahashi, Kwon, Lucas, Abe,
  Brandner, Egner, Feldt, Fujiwara, Goto, Grady, Guyon, Hashimoto, Hayano,
  Hayashi, Hayashi, Hodapp, Ishii, Iye, Knapp, Matsuo, Mayama, Miyama, Morino,
  Nishikawa, Nishimura, Kotani, Kusakabe, Pyo, Serabyn, Suto, Takami, Takato,
  Terada, Tomono, Watanabe, Wisniewski, Yamada, Takami, \&
  Usuda}]{Kuzuhara:2013jz}
Kuzuhara, M., Tamura, M., Kudo, T., {et~al.} 2013, The Astrophysical Journal,
  774, 11

\bibitem[{Kuzuhara {et~al.}(2022)Kuzuhara, Currie, Takarada, Brandt, Sato,
  Uyama, Janson, Chilcote, Tobin, Lawson, Hori, Guyon, Groff, Lozi, Vievard,
  Sahoo, Deo, Jovanovic, Ahn, Martinache, Skaf, Akiyama, Norris, Bonnefoy,
  Hełminiak, Kudo, McElwain, Samland, Wagner, Wisniewski, Knapp, Kwon,
  Nishikawa, Serabyn, Hayashi, \& Tamura}]{Kuzuhara:2022aa}
Kuzuhara, M., Currie, T., Takarada, T., {et~al.} 2022, The Astrophysical
  Journal Letters, 934, L18

\bibitem[{Köhler {et~al.}(2013)Köhler, Ratzka, Petr-Gotzens, \&
  Correia}]{Kohler:2013im}
Köhler, R., Ratzka, T., Petr-Gotzens, M.~G., \& Correia, S. 2013, A\&A, 558,
  A80

\bibitem[{Lacour {et~al.}(2021)Lacour, Wang, Rodet, Nowak, Shangguan, Beust,
  Lagrange, Abuter, Amorim, Asensio-Torres, Benisty, Berger, Blunt, Boccaletti,
  Bohn, Bolzer, Bonnefoy, Bonnet, Bourdarot, Brandner, Cantalloube, Caselli,
  Charnay, Chauvin, Choquet, Christiaens, Clénet, Foresto, Cridland, Dembet,
  Dexter, Zeeuw, Drescher, Duvert, Eckart, Eisenhauer, Gao, Garcia, Lopez,
  Gendron, Genzel, Gillessen, Girard, Haubois, Heißel, Henning, Hinkley,
  Hippler, Horrobin, Houllé, Hubert, Jocou, Kammerer, Keppler, Kervella,
  Kreidberg, Lapeyrère, Bouquin, Léna, Lutz, Maire, Mérand, Mollière,
  Monnier, Mouillet, Nasedkin, Ott, Otten, Paladini, Paumard, Perraut, Perrin,
  Pfuhl, Rickman, Pueyo, Rameau, Rousset, Rustamkulov, Samland, Shimizu, Sing,
  Stadler, Stolker, Straub, Straubmeier, Sturm, Tacconi, Dishoeck, Vigan,
  Vincent, Fellenberg, Ward-Duong, Widmann, Wieprecht, Wiezorrek, Woillez,
  Yazici, Young, \& Collaboration}]{Lacour:2021aa}
Lacour, S., Wang, J.~J., Rodet, L., {et~al.} 2021, Astronomy \& Astrophysics,
  654, L2

\bibitem[{Lafrenière {et~al.}(2008)Lafrenière, Jayawardhana, \&
  Kerkwijk}]{Lafreniere:2008jt}
Lafrenière, D., Jayawardhana, R., \& Kerkwijk, M. H.~v. 2008, The
  Astrophysical Journal, 689, L153

\bibitem[{Lagrange {et~al.}(2019)Lagrange, Meunier, Rubini, Keppler, Galland,
  Chapellier, Michel, Balona, Beust, Guillot, Grandjean, Borgniet, karnia,
  Wilson, Kiefer, Bonnefoy, Lillo-Box, Pantoja, Jones, Iglesias, Rodet, Díaz,
  Zapata, Abe, \& Schmider}]{Lagrange:2019bi}
Lagrange, A.-M., Meunier, N., Rubini, P., {et~al.} 2019, Nature Astronomy, 3,
  1135

\bibitem[{Lagrange {et~al.}(2020)Lagrange, Rubini, Nowak, Lacour, Grandjean,
  Boccaletti, Langlois, Delorme, Gratton, Wang, Flasseur, Galicher, Kral,
  Meunier, Beust, Babusiaux, Coroller, Thebault, Kervella, Zurlo, Maire,
  Wahhaj, Amorim, Asensio-Torres, Benisty, Berger, Bonnefoy, Brandner,
  Cantalloube, Charnay, Chauvin, Choquet, Clénet, Christiaens, Foresto, Zeeuw,
  Desidera, Duvert, Eckart, Eisenhauer, Galland, Gao, Garcia, López, Gendron,
  Genzel, Gillessen, Girard, Hagelberg, Haubois, Henning, Heißel, Hippler,
  Horrobin, Janson, Kammerer, Kenworthy, Keppler, Kreidberg, Lapeyrere,
  Bouquin, Léna, Mérand, Messina, Mollière, Monnier, Ott, Otten, Paumard,
  Paladini, Perraut, Perrin, Pueyo, Pfuhl, Rodet, Rodríguez-Coira, Rousset,
  Samland, Shangguan, Schmidt, Straub, Straubmeier, Stolker, Vigan, Vincent,
  Widmann, Woillez, \& Collaboration}]{Lagrange:2020do}
Lagrange, A.-M., Rubini, P., Nowak, M., {et~al.} 2020, Astronomy and
  Astrophysics, 642, A18

\bibitem[{Lai {et~al.}(2018)Lai, Anderson, \& Pu}]{Lai:2018cb}
Lai, D., Anderson, K.~R., \& Pu, B. 2018, MNRAS, 475, 5231

\bibitem[{Lai {et~al.}(2011)Lai, Foucart, \& Lin}]{Lai:2011bb}
Lai, D., Foucart, F., \& Lin, D. N.~C. 2011, Monthly Notices of the Royal
  Astronomical Society, 412, 2790

\bibitem[{Lambrechts \& Johansen(2012)}]{Lambrechts:2012gr}
Lambrechts, M., \& Johansen, A. 2012, A\&A, 544, A32

\bibitem[{Le Bouquin {et~al.}(2009)Le Bouquin, Absil, Benisty, Massi,
  Mérand, \& Stefl}]{LeBouquin:2009aa}
Le Bouquin, J.-B., Absil, O., Benisty, M., {et~al.} 2009, Astronomy \&
  Astrophysics, 498, L41

\bibitem[{Liebing {et~al.}(2021)Liebing, Jeffers, Reiners, \&
  Zechmeister}]{Liebing:2021gy}
Liebing, F., Jeffers, S.~V., Reiners, A., \& Zechmeister, M. 2021, Astronomy
  and Astrophysics, 654, A168

\bibitem[{Lightkurve Collaboration {et~al.}(2018)Lightkurve Collaboration,
  Miranda, Hedges, Gully-Santiago, Saunders, Cody, Barclay, Hall, Sagear,
  Turtelboom, Zhang, Tzanidakis, Mighell, Coughlin, Bell, Berta-Thompson,
  Williams, Dotson, \& Barentsen}]{LightkurveCollaboration:2018aa}
Lightkurve Collaboration, Miranda, C. V.~d., Hedges, C., {et~al.} 2018,
  Lightkurve: Kepler and TESS time series analysis in Python, Astrophysics
  Source Code Library, ascl:1812.013

\bibitem[{Liu {et~al.}(2016)Liu, Dupuy, \& Allers}]{Liu:2016co}
Liu, M.~C., Dupuy, T.~J., \& Allers, K.~N. 2016, ApJ, 833, 1

\bibitem[{Liu {et~al.}(2002)Liu, Fischer, Graham, Lloyd, Marcy, \&
  Butler}]{Liu:2002fx}
Liu, M.~C., Fischer, D.~A., Graham, J.~R., {et~al.} 2002, The Astrophysical
  Journal, 571, 519

\bibitem[{Liu {et~al.}(2007)Liu, Leggett, \& Chiu}]{Liu:2007cg}
Liu, M.~C., Leggett, S.~K., \& Chiu, K. 2007, The Astrophysical Journal, 660,
  1507

\bibitem[{Louden {et~al.}(2021)Louden, Winn, Petigura, Isaacson, Howard,
  Masuda, Albrecht, \& Kosiarek}]{Louden:2021jz}
Louden, E.~M., Winn, J.~N., Petigura, E.~A., {et~al.} 2021, The Astronomical
  Journal, 161, 68

\bibitem[{Lovis \& Pepe(2007)}]{Lovis:2007cp}
Lovis, C., \& Pepe, F. 2007, A\&A, 468, 1115

\bibitem[{Low \& Lynden-Bell(1976)}]{Low:1976wt}
Low, C., \& Lynden-Bell, D. 1976, Royal Astronomical Society, 176, 367

\bibitem[{Lowrance {et~al.}(1999)Lowrance, McCarthy, Becklin, Zuckerman,
  Schneider, Webb, Hines, Kirkpatrick, Koerner, Low, Meier, Rieke, Smith,
  Terrile, \& Thompson}]{Lowrance:1999ck}
Lowrance, P.~J., McCarthy, C., Becklin, E.~E., {et~al.} 1999, The Astrophysical
  Journal, 512, L69

\bibitem[{Luck(2017)}]{Luck:2017jd}
Luck, R.~E. 2017, The Astronomical Journal, 153, 21

\bibitem[{Luhman {et~al.}(2009)Luhman, Mamajek, Allen, Muench, \&
  Finkbeiner}]{Luhman:2009cx}
Luhman, K.~L., Mamajek, E.~E., Allen, P.~R., Muench, A.~A., \& Finkbeiner,
  D.~P. 2009, The Astrophysical Journal, 691, 1265

\bibitem[{Luhman {et~al.}(2007)Luhman, Patten, Marengo, Schuster, Hora, Ellis,
  Stauffer, Sonnett, Winston, Gutermuth, Megeath, Backman, Henry, Werner, \&
  Fazio}]{Luhman:2007fu}
Luhman, K.~L., Patten, B.~M., Marengo, M., {et~al.} 2007, The Astrophysical
  Journal, 654, 570

\bibitem[{López-Santiago {et~al.}(2010)López-Santiago, Montes, Gálvez-Ortiz,
  Crespo-Chacón, Martínez-Arnáiz, Fernández-Figueroa, Castro, \&
  Cornide}]{LopezSantiago:2010aa}
López-Santiago, J., Montes, D., Gálvez-Ortiz, M.~C., {et~al.} 2010, Astronomy
  \& Astrophysics, 514, A97

\bibitem[{López-Valdivia {et~al.}(2019)López-Valdivia, Mace, Sokal, Hussaini,
  Kidder, Mann, Gosnell, Oh, Kesseli, Muirhead, Johns-Krull, \&
  Jaffe}]{LopezValdivia:2019aa}
López-Valdivia, R., Mace, G.~N., Sokal, K.~R., {et~al.} 2019, The
  Astrophysical Journal, 879, 105

\bibitem[{Mace {et~al.}(2013)Mace, Kirkpatrick, Cushing, Gelino, McLean,
  Logsdon, Wright, Skrutskie, Beichman, Eisenhardt, \& Kulas}]{Mace:2013ku}
Mace, G.~N., Kirkpatrick, J.~D., Cushing, M.~C., {et~al.} 2013, ApJ, 777, 36

\bibitem[{Macintosh {et~al.}(2015)Macintosh, Graham, Barman, Rosa, Konopacky,
  Marley, Marois, Nielsen, \& Pueyo}]{Macintosh:2015fw}
Macintosh, B., Graham, J.~R., Barman, T., {et~al.} 2015, Science, 350, 64

\bibitem[{Macintosh {et~al.}(2001)Macintosh, Max, Zuckerman, Becklin, Kaisler,
  Lowrance, Weinberger, Christou, Schneider, \& Acton}]{Macintosh:2001tr}
Macintosh, B.~A., Max, C., Zuckerman, B., {et~al.} 2001, Young Stars Near
  Earth: Progress and Prospects, 244, 309

\bibitem[{Maire {et~al.}(2018)Maire, Rodet, Lazzoni, Boccaletti, Brandner,
  Galicher, Cantalloube, Mesa, Klahr, Beust, Chauvin, Desidera, Janson,
  Keppler, Olofsson, Augereau, Daemgen, Henning, Thébault, Bonnefoy, Feldt,
  Gratton, Lagrange, Langlois, Meyer, Vigan, D’Orazi, Hagelberg, Coroller,
  Ligi, Rouan, Samland, Schmidt, Udry, Zurlo, Abe, Carle, Delboulbé,
  Feautrier, Magnard, Maurel, Moulin, Pavlov, Perret, Petit, Ramos, Rigal,
  Roux, \& Weber}]{Maire:2018aa}
Maire, A.-L., Rodet, L., Lazzoni, C., {et~al.} 2018, Astronomy \& Astrophysics,
  615, A177

\bibitem[{Maire {et~al.}(2019)Maire, Rodet, Cantalloube, Galicher, Brandner,
  Messina, Lazzoni, Mesa, Melnick, Carson, Samland, Biller, Boccaletti, Wahhaj,
  Beust, Bonnefoy, Chauvin, Desidera, Langlois, Henning, Janson, Olofsson,
  Rouan, Menard, Lagrange, Gratton, Vigan, Meyer, Cheetham, Beuzit, Dohlen,
  Avenhaus, Bonavita, Claudi, Cudel, Daemgen, D'Orazi, Fontanive, Hagelberg,
  Coroller, Perrot, Rickman, Schmidt, Sissa, Udry, Zurlo, Abe, Origné, Rigal,
  Rousset, Roux, \& Weber}]{Maire:2019aa}
Maire, A.~L., Rodet, L., Cantalloube, F., {et~al.} 2019, A\&A, 624, A118

\bibitem[{Maire {et~al.}(2020)Maire, Molaverdikhani, Desidera, Trifonov,
  Mollière, D'Orazi, Frankel, Baudino, Messina, Müller, Charnay, Cheetham,
  Delorme, Ligi, Bonnefoy, Brandner, Mesa, Cantalloube, Galicher, Henning,
  Biller, Hagelberg, Lagrange, Lavie, Rickman, Ségransan, Udry, Chauvin,
  Gratton, Langlois, Vigan, Meyer, Beuzit, Bhowmik, Boccaletti, Lazzoni,
  Perrot, Schmidt, Zurlo, Gluck, Pragt, Ramos, Roelfsema, Roux, \&
  Sauvage}]{Maire:2020iu}
Maire, A.~L., Molaverdikhani, K., Desidera, S., {et~al.} 2020, Astronomy and
  Astrophysics, 639, A47

\bibitem[{Malo {et~al.}(2014)Malo, Artigau, Doyon, LaFreniere, Albert, \&
  Gagné}]{Malo:2014dk}
Malo, L., Artigau, E., Doyon, R., {et~al.} 2014, ApJ, 788, 81

\bibitem[{Mamajek(2012)}]{Mamajek:2012ga}
Mamajek, E.~E. 2012, The Astrophysical Journal, 754, L20

\bibitem[{Marcussen \& Albrecht(2022)}]{Marcussen:2022aa}
Marcussen, M.~L., \& Albrecht, S.~H. 2022, The Astrophysical Journal, 933, 227

\bibitem[{Marino {et~al.}(2015)Marino, Pérez, \& Casassus}]{Marino:2015hg}
Marino, S., Pérez, S., \& Casassus, S. 2015, ApJL, 798, L44

\bibitem[{Marino {et~al.}(2020)Marino, Zurlo, Faramaz, Milli, Henning, Kennedy,
  Matrà, Pérez, Delorme, Cieza, \& Hughes}]{Marino:2020aa}
Marino, S., Zurlo, A., Faramaz, V., {et~al.} 2020, Monthly Notices of the Royal
  Astronomical Society, 498, 1319

\bibitem[{Marois {et~al.}(2008)Marois, Macintosh, Barman, Zuckerman, Song,
  Patience, Lafreniere, \& Doyon}]{Marois:2008ei}
Marois, C., Macintosh, B., Barman, T., {et~al.} 2008, Science, 322, 1348

\bibitem[{Marois {et~al.}(2010)Marois, Zuckerman, Konopacky, Macintosh, \&
  Barman}]{Marois:2010gp}
Marois, C., Zuckerman, B., Konopacky, Q.~M., Macintosh, B., \& Barman, T. 2010,
  Nature, 372, 1

\bibitem[{Marsden {et~al.}(2014)Marsden, Petit, Jeffers, Morin, Fares, Reiners,
  Jr., Aurière, Bouvier, Carter, Catala, Dintrans, Donati, Gastine, Jardine,
  Konstantinova-Antova, Lanoux, Lignières, Morgenthaler, Ramìrez-Vèlez,
  Théado, Grootel, \& Collaboration}]{Marsden:2014bd}
Marsden, S.~C., Petit, P., Jeffers, S.~V., {et~al.} 2014, Monthly Notices of
  the Royal Astronomical Society, 444, 3517

\bibitem[{Martinez \& Kraus(2022)}]{Martinez:2022aa}
Martinez, R.~A., \& Kraus, A.~L. 2022, The Astronomical Journal, 163, 36

\bibitem[{Martínez-Arnáiz {et~al.}(2011)Martínez-Arnáiz, López-Santiago,
  Crespo-Chacón, \& Montes}]{MartinezArnaiz:2011ci}
Martínez-Arnáiz, R., López-Santiago, J., Crespo-Chacón, I., \& Montes, D.
  2011, Monthly Notices of the Royal Astronomical Society, 414, 2629

\bibitem[{Martínez-Arnáiz {et~al.}(2010)Martínez-Arnáiz, Maldonado, Montes,
  Eiroa, \& Montesinos}]{MartinezArnaiz:2010ie}
Martínez-Arnáiz, R., Maldonado, J., Montes, D., Eiroa, C., \& Montesinos, B.
  2010, A\&A, 520, A79

\bibitem[{Masuda(2022)}]{Masuda:2022aa}
Masuda, K. 2022, The Astrophysical Journal, 937, 94

\bibitem[{Masuda {et~al.}(2021)Masuda, Petigura, \& Hall}]{Masuda:2021aa}
Masuda, K., Petigura, E.~A., \& Hall, O.~J. 2021, Monthly Notices of the Royal
  Astronomical Society, 510, 5623

\bibitem[{Masuda \& Winn(2020)}]{Masuda:2020dp}
Masuda, K., \& Winn, J.~N. 2020, The Astronomical Journal, 159, 81

\bibitem[{Mawet {et~al.}(2016)Mawet, Wizinowich, Dekany, Chun, Hall, Cetre,
  Guyon, Wallace, Bowler, Liu, Ruane, Serabyn, Bartos, Wang, Vasisht,
  Fitzgerald, Skemer, Ireland, Fucik, Fortney, Crossfield, Hu, \&
  Benneke}]{Mawet:2016aa}
Mawet, D., Wizinowich, P., Dekany, R., {et~al.} 2016, Proceedings of the SPIE,
  9909, 99090D

\bibitem[{McLaughlin(1924)}]{McLaughlin:1924vz}
McLaughlin, D.~B. 1924, The Astrophysical Journal, 60, 22

\bibitem[{Mellon {et~al.}(2017)Mellon, Mamajek, Oberst, \&
  Pecaut}]{Mellon:2017aa}
Mellon, S.~N., Mamajek, E.~E., Oberst, T.~E., \& Pecaut, M.~J. 2017, The
  Astrophysical Journal, 844, 66

\bibitem[{Meshkat {et~al.}(2015)Meshkat, Bonnefoy, Mamajek, quanz, Chauvin,
  Kenworthy, Rameau, Meyer, Lagrange, Lannier, \& Delorme}]{Meshkat:2015hd}
Meshkat, T., Bonnefoy, M., Mamajek, E.~E., {et~al.} 2015, Monthly Notices of
  the Royal Astronomical Society, 453, 2379

\bibitem[{Messina {et~al.}(2010)Messina, Desidera, Turatto, Lanzafame, \&
  Guinan}]{Messina:2010kx}
Messina, S., Desidera, S., Turatto, M., Lanzafame, A.~C., \& Guinan, E.~F.
  2010, A\&A, 520, A15

\bibitem[{Metchev \& Hillenbrand(2004)}]{Metchev:2004kl}
Metchev, S.~A., \& Hillenbrand, L.~A. 2004, The Astrophysical Journal, 617,
  1330

\bibitem[{Metchev \& Hillenbrand(2006)}]{Metchev:2006bq}
---. 2006, The Astrophysical Journal, 651, 1166

\bibitem[{Metropolis {et~al.}(1953)Metropolis, Rosenbluth, Rosenbluth, Teller,
  \& Teller}]{Metropolis:1953vj}
Metropolis, N., Rosenbluth, A., Rosenbluth, M., Teller, A., \& Teller, E. 1953,
  The Journal of Chemical Physics, 21, 1087

\bibitem[{Meyer {et~al.}(1998)Meyer, Edwards, Hinkle, \& Strom}]{Meyer:1998aa}
Meyer, M.~R., Edwards, S., Hinkle, K.~H., \& Strom, S.~E. 1998, The
  Astrophysical Journal, 508, 397

\bibitem[{Milli {et~al.}(2017)Milli, Hibon, Christiaens, Choquet, Bonnefoy,
  Kennedy, Wyatt, Absil, Gonzalez, Burgo, Matrà, Augereau, Boccaletti,
  Delacroix, Ertel, Dent, Forsberg, Fusco, Girard, Habraken, Huby, Karlsson,
  Lagrange, Mawet, Mouillet, Perrin, Pinte, Pueyo, Reyes, Soummer, Surdej,
  Tarricq, \& Wahhaj}]{Milli:2017fs}
Milli, J., Hibon, P., Christiaens, V., {et~al.} 2017, A\&A, 597, L2

\bibitem[{Mishenina {et~al.}(2008)Mishenina, Soubiran, Bienaymé, Korotin,
  Belik, Usenko, \& Kovtyukh}]{Mishenina:2008aa}
Mishenina, T.~V., Soubiran, C., Bienaymé, O., {et~al.} 2008, Astronomy \&
  Astrophysics, 489, 923

\bibitem[{Mishenina {et~al.}(2012)Mishenina, Soubiran, Kovtyukh, Katsova, \&
  Livshits}]{Mishenina:2012aa}
Mishenina, T.~V., Soubiran, C., Kovtyukh, V.~V., Katsova, M.~M., \& Livshits,
  M.~A. 2012, Astronomy \& Astrophysics, 547, A106

\bibitem[{Mizusawa {et~al.}(2012)Mizusawa, Rebull, Stauffer, Bryden, Meyer, \&
  Song}]{Mizusawa:2012vv}
Mizusawa, T.~F., Rebull, L.~M., Stauffer, J.~R., {et~al.} 2012, The
  Astronomical Journal, 144, 135

\bibitem[{Monin {et~al.}(2013)Monin, Whelan, Lefloch, Dougados, \&
  Oliveira}]{Monin:2013ep}
Monin, J.-L., Whelan, E.~T., Lefloch, B., Dougados, C., \& Oliveira, C. A.~d.
  2013, A\&A, 551, L1

\bibitem[{Montalto {et~al.}(2021)Montalto, Piotto, Marrese, Nascimbeni,
  Prisinzano, Granata, Marinoni, Desidera, Ortolani, Aerts, Alei, Altavilla,
  Benatti, Börner, Cabrera, Claudi, Deleuil, Fabrizio, Gizon, Goupil, Heras,
  Magrin, Malavolta, Mas-Hesse, Pagano, Paproth, Pertenais, Pollacco,
  Ragazzoni, Ramsay, Rauer, \& Udry}]{Montalto:2021aa}
Montalto, M., Piotto, G., Marrese, P.~M., {et~al.} 2021, Astronomy \&
  Astrophysics, 653, A98

\bibitem[{Mora {et~al.}(2020)Mora, Wu, Bowler, \& Sheehan}]{Mora:2020aa}
Mora, A., Wu, Y.-L., Bowler, B.~P., \& Sheehan, P. 2020, RNAAS, 4, 9

\bibitem[{Morton \& Johnson(2011)}]{Morton:2011jc}
Morton, T.~D., \& Johnson, J.~A. 2011, The Astrophysical Journal, 729, 138

\bibitem[{Morton \& Winn(2014)}]{Morton:2014in}
Morton, T.~D., \& Winn, J.~N. 2014, ApJ, 796, 47

\bibitem[{Mugrauer {et~al.}(2006)Mugrauer, Seifahrt, Neuhäuser, \&
  Mazeh}]{Mugrauer:2006iy}
Mugrauer, M., Seifahrt, A., Neuhäuser, R., \& Mazeh, T. 2006, Monthly Notices
  RAS Letters, 373, L31

\bibitem[{Mugrauer {et~al.}(2010)Mugrauer, Vogt, Neuhäuser, \&
  Schmidt}]{Mugrauer:2010cp}
Mugrauer, M., Vogt, N., Neuhäuser, R., \& Schmidt, T. O.~B. 2010, A\&A, 523,
  L1

\bibitem[{Muirhead {et~al.}(2018)Muirhead, Dressing, Mann, Rojas-Ayala,
  Lépine, Paegert, Lee, \& Oelkers}]{Muirhead:2018io}
Muirhead, P.~S., Dressing, C.~D., Mann, A.~W., {et~al.} 2018, The Astronomical
  Journal, 155, 180

\bibitem[{Murphy {et~al.}(2007)Murphy, Tzanavaris, Webb, \&
  Lovis}]{Murphy:2007aa}
Murphy, M.~T., Tzanavaris, P., Webb, J.~K., \& Lovis, C. 2007, PASJ, 378, 221

\bibitem[{Muñoz \& Perets(2018)}]{Munoz:2018jq}
Muñoz, D.~J., \& Perets, H.~B. 2018, The Astronomical Journal, 156, 253

\bibitem[{Müller {et~al.}(2018)Müller, Keppler, Henning, Samland, Chauvin,
  Beust, Maire, Molaverdikhani, Boekel, Benisty, Boccaletti, Bonnefoy,
  Cantalloube, Charnay, Baudino, Gennaro, Long, Cheetham, Desidera, Feldt,
  Fusco, Girard, Gratton, Hagelberg, Janson, Lagrange, Langlois, Lazzoni, Ligi,
  Menard, Mesa, Meyer, Mollière, Mordasini, Moulin, Pavlov, Pawellek, quanz,
  Ramos, Rouan, Sissa, Stadler, Vigan, Wahhaj, Weber, \& Zurlo}]{Muller:2018aa}
Müller, A., Keppler, M., Henning, T., {et~al.} 2018, A\&A, 617, L2

\bibitem[{Nagpal {et~al.}(2022)Nagpal, Blunt, Bowler, Dupuy, Nielsen, \&
  Wang}]{Nagpal:2022aa}
Nagpal, V., Blunt, S., Bowler, B.~P., {et~al.} 2022, arXiv, 2211.02121

\bibitem[{Nakajima {et~al.}(1994)Nakajima, Durrance, Golimowski, \&
  Kulkarni}]{Nakajima:1994ea}
Nakajima, T., Durrance, S.~T., Golimowski, D.~A., \& Kulkarni, S.~R. 1994, The
  Astrophysical Journal, 428, 797

\bibitem[{Nakajima {et~al.}(1995)Nakajima, Oppenheimer, Kulkarni, Golimowski,
  Matthews, \& Durrance}]{Nakajima:1995bb}
Nakajima, T., Oppenheimer, B.~R., Kulkarni, S.~R., {et~al.} 1995, Nature, 378,
  463

\bibitem[{Naud {et~al.}(2014)Naud, Artigau, Malo, Albert, Doyon, LaFreniere,
  Gagné, Saumon, Morley, Allard, Homeier, Beichman, Gelino, \&
  Boucher}]{Naud:2014jx}
Naud, M.-E., Artigau, E., Malo, L., {et~al.} 2014, ApJ, 787, 5

\bibitem[{Nederlander {et~al.}(2021)Nederlander, Hughes, Fehr, Flaherty, Su,
  Moór, Chiang, Andrews, Wilner, \& Marino}]{Nederlander:2021aa}
Nederlander, A., Hughes, A.~M., Fehr, A.~J., {et~al.} 2021, The Astrophysical
  Journal, 917, 5

\bibitem[{Nelder \& Mead(1965)}]{Nelder:1965tk}
Nelder, J.~A., \& Mead, R. 1965, The computer journal, 7, 308

\bibitem[{Neuhäuser {et~al.}(2003)Neuhäuser, Guenther, Alves, Huélamo, Ott,
  \& Eckart}]{Neuhauser:2003tb}
Neuhäuser, R., Guenther, E.~W., Alves, J., {et~al.} 2003, Astronomische
  Nachrichten

\bibitem[{Neuhäuser {et~al.}(2005)Neuhäuser, Guenther, Wuchterl, Mugrauer,
  Bedalov, \& Hauschildt}]{Neuhauser:2005ela}
Neuhäuser, R., Guenther, E.~W., Wuchterl, G., {et~al.} 2005, A\&A, 435, L13

\bibitem[{Newton {et~al.}(2016)Newton, Irwin, Charbonneau, Berta-Thompson,
  Dittmann, \& West}]{Newton:2016ea}
Newton, E.~R., Irwin, J., Charbonneau, D., {et~al.} 2016, ApJ, 821, 93

\bibitem[{Nguyen {et~al.}(2012)Nguyen, Brandeker, Kerkwijk, \&
  Jayawardhana}]{Nguyen:2012aa}
Nguyen, D.~C., Brandeker, A., Kerkwijk, M. H.~v., \& Jayawardhana, R. 2012, The
  Astrophysical Journal, 745, 119

\bibitem[{Nguyen {et~al.}(2021)Nguyen, Rosa, \& Kalas}]{Nguyen:2021aa}
Nguyen, M.~M., Rosa, R. J.~D., \& Kalas, P. 2021, The Astronomical Journal,
  161, 22

\bibitem[{Nielsen {et~al.}(2019)Nielsen, Rosa, Macintosh, Wang, Ruffio, Chiang,
  Marley, Saumon, Savransky, Ammons, Bailey, Barman, Blain, Bulger, Burrows,
  Chilcote, Cotten, Czekala, Doyon, Duchene, Esposito, Fabrycky, Fitzgerald,
  Follette, Fortney, Gerard, Goodsell, Graham, Greenbaum, Hibon, Hinkley,
  Hirsch, Hom, Hung, Dawson, Ingraham, Kalas, Konopacky, Larkin, Lee, Lin,
  Maire, Marchis, Marois, Metchev, Millar-Blanchaer, Morzinski, Oppenheimer,
  Palmer, Patience, Perrin, Poyneer, Pueyo, Rafikov, Rajan, Rameau, Rantakyrö,
  Ren, Schneider, Sivaramakrishnan, Song, Soummer, Tallis, Thomas, Ward-Duong,
  \& Wolff}]{Nielsen:2019cb}
Nielsen, E.~L., Rosa, R. J.~D., Macintosh, B., {et~al.} 2019, The Astronomical
  Journal, 158, 13

\bibitem[{Nordstrom {et~al.}(2004)Nordstrom, Mayor, Andersen, Holmberg, Pont,
  Jorgensen, Olsen, Udry, \& Mowlavi}]{Nordstrom:2004ci}
Nordstrom, B., Mayor, M., Andersen, J., {et~al.} 2004, A\&A, 418, 989

\bibitem[{Nowak {et~al.}(2020)Nowak, Lacour, Lagrange, Rubini, Wang, Stolker,
  Abuter, Amorim, Asensio-Torres, Bauböck, Benisty, Berger, Beust, Blunt,
  Boccaletti, Bonnefoy, Bonnet, Brandner, Cantalloube, Charnay, Choquet,
  Christiaens, Clénet, Foresto, Cridland, Zeeuw, Dembet, Dexter, Drescher,
  Duvert, Eckart, Eisenhauer, Gao, Garcia, López, Gardner, Gendron, Genzel,
  Gillessen, Girard, Grandjean, Haubois, Heißel, Henning, hinkley, Hippler,
  Horrobin, Houllé, Hubert, Jiménez-Rosales, Jocou, Kammerer, Kervella,
  Keppler, Kreidberg, Kulikauskas, Lapeyrere, Bouquin, Léna, Mérand, Maire,
  Mollière, Monnier, Mouillet, Müller, Nasedkin, Ott, Otten, Paumard,
  Paladini, Perraut, Perrin, Pueyo, Pfuhl, Rameau, Rodet, Rodríguez-Coira,
  Rousset, Scheithauer, Shangguan, Stadler, Straub, Straubmeier, Sturm,
  Tacconi, Dishoeck, Vigan, Vincent, Fellenberg, Ward-Duong, Widmann,
  Wieprecht, Wiezorrek, Woillez, \& Collaboration}]{Nowak:2020cc}
Nowak, M., Lacour, S., Lagrange, A.-M., {et~al.} 2020, Astronomy and
  Astrophysics, 642, L2

\bibitem[{Oelkers {et~al.}(2018)Oelkers, Rodriguez, Stassun, Pepper, Somers,
  Kafka, Stevens, Beatty, Siverd, Lund, Kuhn, James, \& Gaudi}]{Oelkers:2018aa}
Oelkers, R.~J., Rodriguez, J.~E., Stassun, K.~G., {et~al.} 2018, The
  Astronomical Journal, 155, 39

\bibitem[{Offner {et~al.}(2016)Offner, Dunham, Lee, Arce, \&
  Fielding}]{Offner:2016gl}
Offner, S. S.~R., Dunham, M.~M., Lee, K.~I., Arce, H.~G., \& Fielding, D.~B.
  2016, The Astrophysical Journal Letters, 827, L11

\bibitem[{Oliveira {et~al.}(2015)Oliveira, Oliveira, Macías, \&
  Antonio}]{Oliveira:2015be}
Oliveira, A., Oliveira, T., Macías, S., \& Antonio. 2015, AIP Conference
  Proceedings, 1648, 840005

\bibitem[{Oppenheimer(2014)}]{Oppenheimer:2014et}
Oppenheimer, B.~R. 2014, in "50 Years of Brown Dwarfs," V. Joergens, ed.,
  Astrophysics and Space Science Library, Volume 401, (Zurich: Springer),
  astro-ph.SR, 1404.4430v1

\bibitem[{Oppenheimer {et~al.}(1995)Oppenheimer, Kulkarni, Matthews, \&
  Nakajima}]{Oppenheimer:1995wl}
Oppenheimer, B.~R., Kulkarni, S.~R., Matthews, K., \& Nakajima, T. 1995,
  Science, 270, 1478

\bibitem[{Ortiz-León {et~al.}(2018)Ortiz-León, Loinard, Dzib, Kounkel, Galli,
  Tobin, II, Hartmann, Rodríguez, Briceño, Torres, \&
  Mioduszewski}]{OrtizLeon:2018aa}
Ortiz-León, G.~N., Loinard, L., Dzib, S.~A., {et~al.} 2018, The Astrophysical
  Journal Letters, 869, L33

\bibitem[{O’Neil {et~al.}(2019)O’Neil, Martinez, Hees, Ghez, Do, Witzel,
  Konopacky, Becklin, Chu, Lu, Matthews, \& Sakai}]{ONeil:2019aa}
O’Neil, K.~K., Martinez, G.~D., Hees, A., {et~al.} 2019, The Astronomical
  Journal, 158, 4

\bibitem[{Pace {et~al.}(2003)Pace, Pasquini, \& Ortolani}]{Pace:2003aa}
Pace, G., Pasquini, L., \& Ortolani, S. 2003, Astronomy \& Astrophysics, 401,
  997

\bibitem[{Palma-Bifani {et~al.}(2022)Palma-Bifani, Chauvin, Bonnefoy, Rojo,
  Petrus, Rodet, Langlois, Allard, Charnay, Desgrange, Homeier, Lagrange,
  Beuzit, Baudoz, Boccaletti, Chomez, Delorme, Desidera, Feldt, Ginski,
  Gratton, Maire, Meyer, Samland, Snellen, Vigan, \&
  Zhang}]{PalmaBifani:2022aa}
Palma-Bifani, P., Chauvin, G., Bonnefoy, M., {et~al.} 2022, arXiv, 2211.01474

\bibitem[{Parker \& Daffern-Powell(2022)}]{Parker:2022aa}
Parker, R.~J., \& Daffern-Powell, E.~C. 2022, Monthly Notices of the Royal
  Astronomical Society: Letters, 516, L91

\bibitem[{Parker \& Quanz(2012)}]{Parker:2012aa}
Parker, R.~J., \& Quanz, S.~P. 2012, Monthly Notices of the Royal Astronomical
  Society, 419, 2448

\bibitem[{Pearce {et~al.}(2019)Pearce, Kraus, Dupuy, Ireland, Rizzuto, Bowler,
  Birchall, \& Wallace}]{Pearce:2019iv}
Pearce, L.~A., Kraus, A.~L., Dupuy, T.~J., {et~al.} 2019, The Astronomical
  Journal, 157, 71

\bibitem[{Pecaut \& Mamajek(2013)}]{Pecaut:2013ej}
Pecaut, M.~J., \& Mamajek, E.~E. 2013, The Astrophysical Journal Supplement
  Series, 208, 9

\bibitem[{Perets \& Kouwenhoven(2012)}]{Perets:2012cv}
Perets, H.~B., \& Kouwenhoven, M. B.~N. 2012, The Astrophysical Journal, 750,
  83

\bibitem[{Petigura {et~al.}(2017)Petigura, Howard, Marcy, Johnson, Isaacson,
  Cargile, Hebb, Fulton, Weiss, Morton, Winn, Rogers, Sinukoff, Hirsch, \&
  Crossfield}]{Petigura:2017bz}
Petigura, E.~A., Howard, A.~W., Marcy, G.~W., {et~al.} 2017, The Astronomical
  Journal, 154, 107

\bibitem[{Phan‐Bao {et~al.}(2008)Phan‐Bao, Bessell, Martín, Simon,
  Borsenberger, Tata, Guibert, Crifo, Forveille, Delfosse, Lim, \&
  Batz}]{PhanBao:2008aa}
Phan‐Bao, N., Bessell, M.~S., Martín, E.~L., {et~al.} 2008, Monthly Notices
  of the Royal Astronomical Society, 383, 831

\bibitem[{Pinfield {et~al.}(2012)Pinfield, Burningham, lodieu, Leggett, Tinney,
  Spaandonk, Marocco, Smart, Gomes, Smith, Lucas, Day-Jones, Murray,
  Katsiyannis, Catalan, Cardoso, Clarke, Folkes, Gálvez-Ortiz, Homeier,
  Jenkins, Jones, \& Zhang}]{Pinfield:2012hm}
Pinfield, D.~J., Burningham, B., lodieu, N., {et~al.} 2012, Monthly Notices of
  the Royal Astronomical Society, 422, 1922

\bibitem[{Potter {et~al.}(2002)Potter, Martín, Cushing, Baudoz, Brandner,
  Guyon, \& Neuhäuser}]{Potter:2002ie}
Potter, D., Martín, E.~L., Cushing, M.~C., {et~al.} 2002, The Astrophysical
  Journal, 567, L133

\bibitem[{Queloz {et~al.}(2000)Queloz, Eggenberger, Mayor, Perrier, Beuzit,
  Naef, Sivan, \& Udry}]{Queloz:2000ui}
Queloz, D., Eggenberger, A., Mayor, M., {et~al.} 2000, A\&A, 359, L13

\bibitem[{Radigan {et~al.}(2008)Radigan, Lafrenière, Jayawardhana, \&
  Doyon}]{Radigan:2008jd}
Radigan, J., Lafrenière, D., Jayawardhana, R., \& Doyon, R. 2008, The
  Astrophysical Journal, 689, 471

\bibitem[{Randich {et~al.}(1993)Randich, Gratton, \&
  Pallavicini}]{Randich:1993aa}
Randich, S., Gratton, R., \& Pallavicini, R. 1993, A\&A, 273, 194

\bibitem[{Rebolo {et~al.}(1998)Rebolo, Osorio, Madruga, Béjar, Arribas, \&
  Licandro}]{Rebolo:1998wx}
Rebolo, R., Osorio, M. R.~Z., Madruga, S., {et~al.} 1998, Science, 282, 1309

\bibitem[{Rebull {et~al.}(2018)Rebull, Stauffer, Cody, Hillenbrand, David, \&
  Pinsonneault}]{Rebull:2018aa}
Rebull, L.~M., Stauffer, J.~R., Cody, A.~M., {et~al.} 2018, The Astronomical
  Journal, 155, 196

\bibitem[{Reid {et~al.}(2008)Reid, Cruz, Kirkpatrick, Allen, Mungall, Liebert,
  Lowrance, \& Sweet}]{Reid:2008fz}
Reid, I.~N., Cruz, K.~L., Kirkpatrick, J.~D., {et~al.} 2008, The Astronomical
  Journal, 136, 1290

\bibitem[{Reid \& Walkowicz(2006)}]{Reid:2006eg}
Reid, I.~N., \& Walkowicz, L.~M. 2006, The Publications of the Astronomical
  Society of the Pacific, 118, 671

\bibitem[{Reiners(2007)}]{Reiners:2007aa}
Reiners, A. 2007, Astronomy \& Astrophysics, 467, 259

\bibitem[{Reiners \& Schmitt(2003)}]{Reiners:2003df}
Reiners, A., \& Schmitt, J. H. M.~M. 2003, Astronomy and Astrophysics, 398, 647

\bibitem[{Reiners {et~al.}(2022)Reiners, Shulyak, Käpylä, Ribas, Nagel,
  Zechmeister, Caballero, Shan, Fuhrmeister, Quirrenbach, Amado, Montes,
  Jeffers, Azzaro, Béjar, Chaturvedi, Henning, Kürster, \&
  Pallé}]{Reiners:2022aa}
Reiners, A., Shulyak, D., Käpylä, P.~J., {et~al.} 2022, Astronomy \&
  Astrophysics, 662, A41

\bibitem[{Reinhold \& Gizon(2015)}]{Reinhold:2015ep}
Reinhold, T., \& Gizon, L. 2015, A\&A, 583, A65

\bibitem[{Reinhold {et~al.}(2013)Reinhold, Reiners, \& Basri}]{Reinhold:2013iz}
Reinhold, T., Reiners, A., \& Basri, G. 2013, A\&A, 560, A4

\bibitem[{Reinhold {et~al.}(2020)Reinhold, Shapiro, Solanki, Montet, Krivova,
  Cameron, \& Amazo-Gómez}]{Reinhold:2020aa}
Reinhold, T., Shapiro, A.~I., Solanki, S.~K., {et~al.} 2020, Science, 368, 518

\bibitem[{Reza \& Pinzon(2004)}]{delaReza:2004aa}
Reza, R. d.~l., \& Pinzon, G. 2004, AJ, 128, 1812

\bibitem[{Rice \& Brewer(2020)}]{Rice:2020fj}
Rice, M., \& Brewer, J.~M. 2020, The Astrophysical Journal, 898, 119

\bibitem[{Ricker {et~al.}(2015)Ricker, Winn, Vanderspek, Latham, Bakos, Bean,
  Berta-Thompson, Brown, Buchhave, Butler, Butler, Chaplin, Charbonneau,
  Christensen-Dalsgaard, Clampin, Deming, Doty, Lee, Dressing, Dunham, Endl,
  Fressin, Ge, Henning, Holman, Howard, Ida, Jenkins, Jernigan, Johnson,
  Kaltenegger, Kawai, Kjeldsen, Laughlin, Levine, Lin, Lissauer, MacQueen,
  Marcy, McCullough, Morton, Narita, Paegert, Pallé, Pepe, Pepper,
  Quirrenbach, Rinehart, Sasselov, Sato, Seager, Sozzetti, Stassun, Sullivan,
  Szentgyorgyi, Torres, Udry, \& Villasenor}]{Ricker:2015fy}
Ricker, G.~R., Winn, J.~N., Vanderspek, R., {et~al.} 2015, Proc. SPIE, 9143,
  914320

\bibitem[{Rodriguez {et~al.}(2017)Rodriguez, Ansdell, Oelkers, Cargile, Gaidos,
  Cody, Stevens, Somers, James, Beatty, Siverd, Lund, Kuhn, Gaudi, Pepper, \&
  Stassun}]{Rodriguez:2017aa}
Rodriguez, J.~E., Ansdell, M., Oelkers, R.~J., {et~al.} 2017, The Astrophysical
  Journal, 848, 97

\bibitem[{Roettenbacher {et~al.}(2016)Roettenbacher, Monnier, Korhonen, Aarnio,
  Baron, Che, Harmon, Kővári, Kraus, Schaefer, Torres, Zhao, Brummelaar,
  Sturmann, \& Sturmann}]{Roettenbacher:2016aa}
Roettenbacher, R.~M., Monnier, J.~D., Korhonen, H., {et~al.} 2016, Nature, 533,
  217

\bibitem[{Rossiter(1924)}]{Rossiter:1924aa}
Rossiter, R.~A. 1924, Astrophysical Journal, 60, 15

\bibitem[{Royer {et~al.}(2002)Royer, Grenier, Baylac, Gomez, \&
  Zorec}]{Royer:2002jd}
Royer, F., Grenier, S., Baylac, M.~O., Gomez, A.~E., \& Zorec, J. 2002,
  Astronomy and Astrophysics, 393, 897

\bibitem[{Ruffio {et~al.}(2019)Ruffio, Macintosh, Konopacky, Barman, Rosa,
  Wang, Hoch, Czekala, \& Marois}]{Ruffio:2019aa}
Ruffio, J.-B., Macintosh, B., Konopacky, Q.~M., {et~al.} 2019, The Astronomical
  Journal, 158, 200

\bibitem[{Ruffio {et~al.}(2021)Ruffio, Konopacky, Barman, Macintosh, Hoch,
  Rosa, Wang, Czekala, \& Marois}]{Ruffio:2021aa}
Ruffio, J.-B., Konopacky, Q.~M., Barman, T., {et~al.} 2021, The Astronomical
  Journal, 162, 290

\bibitem[{Saffe {et~al.}(2021)Saffe, Miquelarena, Alacoria, Flores, Arancibia,
  Calvo, Girardi, Grosso, \& Collado}]{Saffe:2021hr}
Saffe, C., Miquelarena, P., Alacoria, J., {et~al.} 2021, Astronomy and
  Astrophysics, 647, A49

\bibitem[{Santamaría-Miranda {et~al.}(2017)Santamaría-Miranda, Cáceres,
  Schreiber, Hardy, Bayo, Parsons, Gromadzki, \&
  Villegas}]{SantamariaMiranda:2017aa}
Santamaría-Miranda, A., Cáceres, C., Schreiber, M.~R., {et~al.} 2017, Monthly
  Notices of the Royal Astronomical Society, 475, 2994

\bibitem[{Santos {et~al.}(2016)Santos, Meléndez, Nascimento, Bedell, Ramírez,
  Bean, Asplund, Spina, Dreizler, Alves-Brito, \&
  Casagrande}]{dosSantos:2016eq}
Santos, L. A.~d., Meléndez, J., Nascimento, J.-D.~d., {et~al.} 2016, Astronomy
  and Astrophysics, 592, A156

\bibitem[{Savitzky \& Golay(1964)}]{Savitzky:1964aa}
Savitzky, A., \& Golay, M. J.~E. 1964, AnaCh, 36, 1627

\bibitem[{Schaefer {et~al.}(2018)Schaefer, Prato, \& Simon}]{Schaefer:2018aa}
Schaefer, G.~H., Prato, L., \& Simon, M. 2018, The Astronomical Journal, 155,
  109

\bibitem[{Schaefer {et~al.}(2014)Schaefer, Prato, Simon, \&
  Patience}]{Schaefer:2014it}
Schaefer, G.~H., Prato, L., Simon, M., \& Patience, J. 2014, The Astronomical
  Journal, 147, 157

\bibitem[{Schlaufman(2010)}]{Schlaufman:2010db}
Schlaufman, K.~C. 2010, ApJ, 719, 602

\bibitem[{Schlieder {et~al.}(2012)Schlieder, Lépine, \&
  Simon}]{Schlieder:2012gj}
Schlieder, J.~E., Lépine, S., \& Simon, M. 2012, The Astronomical Journal,
  143, 80

\bibitem[{Scholz {et~al.}(2007)Scholz, Coffey, Brandeker, \&
  Jayawardhana}]{Scholz:2007aa}
Scholz, A., Coffey, J., Brandeker, A., \& Jayawardhana, R. 2007, The
  Astrophysical Journal, 662, 1254

\bibitem[{Scholz(2010)}]{Scholz:2010cy}
Scholz, R.-D. 2010, A\&A, 515, A92

\bibitem[{Schröder {et~al.}(2009)Schröder, Reiners, \&
  Schmitt}]{Schroder:2009es}
Schröder, C., Reiners, A., \& Schmitt, J. H. M.~M. 2009, Astronomy and
  Astrophysics, 493, 1099

\bibitem[{Schussler {et~al.}(1996)Schussler, Caligari, Ferriz-Mas, Solanki, \&
  Stix}]{schussler:1996aa}
Schussler, M., Caligari, P., Ferriz-Mas, A., Solanki, S.~K., \& Stix, M. 1996,
  A\&A, 314, 503

\bibitem[{Schwarz {et~al.}(2016)Schwarz, Ginski, Kok, Snellen, Brogi, \&
  Birkby}]{Schwarz:2016fl}
Schwarz, H., Ginski, C., Kok, R. J.~d., {et~al.} 2016, A\&A, 593, A74

\bibitem[{Schweitzer {et~al.}(2019)Schweitzer, Passegger, Cifuentes, Béjar,
  Cortés-Contreras, Caballero, Burgo, Czesla, Kürster, Montes, Osorio, Ribas,
  Reiners, Quirrenbach, Amado, Aceituno, Anglada-Escudé, Bauer, Dreizler,
  Jeffers, Guenther, Henning, Kaminski, Lafarga, Marfil, Morales, Schmitt,
  Seifert, Solano, Tabernero, \& Zechmeister}]{Schweitzer:2019aa}
Schweitzer, A., Passegger, V.~M., Cifuentes, C., {et~al.} 2019, Astronomy \&
  Astrophysics, 625, A68

\bibitem[{Sebastian {et~al.}(2021)Sebastian, Gillon, Ducrot, Pozuelos, Garcia,
  Günther, Delrez, Queloz, Demory, Triaud, Burgasser, Wit, Burdanov,
  Dransfield, Jehin, McCormac, Murray, Niraula, Pedersen, Rackham, Sohy,
  Thompson, \& Grootel}]{Sebastian:2021aa}
Sebastian, D., Gillon, M., Ducrot, E., {et~al.} 2021, Astronomy \&
  Astrophysics, 645, A100

\bibitem[{Seifahrt {et~al.}(2005)Seifahrt, Mugrauer, Wiese, Neuhäuser, \&
  Guenther}]{Seifahrt:2005fz}
Seifahrt, A., Mugrauer, M., Wiese, M., Neuhäuser, R., \& Guenther, E.~W. 2005,
  Astronomische Nachrichten, 326, 974

\bibitem[{Sepulveda \& Bowler(2022)}]{Sepulveda:2022by}
Sepulveda, A.~G., \& Bowler, B.~P. 2022, The Astronomical Journal, 163, 52

\bibitem[{Sepulveda {et~al.}(2023)Sepulveda, Huber, Li, Bedding, Zhang, \&
  Liu}]{Sepulveda:2023aa}
Sepulveda, A.~G., Huber, D., Li, G., {et~al.} 2023, RNAAS, 7, 2

\bibitem[{Sepulveda {et~al.}(2022)Sepulveda, Huber, Zhang, Li, Liu, \&
  Bedding}]{Sepulveda:2022aa}
Sepulveda, A.~G., Huber, D., Zhang, Z., {et~al.} 2022, arXiv, 1

\bibitem[{Shajn \& Struve(1929)}]{Shajn:1929aa}
Shajn, G., \& Struve, O. 1929, MNRAS, 89, 222

\bibitem[{Shan {et~al.}(2021)Shan, Reiners, Fabbian, Marfil, Montes, Tabernero,
  Ribas, Caballero, Quirrenbach, Amado, Aceituno, Béjar, Cortés-Contreras,
  Dreizler, Hatzes, Henning, Jeffers, Kaminski, Kürster, Lafarga, Morales,
  Nagel, Pallé, Passegger, Rodriguez-López, Schweitzer, \&
  Zechmeister}]{Shan:2021aa}
Shan, Y., Reiners, A., Fabbian, D., {et~al.} 2021, Astronomy \& Astrophysics,
  654, A118

\bibitem[{Silva {et~al.}(2020)Silva, Mena, \& Tsantaki}]{Silva:2020cj}
Silva, A. R.~C., Mena, E.~D., \& Tsantaki, M. 2020, Astronomy and Astrophysics,
  634, A136

\bibitem[{Silva {et~al.}(2009)Silva, Torres, Reza, Quast, Melo, \&
  Sterzik}]{daSilva:2009eu}
Silva, L.~d., Torres, C. A.~O., Reza, R. D.~L., {et~al.} 2009, A\&A, 508, 833

\bibitem[{Simon {et~al.}(1987)Simon, Howell, Longmore, Wilking, Peterson, \&
  Chen}]{Simon:1987ga}
Simon, M., Howell, R.~R., Longmore, A.~J., {et~al.} 1987, Astrophysical
  Journal, 320, 344

\bibitem[{Simon \& Schaefer(2011)}]{Simon:2011ix}
Simon, M., \& Schaefer, G.~H. 2011, The Astrophysical Journal, 743, 158

\bibitem[{Smith {et~al.}(2012)Smith, Stumpe, Cleve, Jenkins, Barclay, Fanelli,
  Girouard, Kolodziejczak, McCauliff, Morris, \& Twicken}]{Smith:2012aa}
Smith, J.~C., Stumpe, M.~C., Cleve, J. E.~V., {et~al.} 2012, Publications of
  the Astronomical Society of the Pacific, 124, 1000

\bibitem[{Smith \& Gray(1976)}]{smith:1976aa}
Smith, M.~A., \& Gray, D.~F. 1976, PASP, 88, 809

\bibitem[{Snellen \& Brown(2018)}]{Snellen:2018aa}
Snellen, I. A.~G., \& Brown, A. G.~A. 2018, Nature Astronomy, 2, 883

\bibitem[{Snodgrass \& Ulrich(1990)}]{Snodgrass:1990aa}
Snodgrass, H.~B., \& Ulrich, R.~K. 1990, ApJ, 351, 309

\bibitem[{Soderblom {et~al.}(1998)Soderblom, King, \& Henry}]{Soderblom:1998iy}
Soderblom, D.~R., King, J.~R., \& Henry, T.~J. 1998, The Astronomical Journal,
  116, 396

\bibitem[{Sodor {et~al.}(2014)Sodor, Chené, Cat, Bognár, Wright, Marois,
  Walker, Matthews, Kallinger, Rowe, Kuschnig, Guenther, Moffat, Rucinski,
  Sasselov, \& Weiss}]{Sodor:2014aa}
Sodor, A., Chené, A.-N., Cat, P.~D., {et~al.} 2014, Astronomy \& Astrophysics,
  568, A106

\bibitem[{Soto \& Jenkins(2018)}]{Soto:2018bl}
Soto, M.~G., \& Jenkins, J.~S. 2018, A\&A, 615, A76

\bibitem[{Soubiran {et~al.}(2018)Soubiran, Jasniewicz, Chemin, Zurbach,
  Brouillet, Panuzzo, Sartoretti, Katz, Campion, Marchal, Hestroffer,
  Thévenin, Crifo, Udry, Cropper, Seabroke, Viala, Benson, Blomme,
  Jean-Antoine, Huckle, Smith, Baker, Damerdji, Dolding, Frémat, Gosset,
  Guerrier, Guy, Haigron, Janßen, Plum, Fabre, Lasne, Pailler, Panem, Riclet,
  Royer, Tauran, Zwitter, Gueguen, \& Turon}]{Soubiran:2018fz}
Soubiran, C., Jasniewicz, G., Chemin, L., {et~al.} 2018, Astronomy \&
  Astrophysics, 616, A7

\bibitem[{Spalding(2019)}]{Spalding:2019do}
Spalding, C. 2019, The Astrophysical Journal, 879, 12

\bibitem[{Squicciarini {et~al.}(2022)Squicciarini, Gratton, Janson, Mamajek,
  Chauvin, Delorme, Langlois, Vigan, Ringqvist, Meeus, Reffert, Kenworthy,
  Meyer, Bonnefoy, Bonavita, Mesa, Samland, Desidera, D’Orazi, Engler,
  Alecian, Miglio, Henning, Quanz, Mayer, Flasseur, \&
  Marleau}]{Squicciarini:2022aa}
Squicciarini, V., Gratton, R., Janson, M., {et~al.} 2022, Astronomy \&
  Astrophysics, 664, A9

\bibitem[{Stamatellos \& Whitworth(2009)}]{Stamatellos:2009fw}
Stamatellos, D., \& Whitworth, A.~P. 2009, Monthly Notices of the Royal
  Astronomical Society, 392, 413

\bibitem[{Stassun {et~al.}(2018)Stassun, Oelkers, Pepper, Paegert, Lee, Torres,
  Latham, Charpinet, Dressing, Huber, Kane, Lépine, Mann, Muirhead,
  Rojas-Ayala, Silvotti, Fleming, Levine, \& Plavchan}]{Stassun:2018jr}
Stassun, K.~G., Oelkers, R.~J., Pepper, J., {et~al.} 2018, The Astronomical
  Journal, 156, 102

\bibitem[{Stassun {et~al.}(2019)Stassun, Oelkers, Paegert, Torres, Pepper, Lee,
  Collins, Latham, Muirhead, Chittidi, Rojas-Ayala, Fleming, Rose, Tenenbaum,
  Ting, Kane, Barclay, Bean, Brassuer, Charbonneau, Ge, Lissauer, Mann, McLean,
  Mullally, Narita, Plavchan, Ricker, Sasselov, Seager, Sharma, Shiao,
  Sozzetti, Stello, Vanderspek, Wallace, \& Winn}]{Stassun:2019aa}
Stassun, K.~G., Oelkers, R.~J., Paegert, M., {et~al.} 2019, The Astronomical
  Journal, 158, 138

\bibitem[{Stelzer {et~al.}(2013)Stelzer, Alcalá, Scholz, Natta, Randich, \&
  Covino}]{Stelzer:2013aa}
Stelzer, B., Alcalá, J.~M., Scholz, A., {et~al.} 2013, Astronomy \&
  Astrophysics, 551, A106

\bibitem[{Stelzer {et~al.}(2010)Stelzer, Scholz, Argiroffi, \&
  Micela}]{Stelzer:2010aa}
Stelzer, B., Scholz, A., Argiroffi, C., \& Micela, G. 2010, Monthly Notices of
  the Royal Astronomical Society, 408, 1095

\bibitem[{Stolker {et~al.}(2021)Stolker, Haffert, Kesseli, Holstein, Aoyama,
  Brinchmann, Cugno, Girard, Marleau, Cugno, Meyer, Milli, Quanz, Snellen, \&
  Todorov}]{Stolker:2021aa}
Stolker, T., Haffert, S.~Y., Kesseli, A.~Y., {et~al.} 2021, The Astronomical
  Journal, 162, 286

\bibitem[{Strassmeier(2009)}]{Strassmeier:2009aa}
Strassmeier, K.~G. 2009, The Astronomy and Astrophysics Review, 17, 251

\bibitem[{Strassmeier {et~al.}(1999)Strassmeier, Serkowitsch, \&
  Granzer}]{Strassmeier:1999aa}
Strassmeier, K.~G., Serkowitsch, E., \& Granzer, T. 1999, Astronomy and
  Astrophysics Supplement Series, 140, 29

\bibitem[{Strassmeier {et~al.}(2000)Strassmeier, Washuettl, Granzer, Scheck, \&
  Weber}]{Strassmeier:2000aa}
Strassmeier, K.~G., Washuettl, A., Granzer, T., Scheck, M., \& Weber, M. 2000,
  Astronomy and Astrophysics Supplement Series, 142, 275

\bibitem[{Stumpe {et~al.}(2014)Stumpe, Smith, Catanzarite, Cleve, Jenkins,
  Twicken, \& Girouard}]{Stumpe:2014aa}
Stumpe, M.~C., Smith, J.~C., Catanzarite, J.~H., {et~al.} 2014, Publications of
  the Astronomical Society of the Pacific, 126, 100

\bibitem[{Stumpe {et~al.}(2012)Stumpe, Smith, Cleve, Twicken, Barclay, Fanelli,
  Girouard, Jenkins, Kolodziejczak, McCauliff, \& Morris}]{Stumpe:2012aa}
Stumpe, M.~C., Smith, J.~C., Cleve, J. E.~V., {et~al.} 2012, Publications of
  the Astronomical Society of the Pacific, 124, 985

\bibitem[{Stumpff(1980)}]{stumpff:1980aa}
Stumpff, P. 1980, A\&AS, 41, 1

\bibitem[{Suarez Mascareno {et~al.}(2016)Suarez Mascareno, Rebolo, \&
  Hernández}]{SuarezMascareno:2016aa}
Suarez Mascareno, A., Rebolo, R., \& Hernández, J. I.~G. 2016, Astronomy \&
  Astrophysics, 595, A12

\bibitem[{Swastik {et~al.}(2021)Swastik, Banyal, Narang, Manoj, Sivarani,
  Reddy, \& Rajaguru}]{Swastik:2021jm}
Swastik, C., Banyal, R.~K., Narang, M., {et~al.} 2021, The Astronomical
  Journal, 161, 114

\bibitem[{Sybilski {et~al.}(2018)Sybilski, Pawłaszek, Sybilska, Konacki,
  Hełminiak, Kozłowski, \& Ratajczak}]{Sybilski:2018aa}
Sybilski, P., Pawłaszek, R.~K., Sybilska, A., {et~al.} 2018, Monthly Notices
  of the Royal Astronomical Society, 478, 1942

\bibitem[{Takeda {et~al.}(2010)Takeda, Honda, Kawanomoto, Ando, \&
  Sakurai}]{Takeda:2010aa}
Takeda, Y., Honda, S., Kawanomoto, S., Ando, H., \& Sakurai, T. 2010, Astronomy
  \& Astrophysics, 515, A93

\bibitem[{Takeda {et~al.}(2005)Takeda, Sato, Kambe, \& Masuda}]{Takeda:2005jm}
Takeda, Y., Sato, B., Kambe, E., \& Masuda, S. 2005, PASJ, 57, 13

\bibitem[{Thanathibodee {et~al.}(2020)Thanathibodee, Molina, Calvet, Serna,
  Bae, Reynolds, Hernández, Muzerolle, \& Hernández}]{Thanathibodee:2020fe}
Thanathibodee, T., Molina, B., Calvet, N., {et~al.} 2020, The Astrophysical
  Journal, 892, 81

\bibitem[{Tokunaga(2000)}]{Tokunaga:2000tr}
Tokunaga, A.~T. 2000, in Allen's Astrophysical Quantities, 4th ed., ed. A. N.
  Cox (New York, NY: AIP Press; Springer 2000), 143

\bibitem[{Torres {et~al.}(2006)Torres, Quast, Silva, Reza, Melo, \&
  Sterzik}]{Torres:2006bw}
Torres, C. A.~O., Quast, G.~R., Silva, L.~d., {et~al.} 2006, A\&A, 460, 695

\bibitem[{Tremaine(1991)}]{Tremaine:1991aa}
Tremaine, S. 1991, Icarus, 89, 85

\bibitem[{Tull {et~al.}(1995)Tull, MacQueen, \& Sneden}]{Tull:1995tn}
Tull, R.~G., MacQueen, P.~J., \& Sneden, C. 1995, PASP, 107, 251

\bibitem[{Valenti \& Fischer(2005)}]{Valenti:2005fz}
Valenti, J.~A., \& Fischer, D.~A. 2005, The Astrophysical Journal Supplement
  Series, 159, 141

\bibitem[{Veras {et~al.}(2009)Veras, Crepp, \& Ford}]{Veras:2009br}
Veras, D., Crepp, J.~R., \& Ford, E.~B. 2009, The Astrophysical Journal, 696,
  1600

\bibitem[{Vican(2012)}]{Vican:2012aa}
Vican, L. 2012, The Astronomical Journal, 143, 135

\bibitem[{Vigan {et~al.}(2021)Vigan, Fontanive, Meyer, Biller, Bonavita, Feldt,
  Desidera, Marleau, Emsenhuber, Galicher, Rice, Forgan, Mordasini, Gratton,
  Coroller, Maire, Cantalloube, Chauvin, Cheetham, Hagelberg, Lagrange,
  Langlois, Bonnefoy, Beuzit, Boccaletti, D'Orazi, Delorme, Dominik, Henning,
  Janson, Lagadec, Lazzoni, Ligi, Menard, Mesa, Messina, Moutou, Müller,
  Perrot, Samland, Schmid, Schmidt, Sissa, Turatto, Udry, Zurlo, Abe, Antichi,
  Asensio-Torres, Baruffolo, Baudoz, Baudrand, Bazzon, Blanchard, Bohn,
  Sevilla, Carbillet, Carle, Cascone, Charton, Claudi, Costille, Caprio,
  Delboulbé, Dohlen, Engler, Fantinel, Feautrier, Fusco, Gigan, Girard, Giro,
  Gisler, Gluck, Gry, Hubin, Hugot, Jaquet, Kasper, Mignant, Llored, Madec,
  Magnard, Martinez, Maurel, Möller-Nilsson, Mouillet, Moulin, Origné,
  Pavlov, Perret, Petit, Pragt, Puget, Rabou, Ramos, Rickman, Rigal, Rochat,
  Roelfsema, Rousset, Roux, Salasnich, Sauvage, Sevin, Soenke, Stadler, Suarez,
  Wahhaj, Weber, \& Wildi}]{Vigan:2021dc}
Vigan, A., Fontanive, C., Meyer, M., {et~al.} 2021, Astronomy and Astrophysics,
  651, A72

\bibitem[{Wagner {et~al.}(2019)Wagner, Apai, \& Kratter}]{Wagner:2019iy}
Wagner, K., Apai, D., \& Kratter, K.~M. 2019, The Astrophysical Journal, 877,
  46

\bibitem[{Wahhaj {et~al.}(2011)Wahhaj, Liu, Biller, Clarke, Nielsen, Close,
  Hayward, Mamajek, Cushing, Dupuy, Tecza, Thatte, Chun, Ftaclas, Hartung,
  Reid, Shkolnik, Alencar, Artymowicz, Boss, Pino, Gregorio-Hetem, Ida,
  Kuchner, Lin, \& Toomey}]{Wahhaj:2011by}
Wahhaj, Z., Liu, M.~C., Biller, B.~A., {et~al.} 2011, The Astrophysical
  Journal, 729, 139

\bibitem[{Wang {et~al.}(2018)Wang, Graham, Dawson, Fabrycky, Rosa, Pueyo,
  Konopacky, Macintosh, Marois, Chiang, Ammons, Arriaga, Bailey, Barman,
  Bulger, Chilcote, Cotten, Doyon, Duchêne, Esposito, Fitzgerald, Follette,
  Gerard, Goodsell, Greenbaum, Hibon, Hung, Ingraham, Kalas, Larkin, Maire,
  Marchis, Marley, Metchev, Millar-Blanchaer, Nielsen, Oppenheimer, Palmer,
  Patience, Perrin, Poyneer, Rajan, Rameau, Rantakyrö, Ruffio, Savransky,
  Schneider, Sivaramakrishnan, Song, Soummer, Thomas, Wallace, Ward-Duong,
  Wiktorowicz, \& Wolff}]{Wang:2018aa}
Wang, J.~J., Graham, J.~R., Dawson, R., {et~al.} 2018, The Astronomical
  Journal, 156, 192

\bibitem[{Wang {et~al.}(2020)Wang, Ginzburg, Ren, Wallack, Gao, Mawet, Bond,
  Cetre, Wizinowich, Rosa, Ruane, Liu, Absil, Alvarez, Baranec, Choquet, Chun,
  Defrere, Delorme, Duchene, Forsberg, Ghez, Guyon, Hall, huby, Jolivet,
  Jensen-Clem, Jovanovic, Karlsson, Lilley, Matthews, Menard, Meshkat,
  Millar-Blanchaer, Ngo, Xivry, Pinte, Ragland, Serabyn, Catalán, Wang,
  Wetherell, Williams, Ygouf, \& Zuckerman}]{Wang:2020jb}
Wang, J.~J., Ginzburg, S., Ren, B., {et~al.} 2020, The Astronomical Journal,
  159, 263

\bibitem[{Wang {et~al.}(2021)Wang, Vigan, Lacour, Nowak, Stolker, Rosa,
  Ginzburg, Gao, Abuter, Amorim, Asensio-Torres, Baubck, Benisty, Berger,
  Beust, Beuzit, Blunt, Boccaletti, Bohn, Bonnefoy, Bonnet, Brandner,
  Cantalloube, Caselli, Charnay, Chauvin, Choquet, Christiaens, Clénet,
  Foresto, Cridland, Zeeuw, Dembet, Dexter, Drescher, Duvert, Eckart,
  Eisenhauer, Facchini, Gao, Garcia, Lopez, Gardner, Gendron, Genzel,
  Gillessen, Girard, Haubois, Heißel, Henning, Hinkley, Hippler, Horrobin,
  Houllé, Hubert, Jiménez-Rosales, Jocou, Kammerer, Keppler, Kervella, Meyer,
  Kreidberg, Lagrange, Lapeyrère, Bouquin, Léna, Lutz, Maire, Ménard,
  Mérand, Mollière, Monnier, Mouillet, Müller, Nasedkin, Ott, Otten,
  Paladini, Paumard, Perraut, Perrin, Pfuhl, Pueyo, Rameau, Rodet,
  Rodríguez-Coira, Rousset, Scheithauer, Shangguan, Shimizu, Stadler, Straub,
  Straubmeier, Sturm, Tacconi, Dishoeck, Vincent, Fellenberg, Ward-Duong,
  Widmann, Wieprecht, Wiezorrek, Woillez, \& Collaboration}]{Wang:2021aa}
Wang, J.~J., Vigan, A., Lacour, S., {et~al.} 2021, The Astronomical Journal,
  161, 148

\bibitem[{Ward-Duong {et~al.}(2021)Ward-Duong, Patience, Follette, Rosa,
  Rameau, Marley, Saumon, Nielsen, Rajan, Greenbaum, Lee, Wang, Czekala,
  Duchêne, Macintosh, Ammons, Bailey, Barman, Bulger, Chen, Chilcote, Cotten,
  Doyon, Esposito, Fitzgerald, Gerard, Goodsell, Graham, Hibon, Hom, Hung,
  Ingraham, Kalas, Konopacky, Larkin, Maire, Marchis, Marois, Metchev,
  Millar-Blanchaer, Oppenheimer, Palmer, Perrin, Poyneer, Pueyo, Rantakyro,
  Ren, Ruffio, Savransky, Schneider, Sivaramakrishnan, Song, Soummer, Tallis,
  Thomas, Wallace, Wiktorowicz, \& Wolff}]{WardDuong:2021ev}
Ward-Duong, K., Patience, J., Follette, K., {et~al.} 2021, The Astronomical
  Journal, 161, 5

\bibitem[{Watson {et~al.}(2011)Watson, Littlefair, Diamond, Cameron,
  Fitzsimmons, Simpson, Moulds, \& Pollacco}]{Watson:2011gf}
Watson, C.~A., Littlefair, S.~P., Diamond, C., {et~al.} 2011, Monthly Notices
  RAS Letters, 413, L71

\bibitem[{Webb {et~al.}(1999)Webb, Zuckerman, Platais, Patience, White,
  Schwartz, \& McCarthy}]{Webb:1999kf}
Webb, R.~A., Zuckerman, B., Platais, I., {et~al.} 1999, The Astrophysical
  Journal, 512, L63

\bibitem[{White {et~al.}(2007)White, Gabor, \& Hillenbrand}]{White:2007iz}
White, R.~J., Gabor, J.~M., \& Hillenbrand, L.~A. 2007, The Astronomical
  Journal, 133, 2524

\bibitem[{White \& Ghez(2001)}]{White:2001ic}
White, R.~J., \& Ghez, A.~M. 2001, The Astrophysical Journal, 556, 265

\bibitem[{Wilking {et~al.}(2005)Wilking, Meyer, Robinson, \&
  Greene}]{Wilking:2005aq}
Wilking, B.~A., Meyer, M.~R., Robinson, J.~G., \& Greene, T.~P. 2005, The
  Astronomical Journal, 1733

\bibitem[{Wilner {et~al.}(2018)Wilner, MacGregor, Andrews, Hughes, Matthews, \&
  Su}]{Wilner:2018aa}
Wilner, D.~J., MacGregor, M.~A., Andrews, S.~M., {et~al.} 2018, The
  Astrophysical Journal, 855, 56

\bibitem[{Winn {et~al.}(2010)Winn, Fabrycky, Albrecht, \&
  Johnson}]{Winn:2010dr}
Winn, J.~N., Fabrycky, D., Albrecht, S., \& Johnson, J.~A. 2010, The
  Astrophysical Journal, 718, L145

\bibitem[{Winn \& Fabrycky(2015)}]{Winn:2015jt}
Winn, J.~N., \& Fabrycky, D.~C. 2015, ARA\&A, 53, 409

\bibitem[{Winn {et~al.}(2017)Winn, Petigura, Morton, Weiss, Dai, Schlaufman,
  Howard, Isaacson, Marcy, Justesen, \& Albrecht}]{Winn:2017ip}
Winn, J.~N., Petigura, E.~A., Morton, T.~D., {et~al.} 2017, The Astronomical
  Journal, 154, 270

\bibitem[{Wood {et~al.}(2019)Wood, Boyajian, Braun, Brewer, Crepp, Schaefer,
  Adams, \& White}]{Wood:2019aa}
Wood, C.~M., Boyajian, T., Braun, K.~v., {et~al.} 2019, The Astrophysical
  Journal, 873, 83

\bibitem[{Wright {et~al.}(2011)Wright, Chené, Cat, Marois, Mathias, Macintosh,
  Isaacs, Lehmann, \& Hartmann}]{Wright:2011wu}
Wright, D.~J., Chené, A.-N., Cat, P.~D., {et~al.} 2011, The Astrophysical
  Journal Letters, 728, L20

\bibitem[{Wright {et~al.}(2013)Wright, Skrutskie, Kirkpatrick, Gelino,
  Griffith, Marsh, Jarrett, Nelson, Borish, Mace, Mainzer, Eisenhardt, McLean,
  Tobin, \& Cushing}]{Wright:2013bo}
Wright, E.~L., Skrutskie, M.~F., Kirkpatrick, J.~D., {et~al.} 2013, The
  Astronomical Journal, 145, 84

\bibitem[{Wu \& Sheehan(2017)}]{Wu:2017aa}
Wu, Y.-L., \& Sheehan, P.~D. 2017, The Astrophysical Journal Letters, 846, L26

\bibitem[{Wu {et~al.}(2017)Wu, Sheehan, Males, Close, Morzinski, Teske,
  Haug-Baltzell, Merchant, \& Lyons}]{Wu:2017kd}
Wu, Y.-L., Sheehan, P.~D., Males, J.~R., {et~al.} 2017, ApJ, 836, 1

\bibitem[{Wu {et~al.}(2020)Wu, Bowler, Sheehan, Andrews, Herczeg, Kraus, Ricci,
  Wilner, \& Zhu}]{Wu:2020aa}
Wu, Y.-L., Bowler, B.~P., Sheehan, P.~D., {et~al.} 2020, The Astronomical
  Journal, 159, 229

\bibitem[{Wu {et~al.}(2022)Wu, Bowler, Sheehan, Close, Eisner, Best,
  Ward-Duong, Zhu, \& Kraus}]{Wu:2022aa}
---. 2022, The Astrophysical Journal Letters, 930, L3

\bibitem[{Yadav {et~al.}(2015)Yadav, Gastine, Christensen, \&
  Reiners}]{Yadav:2015aa}
Yadav, R.~K., Gastine, T., Christensen, U.~R., \& Reiners, A. 2015, Astronomy
  \& Astrophysics, 573, A68

\bibitem[{Zechmeister \& Kürster(2009)}]{Zechmeister:2009ii}
Zechmeister, M., \& Kürster, M. 2009, A\&A, 496, 577

\bibitem[{Zhang {et~al.}(2021)Zhang, Liu, Claytor, Best, Dupuy, \&
  Siverd}]{Zhang:2021bb}
Zhang, Z., Liu, M.~C., Claytor, Z.~R., {et~al.} 2021, The Astrophysical Journal
  Letters, 916, L11

\bibitem[{Zhang {et~al.}(2010)Zhang, Pinfield, Day-Jones, Burningham, Jones,
  Yu, Jenkins, Han, Gálvez-Ortiz, Gallardo, García-Pérez, Weights, Tinney,
  \& Pokorny}]{Zhang:2010bq}
Zhang, Z.~H., Pinfield, D.~J., Day-Jones, A.~C., {et~al.} 2010, Monthly Notices
  of the Royal Astronomical Society, 404, 1817

\bibitem[{Zhou {et~al.}(2020)Zhou, Bowler, Morley, Apai, Kataria, Bryan, \&
  Benneke}]{Zhou:2020cl}
Zhou, Y., Bowler, B.~P., Morley, C.~V., {et~al.} 2020, The Astronomical
  Journal, 160, 77

\bibitem[{Zhou {et~al.}(2021)Zhou, Bowler, Wagner, Schneider, Apai, Kraus,
  Close, Herczeg, \& Fang}]{Zhou:2021ky}
Zhou, Y., Bowler, B.~P., Wagner, K.~R., {et~al.} 2021, The Astronomical
  Journal, 161, 244

\bibitem[{Zurlo {et~al.}(2022)Zurlo, Goździewski, Lazzoni, Mesa, Nogueira,
  Desidera, Gratton, Marzari, Langlois, Pinna, Chauvin, Delorme, Girard,
  Hagelberg, Henning, Janson, Rickman, Kervella, Avenhaus, Bhowmik, Biller,
  Boccaletti, Bonaglia, Bonavita, Bonnefoy, Cantalloube, Cheetham, Claudi,
  D’Orazi, Feldt, Galicher, Ghose, Lagrange, Coroller, Ligi, Kasper, Maire,
  Medard, Meyer, Peretti, Perrot, Puglisi, Rossi, Rothberg, Schmidt, Sissa,
  Vigan, \& Wahhaj}]{Zurlo:2022aa}
Zurlo, A., Goździewski, K., Lazzoni, C., {et~al.} 2022, Astronomy \&
  Astrophysics, 666, A133

\bibitem[{Zwintz {et~al.}(2019)Zwintz, Reese, Neiner, Pigulski, Kuschnig,
  Muellner, Zieba, Abe, Guillot, Handler, Kenworthy, Stuik, Moffat, Popowicz,
  Rucinski, Wade, Weiss, Bailey, Crawford, Ireland, Kuhn, Lomberg, Mamajek,
  Mellon, \& Talens}]{Zwintz:2019aa}
Zwintz, K., Reese, D.~R., Neiner, C., {et~al.} 2019, Astronomy \& Astrophysics,
  627, A28

\bibitem[{Zúñiga-Fernández {et~al.}(2021)Zúñiga-Fernández, Bayo, Elliott,
  Zamora, Corvalán, Haubois, Corral-Santana, Olofsson, Huélamo, Sterzik,
  Torres, Quast, \& Melo}]{ZunigaFernandez:2021if}
Zúñiga-Fernández, S., Bayo, A., Elliott, P., {et~al.} 2021, Astronomy and
  Astrophysics, 645, A30

\end{thebibliography}

\end{document}